\documentclass[vecphys]{SVmult}

\usepackage{times}
\usepackage{makeidx}         
\usepackage{graphicx}        
\usepackage{multicol}        
\usepackage[bottom]{footmisc}

\usepackage{epsfig}  
\usepackage{latexsym}    
\usepackage{graphicx}
\usepackage{verbatim}
\usepackage{pstricks} 
\usepackage{diagrams} 
\usepackage{amssymb}
\usepackage{amsmath}
\usepackage[2cell,rotate,ps,dvips,all]{xy}
\usepackage{rotating}
\usepackage{color}  

\newarrow{Is}=====
\newarrow{Maps}|--->

\def\bR{\begin{color}{red}}
\def\bB{\begin{color}{blue}}
\def\bM{\begin{color}{magenta}} 
\def\bC{\begin{color}{cyan}}
\def\bW{\begin{color}{white}}
\def\bBl{\begin{color}{black}}
\def\bG{\begin{color}{green}}
\def\bY{\begin{color}{yellow}}
\def\e{\end{color}}

\DeclareGraphicsExtensions{.eps,.bmp}
\newcommand{\cat}{{\bf C}}

\newcommand{\II}{{\rm I}}

\newcommand{\name}[1]{\ulcorner #1\urcorner}
\newcommand{\coname}[1]{\llcorner #1\lrcorner}
\newcommand{\diag}{{\hspace{1.13 ex}\raisebox{.63 em}[0pt][0pt]{
 \begin{rotate}{180}{$\nabla$}\end{rotate}}}} 
\newcommand{\scdiag}{{\hspace{1.13 ex}\raisebox{.45 em}[0pt][0pt]{
 \begin{rotate}{180}{\scriptsize $\nabla$}\end{rotate}}}}

\newdir{|>}{%
:a(75)@^{|}*:a(80)@^{|}*:a(85)@^{|}*:a(89)@^{|}*:a(-89)@_{|}*:a(-85)@_{|}*:a(-80)@_{|}*:a(-75)@_{|}
} \newcommand{\ars}{\ar@{-|>}}

\UseAllTwocells 

\newcommand{\bit}{\begin{itemize}}
\newcommand{\eit}{\end{itemize}\par\noindent}
\newcommand{\ben}{\begin{enumerate}}
\newcommand{\een}{\end{enumerate}\par\noindent}
\newcommand{\beq}{\begin{equation}}
\newcommand{\eeq}{\end{equation}\par\noindent}
\newcommand{\beqa}{\begin{eqnarray*}}
\newcommand{\eeqa}{\end{eqnarray*}\par\noindent}
\newcommand{\beqn}{\begin{eqnarray}}
\newcommand{\eeqn}{\end{eqnarray}\par\noindent}

\begin{document}

\title*{Categories for the practising physicist} 
\author{Bob Coecke and \'Eric Oliver Paquette}
\authorrunning{Bob Coecke and \'Eric~O.~Paquette}
\institute{OUCL, University of Oxford \texttt{coecke/eop@comlab.ox.ac.uk }}
\maketitle

\begin{abstract}  
In this chapter we survey some particular topics in category theory in a somewhat unconventional manner. Our main focus will be on monoidal categories, mostly symmetric ones, for which we propose a physical interpretation.  Special attention is given to the category which has finite dimensional Hilbert spaces as objects, linear maps as morphisms, and the tensor product as its monoidal structure (${\bf FdHilb}$). We also provide a detailed discussion of the category which has sets as objects, relations as morphisms, and the cartesian product as its monoidal structure (${\bf Rel}$), and thirdly, categories with manifolds  as objects and cobordisms between these as morphisms (${\bf 2Cob}$).  While sets, Hilbert spaces and manifolds do not share any non-trivial common structure, these three categories are in fact structurally very similar.  Shared features are diagrammatic calculus, compact closed structure and particular kinds of internal comonoids which play an important role in each of them.  The categories ${\bf FdHilb}$ and ${\bf Rel}$ moreover admit a categorical matrix calculus.  Together these features guide us towards topological quantum field theories.  We also discuss posetal categories, how group representations are in fact  categorical constructs, and what strictification and coherence of monoidal categories is all about. In our attempt to complement the existing literature we omitted some very basic topics. For these we refer the reader to other available sources.  
\end{abstract}  

\setcounter{section}{-1} 

\section{Prologue: cooking with
vegetables}\label{Sec:Cooking}

Consider a `raw potato'.  Conveniently, we refer to it as $A$.    Raw potato $A$
admits several \em states \em e.g.~`dirty', `clean',  `skinned', ... Since raw 
potatoes don't digest well we need to \em process \em $A$ into `cooked potato'  $B$.  We refer to $A$ and $B$ as \em kinds \em or \em types \em of food. Also $B$ admits several states e.g.~`boiled', `fried', `baked with skin', `baked without skin', ...  Correspondingly, there are several ways to turn raw potato $A$ into cooked potato $B$ e.g.~`boiling', `frying', `baking', to which we respectively refer as $f$,
$f'$ and $f''$.  We make the fact that each of these cooking \em processes \em applies to
raw potato $A$ and  produces cooked potato $B$ explicit via \em labelled arrows\em: 
\[ 
A\rTo^f B\qquad\quad A\rTo^{f'} B\qquad\quad A\rTo^{f''} B\,.  
\]

\par\smallskip
{\bf Sequential composition.}
A plain cooked potato  tastes  dull so we'd like to process it into `spiced
cooked potato'  $C$. We refer to the composite process that consists of
first `boiling' $A\rTo^f B$ \em and then \em `salting' $B\rTo^g C$ as 
\[
A\rTo^{g\circ f} C\,.
\]
%
%
%
We refer to the trivial process of  `doing nothing to vegetable $X$' as 
\[ 
X\rTo^{1_X}X\,.  
\] 
Clearly we have $1_Y\circ \xi=\xi\circ 1_X= \xi$ for all processes $X\rTo^{\xi}Y$.  Note that there is a slight subtlety here: we need to specify what we mean by equality of cooking processes.  We will conceive two cooking processes $X\rTo^{\xi}Y$ and $X\rTo^{\zeta}Y$ as equal, and write $\xi=\zeta$, if the resulting effect on each of the states which $X$ admits is the same.  A stronger notion of equality arises when we also want some additional details of the processes to coincide e.g.~the brand of the cooking pan that we use.

Let  $D$ be a `raw carrot'. Note that it is indeed very important to
explicitly  distinguish our potato and our carrot and any other vegetable such as `lettuce' $L$ in terms of their respective names $A$, $D$ and $L$, since each admits distinct ways of processing. And also a cooked potato admits different ways of processing than a raw one, for example, while we can mash cooked potatoes, we can't mash raw ones.  We denote all processes which turn raw potato $A$ into cooked potato $B$ by ${\bf C}(A,B)$.  Consequently, we can repackage composition of cooking processes as a function
\[
-\circ - : {\bf C}(X,Y)\times {\bf C}(Y,Z)\to {\bf C}(X,Z)\,.
\]

\par\smallskip
{\bf Parallel composition.}
We want to turn `raw potato' $A$  and `raw carrot' $D$ into `carrot-potato mash' $M$.  We refer to the fact that this requires (or consumes) both \em $A$ and $D$ \em as $A\otimes D$. Refer to  `frying the carrot' as $D\rTo^h
E$. Then, by
\[ 
A\otimes D\rTo^{f\otimes h}B\otimes E 
\] 
we mean `boiling potato $A$' \em while \em `frying carrot $D$' and by 
\[ 
C\otimes F\rTo^{x} M
\] 
we mean   `mashing spiced cooked potato $C$ and spiced cooked carrot $F$'. 

\par\smallskip
{\bf Laws.}
The whole process from raw components $A$ and $D$ to  `meal'
$M$ is 
\[ 
A\otimes D\rTo^{f\otimes h}B\otimes E\rTo^{g\otimes k}C\otimes
F\rTo^{x} M\  =\ A\otimes D\rTo^{x\circ(g\otimes k)\circ (f\otimes h)}M\,, 
\]
where `peppering the carrot' is referred to as $E\rTo^k F$.  We refer to the list of the operations that we apply, i.e.~($f$ while $h$, $g$ while $k$, $x$), as a \em recipe\em.  Distinct recipes can yield the same meal.
The reason for this is that the two operations `and then' (i.e.~$-\circ-$) and `while' (i.e.~$-\otimes-$)
which we have at our disposal are not totally independent but \em interact \em
in a certain way.  This is exemplified  by the equality 
\beq\label{food_bifunct1} 
(1_B\otimes h)\circ(f\otimes 1_D)=(f\otimes 1_E)\circ(1_A\otimes h)
\eeq 
on cooking processes, which states that it makes no difference whether `we first boil the potato and then fry the carrot', or, `first fry the carrot and then boil the potato'. 

Eq.(\ref{food_bifunct1}) is in fact a generally valid equational law for
cooking processes, which does not depend on specific properties of $A, B, D, E, f$ nor $h$.  

Of course, chefs do not perform computations
involving eq.(\ref{food_bifunct1}), since their  `intuition' accounts for its content. But, if we were to teach an android how to become a chef, which would require it/him/her to reason  about recipes,  
then we would need to teach it/him/her  the laws governing these 
recipes. 

In fact, there is a more general law governing cooking processes from which
eq.(\ref{food_bifunct1}) can be derived, namely, 
\beq\label{food_bifunct2} 
(g\circ f)\otimes(k\circ h)=(g\otimes k)\circ(f\otimes h)\,.
\eeq 
That is, `boiling the
potato and then salting it, while, frying the carrot and then peppering it', is
equal to  `boiling the potato while frying the carrot, and then, salting the
potato while peppering the carrot'.\footnote{In the light of the previous footnote, note here that this law applies to any reasonable notion of equality for processes.}  A proof of the fact that eq.(\ref{food_bifunct1}) can be derived from eq.(\ref{food_bifunct2}) is in Proposition
\ref{Prop:Bifinct} below.

\par\smallskip
{\bf Logic.}
Eq.(\ref{food_bifunct2}) is indeed a  \em logical \em statement.   
In particular,  note the remarkable similarity, but at the same time also the
essential difference, of eq.(\ref{food_bifunct2}) with the well-known
\em distributive law of classical logic\em,  which states that
\beq\label{food_dist} 
A\ and\ (B\ or\ C)=(A\ and\ B)\ or\ (A\ and\ C)\,.  
\eeq
For simple situations, if one possesses enough brainpower,
`intuition' again accounts for this distributive law.  Ot the other hand,  it needs to be explicitly taught to androids, since this distributive law is key to the resolution method which is the standard implementation of artificial reasoning in AI and robotics \cite{resolution_method}.  Also for complicated sentences we ourselves will need to rely on this method too.

The $(\circ,\otimes)$-logic is a \em logic of interaction\em.  It applies to
cooking processes, physical  processes, biological processes, logical
processes (i.e.~proofs), or computer processes (i.e.~programs).   The theory of \em monoidal
categories\em, the subject of this chapter,  is the mathematical framework that accounts for the common structure of each of these \em theories of processes\em. The framework of monoidal categories moreover enables \em modeling \em and \em axiomatising \em (or `classify')  the  extra structure which certain families of processes may have. For example, how cooking processes differ from physical processes, and how quantum processes differ from classical processes.

\par\smallskip
{\bf Pictures.}
We mentioned that our intuition accounts for
$(\circ,\otimes)$-logic.  Wouldn't it be nice if there would be mathematical
structures which also `automatically' (or `implicitly')  account for the logical mechanisms which we
intuitively perform?  Well, these mathematical structures do exist.  While they are only a fairly recent development, they are becoming more and more prominent in mathematics, including in important `Fields Medal awarding areas' such as algebraic topology and representation theory --- see for example \cite[and references therein]{Morrison}.  Rather than being symbolic, these mathematical structures are purely graphical. Indeed, by far ***the*** coolest thing about monoidal categories is that they admit a purely pictorial calculus, and these pictures automatically account for the logical mechanisms which we
intuitively perform.  As pictures, both sides of
eq.(\ref{food_bifunct2}) become:
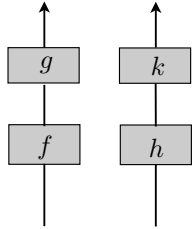
\begin{center} 
\ifx\JPicScale\undefined\def\JPicScale{1}\fi
\psset{unit=\JPicScale mm}
\psset{linewidth=0.3,dotsep=1,hatchwidth=0.3,hatchsep=1.5,shadowsize=1,dimen=middle}
\psset{dotsize=0.7 2.5,dotscale=1 1,fillcolor=black}
\psset{arrowsize=1 2,arrowlength=1,arrowinset=0.25,tbarsize=0.7 5,bracketlength=0.15,rbracketlength=0.15}
\begin{pspicture}(0,0)(25.45,30.44)
\newrgbcolor{userFillColour}{0.8 0.8 0.8}
\pspolygon[linewidth=0.15,fillcolor=userFillColour,fillstyle=solid](15.89,19.55)(25.35,19.55)(25.35,24.25)(15.89,24.25)
\newrgbcolor{userFillColour}{0.8 0.8 0.8}
\pspolygon[linewidth=0.15,fillcolor=userFillColour,fillstyle=solid](16,8.73)(25.45,8.73)(25.45,14)(16,14)
\rput(20.82,10.96){$h$}
\rput(21.02,21.9){$k$}
\psline[linewidth=0.25]{->}(20.72,24.3)(20.72,30.4)
\psline[linewidth=0.25](20.72,0.5)(20.72,8.8)
\psline[linewidth=0.25](20.72,13.8)(20.7,19.5)
\newrgbcolor{userFillColour}{0.8 0.8 0.8}
\pspolygon[linewidth=0.15,fillcolor=userFillColour,fillstyle=solid](1.07,19.59)(10.53,19.59)(10.53,24.29)(1.07,24.29)
\newrgbcolor{userFillColour}{0.8 0.8 0.8}
\pspolygon[linewidth=0.15,fillcolor=userFillColour,fillstyle=solid](1.18,8.77)(10.63,8.77)(10.63,14.04)(1.18,14.04)
\rput(6,11){$f$}
\rput(6.2,21.94){$g$}
\psline[linewidth=0.25]{->}(5.9,24.34)(5.9,30.44)
\psline[linewidth=0.25](5.9,0.54)(5.9,8.84)
\psline[linewidth=0.25](5.9,13.84)(5.9,19.34)
\end{pspicture}
 
\end{center} 
Hence eq.(\ref{food_bifunct2}) becomes
an implicit salient feature of the graphical calculus and needs no
explicit attention anymore. This, as we will see below, substantially
simplifies many computations.    To better understand in which manner these pictures simplify computations note that the differences between the two sides
of eq.(\ref{food_bifunct2})  can be  recovered by introducing
`artificial' brackets within the two pictures: 

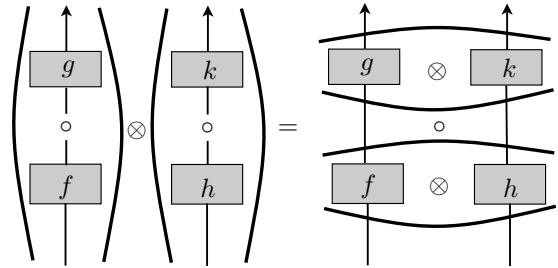
\begin{center}
\ifx\JPicScale\undefined\def\JPicScale{1}\fi
\psset{unit=\JPicScale mm}
\psset{linewidth=0.3,dotsep=1,hatchwidth=0.3,hatchsep=1.5,shadowsize=1,dimen=middle}
\psset{dotsize=0.7 2.5,dotscale=1 1,fillcolor=black}
\psset{arrowsize=1 2,arrowlength=1,arrowinset=0.25,tbarsize=0.7 5,bracketlength=0.15,rbracketlength=0.15}
\begin{pspicture}(0,0)(78.2,35.26)
\psbezier[linewidth=0.45](8.1,34.8)(5.4,19.8)(5.4,17.6)(8.7,1)
\psbezier[linewidth=0.45](26.5,34.5)(23.8,19.5)(23.8,17.3)(27.1,0.7)
\psbezier[linewidth=0.45](17.5,34.6)(21.5,19.4)(21.3,17.4)(18.1,0.8)
\psbezier[linewidth=0.45](36.4,34.5)(40.4,19.3)(40.2,17.3)(37,0.7)
\rput(42.6,18.5){$=$}
\newrgbcolor{userFillColour}{0.8 0.8 0.8}
\pspolygon[linewidth=0.15,fillcolor=userFillColour,fillstyle=solid](8.3,8.53)(17.75,8.53)(17.75,13.8)(8.3,13.8)
\rput(13.12,10.76){$f$}
\psline[linewidth=0.25](13.02,0.3)(13.02,8.6)
\newrgbcolor{userFillColour}{0.8 0.8 0.8}
\pspolygon[linewidth=0.15,fillcolor=userFillColour,fillstyle=solid](48.48,8.57)(57.93,8.57)(57.93,13.84)(48.48,13.84)
\rput(53.3,10.8){$f$}
\psline[linewidth=0.25](53.2,0.34)(53.2,8.64)
\newrgbcolor{userFillColour}{0.8 0.8 0.8}
\pspolygon[linewidth=0.15,fillcolor=userFillColour,fillstyle=solid](27.12,8.49)(36.57,8.49)(36.57,13.76)(27.12,13.76)
\rput(31.94,10.72){$h$}
\psline[linewidth=0.25](31.84,0.26)(31.84,8.56)
\newrgbcolor{userFillColour}{0.8 0.8 0.8}
\pspolygon[linewidth=0.15,fillcolor=userFillColour,fillstyle=solid](67.39,8.49)(76.84,8.49)(76.84,13.76)(67.39,13.76)
\rput(72.21,10.72){$h$}
\psline[linewidth=0.25](72.11,0.26)(72.11,8.56)
\newrgbcolor{userFillColour}{0.8 0.8 0.8}
\pspolygon[linewidth=0.15,fillcolor=userFillColour,fillstyle=solid](47.99,24.41)(57.45,24.41)(57.45,29.11)(47.99,29.11)
\rput(53.12,26.76){$g$}
\psline[linewidth=0.25]{->}(52.82,29.16)(52.82,35.26)
\psline[linewidth=0.25](52.82,13.96)(52.82,24.36)
\newrgbcolor{userFillColour}{0.8 0.8 0.8}
\pspolygon[linewidth=0.15,fillcolor=userFillColour,fillstyle=solid](66.9,24.33)(76.36,24.33)(76.36,29.03)(66.9,29.03)
\rput(72.03,26.68){$k$}
\psline[linewidth=0.25]{->}(71.73,29.08)(71.73,35.18)
\psline[linewidth=0.25](71.72,13.76)(71.62,24.26)
\newrgbcolor{userFillColour}{0.8 0.8 0.8}
\pspolygon[linewidth=0.15,fillcolor=userFillColour,fillstyle=solid](8.3,24.1)(17.76,24.1)(17.76,28.8)(8.3,28.8)
\rput(13.43,26.45){$g$}
\psline[linewidth=0.25]{->}(13.13,28.85)(13.13,34.95)
\psline[linewidth=0.25](13.12,13.66)(13.12,23.96)
\newrgbcolor{userFillColour}{0.8 0.8 0.8}
\pspolygon[linewidth=0.15,fillcolor=userFillColour,fillstyle=solid](27.12,24.06)(36.58,24.06)(36.58,28.76)(27.12,28.76)
\rput(32.25,26.41){$k$}
\psline[linewidth=0.25]{->}(31.95,28.81)(31.95,34.91)
\psline[linewidth=0.25](31.92,13.76)(31.92,24.16)
\rput(22.5,18.3){$\otimes$}
\psbezier[linewidth=0.45](78.2,8.2)(63.7,4.7)(61.3,4.7)(47.4,7.9)
\psbezier[linewidth=0.45](77.9,23.7)(63.4,20.2)(61,20.2)(47.1,23.4)
\psbezier[linewidth=0.45](78,15.3)(63.7,17.4)(61.7,17.5)(47.2,15)
\psbezier[linewidth=0.45](77.5,30.4)(63.2,32.5)(61.2,32.6)(46.7,30.1)
\rput(62.4,26.2){$\otimes$}
\rput(62.7,10.8){$\otimes$}
\pspolygon[linestyle=none,fillcolor=white,fillstyle=solid](11,20.5)(15.4,20.5)(15.4,17)(11,17)
\pspolygon[linestyle=none,fillcolor=white,fillstyle=solid](30,20.4)(34.4,20.4)(34.4,16.9)(30,16.9)
\rput(62.7,18.8){$\circ$}
\rput(31.9,18.6){$\circ$}
\rput(13.1,18.7){$\circ$}
\end{pspicture}
 
\end{center}
A detailed account on  this graphical calculus is in Section
\ref{sec:graphCalc}.   

\par\bigskip
In the remainder of this chapter we provide a formal tutorial on several
kinds of monoidal categories that are relevant to physics.  If you'd rather
stick to the informal story of this prologue you might want to first take a bite of
\cite{C2005c,C2005d}.\footnote{Paper  \cite{C2005c} provided a conceptual
template for setting up the content of this paper. However, here we go in more
detail and provide more examples.}   Section \ref{quiver} introduces categories and Section \ref{muscle} introduces tensor structure.  Section \ref{sec:quantumlike} studies quantum-like tensors and Section \ref{sec:classicallike} studies classical-like tensors. Section \ref{funcNatTQFT} introduces mappings between monoidal categories (= monoidal functors), and natural transformations between these, which enable to concisely define topological quantum field theories. Section \ref{Further_reading} suggests further reading.


\section{The 1D case: New arrows for your quiver}\label{quiver}

The  bulk of the previous section discussed the two manners in which we can compose processes, namely \em sequentially \em and \em in parallel\em, or more physically put, in time and in space.  These are indeed the situations we truly care about in this chapter.  
Historically however, category theoreticians cared mostly about  \em
one-dimensional fragments \em of the \em two-dimensional \em monoidal
categories.  These one-dimensional fragments  are (ordinary) \em categories\em, hence the name \em category theory\em.  Some people will get rebuked by the terminology and particular syntactic
language used in category theory --- which can sound and look like unintelligible
jargon --- resulting in its unfortunate label of {\em generalised abstract
nonsense}. The reader should realise that initially category theory was crafted
as `a theory of mathematical structures'.  Hence substantial effort was made
to avoid any reference to the underlying \em concrete models\em, resulting
in its seemingly idiosyncratic format.  The personalities involved in crafting
category theory, however brilliant minds they had, also did not always help the
cause of making category theory accessible to a broader community.

But this `theory of mathematical structures' view is not the only way to
conceive category theory.  As we argued above, and as is witnessed by its
important use in computer science, in proof theory, and more recently also in
quantum informatics and in quantum foundations, category theory is a theory
which brings the notions of (\em type \em of\,) \em system \em and \em process \em to the forefront,
two notions which are hard to cast within traditional monolithic mathematical
structures.  

We profoundly believe that the fact that the mainstream physics
community has not yet acquired this (type of) systems/process structure as a primal part
of its theories is merely accidental, and temporary, ... and will soon change.  

\subsection{Categories}

We will use the following syntax to denote a function:
\[
f: X\to Y::x\mapsto y
\]
where $X$ is the set of arguments, $Y$ the set of possible values, and 
\[
x\mapsto y
\]
means that argument $x$ is mapped on value $y$.

\begin{definition}\label{cat}\em A {\em category} $\cat$ consists of

\begin{enumerate} 
\item A family\footnote{Typically, `family' will mean a class rather than a set.  While for many constructions the \em size \em of $|\cat|$ is important, it will not play a key role in this paper.} $|\cat|$ of {\em objects}\,; 
\item For any $A,B\in |\cat|$, a set $\cat(A,B)$ of \em morphisms\em, the \em hom-set\em\,; 
\item For any $A,B,C\in |\cat|$, and any $f\in\cat(A,B)$ and $g\in\cat(B,C)$, a \em composite \em $g\circ f\in\cat(A,C)$, i.e., for all $A,B,C\in|\cat|$ there is a {\em composition operation}
\[
-\circ-:\cat(A,B)\times\cat(B,C)\rightarrow\cat(A,C)::(f,g)\mapsto g\circ f\,, 
\] 

and this composition operation is \em
associative \em and has \em units\em, that is,
\begin{itemize} 
\item[i.] for any $f\in\cat(A,B)$, $g\in\cat(B,C)$ and $h\in\cat(C,D)$ we have 
\[ 
h\circ (g\circ f)=(h\circ g)\circ f\,; 
\] 
\item[ii.] for any $A\in |\cat|$, there exists a morphism $1_A\in\cat(A,A)$, called the \em identity\em, which is  such that for any $f\in\cat(A,B)$ we have 
\[ 
f=f\circ
1_A=1_B\circ f\,.  
\] 
\end{itemize}
\end{enumerate} 
\end{definition}	

A shorthand for  $f\in\cat(A,B)$ is  $A\rTo^f B$.  As already mentioned above,
this definition was proposed by Samuel Eilenberg and Saunders Mac Lane in 1945 as part of a framework which intended to unify a variety of mathematical constructions within different areas of mathematics \cite{EilenbergMacLane}.  Consequently, most of the examples of categories that one encounters in the literature encode mathematical structures: the objects will be
examples of this mathematical structure and the morphisms will be the \em structure-preserving \em  maps between these.  This kind of categories is usually referred to as \em concrete categories \em
\cite{Adamek}.  We will also call them \em concrete categorical models\em.  

\subsection{Concrete categories}\label{sec:concretecategories} 

Traditionally,  mathematical structures are defined as a set equipped with some operations and some axioms, for instance:
\begin{itemize}
\item[-] A {\em group} is a set $G$ with an associative binary
operation $-\bullet -:G\times G\rightarrow G$
and with a two-sided identity $1\in G$, relative to which each element is invertible, that is, for all $g\in G$ there exists $g^{-1}\in G$ such that $g\bullet g^{-1}=g^{-1}\bullet g= 1$.
\end{itemize} 
Similarly we define rings and fields.
Slightly more involved but in the same spirit: 
\begin{itemize}
\item[-] A {\em vector space} is a pair $(V,\mathbb{K})$,
respectively a commutative group and a field, and these interact
via the notion of  scalar multiplication, i.e.~a map $V\times \mathbb{K}\to V$ which is subject to a number of axioms.  
\end{itemize} 
It is to these operations and axioms that one usually refers to as \em structure\em.  Functions on the underlying sets which preserve (at least part of) this structure are called \em structure preserving maps\em.  Here are some examples of structure preserving maps:
\begin{itemize} 
\item[-] \em group homomorphisms\em, i.e.~functions which preserve the group multiplication, from which it then also follows that the unit and inverses are preserved; 
\item[-] \em linear maps\em, i.e.~functions from a vector space to a vector space which preserve linear combinations of vectors.
%
\end{itemize}

\begin{example}\label{excat1} 
Let  ${\bf Set}$ be the concrete category with:
\begin{enumerate} 
\item all sets as objects, 
\item all functions between sets as morphisms, that is, more precisely, if $X$ and $Y$ are sets and $f:X\to Y$ is a function between these sets, then $f\in {\bf Set}(X,Y)$,
\item ordinary composition of functions, that is, for $f:X\rightarrow Y$ and $g:Y\rightarrow Z$ we have $(g\circ f)(x):= g(f(x))$ for the composite $g\circ f:X\to Z$, and, \item  the obvious identities i.e.~$1_X(x):=x$.  \end{enumerate} 
${\bf Set}$ is indeed a category since:
\begin{itemize} 
\item[-] function composition is associative, and, 
\item[-] for any function $f:X\rightarrow Y$  we have $(1_Y\circ f)(x)=f(x)=(f\circ 1_X)(x)$\,. 
\end{itemize}
\end{example}

\begin{example}\label{excat2} $\mathbf{FdVect}_\mathbb{K}$ is the concrete
category with: 
\begin{enumerate} 
\item finite dimensional vectors spaces over $\mathbb{K}$ as objects,
\item all linear maps between these vectors spaces as morphisms, and
\item ordinary composition of the underlying functions,  and, 
\item identity functions.
\end{enumerate} 
$\mathbf{FdVect}_\mathbb{K}$  is indeed category since:
\begin{itemize} 
\item[-] the composite of two linear maps is again a linear map, and, 
\item[-] identity functions are linear maps.\end{itemize}
\end{example}

\begin{example}\label{excat3} 
${\bf Grp}$  is the concrete category with:
\begin{enumerate} 
\item groups as objects, 
\item group homomorphisms between these groups as morphisms, and,
\item ordinary function composition, and, 
\item identity functions.
\end{enumerate}
${\bf Grp}$  is indeed category since:
\begin{itemize} 
\item[-] the composite of two group homomorphisms is a group homomorphism, and, 
\item[-] identity functions are group homomorphisms.\end{itemize}
\end{example}

\begin{example}[elements]\label{ex:elements} 
Above we explained that mathematical structures such as groups typically consist of a set with additional structure.  In the case of a category we have a collection of objects, and for each pair of objects a set of morphisms.  The `structure of a category' then consists of the composition operation on morphisms and the identities on objects.  So there is no reference to what the individual objects actually are (e.g.~a set, a vector space, or a group).  Consequently, one would expect that when passing from a mathematical structure (cf.~group) to the corresponding concrete category with these mathematical structures as objects (cf.~${\bf Grp}$), one looses the object's `own' structure.  But fortunately, this happens not to be the case.  The fact that we consider structure preserving maps as morphisms will allow us to recover the mathematical structures that we started from.  In particular, by only relying on categorical concepts  we are still able to identify the `elements' of the objects.  

For the set $X\in |{\bf Set}|$ and some
chosen element $x\in X$ the function 
\[ 
e_x:\{*\}\rightarrow X::*\mapsto x\,,
\] 
where $\{*\}$ is any one-element set, maps the unique element of $\{*\}$ onto the chosen element $x$. If $X$ contains $n$ elements, then there are $n$ such functions  each corresponding to the
element on which $*$ is mapped. Hence the elements of the set $X$  are
now encoded as the set ${\bf Set}(\{*\},X)$. 

In a similar manner we can single out vectors in vectors spaces.  For the
vector space $V\in |\mathbf{FdVect}_\mathbb{K}|$ and some fixed vector $v\in V$
the linear map \[ e_v:\mathbb{K}\rightarrow V::1\mapsto v\,, \] where
$\mathbb{K}$ is now the one-dimensional vector space over itself, maps the
element $1\in\mathbb{K}$ onto  the chosen element $v$. Since $e_v$ is linear, it
is completely characterised by the image of the single element $1$.  Indeed, 
$e_v(\alpha)=e_v(\alpha\cdot 1)=\alpha\cdot e_v( 1)=\alpha\cdot v$,
that is, the element $1$ is a basis for the one-dimensional vector space $\mathbb{K}$.
\end{example}


\begin{example}\label{excat3_5} 
${\bf Pos}$  is the concrete category with:
\begin{enumerate} 
\item partially ordered sets, that is, a set together with a reflexive, anti-symmetric and transitive relation, as objects, 
\item order preserving maps, i.e.~$x\leq y\Rightarrow f(x)\leq f(y)$,  as morphisms, and, 
\item ordinary function composition, and identity functions.  
\end{enumerate} 
An extended version of this category is ${\bf Pre}$ where we
consider arbitrary pre-ordered sets, that is, a set
together with a reflexive and transitive relation.
\end{example} 
\begin{example}\label{excat4} 
${\bf Cat}$  is the concrete category with:\footnote{In order to conceive ${\bf Cat}$ as a concrete category, the family of objects should be restricted to the so-called ``small'' categories i.e., categories for which the family of objects is a set.}
\begin{enumerate} 
\item categories as objects, 
\item so-called \em functors \em between these as morphisms (see Section \ref{Sec:Functors}), and, 
\item functor composition, and identity functors.
\end{enumerate} 
\end{example}

\subsection{Real world categories}\label{sec:real_world}

But viewing category theory as some kind of  \em metatheory about  mathematical
structure \em is  not necessarily the most useful perspective for the
sort of applications that we have in mind.  Indeed, here are a few  examples of the kind of
categories we truly care about, and which are not categories with mathematical
structures as objects and structure preserving maps as morphisms.  

\begin{example}\label{excat11} 
The category $\mathbf{PhysProc}$ with
\begin{enumerate} 
\item all physical systems $A, B, C, \ldots$ as objects, 
\item all physical processes which take a physical system of type $A$ into a physical system of type $B$
as the morphisms of type $A\rTo B$ (these processes typically require some finite amount of time to be
completed), and, 
\item sequential composition of these physical processes as composition, and the process which leaves system $A$ invariant as the identity $1_A$.  
\end{enumerate}
Note that in this case associativity of composition admits a
physical interpretation: if we first have process $f$, then
process $g$, and then process $h$, it doesn't
matter whether we either consider $(g\circ f)$ as a single entity after which we apply $h$, or  whether we consider $(h\circ g)$ as a single entity which we apply after $f$. Hence brackets constitute superfluous
data that can be omitted i.e.~
\[
h\circ g\circ f := h\circ(g\circ f)=(h\circ g)\circ f\,.
\] 
\end{example}

\begin{example}\label{excat12} The category $\mathbf{PhysOpp}$ is an \em
operational variant \em of the above where, rather than general physical
systems such as stars, we focus on systems which can be manipulated in
the lab, and rather than general processes, we consider the operations
which the practising experimenter performs on these systems,
for example, applying force-fields, performing measurements etc. 
\end{example}

\begin{example}\label{excat13} The category $\mathbf{QuantOpp}$ is a
restriction of the above where we restrict ourselves to quantum systems
and operations thereon.  Special processes in $\mathbf{QuantOpp}$ are
\em preparation procedures\em, or \em states\em.   If $Q$ denotes a
qubit, then the type of a preparation procedure would be $\II\rTo Q$
where $\II$ stands for `unspecified'.  Indeed, the point of a
preparation procedure is to provide a qubit in a certain state, and the
resources which we use to produce that state are typically not of
relevance for the remainder of the experimental procedure.  We can further specialise to either pure (or closed) quantum systems or mixed (or open) quantum systems, 
categories to which we respectively refer as $\mathbf{PurQuantOpp}$ and $\mathbf{MixQuantOpp}$. 
\end{example}

Obviously, Example \ref{excat13} is related to the concrete  category which has
Hilbert spaces as objects and certain types of linear mappings (e.g.~completely positive maps) as
morphisms.  The preparation procedures discussed above then correspond to
`categorical elements' in the sense of Example \ref{ex:elements}.  We discuss this
correspondence below.

While to the sceptical reader the above examples still might not seem very useful yet, the next two ones, which are very similar, have become really important for Computer Science and Logic. They are the reason that, for example,  University of Oxford Computing Laboratory  offers category theory to its undergraduates.

\begin{example}\label{excat14} The category $\mathbf{Comp}$ with
\begin{enumerate} 
\item all data types, e.g.~Booleans, integers, reals, as objects, 
\item all programs which take data of type
$A$ as their input and produce data  of type $B$ as
their output  as the morphisms of type $A\rTo B$, and,
\item sequential composition of programs as composition, and the
programs which output their input unaltered as
identities.  
\end{enumerate}
\end{example}

\begin{example}\label{excat15} The category $\mathbf{Prf}$ with
\begin{enumerate} 
\item all propositions as objects, 
\item all proofs which conclude from proposition $A$ that  proposition
$B$ holds as the morphisms of type $A\rTo B$, and,
\item concatenation (or chaining) of proofs as composition, and
the tautologies `from $A$ follows $A$' as identities.
\end{enumerate} 
\end{example}

Computer scientists particularly like category theory because it explicitly
introduces the notion of {\em type}: an arrow $A\rTo^f B$ has type $A\rTo B$.
These types prevent silly mistakes when writing programs, e.g.~the composition
$g\circ f$ makes no sense for $ C\rTo^g  D$ because the output  --- called the {\em
codomain} --- of $f$ doesn't match the input  --- called the {\em domain} --- of $g$.
Computer scientists would say: 
\begin{center}
``types don't match''.
\end{center}

Similar categories ${\bf BioProc}$ and ${\bf ChemProc}$ can be build for
organisms and biological processes, chemicals and chemical reactions,
etc.\footnote{The first time the 1st author heard about categories was in a
Philosophy of Science course, given by a biologist specialised in population
dynamics, who discussed the importance of category theory in the influential
work of Robert Rosen \cite{Rosen}.} The recipe for producing these categories
is obvious: 
\begin{center} 
\begin{tabular}{c|c|c} \ {\bf Name} \   & {\bf
Objects} & {\bf Morphisms} \\ 
\hline \  some area of science \  & \ corresponding systems \  & \ corresponding processes \ \\ 
\end{tabular}
\end{center} 
Composition boils down to `first $f$ and then $g$ \em happens\em'
and identities are just `nothing \em happens\em'.  Somewhat more operationally put,
composition is `first \em do \em $f$ and then \em do \em $g$' and identities
are just `\em doing \em nothing'. The reason for providing both the
`objectivist' (= passive) and `instrumentalist' (= active) perspective is that we both want to appeal to members of the theoretical physics community  and members of the quantum information community.  The first community typically doesn't like instrumentalism since it just doesn't seem to make
sense in the context of theories such as cosmology; on the other hand,
instrumentalism is as important to quantum informatics as it is to ordinary
informatics.  We leave it up to the reader to decide whether it should play a
role in the interpretation of quantum theory.

\subsection{Abstract categorical structures and properties} 
One can treat categories as mathematical structures in their
own right, just as groups and vector spaces are mathematical structures.  In
contrast with concrete categories, abstract categorical structures then arise
by either endowing categories with more \em structure \em or by requiring them
to satisfy certain \em properties\em.  

We are of course aware that this is not a formal definition.  
Our sheepish excuse is that physicists rarely provide precise definitions.  There is however a formal definition which can be found in \cite{Adamek}.  We do provide one below in Example \ref{ConcreteCategory}.

\begin{example}\label{excat7} 
A \emph{monoid} $(M,\bullet,1)$ is a set together with a binary
associative operation 
\[
-\bullet -:M\times M\rightarrow M
\]
which admits a
unit --- i.e.~a `group without inverses'. Equivalently, we can define a monoid as a category ${\bf M}$ with a single object $*$. Indeed, it suffices to identify
\begin{itemize}
	\item the elements of the hom-set ${\bf M}(*,*)$ with those of $M$, 
	\item the associative composition operation 
\[ 
-\circ-:{\bf M}(*,*)\times{\bf M}(*,*)\rightarrow {\bf M}(*,*)
\]
with the associative monoid multiplication $\bullet$, and 
	\item the identity $1_*:*\rightarrow *$ with the unit $1$.   
\end{itemize}
Dually, in any category $\cat$, for any $A\in|\cat|$, the set $\cat(A,A)$ is always a monoid. 
\end{example}

\begin{definition}\label{def:isomorphic} \em 
Two objects $A,B\in |\cat|$ are \em
isomorphic \em if there exists morphisms $f\in\cat(A,B)$ and
$g\in\cat(B,A)$ such that $g\circ f=1_A$ and $f\circ g=1_B$. The
morphism $f$ is called an {\em isomorphism} and $f^{-1}:=g$ is called the \em
inverse \em to $f$.  
\end{definition}

The notion of isomorphism known to the reader is the set-theoretical one,  namely that of a bijection.  
We now show that in the concrete category ${\bf Set}$ the category-theoretical notion of isomorphism coincides  with the notion of bijection. Given functions $f:X\to Y$ and $g:Y\to X$ satisfying $g(f(x))=x$ for all $x \in X$ and $f(g(y))=y$
for all  $y\in Y$  we have: 
\bit 
\item $f(x_1)=f(x_2) \Rightarrow g(f(x_1))=g(f(x_2)) \Rightarrow x_1=x_2$ so $f$ is injective, and, 
\item for all $y\in Y$, setting $x:=g(y)$,  we have $f(x)=y$  so $f$ is surjective, 
\eit
so $f$ is indeed a bijection.   We leave it to the reader to verify that the converse also holds.
For the other concrete categories mentioned above the categorical notion of
isomorphism also coincides with the usual one. 

\begin{example}\label{excat8} 
Since a group $(G,\bullet,1)$ is a monoid with
inverses it can now be equivalently defined as a category with one
object in which each morphism
is an isomorphism.
More generally, a \em groupoid \em is a category in which each morphism
has an inverse. For instance, the category ${\bf Bijec}$ which has sets as objects and bijections as morphisms is such a groupoid. So is  ${\bf FdUnit}$  which has finite dimensional Hilbert spaces as objects and unitary operators as morphisms. Groupoids have  important applications in mathematics, for example, in algebraic topology \cite{Brown}. 
\end{example} 

From this, we see that any group is an example of an abstract categorical
structure.  At the same time, all groups together, with structure preserving
maps between them, constitute a concrete category.  Still following?  That
categories allow several ways of representing mathematical structures might
seem confusing at first, but it is a token of their versatility.
While monoids correspond to categories with only one object, with groups as a
special case, similarly, pre-orders are categories with very few morphisms, with partially
ordered sets as a special case.

\begin{example}\label{excat6}  
Any preordered set $(P,\leq)$ can be seen as a category ${\bf P}$: 
\bit 
\item The elements of $P$ are the objects of ${\bf P}$, 
\item Whenever $a\leq b$ for
$a,b\in P$ then there is a single morphism of type  $a\rTo  b$, that is, ${\bf
P}(a,b)$ is a singleton, and  whenever $a\not\leq b$ then there is no
morphism of type  $a\rTo  b$, that is, ${\bf P}(a,b)$ is empty.  
\item Whenever there is pair of  morphisms of types $a\rTo  b$ and $b\rTo  c$,
that is, whenever $a\leq  b$ and $b\leq  c$, then transitivity of $\leq$ guarantees
the existence of a unique morphism of type $a\rTo  c$,  which we take to be the
composite of the morphisms of type $a\rTo  b$ and $b\rTo  c$.  
\item Reflexivity
guarantees the existence of a unique morphism of type $a\rTo  a$, which we take
to be the identity on the object $a$.  
\eit

Conversely, a category ${\bf C}$ of which the objects constitute a set, and in which there is at most one morphism of any
type i.e., hom-sets are either singletons or empty,  is in fact  a preordered set. Concretely:
\bit 
\item  The set $|{\bf C}|$ are the elements of the preordered set, 
\item We set $A\leq B$ if and only if ${\bf C}(A,B)$ is non-empty,
\item Since ${\bf C}$ is a category, whenever there exist morphisms $f\in{\bf C}(A,B)$ and $g\in{\bf C}(B,C)$, that is, whenever both ${\bf C}(A,B)$ and ${\bf C}(B,C)$ are non-empty,  then there exist a morphism $g\circ f\in{\bf C}(A,C)$, so ${\bf C}(A,C)$ is also non-empty. Hence,  $A\leq B$ and $B\leq C$ yields $A\leq C$, so $\leq$ is transitive. 
\item Since $1_A\in{\bf C}(A,A)$ we also have $A\leq A$, so $\leq$ is reflexive.  
\eit 
Hence, preordered sets indeed constitute an abstract category: its defining property is that every hom-set contains at most
one morphism. Such categories are sometimes called \em thin categories\em.
Conversely, categories with non-trivial hom-sets are called \em thick\em.
Partially ordered sets also constitute an abstract category, namely one in which:
\bit
\item every hom-set contains at most one morphism\,;
\item whenever two objects are  isomorphic then they must be equal\,.
\eit
This second condition imposes anti-symmetry on the partial order.
\end{example}

Let $\{*\}$ and $\emptyset$ denote a singleton set and the
empty set respectively. Then for any set $A\in|{\bf Set}|$, the set ${\bf Set}(A,\{*\})$ of all functions of type
$A\rightarrow\{*\}$ is itself a singleton, since there is only one function which maps all $a\in A$ on $*$, the single element of $\{*\}$. This concept can be \emph{dualised}.
The set ${\bf Set}(\emptyset,A)$ of functions of type $\emptyset\rightarrow A$ is again a singleton consisting of the `empty function'. Due to these special properties, we call $\{*\}$ and $\emptyset$  respectively the {\em terminal} object and the {\em initial} object in ${\bf Set}$. All this can be generalised to arbitrary categories as follows:


\begin{definition} \em An object $\top\in|\cat|$ is {\em terminal} in $\cat$
if, for any $A\in |\cat|$, there is only one morphism of type
$A\rTo\top$. Dually, an object $\bot\in|\cat|$ is {\em initial} in $\cat$ if,
for any $A\in|\cat|$, there is only one morphism of type $\bot\rTo A$.
\end{definition}

\begin{proposition} If a category $\cat$ has two initial objects then they
are isomorphic. The same property holds for terminal objects.
\rm \end{proposition}

\noindent Indeed.  Let $\bot$ and $\bot'$ both be initial objects in $\cat$. Since $\bot$ is initial, there is a unique morphism $f$ such that $\cat(\bot,\bot')=\{f\}$. Analogously, there is a unique morphism $g$ such that $\cat(\bot',\bot)=\{g\}$. Now, since $\cat$ is a category and relying again on the fact that $\bot$ is initial, it follows that $g\circ f\in\cat(\bot,\bot)=\{1_\bot\}$. Similarly, 
 $g\circ f\in\cat(\bot',\bot')=\{1_{\bot'}\}$.
Hence, $\bot\simeq\bot'$ as claimed. Similarly we show that $\top\simeq\top'$. 


\begin{example}\label{excat6_5} A partially ordered set $P$  is \em bounded \em
if there exist two elements $\top$ and $\bot$ such that for all $a\in
P$ we have $\bot\leq a\leq \top$.  Hence, when $P$ is viewed as a
category, this means that it has both a terminal and an initial object.
\end{example}

The next example of an abstract categorical structure is the most important one
in this paper. Therefore, we state it as a definition.  Among many (more important) things, it axiomatises `cooking with vegetables'. 

\begin{definition}\label{excat10}\em 
A \em strict monoidal category \em is a category for which: 
\ben 
\item objects come with monoid structure $(|{\bf C}|, \otimes, \II)$ i.e, for all $A,B,C\in |\cat|$, 
\[ 
A\otimes (B\otimes C)= (A\otimes B)\otimes C\quad\mbox{and}\quad \II\otimes A=A=A\otimes \II\,, 
\] 
\item for all objects $A,B,C,D\in |{\bf C}|$ there exists an operation 
\[ 
-\otimes -:{\bf C}(A,B)\times {\bf C}(C,D)\to{\bf C}(A\otimes C,B\otimes D):: (f,g)\mapsto f\otimes g 
\] 
which is associative and has $1_\II$ as its unit, that is,\footnote{Note that
this operation on morphisms is a \em typed \em variant of the notion of
monoid.} 
\[ 
f\otimes (g\otimes h)= (f\otimes g)\otimes h\quad\mbox{and}\quad 1_\II\otimes f=f=f\otimes 1_\II\,, 
\] 
\item for all morphisms $f, g, h, k$ with matching types we have 
\beq\label{food_bifunct2bis}
(g\circ f)\otimes(k\circ h)=(g\otimes k)\circ(f\otimes h)\,,
\eeq
%
\item for all objects $A,B\in |{\bf C}|$ we have
\beq\label{eq:tensoridentity} 
1_A\otimes 1_B=1_{A\otimes B}\,.  
\eeq
\een \end{definition}

As we will see in Section \ref{sec:BifunctAsFunct},  
the two equational constraints eq.(\ref{food_bifunct2bis}) and
eq.(\ref{eq:tensoridentity}) can be conceived as a single principle.

The symbol $\otimes$ is sometimes called the tensor.  We will also use this terminology, since `tensor' is shorter than `monoidal product'.  However, the reader should not deduce from this that the above definition necessitates $\otimes$ to be anything like a tensor product, since this is not at all the case.

The categories of systems and processes discussed in Section \ref{sec:real_world} are all examples of strict monoidal categories.  We already explained in Section \ref{Sec:Cooking} what
$-\otimes-$ stands for: it enables dealing with situations where several
systems are involved.  To a certain extent $-\otimes-$ can be interpreted as a
logical conjunction:  
\[ 
A\otimes B := \mbox{system}\ A\ \mbox{\underline{and}\
system}\ B 
\]
\[
\ f\otimes g := \mbox{process}\ f\ \mbox{\underline{and}\
process}\ g\,.  
\] 
There is however considerable care required with this view:
while 
\[
A\wedge A=A\,, 
\]
in general
\[
A\otimes A\not=A\,.
\]
This is where
the so-called \em linear logic \em \cite{Girard,Seely} kicks in, which is
discussed in substantial detail in  \cite{AbramskyT}. 

For the special object $\II$ we have 
\[
A\otimes\II=A=\II\otimes A
\]
since it is
the unit for the monoid. Hence, it refers to a system which leaves any system
invariant when adjoined to it.  In short, it stands for `unspecified', for `no system', or even
for `nothing'.  We already made reference to it in Example \ref{excat13} when
discussing preparation procedures.  Similarly, $1_\II$ is the operation which
`does nothing to nothing'.  The system $\II$ will allow us to encode a notion of
state within arbitrary monoidal categories, and also a notion of number and
probabilistic weight --- see Example \ref{ex:stateseffects} below. 

\begin{example}\label{excat7_5} 
Now, a monoid $(M,\bullet, 1)$ can also be
conceived as a strict monoidal category in which all morphisms are
identities. Indeed, take $M$ to be the objects, $\bullet$ to be the
tensor and $1$ to be the unit for the tensor.  By taking identities to
be the only morphisms, we can equip these with the same monoid
structure as the monoid structure on the objects.
Hence it satisfies eq.(\ref{eq:tensoridentity}). By 
\[
(1_A\circ 1_A)\otimes (1_B\circ 1_B)=1_A\otimes 1_B=1_{A\otimes B}=
1_{A\otimes B}\circ 1_{A\otimes B}=(1_A\otimes 1_B)\circ(1_A\otimes
1_B) 
\] 
eq.(\ref{food_bifunct2}) is also satisfied.

\end{example}

\subsection{Categories in physics}\label{Cats_physics}

In the previous section, we saw how groups and partial orders, both of massive
importance for physics, are themselves abstract categorical structures.
\bit 
\item While there is no need to argue for
the importance of group theory to physics here, it is worth mentioning that
John Slater (cf.~Slater determinant in quantum chemistry) referred to Weyl,
Wigner and others' use of group theory in quantum physics as \em der
Gruppenpest\em, what translates as the `plague of groups'. Even in 1975 he wrote:
As soon as [my] paper became known, it was obvious that a great many
other physicists were as \underline{disgusted} as I had been with the
group-theoretical approach to the problem.  As I heard later, there were
remarks made such as `Slater has slain the Gruppenpest'. I believe that no
other piece of work I have done was so universally popular.''  
Similarly, we may wonder whether it are the category theoreticians or their opponents which are the true aliens.
\item 
Partial orders model spatio-temporal causal structure \cite{Penrose72,Sorkin91}.
Roughly speaking, if $a\leq b$ then events $a$ and $b$ are causally related, if
$a< b$ then they are time-like separated, and if $a$ and $b$ don't compare then
they are space-like separated.  This theme is discussed in great detail in
\cite{MartinPanangaden}.  
\item 
The degree of bipartite quantum entanglement gives rise to a preorder on bipartite quantum states
\cite{Nielsen}.  The relevant preorder is Muirheads' majorization order
\cite{Muirhead}.  However,  multipartite quantum
entanglement and mixed state quantum entanglement are not well understood yet. We strongly believe that category theory provides the key to the
solution, in the following sense: 
\[ 
{{\mbox{\rm bipartite entanglement}} \over{\mbox{\rm some preorder}}} = {{\mbox{\rm
multipartite entanglement}} \over{\mbox{\rm some thick category}}} 
\]
\eit 
We also acknowledge the use of category theory in several involved
subjects in mathematical physics ranging from topological quantum field
theories (TQFTs) to proposals for a theory of quantum gravity; here the motivation to use category theory is of a mathematical nature.  We discuss one such topic, namely TQFT, in
Section \ref{sec:TQFT}.


But the particular perspective
which we would like to promote here is \em categories as physical theories\em.
Above we discussed three kinds of categories: 
\bit 
\item \em Concrete categories \em have mathematical structures as objects, and structure
preserving maps between these as morphisms.  
\item \em Real world categories \em have some notion of system as objects, and corresponding processes thereoff as morphisms.  
\item \em Abstract categorical structures \em are mathematical structures in their own right; they are defined in terms of  additional structure and/or
certain properties.  
\eit 
The real world categories constitute the \em area \em of our focus (e.g.~quantum physics, proof theory, computation,
organic chemistry, ...), the concrete categories constitute the formal
mathematical \em models \em for these (e.g., in the case of quantum physics,
Hilbert spaces as objects, certain types of linear maps as morphisms, and the
tensor product as the monoidal structure), while the abstract categorical
structures constitute \em axiomatisations \em of these.  

The latter is the obvious place to start when one is interested in comparing
theories. We can study which axioms and/or structural properties give rise to certain physical phenomena,  for example, which tensor structures give rise to teleportation (e.g.~\cite{AC2004}), or to non-local
quantum-like behavior \cite{BES}.   Or, we can study which
structural features distinguish  classical from quantum theories
(e.g.~\cite{CPav2006,CPP2008}).  

Quantum theory is subject to the so-called
No-Cloning, No-Deleting and No-Broadcasting theorems \cite{Broadcast, Pati,
WZ}, which impose key constraints on our capabilities to process quantum states.
Expressing these clearly requires a formalism that allows to vary types from a
single to multiple systems, as well as one which explicitly  accommodates processes
(cf.~copying/deleting process).  Monoidal categories provide the appropriate
mathematical arena for this on-the-nose.





\begin{example} Why does a tiger have stripes and a lion doesn't?
One might expect that the explanation  is written within the fundamental building
blocks which these animals are made up from, so one could take a big knife and open the lion's and the tiger's bellies.  One finds intestines, but these are the same for both animals.  So maybe the answer is hidden in even smaller constituents.   With a tiny knife we keep cutting and identify a smaller kind of building block, namely the cell.  Again, there is no obvious difference between tigers and lions at this level.  So we need to go even smaller.  After a century of advancing `small knife technology' we  discover DNA and this constituent truly reveals the difference.  So yes, now we know why tigers have stripes and lions don't!  Do we really? No, of course not.    Following in the footsteps of Charles Darwin, your favorite nature channel would tell you that the explanation is given by a process of type
\[ 
prey\otimes predator\otimes environment\rTo^{} 
dead\ prey\otimes eating\ predator 
\] 
which represents the successful challenge of  a predator, operating within some environment, on some prey. Key to  the success of such a  challenge is the predator's camouflage. Sandy savanna is the lion's habitat while forests constitute the tiger's habitat, so their respective  coat blends them within their natural habitat.  Any \mbox{(neo-)}Darwinist biologist will tell you that the fact that this is encoded in the animal's DNA is not a cause, but rather a consequence, via the process of natural selection.

This example
illustrates how monoidal categories enable to shift  the focus from an
\em atomistic \em or \em reductionist \em attitude to one where systems are studied in terms of their interactions with other systems, rather than in terms of their constituents.  Clearly, in recent history, physics has solely focused on chopping down things into smaller things.  Focussing on interactions might provide us  with a complementary  understanding of the fundamental theories of nature.
\end{example}

\subsection{Structure preserving maps for categories}\label{Sec:Functors} 


The notion of structure preserving map between categories ---  which we referred to
in Example \ref{excat4} --- wasn't made explicit yet.  These `maps which preserve
categorical structure', the  so-called  \em functors\em,  must preserve the \em
structure of a category\em, that is, \em composition \em and \em identities\em.   An
example of a functor that might be known to the reader because of its
applications in physics, is the linear representation of a group. A
representation of a group $G$ on a vector space $V$ is a group homomorphism
from $G$ to $\mbox{GL}(V)$, the general linear group on $V$, i.e., a map
$\rho: G\rightarrow\mbox{GL}(V)$ such that 
\[ 
\rho(g_1\bullet g_2)=\rho(g_1)\circ\rho(g_2)\quad \mbox{for all}\quad g_1,g_2\in G\,,
\quad\mbox{and}\,, \quad\rho(1)=1_V\,. 
\]
Consider $G$ as a category ${\bf G}$ as in Example \ref{excat8}.  We also have
that $\mbox{GL}(V)\subset\mathbf{FdVect}_\mathbb{K}(V,V)$ (cf.~Example
\ref{excat2}).  Hence,  a group representation $\rho$ from $G$ to
$\mbox{GL}(V)$ induces `something' from ${\bf G}$ to
$\mathbf{FdVect}_\mathbb{K}$: 
\[ 
\rho:G\stackrel{}{\rightarrow}\mbox{GL}(V)\ \ \ \ \leadsto\ \ \ \ {\bf
G}\stackrel{R_\rho}{\longrightarrow}\mathbf{FdVect}_\mathbb{K}\,.  
\] 
However, specifying ${\bf G}\stackrel{R_\rho}{\longrightarrow}\mathbf{FdVect}_\mathbb{K}$ requires some care: 
\bit 
\item
Firstly, we need to specify that we are representing on the general linear
group of the vector space $V\in\mathbf{FdVect}_\mathbb{K}$.  We do this by
mapping the unique object $*$ of ${\bf G}$ on $V$, thus defining  a map from
objects to objects 
\[ 
R_\rho:|{\bf G}|\rightarrow |\mathbf{FdVect}_\mathbb{K}|:: *\mapsto V\,.  
\] 
\item Secondly, we need to specify to which 
linear map in 
\[
\mbox{GL}(R_\rho(*))\subset {\bf FdVect}_\mathbb{K}(R_\rho(*),R_\rho(*)) 
\]
a group element
\[
g\in{\bf G}(*,*)=G
\]
is mapped.  This defines a map from a hom-set to
a hom-set, namely
\[ 
R_\rho:{\bf G}(*,*)\rightarrow \mathbf{FdVect}_\mathbb{K}(R_\rho(*),R_\rho(*))::g\mapsto \rho(g)\,. 
\] 
The fact that $\rho$ is a group homomorphism implies in our category-theoretic context that $R_\rho$ preserves composition of morphisms as well as identities, that is,  $R_\rho$ preserves the categorical structure.
\eit 
Having this example in mind, we infer that a
functor must consist not of a single but of \em two \em kinds of mappings: one
map on the objects, and a family of maps on the hom-sets which preserve
identities and composition. 

\begin{definition} \em Let $\cat$ and ${\bf D}$ be categories. A {\em functor}
\[
F:\cat\longrightarrow{\bf D}
\]
consists of: \begin{enumerate} \item A
mapping 
\[ 
F: |\cat|\rightarrow |{\bf D}|::A\mapsto
F(A)\,; 
\] 
\item For any $A,B\in |\cat|$, a mapping 
\[
F:\cat(A,B)\rightarrow{\bf D}(F(A),F(B))::f\mapsto F(f)
\] 
which preserves identities and composition, i.e.,
\begin{itemize} 
\item[i.] for any $f\in\cat(A,B)$ and $g\in\cat(B,C)$ we have 
\[ 
F(g\circ
f)=F(g)\circ F(f)\,, 
\] 
\item[ii.] and, for any $A\in|\cat|$ we have 
\[
F(1_A)=1_{F(A)}\,.  
\] 
\end{itemize}
\end{enumerate}
\end{definition}

Typically one drops the parentheses unless they are necessary. For instance,
$F(A)$ and $F(f)$ will be denoted simply as $FA$ and $Ff$.

Consider the category ${\bf PhysProc}$ of Example \ref{excat11}  and a concrete
category ${\bf Mod}$ in which we wish to model these
mathematically by assigning to each process a morphism in the concrete
category ${\bf Mod}$. Functoriality of 
\[ 
F:{\bf PhysProc}\longrightarrow{\bf Mod} 
\] means
that sequential composition of physical processes is mapped on composition of
morphisms in ${\bf Mod}$, and that void processes are mapped on the identity
morphisms.  From this, we see that functoriality is an obvious requirement when designing
mathematical models for physical processes.

\begin{example}\label{excat2_5} Define the category $\mathbf{Mat}_\mathbb{K}$
with 
\begin{enumerate}
\item the set of natural numbers $\mathbb{N}$ as objects, 
\item all  $m\times n$-matrices with entries
in $\mathbb{K}$ as morphisms of type $n\rTo m$, and
\item matrix composition, and identity matrices.
\end{enumerate} 
This example is closely related to Example \ref{excat2}.  
However, it strongly emphasizes that objects are but
labels with no internal structure.  Strictly speaking this is not a
concrete category in the sense of Section \ref{sec:concretecategories}.
However, for all practical purposes, it can serve as well as a model as
any other concrete category.  Therefore, we can relax our conception of concrete
categories to accommodate such models.  

Assume now that for each
vector space $V\in|\mathbf{FdVect}_\mathbb{K}|$, we pick a fixed basis.
Then any linear function $f\in\mathbf{FdVect}_\mathbb{K}(V,W)$ admits a
matrix in these bases.  This `assigning of matrices' to linear maps is
described by the functor 
\[
F:\mathbf{FdVect}_\mathbb{K}\longrightarrow\mathbf{Mat}_\mathbb{K} 
\] 
which maps vector spaces  on their respective dimension,  and which maps linear maps on their matrices
in the chosen bases. Importantly, note that it is the functor $F$ which encodes the choices of bases, and not the categorical structure of $\mathbf{FdVect}_\mathbb{K}$.
\end{example}

\begin{example}\label{excat2_7} In $\mathbf{Mat}_\mathbb{C}$, if we map each
natural number on itself and conjugate all the entries of each matrix
we also obtain a functor. 
\end{example}

We now introduce the concept of \em
duality \em which we already hinted at above. Simply put, it means reversal of the
arrows in a given category $\cat$. We illustrate this notion in term of an example.  Transposition of matrices, just like a functor, is a mapping on both objects and  morphisms which: 
\begin{itemize} 
\item[i.] preserves objects and  identities, 
\item[ii.] reverses the direction of the morphisms since  when the matrix  $M$ has type $n\rTo m$, then the matrix  $M^T$ has type $m\rTo n$, and
\item[iii.] preserves the composition `up to this
reversal of the arrows', i.e.~for any pair of matrices  $N$ and $M$ for which types match we have 
\[ 
(N\circ M)^T=M^T\circ N^T\,. 
\]
\end{itemize} 
So transposition is a functor {\em up to reversal of the arrows}.   

\begin{definition} \em 
A {\em contravariant functor} $F:\cat\rightarrow{\bf D}$
consists of the same data as a functor, it also preserves identities, but
reverses composition that is: 
\[ 
F(g\circ f)=Ff\circ Fg\,, 
\]
\end{definition}

In contrast to contravariant functors, ordinary functors are often referred to as {\em covariant functors}. 

\begin{definition} \em 
The {\em opposite category} $\cat^{\scriptsize \mbox{\em
op}}$ of a category $\cat$ is the category with
\begin{itemize}
\item the same objects as $\cat$, 
\item in which morphisms are `reversed', that is,  
\[
f\in\cat(A,B)
 \Leftrightarrow  f\in\cat^{\scriptsize \mbox{\em op}}(B,A)\,,
 \] 
where to avoid confusion from now on we denote $f\in\cat^{\scriptsize \mbox{\em
op}}(B,A)$ by $f^{\scriptsize \mbox{\em op}}$,
\item identities in $\cat^{\scriptsize \mbox{\em op}}$ are those of $\cat$, and 
\[
f^{\scriptsize \mbox{\em op}}\circ g^{\scriptsize \mbox{\em op}}=(g\circ f)^{\scriptsize
\mbox{\em op}}\,.
\]  
\end{itemize}
\end{definition}

Contravariant functors of type $\cat\rightarrow{\bf D}$ can now be defined as
functors of type $\cat^{\scriptsize \mbox{\em op}}\rightarrow{\bf D}$.  
Of course, the operation $(-)^{\scriptsize \mbox{\em op}}$ on categories
is involutive: reversing the arrows twice is the same as doing nothing.   The
process of reversing the arrow is sometimes indicated by the prefix `co', 
indicating that the defining equations for those structures are the same as
the defining equations for the original structure, 
but with arrows reversed.  

\begin{example}\label{ex:transpose} The \em transpose \em is the involutive
contravariant functor 
\[ 
T:\mathbf{FdVect}_\mathbb{K}^{\scriptsize \mbox{\em op}}\rightarrow\mathbf{FdVect}_\mathbb{K} 
\] 
which maps each vector space on the corresponding dual vector space, and which maps each
linear map $f$ on its transpose $f^T$. 
\end{example}


\begin{example}\label{ex:dagger} 
A Hilbert space is a vector space over
$\mathbb{C}$ with an  inner-product
\[ 
\langle-,-\rangle:{\cal H}\times{\cal H}\to\mathbb{C}\,.
\] 
Let ${\bf FdHilb}$ be the category with finite
dimensional Hilbert spaces as objects and with  linear maps as morphisms.  
Of course, one could define other categories with Hilbert
spaces as objects, for example, the groupoid ${\bf FdUnit}$ of Example \ref{excat8}.  But as we will see below in Section \ref{sec:Dirac}, the category ${\bf FdHilb}$ as defined here comes with enough extra structure to extract all unitary maps from it.  Hence,  ${\bf FdHilb}$ subsumes ${\bf FdUnit}$. This 
extra structure comes as a functor, whose action is \em taking the adjoint \em or \em
hermitian transpose\em.  This is the contravariant functor 
\[
\dagger:{\bf FdHilb}^{\scriptsize \mbox{\em op}}\longrightarrow{\bf FdHilb}
\] 
which: 
\begin{enumerate} 
\item is \em identity-on-object\em, that is, 
\[ 
\dagger:|{\bf FdHilb}^{\scriptsize \mbox{\em op}}|\rightarrow|{\bf FdHilb}|::{\cal H}\mapsto {\cal H}\,, 
\]
\item and assigns morphisms to their adjoints, that is, 
\[ 
\dagger:{\bf FdHilb}^{\scriptsize \mbox{\em op}}({\cal H},{\cal K})\to {\bf FdHilb}({\cal K},{\cal H}):: f\mapsto f^\dagger\,.  
\] 
\end{enumerate} 
Since for $f\in{\bf FdHilb}({\cal H},{\cal K})$ and $g\in{\bf FdHilb}({\cal K},{\cal L})$
we have: 
\[ 
1^\dagger_{\cal H}=1_{\cal H}\qquad\mbox{\rm and}\qquad
(g\circ f)^\dagger=f^\dagger\circ g^\dagger 
\] 
we indeed obtain an identity-on-object contravariant functor. This functor is moreover \em involutive\em, that is, for all morphisms $f$ we have 
\[ 
f^{\dagger\dagger}=f\,.  
\] 
While the morphisms of ${\bf FdHilb}$ do
not reflect the inner-product structure, the latter is required to
specify the adjoint.  In turn, this adjoint will allow us to recover the inner-product in purely
category-theoretic terms, as we shall see in Section
\ref{sec:Dirac}. 
\end{example}

\begin{example}\label{excat5} Define the category ${\bf Funct}_{{\bf C},{\bf D}}$ with 
\begin{enumerate} 
\item all functors from ${\bf C}$ to ${\bf
D}$ as objects, 
\item  \em natural transformations \em between these as morphisms (cf.~Section
\ref{sec:naturaltransformations}), and, 
\item
composition of natural transformations and
corresponding identities.  
\end{enumerate}
\end{example}

\begin{example}\label{exsmcb} The defining equations of strict monoidal categories, that is,
\beq\label{def:bifunceq} (g\circ f)\otimes(k\circ h)=(g\otimes
k)\circ(f\otimes h)\quad\mbox{\rm and} \quad 1_A\otimes 1_B=1_{A\otimes
B}\,, 
\eeq 
to which we from now on refer as \em bifunctoriality\em, is nothing but
functoriality of a certain functor.   We will discuss this in detail in Section
\ref{sec:BifunctAsFunct}. 
\end{example}

\begin{example}\label{ConcreteCategory}
A \em concrete category\em, or even better, a ${\bf Set}$-concrete category, is a category ${\bf C}$ together 
with a functor $U: {\bf C}\longrightarrow{\bf Set}$.  The way in which we construct this functor for categories with mathematical structures as objects is by sending each object to the underlying set, and morphisms to the underlying functions.  So we forget the extra structure the object has.  Therefore the functor $U$ is typically called \em forgetful\em.  For example, the category ${\bf Grp}$ is a concrete category for the functor
\[
U: {\bf Grp}\longrightarrow{\bf Set}::\left\{\begin{array}{l} 
(G,\bullet, 1)\mapsto G\vspace{1.5mm}\\
f \mapsto f
\end{array}\right.
\]
which `forgets' the group's multiplication and unit, and morphisms are mapped on their underlying functions.  More generally, a ${\bf D}$-concrete category is a category ${\bf C}$ with a functor $U: {\bf C}\longrightarrow{\bf D}$.
\end{example}

\begin{example} The TQFTs of Section \ref{sec:TQFT} are special kinds of
functors.  \end{example}

\section{The 2D case: Muscle power}\label{muscle}

We now genuinely start to study the interaction of the parallel and the sequential modes of composing systems, and operations thereon.

\subsection{Strict symmetric monoidal categories}

The starting point of this Section is the notion of a strict monoidal category as given in
Definition \ref{excat10}.  Such categories enable us to give formal meaning to physical
processes which involve several types, e.g.~classical and quantum as the
following example clearly demonstrates. 

\begin{example}\label{excatCQOpp} Define ${\bf CQOpp}$ to be the strict
monoidal category containing both classical and quantum systems, with
operations thereon as morphisms, and with the obvious notion of monoidal
tensor, that is, a physical analogue of the tensor for vegetables that we saw in the prologue.  Concretely, by $A\otimes B$ we mean that we have \em both $A$ and $B$ \em available to operate on.  Note in particular that at this stage of the discussion there are no Hilbert spaces involved, so $\otimes$ cannot stand for the tensor product, but this does not exclude that we may want to model it by the tensor product at a later stage.  In this category, non-destructive (projective) measurements
have type 
\[
Q\rTo X\otimes Q
\]
where $Q$ is a quantum system and $X$ is
the classical data produced by the measurement.  Obviously, the hom-sets
\[
{\bf CQOpp}(Q,Q)\qquad \mbox{and} \qquad {\bf CQOpp}(X,X)
\]
have a very different
structure since ${\bf CQOpp}(Q,Q)$ stands for the operations we can
perform on a quantum system while ${\bf CQOpp}(X,X)$ stands for the
classical operations (e.g.~classical computations) which we can perform
on classical systems.  But all of these now live within a single
mathematical entity ${\bf CQOpp}$.  \end{example}

The structure of a strict monoidal category does not yet capture certain
important properties of cooking with vegetables.  Denote the strict monoidal
category constructed in the Prologue by ${\bf Cook}$.  

Clearly `boil the potato while fry the carrot' is very much the same thing as
`fry the carrot while  boil the potato'. But we cannot just bluntly say that in the category ${\bf Cook}$ the equality 
\[
h\otimes f=f\otimes h
\]
holds.  By plain set theory, for this equality to be meaningful, the two morphisms $h\otimes f$ and $f\otimes h$ need to live in the same set. That is, respecting the structure of a category, within the same
hom-set.  So  
\[
A\otimes D\rTo^{f\otimes h}B\otimes F\qquad \mbox{and}\qquad D\otimes
A\rTo^{h\otimes f}F\otimes B
\] 
need to have the same type, which implies that
\beq\label{eq:degentypes}
A\otimes D=D\otimes A \qquad \mbox{and}\qquad B\otimes F=F\otimes B
\eeq
must hold. 
But this completely blurs the distinction between a carrot and a potato.  For example, we cannot distinguish anymore between `boil the potato while fry the carrot', which we denoted by 
\[ 
A\otimes D\rTo^{f\otimes h}B\otimes F\,, 
\] 
and  `fry the potato while boil the carrot', which given eqs.(\ref{eq:degentypes}), we can
write as 
\[ 
A\otimes D=D\otimes A\rTo^{h\otimes f}F\otimes B=B\otimes F\,. 
\]
So we basically threw out the child with the bath water.

The solution to this problem 
is to introduce  an operation 
\[ 
\sigma_{A,D}: A\otimes D\rTo D\otimes A
\]
which \em swaps \em the role of the potato and the carrot relative to
the monoidal tensor. The fact that `boil the potato  while fry the carrot' is
essentially the same thing as `fry the carrot while boil the potato' can now be
expressed as 
\[ 
\sigma_{B,F}\circ(f\otimes h)=(h\otimes f)\circ\sigma_{A,D}\,.
\] 
In our `real world example' of cooking this operation can be
interpreted as physically swapping the vegetables \cite{C2005d}.  An equational law governing `swapping' is:
\[
\sigma_{B,A}\circ\sigma_{A,B}=1_{A\otimes B}\,.
\]

\begin{definition}\label{excat10_5}\em 
A \em strict symmetric monoidal category
\em is a strict monoidal category ${\bf C}$ which  moreover comes with a family
of isomorphisms 
\[ 
\left\{A\otimes B\rTo^{\sigma_{A,B}} B\otimes A\Bigm| A,B\in|{\bf C}|\right\}
\] 
called {\em symmetries}, and which are such that:
\begin{itemize}
\item for all $A,B\in|{\bf C}|$  we have $\sigma_{A,B}^{-1}=\sigma_{B,A}$,  and 
\item for all $A,B,C,D\in|{\bf C}|$ and all $f,g$ of appropriate type we have
\beq\label{eq:symmetry} 
\sigma_{C,D}\circ(f\otimes g)=(g\otimes f)\circ\sigma_{A,B}\,.  
\eeq 
\end{itemize}
\end{definition}

All Examples of Section \ref{sec:real_world} are strict symmetric monoidal
categories for the obvious notion of symmetry in terms of `swapping'.  

We can rewrite eq.(\ref{eq:symmetry}) in a form which makes the types explicit: 
\beq\label{Eq:symdiag} 
\xymatrix@=0.6in{%
A\otimes B\ar[d]_{f\otimes g}\ar[r]^{\sigma_{A,B}} & B\otimes A\ar[d]^{g\otimes f}\\%
C\otimes D\ar[r]_{\sigma_{C,D}} & D\otimes C} 
\eeq 
This representation is referred to as \em commutative diagrams\em.

\begin{proposition}\label{Prop:Bifinct} In any strict monoidal category we have
\beq\label{Eq:bifunctdiag} 
\xymatrix@=0.6in{%
A\otimes
B\ar[d]_{f\otimes 1_B}\ar[r]^{1_A\otimes g} & A\otimes
D\ar[d]^{f\otimes 1_D}\\%
C\otimes B\ar[r]_{1_C\otimes g} & C\otimes D}
\eeq 
\end{proposition}
Indeed, relying on bifunctoriality we have:
\beqa 
(f\otimes 1_D)\circ(1_A\otimes g)=(f\circ 1_A)\otimes(1_D\circ
g)\hspace{-1.95cm}&&\\ &||&\\ &f\otimes g&\\ &||&\\
&&\hspace{-1.95cm}(1_C\circ f)\otimes(g\circ 1_B)=(1_C\otimes
g)\circ(f\otimes 1_B)\,.  
\eeqa 
The reader can easily verify that,
given a connective $-\otimes-$ defined both on objects and morphisms as in
items 1 \& 2 of Definition \ref{excat10}, the  four equations	
\beq\label{eq:twice_functorial_pre} 
(f\circ 1_A)\otimes(1_D\circ g)=f\otimes g=(1_B\circ f)\otimes(g\circ 1_C)
\eeq
\beq\label{eq:twice_functorial} 
(g\otimes 1_B)\circ(f\otimes 1_B)=(g\circ f)\otimes 1_B
\eeq
\beq\label{eq:twice_functorialBis} 
\ (1_A\otimes g)\circ(1_A\otimes f)=1_A \otimes (g\circ f)\,, 
\eeq 
when varying over all objects $A,B,C,D\in|{\bf C}|$ and all  morphisms $f$ and $g$ of appropriate
type, are equivalent to the single equation
\beq\label{eq:twice_functorial_post} 
(g\circ f)\otimes(k\circ h)=(g\otimes k)\circ(f\otimes h)\, 
\eeq 
when varying over $f,g,h,k$. Eqs.(\ref{eq:twice_functorial},\ref{eq:twice_functorialBis}) together with 
\[
1_A\otimes 1_B=1_{A\otimes B}
\]
is usually referred to as $-\otimes-$ being \em functorial in both arguments\em. They are indeed equivalent to the mappings on objects and morphisms   
\[ 
(1_A\otimes -):{\bf C}\longrightarrow {\bf C}
\quad\qquad\mbox{\rm and}\qquad\quad
(-\otimes 1_B):{\bf C}\longrightarrow {\bf C}
\] 
both being functors, for all objects $A,B\in|{\bf C}|$ --- their action on objects  is   
\[
(1_A\otimes -)::X\mapsto A\otimes X
\qquad\mbox{\rm and}\qquad
(-\otimes 1_B)::X\mapsto X\otimes B\,. 
\]
Hence, functoriality in both arguments is strictly weaker than bifunctoriality (cf.~Example~\ref{exsmcb}), since the latter also requires eqs(\ref{eq:twice_functorial_pre}). 

\subsection{Graphical calculus for symmetric monoidal categories}\label{sec:graphCalc}

The most attractive, and at the same time, also the most powerful feature of
strict symmetric monoidal categories, is that they admit a purely diagrammatic
calculus.  Such a graphical language is subject to the following characteristics:
\bit 
\item The symbolic ingredients in the definition  of strict symmetric
monoidal structure, e.g.~$\otimes$, $\circ$, $A$, $\II$, $f$ etc., or any other abstract categorical structure which
refines it,  all have a purely diagrammatic counterpart\,; 
\item The corresponding axioms become very intuitive graphical manipulations\,; 
\item And crucially, an equational statement is derivable in the graphical language \underline{if and
only if} it is symbolically derivable from the axioms of the theory.  
\eit
For a more formal presentation of what we precisely mean by a graphical calculus we
refer the reader to Peter Selinger's marvelous paper \cite{Selinger} in these volumes.

These diagrammatic calculi trace back to Penrose's work in the early 1970s, and
have been given rigorous formal treatments in \cite{FY, JS, JSV, SelingerPre}.
Some examples of possible  elaborations and corresponding applications of the
graphical language presented in this paper are in \cite{CP2006, CPP2008,
CoeckeDuncan, Lauda,Selinger, StreetBook,Vicary, YetterBook}. 

The graphical counterparts to the axioms are typically much simpler then their
formal counterparts. For example, in the Prologue we mentioned that
bifunctoriality becomes a tautology in this context. Therefore such a graphical language
radically simplifies algebraic manipulations, and in many cases trivialises
something very complicated.  Also the physical interpretation of the axioms,
something which is dear to the authors of this paper, becomes very direct.  

The graphical counterparts to  strict symmetric monoidal structure are:
\begin{itemize} 
\item[-] The {\em identity} $1_\II$ is the empty picture (= it is not depicted).  
\item[-] The {\em identity} $1_A$ for and object $A$ different of $\II$ is depicted as 
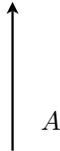
\begin{center} 
\ifx\JPicScale\undefined\def\JPicScale{1}\fi
\psset{unit=\JPicScale mm}
\psset{linewidth=0.3,dotsep=1,hatchwidth=0.3,hatchsep=1.5,shadowsize=1,dimen=middle}
\psset{dotsize=0.7 2.5,dotscale=1 1,fillcolor=black}
\psset{arrowsize=1 2,arrowlength=1,arrowinset=0.25,tbarsize=0.7 5,bracketlength=0.15,rbracketlength=0.15}
\begin{pspicture}(0,0)(11,19.8)
\psline{->}(5.9,0)(5.9,19.8)
\rput(11,4){$A$}
\end{pspicture}

\end{center} 
\item[-] A {\em morphism} $f:A\rTo B$ is
depicted as 
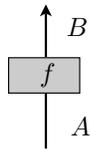
\begin{center} 
\ifx\JPicScale\undefined\def\JPicScale{1}\fi
\psset{unit=\JPicScale mm}
\psset{linewidth=0.3,dotsep=1,hatchwidth=0.3,hatchsep=1.5,shadowsize=1,dimen=middle}
\psset{dotsize=0.7 2.5,dotscale=1 1,fillcolor=black}
\psset{arrowsize=1 2,arrowlength=1,arrowinset=0.25,tbarsize=0.7 5,bracketlength=0.15,rbracketlength=0.15}
\begin{pspicture}(0,0)(10.6,19.2)
\psline(6,0)(6,7.5)
\rput(10.6,2.9){$A$}
\newrgbcolor{userFillColour}{0.8 0.8 0.8}
\pspolygon[linewidth=0.15,fillcolor=userFillColour,fillstyle=solid](1.05,7.4)(10.5,7.4)(10.5,12.1)(1.05,12.1)
\rput(10.2,16.2){$B$}
\rput(6.25,9.7){$f$}
\psline{->}(6,12.2)(6,19.2)
\end{pspicture}
 
\end{center}
\item[-] The {\em composition} of morphisms $f:A\rTo B$
and $g:B\rTo C$ is depicted by locating $g$
above $f$ and by connecting the output of $f$ to the
input of $g$, i.e.
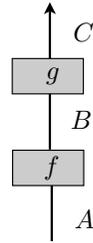
\begin{center} 
\ifx\JPicScale\undefined\def\JPicScale{1}\fi
\psset{unit=\JPicScale mm}
\psset{linewidth=0.3,dotsep=1,hatchwidth=0.3,hatchsep=1.5,shadowsize=1,dimen=middle}
\psset{dotsize=0.7 2.5,dotscale=1 1,fillcolor=black}
\psset{arrowsize=1 2,arrowlength=1,arrowinset=0.25,tbarsize=0.7 5,bracketlength=0.15,rbracketlength=0.15}
\begin{pspicture}(0,0)(10.6,31.8)
\psline(6.25,0)(6.25,7.5)
\rput(10.6,2.9){$A$}
\newrgbcolor{userFillColour}{0.8 0.8 0.8}
\pspolygon[linewidth=0.15,fillcolor=userFillColour,fillstyle=solid](1.05,7.4)(10.5,7.4)(10.5,12.1)(1.05,12.1)
\rput(10.2,16.2){$B$}
\rput(6.25,9.7){$f$}
\psline(6,12.2)(6,19.8)
\newrgbcolor{userFillColour}{0.8 0.8 0.8}
\pspolygon[linewidth=0.15,fillcolor=userFillColour,fillstyle=solid](1,19.6)(10.45,19.6)(10.45,24.3)(1,24.3)
\psline{->}(6,24.3)(6,31.8)
\rput(6.4,22){$g$}
\rput(10.5,27.8){$C$}
\end{pspicture}

\end{center} 
\item[-] The {\em tensor product} of
morphisms $f:A\rTo B$ and $g:C\rTo D$ is
depicted by aligning the graphical representation of
$f$ and $g$ side by side in the order they occur
within the expression $f\otimes g$, i.e.  
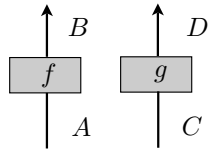
\begin{center}
\ifx\JPicScale\undefined\def\JPicScale{1}\fi
\psset{unit=\JPicScale mm}
\psset{linewidth=0.3,dotsep=1,hatchwidth=0.3,hatchsep=1.5,shadowsize=1,dimen=middle}
\psset{dotsize=0.7 2.5,dotscale=1 1,fillcolor=black}
\psset{arrowsize=1 2,arrowlength=1,arrowinset=0.25,tbarsize=0.7 5,bracketlength=0.15,rbracketlength=0.15}
\begin{pspicture}(0,0)(26,19.3)
\psline(6,0)(6,7.6)
\rput(10.6,2.9){$A$}
\newrgbcolor{userFillColour}{0.8 0.8 0.8}
\pspolygon[linewidth=0.15,fillcolor=userFillColour,fillstyle=solid](1.05,7.4)(10.5,7.4)(10.5,12.1)(1.05,12.1)
\rput(10.2,16.2){$B$}
\rput(6.25,9.7){$f$}
\psline{->}(6,12.2)(6,19.2)
\newrgbcolor{userFillColour}{0.8 0.8 0.8}
\pspolygon[linewidth=0.15,fillcolor=userFillColour,fillstyle=solid](15.8,7.5)(25.25,7.5)(25.25,12.2)(15.8,12.2)
\psline{->}(20.75,12.3)(20.75,19.3)
\rput(21.2,9.9){$g$}
\rput(25.3,3){$C$}
\psline(20.8,0.1)(20.8,7.6)
\rput(26,16.1){$D$}
\end{pspicture}
 
\end{center}
\item[-] {\em Symmetry} 
\[
\sigma_{AB}: A\otimes B\rTo
B\otimes A
\]
is depicted as 
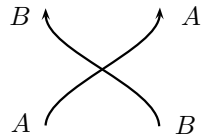
\begin{center}
\psset{xunit=1mm,yunit=1mm,runit=1mm}
\begin{pspicture}(0,0)(26.40,16.70)
\rput(3.80,1.30){$A$}
\psbezier[linewidth=0.30,linecolor=black]{->}(7.10,1.00)(7.30,6.00)(22.10,11.10)(22.30,16.70)
\rput(26.40,15.60){$A$}
\psbezier[linewidth=0.30,linecolor=black]{->}(22.20,0.90)(22.00,6.20)(7.20,10.90)(7.10,16.60)
\rput(25.70,1.10){$B$}
\rput(3.70,15.60){$B$}
\end{pspicture}

\end{center}

\item[-] Morphisms 
\[
\psi:\II\rTo A\ \ \ , \ \ \ \phi:A\rTo \II\ \quad  \mbox{and}\quad \ s: \II\rTo \II
\]
are respectively depicted as 
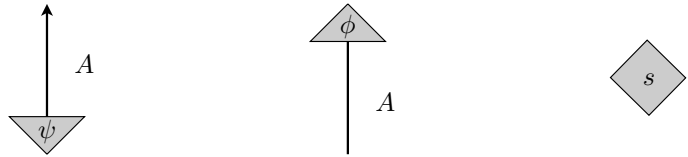
\begin{center}
\ifx\JPicScale\undefined\def\JPicScale{1}\fi
\psset{unit=\JPicScale mm}
\psset{linewidth=0.3,dotsep=1,hatchwidth=0.3,hatchsep=1.5,shadowsize=1,dimen=middle}
\psset{dotsize=0.7 2.5,dotscale=1 1,fillcolor=black}
\psset{arrowsize=1 2,arrowlength=1,arrowinset=0.25,tbarsize=0.7 5,bracketlength=0.15,rbracketlength=0.15}
\begin{pspicture}(0,0)(89.91,20)
\newrgbcolor{userFillColour}{0.8 0.8 0.8}
\psline[linewidth=0.15,fillcolor=userFillColour,fillstyle=solid](5,0)
(10,5)
(0,5)(5,0)
\psline{->}(5,5)(5,20)
\rput(10,12){$A$}
\newrgbcolor{userFillColour}{0.8 0.8 0.8}
\pspolygon[linewidth=0.15,fillcolor=userFillColour,fillstyle=solid](89.91,10.16)(85.03,5.18)(79.91,10.2)(84.79,15.18)
\rput(85,10){$s$}
\rput(5,3){$\psi$}
\psline(45,0)(45,15)
\rput(50,7){$A$}
\newrgbcolor{userFillColour}{0.8 0.8 0.8}
\psline[linewidth=0.15,fillcolor=userFillColour,fillstyle=solid](45,20)
(50,15)
(40,15)(45,20)
\rput(45,17){$\phi$}
\end{pspicture}
 
\end{center} 
\end{itemize}

\noindent The diamond shape of the morphisms of type $\II\rTo\II$ indicates that
they arise when composing two triangles:

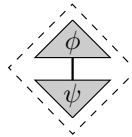
\begin{center} 
\ifx\JPicScale\undefined\def\JPicScale{1}\fi
\psset{unit=\JPicScale mm}
\psset{linewidth=0.3,dotsep=1,hatchwidth=0.3,hatchsep=1.5,shadowsize=1,dimen=middle}
\psset{dotsize=0.7 2.5,dotscale=1 1,fillcolor=black}
\psset{arrowsize=1 2,arrowlength=1,arrowinset=0.25,tbarsize=0.7 5,bracketlength=0.15,rbracketlength=0.15}
\begin{pspicture}(0,0)(17.18,17.5)
\newrgbcolor{userFillColour}{0.8 0.8 0.8}
\psline[linewidth=0.15,fillcolor=userFillColour,fillstyle=solid](8.7,2.4)
(13.7,7.4)
(3.7,7.4)(8.7,2.4)
\psline(8.7,7.4)(8.7,12.4)
\rput(8.7,5.4){$\psi$}
\newrgbcolor{userFillColour}{0.8 0.8 0.8}
\psline[linewidth=0.15,fillcolor=userFillColour,fillstyle=solid](8.7,15.4)
(13.7,10.4)
(3.7,10.4)(8.7,15.4)
\rput(8.7,12.4){$\phi$}
\pspolygon[linewidth=0.15,linestyle=dashed,dash=1 1](17.18,8.66)(8.72,0.22)(0.2,9.06)(8.66,17.5)
\end{pspicture}
 
\end{center}

\begin{example}\label{ex:stateseffects} 
In the category ${\bf QuantOpp}$ the triangles of respective types $\II\rTo A$ and $A\rTo \II$
represent states and effects, and the diamonds of type $\II\rTo\II$ can
be interpreted as probabilistic weights: they give the likeliness of a
certain effect to occur when the system is in a certain state. In the usual quantum formalism these values are obtained when computing  the Born rule or Luders' rule.  In appropriate categories, we find these exact values back as one of these diamonds, by composing a state and an effect \cite{deLL,Selinger}.
\end{example}

The equation 
\beq\label{eq:SMCsliding} 
f\otimes g=(f\otimes 1_D)\circ (1_A\otimes g)=(1_B\otimes g)\circ(f\otimes 1_C)  
\eeq established in Proposition \ref{Prop:Bifinct} is depicted as: 
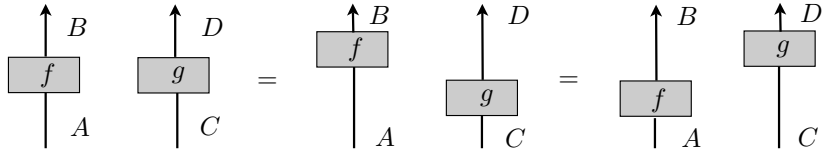
\begin{center} 
\ifx\JPicScale\undefined\def\JPicScale{1}\fi
\psset{unit=\JPicScale mm}
\psset{linewidth=0.3,dotsep=1,hatchwidth=0.3,hatchsep=1.5,shadowsize=1,dimen=middle}
\psset{dotsize=0.7 2.5,dotscale=1 1,fillcolor=black}
\psset{arrowsize=1 2,arrowlength=1,arrowinset=0.25,tbarsize=0.7 5,bracketlength=0.15,rbracketlength=0.15}
\begin{pspicture}(0,0)(108.3,19.35)
\psline(6,0)(6,7.5)
\rput(10.6,2.9){$A$}
\newrgbcolor{userFillColour}{0.8 0.8 0.8}
\pspolygon[linewidth=0.15,fillcolor=userFillColour,fillstyle=solid](1.05,7.4)(10.5,7.4)(10.5,12.1)(1.05,12.1)
\rput(10.2,16.2){$B$}
\rput(6.25,9.7){$f$}
\psline{->}(6,12.2)(6,19.2)
\psline(23.5,0)(23.5,7.5)
\newrgbcolor{userFillColour}{0.8 0.8 0.8}
\pspolygon[linewidth=0.15,fillcolor=userFillColour,fillstyle=solid](18.27,7.35)(27.72,7.35)(27.72,12.05)(18.27,12.05)
\psline{->}(23.22,12.15)(23.22,19.15)
\rput(23.67,9.75){$g$}
\rput(27.9,3){$C$}
\rput(28.2,16.2){$D$}
\rput(35.5,8.8){$=$}
\psline(47,0)(47,11)
\rput(51.1,1.5){$A$}
\newrgbcolor{userFillColour}{0.8 0.8 0.8}
\pspolygon[linewidth=0.15,fillcolor=userFillColour,fillstyle=solid](42.05,10.8)(51.5,10.8)(51.5,15.5)(42.05,15.5)
\rput(50.4,17.7){$B$}
\rput(47.25,13.1){$f$}
\psline{->}(46.9,15.4)(46.83,19.25)
\psline(64,0)(64,4.5)
\newrgbcolor{userFillColour}{0.8 0.8 0.8}
\pspolygon[linewidth=0.15,fillcolor=userFillColour,fillstyle=solid](59.27,4.35)(68.72,4.35)(68.72,9.05)(59.27,9.05)
\psline{->}(64.1,9.1)(64.05,19.2)
\rput(64.67,6.75){$g$}
\rput(68.5,1.4){$C$}
\rput(68.9,17.7){$D$}
\rput(75.55,8.9){$=$}
\psline(103.5,0)(103.5,11)
\rput(91.9,1.4){$A$}
\newrgbcolor{userFillColour}{0.8 0.8 0.8}
\pspolygon[linewidth=0.15,fillcolor=userFillColour,fillstyle=solid](98.85,10.9)(108.3,10.9)(108.3,15.6)(98.85,15.6)
\rput(91.3,17.6){$B$}
\psline{->}(103.7,15.5)(103.63,19.35)
\psline(87,0)(87,4)
\newrgbcolor{userFillColour}{0.8 0.8 0.8}
\pspolygon[linewidth=0.15,fillcolor=userFillColour,fillstyle=solid](82.38,4.25)(91.83,4.25)(91.83,8.95)(82.38,8.95)
\psline{->}(87.2,9)(87.15,19.1)
\rput(104.1,13.3){$g$}
\rput(107.4,1.5){$C$}
\rput(107.8,18){$D$}
\rput(87.5,6.4){$f$}
\end{pspicture}

\end{center} 
In words: we can `slide' boxes along their wires.  

The first defining equation of symmetry, i.e.~eq.(\ref{Eq:symdiag}), depicts as:
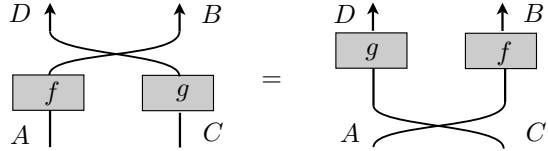
\begin{center} 
\ifx\JPicScale\undefined\def\JPicScale{1}\fi
\psset{unit=\JPicScale mm}
\psset{linewidth=0.3,dotsep=1,hatchwidth=0.3,hatchsep=1.5,shadowsize=1,dimen=middle}
\psset{dotsize=0.7 2.5,dotscale=1 1,fillcolor=black}
\psset{arrowsize=1 2,arrowlength=1,arrowinset=0.25,tbarsize=0.7 5,bracketlength=0.15,rbracketlength=0.15}
\begin{pspicture}(0,0)(70.67,19.25)
\psline(6,0)(6,5.5)
\rput(2,1.7){$A$}
\newrgbcolor{userFillColour}{0.8 0.8 0.8}
\pspolygon[linewidth=0.15,fillcolor=userFillColour,fillstyle=solid](1.05,4.9)(10.5,4.9)(10.5,9.6)(1.05,9.6)
\rput(27.5,17.7){$B$}
\rput(6.25,7.2){$f$}
\psline{->}(6,15.43)(6,19.2)
\psline(23.27,-0.05)(23.27,4.57)
\newrgbcolor{userFillColour}{0.8 0.8 0.8}
\pspolygon[linewidth=0.15,fillcolor=userFillColour,fillstyle=solid](18.28,4.85)(27.73,4.85)(27.73,9.55)(18.28,9.55)
\psline{->}(23.22,15.4)(23.22,19.17)
\rput(23.67,7.25){$g$}
\rput(27.7,2){$C$}
\rput(2,17.8){$D$}
\rput(35.5,8.8){$=$}
\psbezier[linewidth=0.25](5.9,9.6)(6,12.9)(23.2,11.7)(23.2,15.4)
\psbezier[linewidth=0.25](23.2,9.4)(23.3,12.7)(6.1,12.1)(6,15.4)
\psline(49,5.5)(49,10.5)
\rput(45.9,1.9){$A$}
\newrgbcolor{userFillColour}{0.8 0.8 0.8}
\pspolygon[linewidth=0.15,fillcolor=userFillColour,fillstyle=solid](44,10.5)(53.45,10.5)(53.45,15.2)(44,15.2)
\rput(70.3,17.8){$B$}
\psline{->}(48.93,15.48)(48.93,19.25)
\psline(66.5,6)(66.5,10.5)
\newrgbcolor{userFillColour}{0.8 0.8 0.8}
\pspolygon[linewidth=0.15,fillcolor=userFillColour,fillstyle=solid](61.22,10.45)(70.67,10.45)(70.67,15.15)(61.22,15.15)
\psline{->}(66.15,15.45)(66.15,19.22)
\rput(48.8,12.7){$g$}
\rput(70.63,2.05){$C$}
\rput(45.2,17.7){$D$}
\psbezier[linewidth=0.25](49,0)(49.1,3.3)(66.5,2.3)(66.5,6)
\psbezier[linewidth=0.25](66.3,-0.2)(66.4,3.1)(49.3,2.6)(49,5.5)
\rput(66.4,12.5){$f$}
\end{pspicture}
 
\end{center} 
i.e., we can still `slide' boxes along crossings of wires. 
The equation 
\beq\label{eq:doubleswap} 
\sigma_{B,A}\circ\sigma_{A,B}=1_{A,B}\,,
\eeq
which when varying $A,B\in|{\bf C}|$ states that 
\[
\sigma_{A,B}^{-1}=\sigma_{B,A}\,, 
\]
depicts as 
\vspace{-6.5mm}\begin{center} 
\ifx\JPicScale\undefined\def\JPicScale{1}\fi
\psset{unit=\JPicScale mm}
\psset{linewidth=0.3,dotsep=1,hatchwidth=0.3,hatchsep=1.5,shadowsize=1,dimen=middle}
\psset{dotsize=0.7 2.5,dotscale=1 1,fillcolor=black}
\psset{arrowsize=1 2,arrowlength=1,arrowinset=0.25,tbarsize=0.7 5,bracketlength=0.15,rbracketlength=0.15}
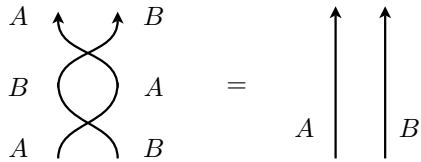
\begin{pspicture}(0,0)(55.02,29)
\psbezier(8.29,10.63)(8.29,4.75)(16.2,5.59)(16.2,0.55)
\psbezier(16.2,10.63)(16.2,4.75)(8.29,5.59)(8.29,0.55)
\psbezier{<-}(8.29,20.08)(8.29,14.2)(16.2,15.04)(16.2,10)
\psbezier{<-}(16.2,20.08)(16.2,14.2)(8.29,15.04)(8.29,10)
\rput(3.02,19.45){$A$}
\rput(21.02,10.1){$A$}
\rput(3.02,1.9){$A$}
\rput(54.39,22.15){}
\rput(53.05,24.95){}
\rput(21.02,19.55){$B$}
\rput(51.73,29){}
\rput(21.02,2){$B$}
\rput(3.02,10){$B$}
\rput(32,10){$=$}
\psline{->}(45.17,0.55)(45.17,20.8)
\psline{->}(51.76,0.55)(51.76,20.8)
\rput(41.02,4.55){$A$}
\rput(55.02,4.55){$B$}
\end{pspicture}

\end{center} 

Suppose now that for any three arbitrary morphisms 
\[
f:A\rTo A'\quad,\quad g:B\rTo B'\qquad \mbox{and} \qquad h:C\rTo C'
\]
in any strict symmetric monoidal category, one intends to prove that
\begin{equation}\nonumber 
(\sigma_{B',C'}\otimes f)\circ(g\otimes\sigma_{A,C'})\circ(\sigma_{A,B}\otimes h)\hspace{4.2cm} 
\end{equation}
\begin{equation}\nonumber 
\hspace{1.2cm}
=(h\otimes\sigma_{A',B'})\circ(\sigma_{A',C}\otimes 1_{B'})\circ(1_{A'}\otimes\sigma_{B',C})\circ(f\otimes
g\otimes 1_C) 
\end{equation} 
always holds. Then, the typical textbook proof  proceeds by \em diagram chasing\em: 
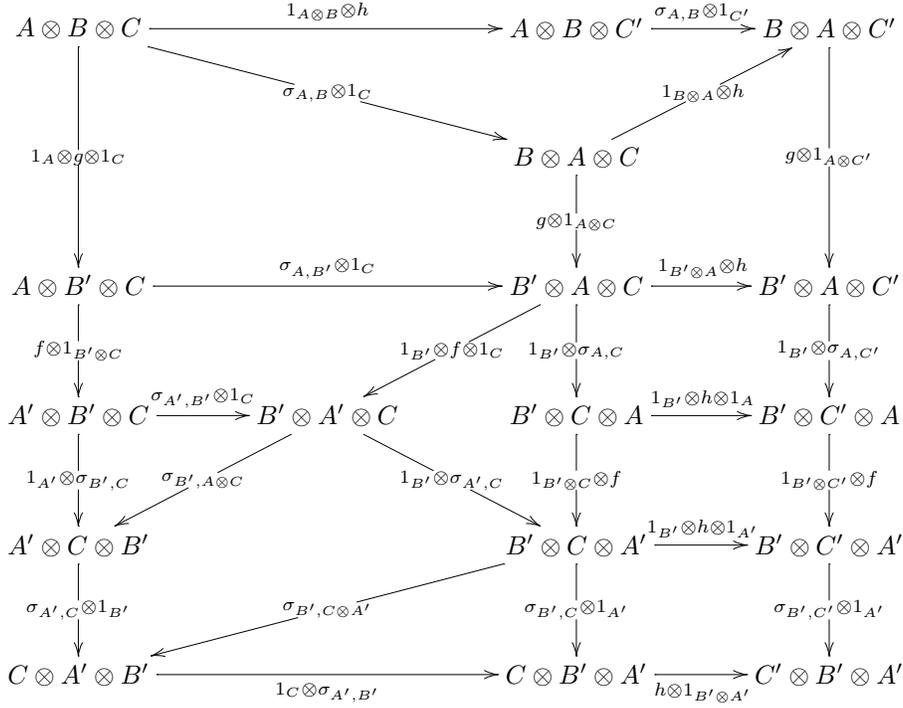
\[ \!\!\xymatrix@=.48in{%
A\otimes B\otimes C \ar[dd]|{1_A\otimes g\otimes 1_C}\ar[drr]|{\sigma_{A,B}\otimes 1_C}\ar[rr]^{1_{A\otimes B}\otimes h} & & A\otimes B\otimes C' \ar[r]^{\sigma_{A, B}\otimes 1_{C'}} & B\otimes A\otimes C' \ar[dd]|{g\otimes 1_{A\otimes C'}}\\%
& & B\otimes A\otimes C\ar[ur]|{1_{B\otimes A}\otimes h} \ar[d]|{g\otimes 1_{A\otimes C}} &\\%
A\otimes B'\otimes C\ar[rr]^{\sigma_{A,B'}\otimes 1_C}\ar[d]|{f\otimes 1_{B'\otimes C}} & & B'\otimes A\otimes C\ar[r]^{1_{B'\otimes A}\otimes h} \ar[d]|{1_{B'}\otimes\sigma_{A,C}} \ar[dl]|{1_{B'}\otimes f\otimes 1_{C}} & B'\otimes A\otimes C' \ar[d]|{1_{B'}\otimes\sigma_{A,C'}}\\%
A'\otimes B'\otimes C \ar[r]^{\sigma_{A',B'}\otimes 1_C}\ar[d]|{1_{A'}\otimes \sigma_{B',C}}& B'\otimes A'\otimes C\ar[dl]|{\sigma_{B',A\otimes C}}\ar[dr]|{1_{B'}\otimes\sigma_{A',C}} & B'\otimes C\otimes A\ar[r]^{1_{B'}\otimes h\otimes 1_A}\ar[d]|{1_{B'\otimes C}\otimes f}& B'\otimes C'\otimes A\ar[d]|{1_{B'\otimes C'}\otimes f}\\%
A'\otimes C\otimes B'\ar[d]|{\sigma_{A',C}\otimes 1_{B'}}& & B'\otimes C\otimes A'\ar[dll]|{\sigma_{B',C\otimes A'}}\ar[r]^{1_{B'}\otimes h\otimes 1_{A'}} \ar[d]|{\sigma_{B',C}\otimes 1_{A'}} & B'\otimes C'\otimes A'\ar[d]|{\sigma_{B',C'}\otimes 1_{A'}}\\%
C\otimes A'\otimes B'\ar[rr]_{1_C\otimes\sigma_{A',B'}} & & C\otimes B'\otimes A'\ar[r]_{h\otimes 1_{B'\otimes A'}} & C'\otimes B'\otimes A'%
}\]
One needs to read this `dragon' as follows.  The two outer paths both going from the left-upper-corner to the right-lower-corner represent the two sides of the equality we want to prove. Then, we do what category-theoreticians call diagram chasing, that is, `pasting'  together several commutative diagrams, which connect one of the outer paths to the other.  For example, the triangle at the top of the diagram expresses that 
\[ 
(\sigma_{A, B}\otimes 1_{C'})\circ(1_{A\otimes B}\otimes h)=(1_{B\otimes A}\otimes h)\circ(\sigma_{A, B}\otimes 1_{C})\,, 
\] 
that is, an instance of bifunctoriality. Using properties of strict symmetric monoidal categories, namely bifunctoriality and  eq.(\ref{Eq:symdiag})  expressed as commutative diagrams, we can pass from the outer path at the top and the right to the outer path on the left and the bottom. This is clearly a very tedious task and getting these diagrams into LaTeX becomes a time-consuming activity. 

On the other hand, when using the graphical calculus, one immediately sees that 
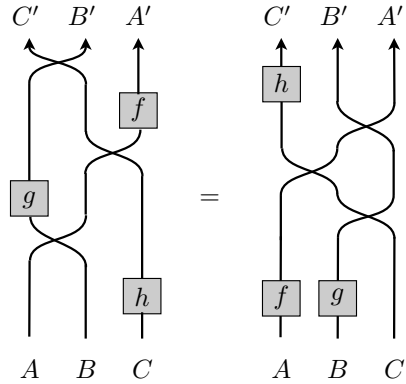
\begin{center}
\ifx\JPicScale\undefined\def\JPicScale{1}\fi
\psset{unit=\JPicScale mm}
\psset{linewidth=0.3,dotsep=1,hatchwidth=0.3,hatchsep=1.5,shadowsize=1,dimen=middle}
\psset{dotsize=0.7 2.5,dotscale=1 1,fillcolor=black}
\psset{arrowsize=1 2,arrowlength=1,arrowinset=0.25,tbarsize=0.7 5,bracketlength=0.15,rbracketlength=0.15}
\begin{pspicture}(0,0)(51.05,49)
\newrgbcolor{userFillColour}{0.8 0.8 0.8}
\pspolygon[linewidth=0.15,fillcolor=userFillColour,fillstyle=solid](-0.04,22.3)(4.96,22.3)(4.96,27)(-0.04,27)
\newrgbcolor{userFillColour}{0.8 0.8 0.8}
\pspolygon[linewidth=0.15,fillcolor=userFillColour,fillstyle=solid](14.96,9.3)(19.96,9.3)(19.96,14)(14.96,14)
\newrgbcolor{userFillColour}{0.8 0.8 0.8}
\pspolygon[linewidth=0.15,fillcolor=userFillColour,fillstyle=solid](14.46,34)(19.46,34)(19.46,38.7)(14.46,38.7)
\rput(2.5,2){$A$}
\rput(10,2){$B$}
\rput(17.5,2){$C$}
\rput(17,49){$A'$}
\rput(9.5,49){$B'$}
\rput(2,49){$C'$}
\rput(36,2){$A$}
\rput(43.5,2){$B$}
\rput(51,2){$C$}
\rput(50.5,49){$A'$}
\rput(43,49){$B'$}
\rput(35.5,49){$C'$}
\rput(26.46,24.5){$=$}
\psline(35.86,6.18)(35.9,19.6)
\psbezier(43.5,19.5)(43.54,22.8)(50.9,21.4)(50.9,25.1)
\psbezier(51,19.3)(51.05,22.6)(43.44,22.1)(43.4,25.4)
\psline(43.46,6)(43.5,19.5)
\psbezier(43.4,31.3)(43.44,34.6)(51,33.7)(51,37.4)
\psbezier(50.9,30.9)(50.95,34.2)(43.44,34.3)(43.4,37.6)
\psline(50.96,6)(51,19.5)
\newrgbcolor{userFillColour}{0.8 0.8 0.8}
\pspolygon[linewidth=0.15,fillcolor=userFillColour,fillstyle=solid](40.96,9)(45.96,9)(45.96,13.7)(40.96,13.7)
\psline(35.9,19.1)(35.9,25.1)
\psline(36,31.3)(36,37.5)
\psline{->}(51,37.3)(50.96,46)
\newrgbcolor{userFillColour}{0.8 0.8 0.8}
\pspolygon[linewidth=0.15,fillcolor=userFillColour,fillstyle=solid](33.46,9)(38.46,9)(38.46,13.7)(33.46,13.7)
\psbezier(35.9,25.1)(35.94,28.4)(43.4,27.6)(43.4,31.3)
\psbezier(43.4,25.3)(43.44,28.6)(36.13,28)(36.08,31.3)
\psline(50.9,25.1)(50.9,31.1)
\psline{->}(43.46,37.5)(43.46,46)
\psline{->}(35.96,37.5)(35.96,46)
\rput(35.96,11.5){$f$}
\rput(43.46,11.5){$g$}
\rput(17.46,11.5){$h$}
\rput(16.96,36.5){$f$}
\rput(2.46,24.5){$g$}
\newrgbcolor{userFillColour}{0.8 0.8 0.8}
\pspolygon[linewidth=0.15,fillcolor=userFillColour,fillstyle=solid](33.5,37.5)(38.5,37.5)(38.5,42.2)(33.5,42.2)
\rput(36,40){$h$}
\psline(9.96,22.5)(9.96,28)
\psline(2.46,6.26)(2.5,16.4)
\psbezier(2.56,16.4)(2.6,19.7)(10,18.7)(10,22.4)
\psbezier(10,15.5)(10.05,18.8)(2.64,18.9)(2.6,22.2)
\psline(9.96,6)(9.96,15.86)
\psline(17.46,14)(17.46,27.62)
\psbezier(9.9,27.9)(9.94,31.2)(17.23,30.1)(17.23,33.8)
\psbezier(17.5,27.6)(17.54,30.9)(10.05,30.3)(10,33.6)
\psline(17.46,6)(17.46,9.5)
\psline(2.46,27)(2.46,40)
\psline(9.96,33.5)(9.96,39.81)
\psline{->}(16.96,38.5)(16.96,46)
\psbezier{->}(2.46,40)(2.5,43.3)(9.96,42.3)(9.96,46)
\psbezier{->}(10,39.8)(10.05,43.1)(2.5,42.7)(2.46,46)
\end{pspicture}
 
\end{center} 
must hold.  We pass from one picture to
the other by sliding the boxes along wires and then by rearranging
these wires.  In terms of the underlying equations of strict symmetric
monoidal structure, `sliding the boxes along wires' uses
eq.(\ref{Eq:symdiag})   and eq.(\ref{eq:SMCsliding}), while
`rearranging these wires' means that we used
eq.(\ref{Eq:symdiag}) as follows: 
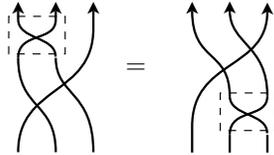
\begin{center}
\ifx\JPicScale\undefined\def\JPicScale{1}\fi
\psset{unit=\JPicScale mm}
\psset{linewidth=0.3,dotsep=1,hatchwidth=0.3,hatchsep=1.5,shadowsize=1,dimen=middle}
\psset{dotsize=0.7 2.5,dotscale=1 1,fillcolor=black}
\psset{arrowsize=1 2,arrowlength=1,arrowinset=0.25,tbarsize=0.7 5,bracketlength=0.15,rbracketlength=0.15}
\begin{pspicture}(0,0)(35.62,20.1)
\rput(16.88,11.25){$=$}
\psbezier(6.25,0.05)(6.25,6.88)(1.25,5.62)(1.25,12.6)
\psbezier(11.25,0.05)(11.25,8.28)(6.25,7.5)(6.25,12.6)
\psbezier(1.25,0.05)(1.25,9.07)(11.25,7.5)(11.25,16.52)
\psbezier(1.25,12.6)(1.28,16.13)(6.25,14.65)(6.25,17.7)
\psbezier(6.22,12.76)(6.25,15.48)(1.28,14.98)(1.25,17.7)
\psline{->}(1.25,17.7)(1.25,20.05)
\psline{->}(6.25,17.7)(6.25,20.05)
\psline{->}(11.25,16.52)(11.25,20.05)
\pspolygon[linewidth=0.15,linestyle=dashed,dash=1 1](0,18.12)(7.5,18.12)(7.5,13.12)(0,13.12)
\psbezier{->}(29.38,7.5)(29.38,13.75)(24.38,11.88)(24.38,20.05)
\psbezier{->}(34.38,7.55)(34.38,14.38)(29.38,14.38)(29.38,20.1)
\psbezier{->}(24.38,3.48)(24.38,12.5)(34.38,10.93)(34.38,19.95)
\psbezier(29.38,2.4)(29.4,5.93)(34.38,4.46)(34.38,7.5)
\psbezier(34.35,2.57)(34.38,5.28)(29.41,4.79)(29.38,7.5)
\psline(29.38,0)(29.38,2.35)
\psline(34.38,0)(34.38,2.35)
\psline(24.38,0)(24.38,3.53)
\pspolygon[linewidth=0.15,linestyle=dashed,dash=1 1](28.12,7.93)(35.62,7.93)(35.62,2.93)(28.12,2.93)
\end{pspicture}
 
\end{center} 
Indeed, since symmetry is a morphism it can be conceived as a box,  and hence we can `slide it along wires'. 

In a broader historical perspective, we are somewhat unfair here.  Writing equational reasoning down in terms of these commutative diagrams rather than long lists of equalities was an important step towards a better geometrical understanding of the structure of proofs.

\subsection{Extended Dirac notation}\label{sec:Dirac}

\begin{definition}\em A \em strict dagger monoidal category \em ${\bf C}$ is a
strict monoidal category equipped with an involutive identity-on-objects
contravariant functor 
\[
\dagger:{\bf C}^{op}{\longrightarrow}{\bf C}\,,
\]
that is, 
\begin{itemize}
	\item $A^\dagger=A$ for all $A\in|{\bf C}|$, and
	\item $f^{\dagger\dagger}=f$ for all morphisms $f$, 
\end{itemize}
and this functor preserves the tensor, that is, 
\beq\label{eq:daggermonoidal} 
(f\otimes g)^\dagger=f^\dagger\otimes g^\dagger\,.  
\eeq
We will refer to $B\rTo^{f^\dagger} A$ as the \em
adjoint \em to $A\rTo^f B$.  A \em strict dagger symmetric monoidal
category \em ${\bf C}$ is both a strict dagger monoidal category and a
strict symmetric monoidal category such that
\[
\sigma_{A,B}^\dagger=\sigma_{A,B}^{-1}. 
\] \end{definition}

\begin{definition}{\rm\cite{AC2004}}\label{def:unitarity} 
\em A morphism $U:A\rTo B$ in a
strict dagger monoidal category $\cat$ is called \em unitary \em if its
inverse and its adjoint coincide, that is, if
\[
U^\dagger=U^{-1}\,.
\]
Let $\psi,\phi:\II\rTo A$ be `elements' in $\cat$.  Their \em inner-product \em is the `scalar'
\[
\langle \phi\, |\,\psi\rangle:= \phi^\dagger\circ\psi:\II\rTo\II\,.
\] 
\end{definition}

So in any strict monoidal category we refer to morphisms of type 
\[
\II\rTo A 
\]
as \em elements \em (cf.~Example \ref{ex:elements}), to those of type 
\[
A\rTo\II
\]  
as \emph{co-elements}, and to those of type 
\[
\II\rTo\II
\]
as \em scalars\em.  As already discussed in Example   \ref{ex:stateseffects} in the category ${\bf QuantOpp}$ these corresponds respectively to states, effects and probabilistic weights.

Even at this abstract level, many familiar things follow from Definition \ref{def:unitarity}.  For example, we recover the defining property of adjoints for any dagger functor:
\begin{eqnarray*} \langle f^\dagger\circ \psi\, |\,\phi\rangle 
&=& (f^\dagger\circ \psi)^\dagger\circ\phi\\
&=& (\psi^\dagger\circ f)\circ \phi\\
&=& \psi^\dagger\circ (f\circ \phi)\\
&=& \langle\psi\, |\,f\circ\phi\rangle.
\end{eqnarray*}
From this it follows that unitary morphisms preserve the inner-product:
\begin{eqnarray*}
\langle U\circ \psi \,|\,U\circ\phi\rangle 
&=& \langle U^\dagger\circ (U\circ\psi) \,|\, \phi\rangle\\
&=& \langle (U^\dagger\circ U)\circ\psi \,|\, \phi\rangle\\
&=& \langle\psi \,|\,\phi\rangle. 
\end{eqnarray*}

Importantly, the graphical calculus of the previous section extends to strict dagger
symmetric monoidal categories.  Following Selinger \cite{Selinger}, we introduce
an asymmetry in the graphical notation of the morphisms $A\rTo{f} B$ as
follows:\vspace{-6mm}
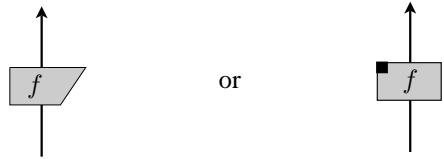
\begin{center} 
\ifx\JPicScale\undefined\def\JPicScale{1}\fi
\psset{unit=\JPicScale mm}
\psset{linewidth=0.3,dotsep=1,hatchwidth=0.3,hatchsep=1.5,shadowsize=1,dimen=middle}
\psset{dotsize=0.7 2.5,dotscale=1 1,fillcolor=black}
\psset{arrowsize=1 2,arrowlength=1,arrowinset=0.25,tbarsize=0.7 5,bracketlength=0.15,rbracketlength=0.15}
\begin{pspicture}(0,0)(58.12,29)
\psline{->}(54,12.5)(54,20.62)
\psline(54,0.62)(54,7.5)
\psline{->}(5,11.25)(5,20)
\rput(54.39,22.15){}
\rput(53.05,24.95){}
\rput(51.73,29){}
\psline(5,0)(5,6.88)
\newrgbcolor{userFillColour}{0.8 0.8 0.8}
\psline[linewidth=0.1,fillcolor=userFillColour,fillstyle=solid](0.77,11.86)
(0.77,6.88)
(7.5,6.88)
(10.86,11.86)(0.77,11.86)
\rput(4.14,9.36){$f$}
\newrgbcolor{userFillColour}{0.8 0.8 0.8}
\pspolygon[linewidth=0.15,fillcolor=userFillColour,fillstyle=solid](49.62,12.5)(58.12,12.5)(58.12,7.5)(49.62,7.5)
\rput(54,10){$f$}
\pspolygon[fillstyle=solid](49.62,12.5)(50.88,12.5)(50.88,11.24)(49.62,11.24)
\rput(30,10){or}
\rput(-0.9,18.4){}
\rput(45.62,10.62){}
\end{pspicture}
 
\end{center}
Then we depict the adjoint $B\rTo{f^\dagger} A$ of $A\rTo{f} B$ as follows:
\vspace{-5mm}
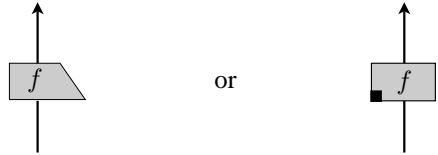
\begin{center} 
\ifx\JPicScale\undefined\def\JPicScale{1}\fi
\psset{unit=\JPicScale mm}
\psset{linewidth=0.3,dotsep=1,hatchwidth=0.3,hatchsep=1.5,shadowsize=1,dimen=middle}
\psset{dotsize=0.7 2.5,dotscale=1 1,fillcolor=black}
\psset{arrowsize=1 2,arrowlength=1,arrowinset=0.25,tbarsize=0.7 5,bracketlength=0.15,rbracketlength=0.15}
\begin{pspicture}(0,0)(57.88,29)
\psline{->}(53.75,11.88)(53.75,20)
\psline(53.75,0)(53.75,6.88)
\psline{->}(5,11.25)(5,20)
\rput(54.39,22.15){}
\rput(53.05,24.95){}
\rput(51.73,29){}
\psline(5,0)(5,6.88)
\newrgbcolor{userFillColour}{0.8 0.8 0.8}
\pspolygon[linewidth=0.15,fillcolor=userFillColour,fillstyle=solid](49.38,11.88)(57.88,11.88)(57.88,6.88)(49.38,6.88)
\rput(53.75,9.38){$f$}
\pspolygon[fillstyle=solid](49.38,8.12)(50.62,8.12)(50.62,6.88)(49.38,6.88)
\rput(30,9.38){or}
\rput(-0.9,18.4){}
\rput(31.88,10){}
\newrgbcolor{userFillColour}{0.8 0.8 0.8}
\psline[linewidth=0.1,fillcolor=userFillColour,fillstyle=solid](1.25,7.05)
(1.25,11.88)
(7.98,11.88)
(11.34,7.05)(1.25,7.05)
\rput(4.62,9.54){$f$}
\end{pspicture}

\end{center} 
\noindent that is, we turn the box
representing $f$ upside-down. All this enables interpreting Dirac
notation \cite{Dirac} in terms of strict dagger symmetric monoidal categories, and in
particular, in terms of the corresponding graphical calculus:
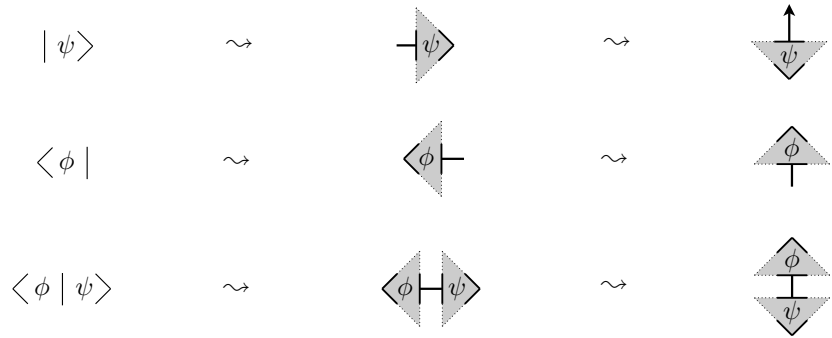
\begin{center}
\ifx\JPicScale\undefined\def\JPicScale{1}\fi
\psset{unit=\JPicScale mm}
\psset{linewidth=0.3,dotsep=1,hatchwidth=0.3,hatchsep=1.5,shadowsize=1,dimen=middle}
\psset{dotsize=0.7 2.5,dotscale=1 1,fillcolor=black}
\psset{arrowsize=1 2,arrowlength=1,arrowinset=0.25,tbarsize=0.7 5,bracketlength=0.15,rbracketlength=0.15}
\begin{pspicture}(0,0)(109.3,12.5)
\newrgbcolor{userFillColour}{0.8 0.8 0.8}
\psline[linewidth=0.15,linestyle=none,fillcolor=userFillColour,fillstyle=solid](104.3,2.5)
(109.3,7.5)
(99.3,7.5)(104.3,2.5)
\psline{->}(104.3,7.5)(104.3,12.5)
\rput(104.3,5.5){$\psi$}
\psline[linewidth=0.25](102.3,7.5)(106.3,7.5)
\psline[linewidth=0.25](102.3,4.5)(104.3,2.5)
\psline[linewidth=0.25](104.3,2.5)(106.3,4.5)
\psline[linewidth=0.15,linestyle=dotted,dotsep=0.3](102.3,7.5)(99.3,7.5)
\psline[linewidth=0.15,linestyle=dotted,dotsep=0.3](99.3,7.5)(102.3,4.5)
\psline[linewidth=0.15,linestyle=dotted,dotsep=0.3](106.3,4.5)(109.3,7.5)
\psline[linewidth=0.15,linestyle=dotted,dotsep=0.3](109.3,7.5)(106.3,7.5)
\newrgbcolor{userFillColour}{0.8 0.8 0.8}
\psline[linewidth=0.15,linestyle=none,fillcolor=userFillColour,fillstyle=solid](59.7,7)
(54.7,12)
(54.7,2)(59.7,7)
\psline(54.7,7)(52.1,7)
\rput(56.7,7){$\psi$}
\psline[linewidth=0.25](54.7,5)(54.7,9)
\psline[linewidth=0.25](57.7,5)(59.7,7)
\psline[linewidth=0.25](59.7,7)(57.7,9)
\psline[linewidth=0.15,linestyle=dotted,dotsep=0.3](54.7,5)(54.7,2)
\psline[linewidth=0.15,linestyle=dotted,dotsep=0.3](54.7,2)(57.7,5)
\psline[linewidth=0.15,linestyle=dotted,dotsep=0.3](57.7,9)(54.7,12)
\psline[linewidth=0.15,linestyle=dotted,dotsep=0.3](54.7,12)(54.7,9)
\rput(8.2,6.9){$\psi$}
\psline[linewidth=0.15](5.3,4.7)(5.3,9)
\psline[linewidth=0.15](9.7,4.9)(11.7,6.9)
\psline[linewidth=0.15](11.7,6.9)(9.7,8.9)
\rput(81.5,7.3){$\leadsto$}
\rput(31.3,7){$\leadsto$}
\end{pspicture}
 
\end{center}	

\begin{center}
\ifx\JPicScale\undefined\def\JPicScale{1}\fi
\psset{unit=\JPicScale mm}
\psset{linewidth=0.3,dotsep=1,hatchwidth=0.3,hatchsep=1.5,shadowsize=1,dimen=middle}
\psset{dotsize=0.7 2.5,dotscale=1 1,fillcolor=black}
\psset{arrowsize=1 2,arrowlength=1,arrowinset=0.25,tbarsize=0.7 5,bracketlength=0.15,rbracketlength=0.15}
\begin{pspicture}(0,0)(110,12.4)
\psline(105,3.7)(105,8.7)
\newrgbcolor{userFillColour}{0.8 0.8 0.8}
\psline[linewidth=0.15,linestyle=none,fillcolor=userFillColour,fillstyle=solid](105,11.7)
(110,6.7)
(100,6.7)(105,11.7)
\rput(105,8.7){$\phi$}
\psline[linewidth=0.25](105,11.7)(103,9.7)
\psline[linewidth=0.25](105,11.7)(107,9.7)
\psline[linewidth=0.25](103,6.7)(107,6.7)
\psline[linewidth=0.15,linestyle=dotted,dotsep=0.3](107,6.7)(110,6.7)
\psline[linewidth=0.15,linestyle=dotted,dotsep=0.3](110,6.7)(107,9.7)
\psline[linewidth=0.15,linestyle=dotted,dotsep=0.3](103,6.7)(100,6.7)
\psline[linewidth=0.15,linestyle=dotted,dotsep=0.3](100,6.7)(103,9.7)
\psline(61.4,7.4)(56.4,7.4)
\newrgbcolor{userFillColour}{0.8 0.8 0.8}
\psline[linewidth=0.15,linestyle=none,fillcolor=userFillColour,fillstyle=solid](53.4,7.4)
(58.4,12.4)
(58.4,2.4)(53.4,7.4)
\rput(56.4,7.4){$\phi$}
\psline[linewidth=0.25](53.4,7.4)(55.4,5.4)
\psline[linewidth=0.25](53.4,7.4)(55.4,9.4)
\psline[linewidth=0.25](58.4,5.4)(58.4,9.4)
\psline[linewidth=0.15,linestyle=dotted,dotsep=0.3](58.4,9.4)(58.4,12.4)
\psline[linewidth=0.15,linestyle=dotted,dotsep=0.3](58.4,12.4)(55.4,9.4)
\psline[linewidth=0.15,linestyle=dotted,dotsep=0.3](58.4,5.4)(58.4,2.4)
\psline[linewidth=0.15,linestyle=dotted,dotsep=0.3](58.4,2.4)(55.4,5.4)
\rput(8.6,6.9){$\phi$}
\psline[linewidth=0.15](4.7,6.9)(6.7,4.9)
\psline[linewidth=0.15](4.7,6.9)(6.7,8.9)
\psline[linewidth=0.15](11.3,4.7)(11.3,9)
\rput(81.5,7.3){$\leadsto$}
\rput(31.3,7){$\leadsto$}
\end{pspicture}
 
\end{center}	

\begin{center}
\ifx\JPicScale\undefined\def\JPicScale{1}\fi
\psset{unit=\JPicScale mm}
\psset{linewidth=0.3,dotsep=1,hatchwidth=0.3,hatchsep=1.5,shadowsize=1,dimen=middle}
\psset{dotsize=0.7 2.5,dotscale=1 1,fillcolor=black}
\psset{arrowsize=1 2,arrowlength=1,arrowinset=0.25,tbarsize=0.7 5,bracketlength=0.15,rbracketlength=0.15}
\begin{pspicture}(0,0)(110.1,13.8)
\newrgbcolor{userFillColour}{0.8 0.8 0.8}
\psline[linewidth=0.15,linestyle=none,fillcolor=userFillColour,fillstyle=solid](105.1,0.8)
(110.1,5.8)
(100.1,5.8)(105.1,0.8)
\psline(105.1,5.8)(105.1,10.8)
\rput(105.1,3.8){$\psi$}
\newrgbcolor{userFillColour}{0.8 0.8 0.8}
\psline[linewidth=0.15,linestyle=none,fillcolor=userFillColour,fillstyle=solid](105.1,13.8)
(110.1,8.8)
(100.1,8.8)(105.1,13.8)
\rput(105.1,10.8){$\phi$}
\psline[linewidth=0.25](105.1,13.8)(103.1,11.8)
\psline[linewidth=0.25](105.1,13.8)(107.1,11.8)
\psline[linewidth=0.25](103.1,8.8)(107.1,8.8)
\psline[linewidth=0.25](103.1,5.8)(107.1,5.8)
\psline[linewidth=0.25](103.1,2.8)(105.1,0.8)
\psline[linewidth=0.25](105.1,0.8)(107.1,2.8)
\psline[linewidth=0.15,linestyle=dotted,dotsep=0.3](107.1,8.8)(110.1,8.8)
\psline[linewidth=0.15,linestyle=dotted,dotsep=0.3](110.1,8.8)(107.1,11.8)
\psline[linewidth=0.15,linestyle=dotted,dotsep=0.3](103.1,8.8)(100.1,8.8)
\psline[linewidth=0.15,linestyle=dotted,dotsep=0.3](100.1,8.8)(103.1,11.8)
\psline[linewidth=0.15,linestyle=dotted,dotsep=0.3](103.1,5.8)(100.1,5.8)
\psline[linewidth=0.15,linestyle=dotted,dotsep=0.3](100.1,5.8)(103.1,2.8)
\psline[linewidth=0.15,linestyle=dotted,dotsep=0.3](107.1,2.8)(110.1,5.8)
\psline[linewidth=0.15,linestyle=dotted,dotsep=0.3](110.1,5.8)(107.1,5.8)
\newrgbcolor{userFillColour}{0.8 0.8 0.8}
\psline[linewidth=0.15,linestyle=none,fillcolor=userFillColour,fillstyle=solid](63.6,7)
(58.6,12)
(58.6,2)(63.6,7)
\psline(58.6,7)(53.6,7)
\rput(60.6,7){$\psi$}
\newrgbcolor{userFillColour}{0.8 0.8 0.8}
\psline[linewidth=0.15,linestyle=none,fillcolor=userFillColour,fillstyle=solid](50.6,7)
(55.6,12)
(55.6,2)(50.6,7)
\rput(53.6,7){$\phi$}
\psline[linewidth=0.25](50.6,7)(52.6,5)
\psline[linewidth=0.25](50.6,7)(52.6,9)
\psline[linewidth=0.25](55.6,5)(55.6,9)
\psline[linewidth=0.25](58.6,5)(58.6,9)
\psline[linewidth=0.25](61.6,5)(63.6,7)
\psline[linewidth=0.25](63.6,7)(61.6,9)
\psline[linewidth=0.15,linestyle=dotted,dotsep=0.3](55.6,9)(55.6,12)
\psline[linewidth=0.15,linestyle=dotted,dotsep=0.3](55.6,12)(52.6,9)
\psline[linewidth=0.15,linestyle=dotted,dotsep=0.3](55.6,5)(55.6,2)
\psline[linewidth=0.15,linestyle=dotted,dotsep=0.3](55.6,2)(52.6,5)
\psline[linewidth=0.15,linestyle=dotted,dotsep=0.3](58.6,5)(58.6,2)
\psline[linewidth=0.15,linestyle=dotted,dotsep=0.3](58.6,2)(61.6,5)
\psline[linewidth=0.15,linestyle=dotted,dotsep=0.3](61.6,9)(58.6,12)
\psline[linewidth=0.15,linestyle=dotted,dotsep=0.3](58.6,12)(58.6,9)
\rput(11,7){$\psi$}
\rput(5.4,7){$\phi$}
\psline[linewidth=0.15](1.5,7)(3.5,5)
\psline[linewidth=0.15](1.5,7)(3.5,9)
\psline[linewidth=0.15](8.1,4.8)(8.1,9.1)
\psline[linewidth=0.15](12.5,5)(14.5,7)
\psline[linewidth=0.15](14.5,7)(12.5,9)
\rput(81.5,7.3){$\leadsto$}
\rput(31.3,7){$\leadsto$}
\end{pspicture}

\end{center}
\noindent The latter notation merely requires closing the bra's and ket's and performing a $90^\circ$ rotation.\footnote{This $90^\circ$ rotation is merely a consequence of our convention to read pictures from bottom-to-top.  Other authors obey different conventions e.g.~top-to-bottom or left-to-right.} Summarising we
now have: 
\begin{center} 
\begin{tabular}{|c|c|c|c|} \hline \ {\bf Dirac} \  &
{\bf matrix}  & {\bf strict $\dagger$-SMC} & \ {\bf picture} \ \\ 
\hline\hline $|\psi\rangle$  & $\left(\begin{array}{c}\psi_1\\ \vdots \\ \psi_n\end{array}\right)$ & $\II\rTo{\psi}A$ & \ensuremath{\vcenter{\hbox{
\ifx\JPicScale\undefined\def\JPicScale{1}\fi
\psset{unit=\JPicScale mm}
\psset{linewidth=0.3,dotsep=1,hatchwidth=0.3,hatchsep=1.5,shadowsize=1,dimen=middle}
\psset{dotsize=0.7 2.5,dotscale=1 1,fillcolor=black}
\psset{arrowsize=1 2,arrowlength=1,arrowinset=0.25,tbarsize=0.7 5,bracketlength=0.15,rbracketlength=0.15}
\begin{pspicture}(0,0)(10,10)
\newrgbcolor{userFillColour}{0.8 0.8 0.8}
\psline[linewidth=0.15,fillcolor=userFillColour,fillstyle=solid](5,-0)
(10,5)
(-0,5)(5,-0)
\psline{->}(5,5)(5,10)
\rput(5,3){$\psi$}
\psline[linewidth=0.15](3,5)(7,5)
\psline[linewidth=0.15](3,2)(5,-0)
\psline[linewidth=0.15](5,-0)(7,2)
\psline[linewidth=0.15](3,5)(-0,5)
\psline[linewidth=0.15](-0,5)(3,2)
\psline[linewidth=0.15](7,2)(10,5)
\psline[linewidth=0.15](10,5)(7,5)
\end{pspicture}
}}} \\ 
\hline $\langle\phi|$ & $\left(\begin{array}{ccc}\bar{\phi}_1 & \ldots & \bar{\phi}_n\end{array}\right)$ & $A\rTo{\phi} \II$ & \ensuremath{\vcenter{\hbox{
\ifx\JPicScale\undefined\def\JPicScale{1}\fi
\psset{unit=\JPicScale mm}
\psset{linewidth=0.3,dotsep=1,hatchwidth=0.3,hatchsep=1.5,shadowsize=1,dimen=middle}
\psset{dotsize=0.7 2.5,dotscale=1 1,fillcolor=black}
\psset{arrowsize=1 2,arrowlength=1,arrowinset=0.25,tbarsize=0.7 5,bracketlength=0.15,rbracketlength=0.15}
\begin{pspicture}(0,0)(10,9.7)
\newrgbcolor{userFillColour}{0.8 0.8 0.8}
\psline[linewidth=0.15,fillcolor=userFillColour,fillstyle=solid](5,9.7)
(10,4.85)
(-0,4.85)(5,9.7)
\psline(5,4.85)(5,0)
\rput(5,6.79){$\phi$}
\psline[linewidth=0.15](3,4.85)(7,4.85)
\psline[linewidth=0.15](3,7.76)(5,9.7)
\psline[linewidth=0.15](5,9.7)(7,7.76)
\psline[linewidth=0.15](3,4.85)(-0,4.85)
\psline[linewidth=0.15](-0,4.85)(3,7.76)
\psline[linewidth=0.15](7,7.76)(10,4.85)
\psline[linewidth=0.15](10,4.85)(7,4.85)
\end{pspicture}
}}} \\ 
\hline $\langle\phi|\psi\rangle$  & $\left(\begin{array}{ccc}\bar{\phi}_1 & \ldots & \bar{\phi}_n\end{array}\right) \left(\begin{array}{c}\psi_1\\ \vdots\\ \psi_n\end{array}\right)$ & $\II\rTo{\psi}A\rTo{\phi^\dagger} \II$ & \ensuremath{\vcenter{\hbox{
\ifx\JPicScale\undefined\def\JPicScale{1}\fi
\psset{unit=\JPicScale mm}
\psset{linewidth=0.3,dotsep=1,hatchwidth=0.3,hatchsep=1.5,shadowsize=1,dimen=middle}
\psset{dotsize=0.7 2.5,dotscale=1 1,fillcolor=black}
\psset{arrowsize=1 2,arrowlength=1,arrowinset=0.25,tbarsize=0.7 5,bracketlength=0.15,rbracketlength=0.15}
\begin{pspicture}(0,0)(10,13)
\newrgbcolor{userFillColour}{0.8 0.8 0.8}
\psline[linewidth=0.15,fillcolor=userFillColour,fillstyle=solid](5,0)
(10,5)
(0,5)(5,0)
\psline(5,5)(5,10)
\rput(5,3){$\psi$}
\newrgbcolor{userFillColour}{0.8 0.8 0.8}
\psline[linewidth=0.15,fillcolor=userFillColour,fillstyle=solid](5,13)
(10,8)
(0,8)(5,13)
\rput(5,10){$\phi$}
\psline[linewidth=0.15](5,13)(3,11)
\psline[linewidth=0.15](5,13)(7,11)
\psline[linewidth=0.15](3,8)(7,8)
\psline[linewidth=0.15](3,5)(7,5)
\psline[linewidth=0.15](3,2)(5,0)
\psline[linewidth=0.15](5,0)(7,2)
\psline[linewidth=0.15](7,8)(10,8)
\psline[linewidth=0.15](10,8)(7,11)
\psline[linewidth=0.15](3,8)(0,8)
\psline[linewidth=0.15](0,8)(3,11)
\psline[linewidth=0.15](3,5)(0,5)
\psline[linewidth=0.15](0,5)(3,2)
\psline[linewidth=0.15](7,2)(10,5)
\psline[linewidth=0.15](10,5)(7,5)
\end{pspicture}
}}} \\ 
\hline $|\psi\rangle\langle\phi|$ & $\left(\begin{array}{c}\psi_1\\ \vdots\\ \psi_n\end{array}\right) \left(\begin{array}{ccc}\bar{\phi}_1 & \ldots & \bar{\phi}_n\end{array}\right)$ & $A\rTo{\phi^\dagger}\II\rTo{\psi} A$ & \ensuremath{\vcenter{\hbox{
\ifx\JPicScale\undefined\def\JPicScale{1}\fi
\psset{unit=\JPicScale mm}
\psset{linewidth=0.3,dotsep=1,hatchwidth=0.3,hatchsep=1.5,shadowsize=1,dimen=middle}
\psset{dotsize=0.7 2.5,dotscale=1 1,fillcolor=black}
\psset{arrowsize=1 2,arrowlength=1,arrowinset=0.25,tbarsize=0.7 5,bracketlength=0.15,rbracketlength=0.15}
\begin{pspicture}(0,0)(10,16)
\psline{->}(5,11)(5,16)
\newrgbcolor{userFillColour}{0.8 0.8 0.8}
\psline[linewidth=0.15,fillcolor=userFillColour,fillstyle=solid](5,8)
(10,13)
(0,13)(5,8)
\psline(5,0)(5,5)
\rput(5,11){$\psi$}
\newrgbcolor{userFillColour}{0.8 0.8 0.8}
\psline[linewidth=0.15,fillcolor=userFillColour,fillstyle=solid](5,8)
(10,3)
(0,3)(5,8)
\rput(5,5){$\phi$}
\psline[linewidth=0.15](5,8)(3,6)
\psline[linewidth=0.15](5,8)(7,6)
\psline[linewidth=0.15](3,3)(7,3)
\psline[linewidth=0.15](3,13)(7,13)
\psline[linewidth=0.15](3,10)(5,8)
\psline[linewidth=0.15](5,8)(7,10)
\psline[linewidth=0.15](7,3)(10,3)
\psline[linewidth=0.15](10,3)(7,6)
\psline[linewidth=0.15](3,3)(0,3)
\psline[linewidth=0.15](0,3)(3,6)
\psline[linewidth=0.15](3,13)(0,13)
\psline[linewidth=0.15](0,13)(3,10)
\psline[linewidth=0.15](7,10)(10,13)
\psline[linewidth=0.15](10,13)(7,13)
\end{pspicture}
}}} \\ \hline 
\end{tabular}
\end{center} 
In particular, note that in the language of strict dagger symmetric monoidal categories both a bra-ket and a ket-bra are compositions of morphisms, namely  $\phi^\dagger\circ\psi$ and $\psi\circ\phi^\dagger$ respectively.  What the diagrammatic calculus adds to standard Dirac notation is a second dimension to accommodate the monoidal composition: 
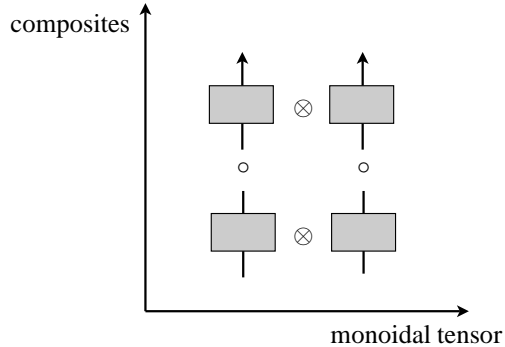
\begin{center}
\ifx\JPicScale\undefined\def\JPicScale{1}\fi
\psset{unit=\JPicScale mm}
\psset{linewidth=0.3,dotsep=1,hatchwidth=0.3,hatchsep=1.5,shadowsize=1,dimen=middle}
\psset{dotsize=0.7 2.5,dotscale=1 1,fillcolor=black}
\psset{arrowsize=1 2,arrowlength=1,arrowinset=0.25,tbarsize=0.7 5,bracketlength=0.15,rbracketlength=0.15}
\begin{pspicture}(0,0)(65.25,45.5)
\psline(25.25,17.5)(25.25,20.5)
\psline(25.25,9)(25.25,12.5)
\rput(51.98,16.5){}
\newrgbcolor{userFillColour}{0.8 0.8 0.8}
\pspolygon[linewidth=0.15,fillcolor=userFillColour,fillstyle=solid](21,17.5)(29.5,17.5)(29.5,12.5)(21,12.5)
\rput(45.87,32.62){}
\psline{->}(41.13,34.5)(41.13,39)
\psline(41.13,26)(41.13,29.5)
\newrgbcolor{userFillColour}{0.8 0.8 0.8}
\pspolygon[linewidth=0.15,fillcolor=userFillColour,fillstyle=solid](36.75,34.5)(45.25,34.5)(45.25,29.5)(36.75,29.5)
\psline(41.25,17.5)(41.25,20.5)
\psline(41.25,9.5)(41.25,13)
\newrgbcolor{userFillColour}{0.8 0.8 0.8}
\pspolygon[linewidth=0.15,fillcolor=userFillColour,fillstyle=solid](37,17.5)(45.5,17.5)(45.5,12.5)(37,12.5)
\psline{->}(25.13,34.5)(25.13,39)
\psline(25.13,26)(25.13,29.5)
\newrgbcolor{userFillColour}{0.8 0.8 0.8}
\pspolygon[linewidth=0.15,fillcolor=userFillColour,fillstyle=solid](20.75,34.5)(29.25,34.5)(29.25,29.5)(20.75,29.5)
\rput(41.25,23.5){$\circ$}
\rput(25.25,23.5){$\circ$}
\rput(65.25,13.5){}
\rput(33.25,14.5){$\otimes$}
\rput(60,8){}
\rput(33.25,31.5){$\otimes$}
\psline{->}(12.25,4.5)(55.25,4.5)
\psline{->}(12.25,4.5)(12.25,45.5)
\rput(2.25,42.5){composites}
\rput(48.25,1.5){monoidal tensor}
\end{pspicture}

\end{center} 
\noindent The advantages of this have already been made clear in the previous section and will even become clearer in Section \ref{sec:compact}.

Concerning the types of the morphisms in the third column of the above
table, recall that in Example \ref{ex:elements} we showed that the
vectors in Hilbert spaces ${\cal H}$ can be faithfully represented by linear
maps of type $\mathbb{C}\to{\cal H}$.  Similarly, complex numbers
$c\in\mathbb{C}$, that is, equivalently, vectors in the `one-dimensional
Hilbert space $\mathbb{C}$', can be faithfully represented by linear maps 
\[
s_c:\mathbb{C}\to\mathbb{C}::1\mapsto c\,,
\] 
since by linearity the image of $1$
fully specifies this map.


However, by making explicit reference to ${\bf FdHilb}$ and hence also by
having matrices (morphisms in ${\bf FdHilb}$ expressed relative to
some bases)  in the above table, we are actually cheating. The fact that Hilbert
spaces and linear maps are set-theoretic based mathematical structures has
non-trivial `unpleasant' implications.  In particular, while the $\otimes$-notation for the
monoidal structure of strict monoidal categories insinuates that the tensor
product would turn ${\bf FdHilb}$ into a strict symmetric monoidal category,
this turns out not to be  true in the `strict' sense of the word true.

\subsection{The set-theoretic verdict on strictness}

As outlined in Section \ref{Cats_physics}, we `model' real world
categories in terms of concrete categories.  While the real world categories
\em are \em indeed strict monoidal categories, their corresponding models
typically \em aren't\em.  

What goes wrong is the following: for set-theory
based mathematical structures such as groups, topological spaces, partial
orders and vector spaces, neither 
\[ 
A\otimes (B\otimes C)= (A\otimes B)\otimes C\quad\mbox{nor}\quad \II\otimes A=A=A\otimes \II 
\] 
hold. This is due to the
fact that  for the underlying sets $X,Y,Z$ we have that 
\[
(x,(y,z))\not=((x,y),z)
\qquad\mbox{and}\qquad(*,x)\not=x\not=(x,*)
\]
so, as a consequence,  neither 
\[ X\times (Y\times Z)= (X\times Y)\times Z\quad\mbox{nor}\quad \{*\}\times X=X=X\times
\{*\} 
\] 
hold.  We do have something very closely related to this, namely 
\[
X\times (Y\times Z)\simeq(X\times Y)\times Z\quad\mbox{and}\quad \{*\}\times
X\simeq X\simeq X\times \{*\}\,.  
\] 
That is, we have \em isomorphisms \em
rather than \em strict \em equations. But these isomorphisms are not just
ordinary isomorphisms but so-called \em natural \em isomorphisms.  They are an instance of the more general  \em natural transformations \em which we will discuss in Section
\ref{sec:naturaltransformations}.\footnote{Naturality is one of the most
important concepts of formal category theory. In fact, in the founding paper 
\cite{EilenbergMacLane} Eilenberg and MacLane argue that \em their \em main
motivation for introducing the notion of a category is to introduce the notion
of a functor, and that \em their \em main motivation  for introducing the
notion of a functor is to introduce the notion of a natural transformation.}
Meanwhile we introduce a  restricted version of this general notion of natural
transformation, one which comes with a clear interpretation.

Consider a category ${\bf C}$ that comes with an operation on objects
\beq\label{bifunctonob} 
-\otimes-:|{\bf C}|\times|{\bf C}|\to |{\bf C}|::(A,B)\mapsto A\otimes B\,, 
\eeq 
and with for all objects $A,B,C,D\in |{\bf C}|$ we also have an operation on hom-sets 
\beq\label{bifunctonmor} 
-\otimes -:{\bf C}(A,B)\times {\bf C}(C,D)\to{\bf C}(A\otimes C,B\otimes D):: (f,g)\mapsto
f\otimes g. 
\eeq 
Let 
\[ 
\Lambda(x_1,\mbox{\scriptsize\ldots},x_n, C_1, \mbox{\scriptsize\ldots},
C_m)\quad\mbox{and}\quad\Xi(x_1,\mbox{\scriptsize\ldots},x_n, C_1,
\mbox{\scriptsize\ldots}, C_m) 
\] 
be two well-formed expressions built from:
\bit 
\item $-\otimes -$, 
\item brackets, 
\item variables $x_1,\ldots,x_n$, 
\item  and constants $C_1, \ldots, C_m\in|\cat|$. 
\eit
Then a natural transformation is a family 
\[
\hspace{-1.8mm}\bigl\{\Lambda(A_1,\mbox{\scriptsize\ldots},A_n, C_1,
\mbox{\scriptsize\ldots}, C_m)\rTo^{\xi_{A_1,\ldots\!,A_n}}
\Xi(A_1,\mbox{\scriptsize\ldots},A_n, C_1, \mbox{\scriptsize\ldots}, C_m)\mid
A_1,\mbox{\scriptsize\ldots},A_n\in{\bf C} \bigr\} 
\] 
of morphisms which are such that for all objects $A_1,\ldots,A_n, B_1,\ldots,B_n\in |\cat|$ and all morphisms $A_1\rTo^{f_1} B_1\,,\,\ldots,\,A_n\rTo^{f_n} B_n$ we have: 
\[ 
\begin{diagram} 
\Lambda(A_1,\mbox{\scriptsize\ldots},A_n, C_1,
\mbox{\scriptsize\ldots}, C_m)  &\rTo^{\xi_{A_1,\ldots,A_n}}
&\Xi(A_1,\ldots,A_n, C_1, \ldots, C_m) \\ 
\dTo^{\Lambda(f_1,\ldots,f_n,
1_{C_1}, \ldots, 1_{C_m})}&&\dTo_{\Xi(f_1,\ldots,f_n, 1_{C_1}, \ldots,
1_{C_m})}\\
\Lambda(B_1,\ldots,B_n, C_1, \ldots, C_m)
&\rTo_{\xi_{B_1,\ldots,B_n}} &\Xi(B_1,\mbox{\scriptsize\ldots},B_n,
C_1, \mbox{\scriptsize\ldots}, C_m) 
\end{diagram} 
\]

A natural transformation is a \em natural  isomorphism \em if, in addition, all these
morphisms  $\xi_{A_1,\ldots\!,A_n}$  are isomorphisms in the sense of
Definition \ref{def:isomorphic}.

Examples of such well-formed expressions  are 
\[ 
 x\otimes(y\otimes z)
 \qquad\mbox{\rm and}\qquad 
(x\otimes y)\otimes z
\] 
and the corresponding  constraint on the morphims is 
\beq\label{assocdiag} 
\xymatrix@=0.6in{%
A\otimes (B\otimes C)\ar[d]_{f\otimes (g\otimes h)}\ar[r]^{\alpha_{A,B,C}} & (A\otimes B)\otimes
C\ar[d]^{(f\otimes g)\otimes h}\\%
A'\otimes (B'\otimes
C')\ar[r]_{\alpha_{A',B',C'}} & (A'\otimes B')\otimes C'} 
\eeq
If diagram (\ref{assocdiag}) commutes for all $A,B,C,A',B',C',f,g,h$ and the morphisms
\[ 
\alpha:=\{\alpha_{A,B,C}\mid A,B,C\in{\bf C}\} 
\] 
are all isomorphisms, then this natural isomorphism is called \em associativity\em.   Its name refers to the fact that this natural isomorphism embodies a \em weaker \em form of the strict
associative law $A\otimes (B\otimes C)= (A\otimes B)\otimes C$.  A better name
would actually be \em re-bracketing\em, since that is what it truly does: it is
a morphism ---which we like to think of as a \em process\em--- which transforms
type $A\otimes (B\otimes C)$ into type $(A\otimes B)\otimes C$.  In other
words, it provides a \em formal witness \em to the actual \em processes of
re-bracketing  a mathematical expression\em. The \em naturality condition \em
in diagram (\ref{assocdiag}) formally states that re-bracketing  \em commutes \em with any triple of operations $f,g,h$ we apply to the systems, and hence it tells us that the process of re-bracketing \em does not interfere \em with any  non-trivial processes $f,g,h$ --- almost as if it wasn't there.

Other important pairs of well-formed formal expressions are
\[
x\quad\mbox{\rm and}\quad c\otimes  x \qquad\qquad\qquad\qquad x\quad\mbox{\rm
and}\quad x\otimes c 
\] 
and, if $\II$ is taken to be the constant object, the corresponding  naturality constraint  is
\beq\label{unitdiag} 
\xymatrix@=0.6in{%
A\ar[r]^{\lambda_A}\ar[d]_{f} & \II\otimes A\ar[d]^{1_\II\otimes f} & & A\ar[r]^{\rho_A}\ar[d]_{f} & A\otimes\II\ar[d]^{f\otimes 1_\II}\\ 
B\ar[r]_{\lambda_B} & \II\otimes B & & B\ar[r]_{\rho_B} & B\otimes\II } 
\eeq 
The natural isomorphisms $\lambda$ and $\rho$ in  diagrams (\ref{unitdiag}) are called \em left- \em and \em right unit\em.
In this case, a better name would have been \em left- \em and \em right introduction \em since they correspond to the process of introducing a new object relative
to an existing one. 

We encountered a fourth important example in Definition \ref{excat10_5}, namely 
\[ 
x\otimes y\qquad\mbox{\rm and}\qquad y\otimes x\,,
\] 
for which  diagram (\ref{Eq:symdiag}) is the naturality condition.  The isomorphism $\sigma$ is called \em symmetry \em but a better name could have been \em exchange \em or \em swapping\em.

\begin{example}\label{Set_cart_natural_isos} 
The category ${\bf Set}$ has
associativity, left- and right unit, and symmetry natural isomorphisms
relative to the Cartesian product, with the singleton set $\{*\}$ as the monoidal unit.
Explicitly,  setting  
\[
f\times f': X\times X'\to Y\times Y'::(x,x')\mapsto (f(x), f'(x'))
\]
for $f:X\to Y$ and $f':X'\to Y'$, these natural isomorphisms are 
\[ 
\alpha_{X,Y,Z}: X\times (Y\times Z)\to  (X\times Y)\times Z:: (x,(y,z))\mapsto ((x,y),z) \] \[ \lambda_X:X\to\{*\}\times
X::x\mapsto(*,x) \qquad\qquad \rho_X:X\to X\times \{*\}::x\mapsto (x,*)
\] 
\[ 
\sigma_{X,Y}:X\times Y\to Y\times X::(x,y)\mapsto(y,x) 
\] 
The reader can easily verify  that diagrams (\ref{Eq:symdiag}),  (\ref{assocdiag}) and (\ref{unitdiag}) all commute.  Showing that bifunctoriality holds is somewhat more tedious.   
\end{example}

\begin{definition}\label{moncat} \em A {\em monoidal category} consists of the
following data: 
\begin{enumerate} 
\item a category $\cat$\,{\rm,} 
\item an object ${\rm I}\in|\cat|$\,,
\item a \em bifunctor \em $-\otimes-$, that is, an operation  both on objects
and on morphisms as in prescriptions (\ref{bifunctonob}) and (\ref{bifunctonmor}) above,
which moreover satisfies 
\[ 
(g\circ f)\otimes(k\circ h)=(g\otimes k)\circ(f\otimes h)\quad\mbox{\rm and}
\quad 1_A\otimes 1_B=1_{A\otimes B} 
\] 
for all $A,B\in|\cat|$ and all morphisms $f,g,h,k$ of
appropriate type\,, and
\item three natural isomorphisms
\[
\alpha=\{ A\otimes (B\otimes C)\rTo^{\alpha_{A,B,C}}
(A\otimes B)\otimes C \mid A, B, C\in|\cat|\}\,,
\]
\[
\lambda=\{A\rTo^{\lambda_A} \II\otimes A \mid
A\in|\cat|\}\quad\mbox{\rm and}
\quad\rho=\{A\rTo^{\rho_A}
A\otimes\II \mid A\in|\cat| \}\,,
\] 
hence satisfying eq.(\ref{assocdiag}) and eq.(\ref{unitdiag}), and such
that the \em Mac Lane pentagon \em
\beq\label{cohere1}
\xymatrix@=0.04in{
&(A\otimes B)\otimes(C\otimes D) \ar[rddd]^{\alpha_{-}}&\\ \\ \\
A\otimes (B\otimes (C\otimes D)) \ar[ruuu]^{\alpha_{-}}
\ar[dddd]_{1_A\otimes \alpha_{-}}&& ((A\otimes B)\otimes
C)\otimes D\\ \\ \\ \\
A\otimes ((B\otimes C)\otimes D)
\ar[rr]_{\alpha_{-}}& & (A\otimes (B\otimes C))\otimes
D \ar[uuuu]_{\alpha_{-}\otimes 1_D}} 
\eeq 

commutes for all $A,B,C,D\in|\cat|$, that also  
\beq\label{cohere2}
\xymatrix{A\otimes
B\ar[r]^{1_A\otimes\lambda_B\ \ \
}\ar[dr]_{\rho_A\otimes 1_B} & A\otimes (\II\otimes
B)\ar[d]^{\alpha_{A,\II,B}}\\
& (A\otimes\II)\otimes B
} 
\eeq 
commutes for all $A,B\in|\cat|$, and that
\beq\label{cohere2_5}
\lambda_\II=\rho_\II\,.  \eeq
\end{enumerate}
A monoidal category is moreover {\em symmetric} if there is a
fourth natural isomorphism 
\[ 
\sigma=\{A\otimes B\rTo^{\sigma_{A,B}} B\otimes A\mid A,B\in|\cat|\}\,, 
\] 
satisfying eq.(\ref{Eq:symdiag}), and such that  
\beq\label{cohere3} 
\xymatrix{%
A\otimes B\ar[r]^{\sigma_{A,B}} \ar[dr]_{1_{A\otimes B}} & B\otimes A \ar[d]^{\sigma_{B,A}}\\%
& A\otimes B} 
\eeq 
commutes for all $A,B\in|\cat|$, that
\beq\label{cohere3_5}
\hspace{-3mm}\xymatrix{%
A\ar[r]^{\lambda_A} \ar[dr]_{\rho_A}& \II\otimes A\ar[d]^{\sigma_{\II,A}}\\ %
& A\otimes
\II} 
\eeq 
commutes for all $A\in|\cat|$, and that
\beq\label{cohere4}
\xymatrix@C=.5in{A\otimes (B\otimes
C)\ar[r]^{\alpha_{-}}\ar[d]_{1_A\otimes\sigma_{B,C}} & (A\otimes B)\otimes C\ar[r]^{\sigma_{(A\otimes B),C}} & C\otimes (A\otimes B) \ar[d]^{\alpha_{-}}\\%
A\otimes (C\otimes B)\ar[r]_{\alpha_{-}} & (A\otimes C)\otimes B\ar[r]_{\sigma_{A,C}\otimes 1_B} & (C\otimes A)\otimes B} 
\eeq 
commutes for all $A,B,C\in|\cat|$\,.  
\end{definition}

The set-theoretic verdict on strictness is very hard!  The punishment is grave: a definition which stretches over two pages, since  we need to carry along associativity and unit natural isomorphisms,  which, on top of that,  are subject to a formal overdose of \em coherence conditions\em,
that is, eqs.(\ref{cohere1},\ref{cohere2},\ref{cohere2_5},\ref{cohere3},\ref{cohere4}).
They embody rules which should be obeyed when natural ismorphisms interact with
each other, in addition to the naturality conditions which state how natural
isomorphisms  interact with other morphisms in the category. For example,
eq.(\ref{cohere3_5}) tells us that if we introduce $\II$ on the left of $A$, and
then swap $\II$ and  $A$, that this should be the same as introducing $\II$ on
the right of $A$.  Eq.(\ref{cohere3_5}) tells us that the two ways of
re-bracketing the four variable expressions involved should be the same.  

The idea behind coherence conditions is as follows:  if for formal expressions
$\Lambda(A_1,\mbox{\scriptsize\ldots},A_n, C_1, \mbox{\scriptsize\ldots}, C_m)$ and
 $\Xi(A_1,\mbox{\scriptsize\ldots},A_n, C_1,
\mbox{\scriptsize\ldots}, C_m)$ there are two morphisms
\[
\Lambda(A_1,\mbox{\scriptsize\ldots},A_n, C_1, \mbox{\scriptsize\ldots}, C_m)\rTo^{f,g}\Xi(A_1,\mbox{\scriptsize\ldots},A_n, C_1,
\mbox{\scriptsize\ldots}, C_m)
\]
which are obtained by composing the natural isomorphisms $\alpha$, $\sigma$, $\lambda$, $\rho$  and $1$ both with $-\otimes-$ and $-\circ-$, then $f=g$ --  
identities are indeed natural isomorphisms, for the formal expressions
$\Lambda(A)=\Xi(A)=A$. That
eqs.(\ref{cohere1},\ref{cohere2},\ref{cohere2_5},\ref{cohere3},\ref{cohere4})
suffice for this purpose is in itself remarkable.  This is a the consequence of MacLane's highly non-trivial
coherence theorem for symmetric monoidal categories \cite{SML}, which states that from this set of equations we can derive any other one.  

If it wasn't for this theorem, things could have been even worse, potentially involving equations with an unbounded number of symbols.\\

Pf\/f\/f\/f\/f\/f\/f\/f\/f\/f\/f\/f\/f\/f\/f\/f\/f\/f\/f\/f\/f\/f\/f . . . 

\bigskip . . .\ \ sometimes miracles do happen:

\begin{theorem}[Strictification \cite{SML} p.257] Any monoidal category ${\bf
C}$ is categorically equivalent, via a pair of strong monoidal
functors $G:{\bf C}\longrightarrow {\bf D}$ and $F:{\bf D}\longrightarrow {\bf C}$, to a
strict monoidal category ${\bf D}$. 
\end{theorem}

The definitions of categorical equivalence and strong monoidal functor can be found below in  Section \ref{sec:monoidal_functors}.  In words, what this means is that for practical purposes, arbitrary monoidal categories behave the same as strict monoidal categories.  In
particular, the connection between diagrammatic reasoning (incl.~Dirac notation)
and axiomatic reasoning for strict monoidal categories extends to arbitrary
monoidal categories.  The essence of the above theorem is that the unit and
associativity isomorphims are so well-behaved that they don't affect this
correspondence. In the graphical calculus, the associativity natural isomorphisms becomes
implicit when we write 
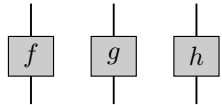
\begin{center} 
\ifx\JPicScale\undefined\def\JPicScale{1}\fi
\psset{unit=\JPicScale mm}
\psset{linewidth=0.3,dotsep=1,hatchwidth=0.3,hatchsep=1.5,shadowsize=1,dimen=middle}
\psset{dotsize=0.7 2.5,dotscale=1 1,fillcolor=black}
\psset{arrowsize=1 2,arrowlength=1,arrowinset=0.25,tbarsize=0.7 5,bracketlength=0.15,rbracketlength=0.15}
\begin{pspicture}(0,0)(28,13.37)
\psline[linewidth=0.25](3,9)(3,13.37)
\psline[linewidth=0.25](3,-0)(3,3.75)
\newrgbcolor{userFillColour}{0.8 0.8 0.8}
\pspolygon[linewidth=0.15,fillcolor=userFillColour,fillstyle=solid](0,3.72)(6,3.72)(6,9)(0,9)
\rput(3,6.5){$f$}
\psline[linewidth=0.25](14,9)(14,13.37)
\psline[linewidth=0.25](14,-0)(14,3.75)
\newrgbcolor{userFillColour}{0.8 0.8 0.8}
\pspolygon[linewidth=0.15,fillcolor=userFillColour,fillstyle=solid](11,3.72)(17,3.72)(17,9)(11,9)
\rput(14,6.5){$g$}
\psline[linewidth=0.25](25,9)(25,13.37)
\psline[linewidth=0.25](25,-0)(25,3.75)
\newrgbcolor{userFillColour}{0.8 0.8 0.8}
\pspolygon[linewidth=0.15,fillcolor=userFillColour,fillstyle=solid](22,3.72)(28,3.72)(28,9)(22,9)
\rput(25,6.5){$h$}
\end{pspicture}
 \end{center} 
\noindent The absence of any brackets means that we can interpret this picture either as

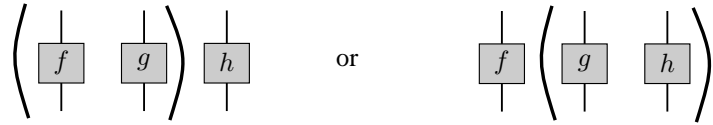
\begin{center} 
\ifx\JPicScale\undefined\def\JPicScale{1}\fi
\psset{unit=\JPicScale mm}
\psset{linewidth=0.3,dotsep=1,hatchwidth=0.3,hatchsep=1.5,shadowsize=1,dimen=middle}
\psset{dotsize=0.7 2.5,dotscale=1 1,fillcolor=black}
\psset{arrowsize=1 2,arrowlength=1,arrowinset=0.25,tbarsize=0.7 5,bracketlength=0.15,rbracketlength=0.15}
\begin{pspicture}(0,0)(93.88,16.27)
\psline[linewidth=0.25](7,10.27)(7,14.65)
\psline[linewidth=0.25](7,1.27)(7,5.02)
\newrgbcolor{userFillColour}{0.8 0.8 0.8}
\pspolygon[linewidth=0.15,fillcolor=userFillColour,fillstyle=solid](4,5)(10,5)(10,10.27)(4,10.27)
\rput(7,7.77){$f$}
\psline[linewidth=0.25](18,10.27)(18,14.65)
\psline[linewidth=0.25](18,1.27)(18,5.02)
\newrgbcolor{userFillColour}{0.8 0.8 0.8}
\pspolygon[linewidth=0.15,fillcolor=userFillColour,fillstyle=solid](15,5)(21,5)(21,10.27)(15,10.27)
\rput(18,7.77){$g$}
\psline[linewidth=0.25](29,10.27)(29,14.65)
\psline[linewidth=0.25](29,1.27)(29,5.02)
\newrgbcolor{userFillColour}{0.8 0.8 0.8}
\pspolygon[linewidth=0.15,fillcolor=userFillColour,fillstyle=solid](26,5)(32,5)(32,10.27)(26,10.27)
\rput(29,7.77){$h$}
\psbezier[linewidth=0.45](2.23,15.53)(0.29,8.78)(0.29,7.8)(2.66,0.33)
\psbezier[linewidth=0.45](21,15.27)(23.88,8.43)(23.74,7.53)(21.43,0.07)
\rput(45,8){or}
\rput(35,-1){}
\rput(5,16.27){}
\psline[linewidth=0.25](65.57,10.2)(65.57,14.58)
\psline[linewidth=0.25](65.57,1.2)(65.57,4.95)
\newrgbcolor{userFillColour}{0.8 0.8 0.8}
\pspolygon[linewidth=0.15,fillcolor=userFillColour,fillstyle=solid](62.57,4.93)(68.57,4.93)(68.57,10.2)(62.57,10.2)
\rput(65.57,7.7){$f$}
\psline[linewidth=0.25](76.57,10.2)(76.57,14.58)
\psline[linewidth=0.25](76.57,1.2)(76.57,4.95)
\newrgbcolor{userFillColour}{0.8 0.8 0.8}
\pspolygon[linewidth=0.15,fillcolor=userFillColour,fillstyle=solid](73.57,4.93)(79.57,4.93)(79.57,10.2)(73.57,10.2)
\rput(76.57,7.7){$g$}
\psline[linewidth=0.25](87.57,10.2)(87.57,14.58)
\psline[linewidth=0.25](87.57,1.2)(87.57,4.95)
\newrgbcolor{userFillColour}{0.8 0.8 0.8}
\pspolygon[linewidth=0.15,fillcolor=userFillColour,fillstyle=solid](84.57,4.93)(90.57,4.93)(90.57,10.2)(84.57,10.2)
\rput(87.57,7.7){$h$}
\psbezier[linewidth=0.45](72.23,15.26)(70.29,8.51)(70.29,7.53)(72.66,0.06)
\psbezier[linewidth=0.45](91,15)(93.88,8.16)(93.74,7.26)(91.43,-0.2)
\rput(63.57,16.2){}
\end{pspicture}
 \end{center}

\noindent That is, it does not matter whether in first order we want to associate $f$ with $g$, and then in second order this pair as a whole with $h$, or whether in first order we want to associate $g$ with $h$, and then in second order this pair as  a whole with $f$.

So things turn out not to be as bad as they looked at first sight! 

\begin{example}\label{loads_of_examples} 
The category ${\bf Set}$ admits two
important symmetric monoidal structures.  We discussed the Cartesian product in Example \ref{Set_cart_natural_isos}. The other one is the \em disjoint union\em. Given two sets $X$ and $Y$ their disjoint union 
is the set 
\begin{equation}\nonumber
X+Y:=\{(x,1)\ |\ x\in X\}\cup\{(y,2)\ |\ y\in Y\}.
\end{equation}
This set can be thought of as the set of all elements both of $X$ and $Y$, but where the elements of $X$ are ``coloured'' with 1 while those of $Y$ are ``coloured'' with 2.  This guarantees  that, when the same element occurs both in $X$ and $Y$, it is twice accounted for in $X+Y$ since the ``colours'' 1 and 2 recall whether the elements in $X+Y$ either originated in $X$ or in $Y$.  As a consequence, the intersection of $\{(x,1)\ |\ x\in X\}$ and $\{(y,2)\ |\ y\in Y\}$ is empty, hence the name `disjoint' union. 

For the disjoint union, we take the empty set $\emptyset$ as the monoidal unit and set
\[
f+ f': X + X'\to Y + Y'::
\left\{\begin{array}{ll}
(x,1)\mapsto(f(x),1)\\
(x,2)\mapsto(f'(x),2)
\end{array}\right.
\]
for $f:X\to Y$ and $f':X'\to Y'$. The natural isomorphisms of the symmetric monoidal structure are
\[ 
\alpha_{X,Y,Z}: X + (Y + Z)\to  (X + Y) + Z:: 
\left\{\begin{array}{ll}
(x,1)\mapsto((x,1),1)\\
((x,1),2)\mapsto ((x,2),1)\\
((x,2),2)\mapsto (x,2)
\end{array}\right.\,.
\] 
\[ 
\lambda_X:X\to\emptyset+X::x\mapsto (x,2) \qquad\qquad \rho_X:X\to X+\emptyset::x\mapsto (x,1)
\] 
\[ 
\sigma_{X,Y}:X+ Y\to Y+ X::(x,i)\mapsto(x,3-i) 
\] 
One again easily verifies  that diagrams (\ref{assocdiag}), (\ref{unitdiag}) and (\ref{Eq:symdiag}) all commute.  Showing that bifunctoriality holds is again somewhat more tedious.   
\end{example}

\begin{example} 
The category ${\bf FdVect}_\mathbb{K}$ also admits two symmetric monoidal
structures, provided respectively by the tensor product $\otimes$ and by the
direct sum $\oplus$.  

For the tensor product, the monoidal unit is the underlying field
$\mathbb{K}$, while the natural isomorphisms of the monoidal structure are given by
\[ 
\alpha_{V_1,V_2,V_3}:V_1\otimes (V_2\otimes V_3)\to (V_1\otimes V_2)\otimes V_3:: v'\otimes (v''\otimes v''')\mapsto (v'\otimes v'')\otimes v'''
\] 
\[ 
\lambda_V: V\to\mathbb{K}\otimes V::v\mapsto 1\otimes v \qquad\qquad \rho_V: V\to V\otimes \mathbb{K}::v\mapsto v\otimes 1
\] 
\[ 
\sigma_{V_1,V_2}:V_1\otimes V_2\to V_2\otimes V_1::v'\otimes v''\mapsto v''\otimes v'\,.
\] 
Note that  the inverse to $\lambda_V$ is 
\[
\lambda_V^{-1}:\mathbb{K}\otimes V\to V:: k\otimes v\mapsto k\cdot v\,.
\]
The `scalars'  are provided by the field $\mathbb{K}$ itself, since it is in bijective correspondence with the linear maps from $\mathbb{K}$ to itself.  We leave it to the reader to verify that this defines a monoidal structure.  

On the other hand, the monoidal unit for the direct sum is the $0$-dimensional vector space. Hence this monoidal structure only  admits a single `scalar'.  The following subsection discusses scalars in more detail.
\end{example} 

\begin{definition}\em 
A \em dagger monoidal category \em ${\bf C}$ is a
monoidal category which comes with an identity-on-objects contravariant involutive
functor 
\[
\dagger:{\bf C}^{op}{\longrightarrow}{\bf C}
\]
satisfying eq.(\ref{eq:daggermonoidal}), and for which all unit and associativity
natural isomorphisms are unitary. A \em dagger symmetric monoidal
category \em ${\bf C}$ is both a dagger monoidal category and a symmetric monoidal category, in which the symmetry natural isomorphism is also unitary.
\end{definition}

\begin{example} 
The category ${\bf FdHilb}$ admits two dagger symmetric monoidal structures, respectively provided by the tensor product and by the direct sum. In both cases,  the adjoint of Example \ref{ex:dagger} is the dagger functor.    
\end{example} 


\begin{example} 
As we will see in great detail in Sections \ref{sect:rel} and \ref{sec:biprod}, the category ${\bf Rel}$ which has sets as objects and relations as morphisms also admits two symmetric monoidal structures, just like ${\bf Set}$:  these are again the Cartesian product and the disjoint union.  Moreover, ${\bf Rel}$ is  dagger symmetric monoidal relative to both  monoidal structures with the relational converse as the dagger functor.  This is a first very important difference between ${\bf Rel}$ and ${\bf Set}$, since the latter does not admit a dagger functor for either of the monoidal structures we identified on it.
\end{example}

\begin{example} 
The category ${\bf 2Cob}$ has  1-dimensional closed manifolds as objects, and 2-dimensional cobordisms between these as morphisms, it is dagger symmetric monoidal with the disjoint union of manifolds as its monoidal product and with the reversal of cobordisms as the dagger. This category will be discussed in great detail in Section~\ref{sect:2cob}. 
\end{example} 




Of course, in ${\bf FdHilb}$ the tensor product $\otimes$ and the direct sum
$\oplus$ are very different monoidal structures as exemplified by the
particular role each of these plays within quantum theory.  In particular, as
pointed out by Schr\"odinger in the 1930's \cite{Schrodinger}, the tensor product description of
compound quantum systems is what makes quantum physics so different from
classical physics.  We will refer to monoidal structures which are somewhat
like $\otimes$ in ${\bf FdHilb}$ as \em quantum-like\em, and to those that
are rather like $\oplus$ in ${\bf FdHilb}$ as \em classical-like\em.  As we will
see below, the quantum-like tensors allow for correlations between subsystems,
so the joint state can in general not be decomposed into states of the individual
subsystems. In contrast, the classical-like tensors can only describe
`separated' systems, that is, the state of a joint system can always be
faithfully represented by states of the individual subsystems. 

The tensors considered in this paper have the following nature:
\begin{center}
\begin{tabular}{|c|c|c|c|} \hline \ {\bf category} \  & \ {\bf classical-like}
\ & \ {\bf quantum-like} \ & \ {\bf other} (see \S
\ref{sec:coprod}) \ \\ \hline\hline {\bf Set} & $\times$ & &
$+$ \\ \hline {\bf Rel} & $+$ & $\times$ & \\ \hline {\bf
FdHilb} & $\oplus$ & $\otimes$ & \\ \hline {\bf nCob} &  & $+$
& \\ \hline 
\end{tabular}
\end{center}  

Observe the following remarkable facts: 
\bit 
\item While  $\times$ behaves `classical-like' in ${\bf Set}$, it behaves `quantum-like' in  ${\bf Rel}$, and this
despite the fact that ${\bf Rel}$ contains ${\bf Set}$ as a subcategory with the same objects as ${\bf Rel}$, and which inherits its monoidal structures from ${\bf Rel}$.  
\item There is a remarkable parallel between the role that the pair $(\oplus,\otimes)$ plays for ${\bf FdHilb}$ and the role that the pair
$(+,\times)$ plays for ${\bf Rel}$.  
\item In ${\bf nCob}$ the direct sum even becomes `quantum-like' --- a point which has been strongly
emphasized for a while by John Baez \cite{B2004}.  
\eit 
All of this clearly indicates that being either quantum-like and classical-like is something that involves not just the objects, but also the tensor and the morphism structure.

Sections \ref{sec:quantumlike} and \ref{sec:classicallike} provide a detailed
discussion of  these two very distinct kinds of monoidal structures,
which will shed more light on the above table.  

To avoid confusion concerning which monoidal structure on a category we are considering, we may specify it e.g., $({\bf FdHilb},\otimes,\mathbb{C})$.

\subsection{Scalar valuation and multiples}\label{sec:scalars}

In any monoidal category ${\bf C}$ the hom-set $\mathbb{S}_{\bf C}:={\bf
C}(\II,\II)$ is always a monoid with categorical composition as monoid
multiplication.  Therefore we call $\mathbb{S}_{\bf C}$ the \em scalar monoid
\em of the monoidal category $\cat$.   Such a monoid equips any monoidal category with explicit
quantitative content. For instance, if $\cat$ is dagger monoidal, scalars can
be produced  in terms of  the inner-product of Definition \ref{def:unitarity}.

The following is a fascinating fact discovered by Kelly and Laplaza in \cite{KellyLaplaza}:  even for ``non-symmetric'' monoidal categories, the scalar
monoid is always commutative. The proof is given by the following
commutative diagram: 
\beq\label{KellyDiagram}
\raisebox{2.1cm}{\xymatrix @=.6in{%
\II \ars[r]^{\simeq} & \II\otimes\II \ar@{=}[r] & \II\otimes\II\ar@{=}[r] & \II\otimes\II\ar@{-|>}[r]^{\simeq} & \II \\ 
\II\ars[u]^t \ar@{-|>}[r]^{\simeq} & \II\otimes\II\ar@{|>}[u]|{1_{\II}\otimes t} & & \II\otimes\II\ar@{-|>}[u]|{s\otimes 1_\II}\ar@{-|>}[r]^{\simeq} & \II \ars[u]_{s}\\ 
\II \ars[r]_{\simeq} \ars[u]^s & \II\otimes\II \ar@{=}[r] \ars[u]|{s\otimes 1_\II}& \II\otimes\II\ar@{=}[r] \ars[uu]|{s\otimes t}&
\II\otimes\II\ars[r]_{\simeq} \ars[u]|{1_\II\otimes t}& \II \ars[u]_t }}
\eeq
Equality of the two outer paths both going from the left-lower-corner to the
right-upper-corner boils down to equality between: 
\bit 
\item the outer left/upper path which consists of $t\circ s$, and the composite of
isomorphism $\II\simeq\II\otimes\II$ with its inverse,  so nothing but $1_\II$, giving all
together $t\circ s$, and
\item the outer lower/right path, giving all together  $s\circ t$.  
\eit
Their equality relies on bifunctoriality (cf.~middle two rectangles) and
naturality of the left- and right-unit isomorphisms (cf.~the four squares). 

Diagrammatically commutativity  is subsumed by the fact that scalars do not
have wires, and hence can `move freely around in the picture': 

\begin{center} 
\ifx\JPicScale\undefined\def\JPicScale{1}\fi
\psset{unit=\JPicScale mm}
\psset{linewidth=0.3,dotsep=1,hatchwidth=0.3,hatchsep=1.5,shadowsize=1,dimen=middle}
\psset{dotsize=0.7 2.5,dotscale=1 1,fillcolor=black}
\psset{arrowsize=1 2,arrowlength=1,arrowinset=0.25,tbarsize=0.7 5,bracketlength=0.15,rbracketlength=0.15}
\begin{pspicture}(0,0)(42.41,18)
\newrgbcolor{userFillColour}{0.8 0.8 0.8}
\pspolygon[linewidth=0.15,fillcolor=userFillColour,fillstyle=solid](7.41,15.08)(5.01,12.5)(2.5,15.1)(4.9,17.68)
\rput(5,15){$s$}
\newrgbcolor{userFillColour}{0.8 0.8 0.8}
\pspolygon[linewidth=0.15,fillcolor=userFillColour,fillstyle=solid](7.41,7.08)(5.01,4.5)(2.5,7.1)(4.9,9.68)
\rput(5,7){$t$}
\rput(12,11){$=$}
\rput(1,18){}
\newrgbcolor{userFillColour}{0.8 0.8 0.8}
\pspolygon[linewidth=0.15,fillcolor=userFillColour,fillstyle=solid](42.41,7.08)(40.01,4.5)(37.5,7.1)(39.9,9.68)
\rput(40,7){$s$}
\newrgbcolor{userFillColour}{0.8 0.8 0.8}
\pspolygon[linewidth=0.15,fillcolor=userFillColour,fillstyle=solid](42.41,15.08)(40.01,12.5)(37.5,15.1)(39.9,17.68)
\rput(40,15){$t$}
\rput(33,11){$=$}
\newrgbcolor{userFillColour}{0.8 0.8 0.8}
\pspolygon[linewidth=0.15,fillcolor=userFillColour,fillstyle=solid](22.41,11.08)(20.01,8.5)(17.5,11.1)(19.9,13.68)
\rput(20,11){$s$}
\newrgbcolor{userFillColour}{0.8 0.8 0.8}
\pspolygon[linewidth=0.15,fillcolor=userFillColour,fillstyle=solid](28.41,11.08)(26.01,8.5)(23.5,11.1)(25.9,13.68)
\rput(26,11){$t$}
\end{pspicture}
 \end{center}

This result has physical consequences.  Above we argued that strict monoidal
categories model physical systems and processes thereon.  We now discovered
that a strict monoidal category ${\bf C}$ always has a commutative
endomorphism monoid $\mathbb{S}_{\bf C}$.  
So when varying
quantum theory by changing the underlying field $\mathbb{K}$ of the vector
space, we need to restrict ourselves to commutative fields, hence excluding
things like `quaternionic quantum mechanics' \cite{Finkelstein}. 

\begin{example} 
We already saw that the elements of $\mathbb{S}_{({\bf FdHilb},\otimes, \mathbb{C})}$ are in bijective correspondence  with those of $\mathbb{C}$, in short, 
\[
\mathbb{S}_{({\bf FdHilb},\otimes, \mathbb{C})}\simeq \mathbb{C}\,.
\]
In ${\bf Set}$ however, since there is only one
function of type $\{*\}\to\{*\}$, namely the identity, $\mathbb{S}_{({\bf Set},\times,\{*\})}$ is a singleton, in short,
\[
\mathbb{S}_{({\bf Set},\times,\{*\})}\simeq\{*\}\,.
\]
 Thus, the scalar
structure on $({\bf Set},\times,\{*\})$ is trivial. On the other hand, in ${\bf Rel}$
there are two relations of type $\{*\}\to\{*\}$, the identity and the
empty relation,  so 
\[
\mathbb{S}_{({\bf Rel},\times,\{*\})}\simeq\mathbb{B}\,, 
\]
where $\mathbb{B}$ are the Booleans.  Hence, the  scalar
structure on $({\bf Rel},\times,\{*\})$ is non-trivial as it is that of
Boolean logic.  Operationally, we can interpret these two scalars as `possible' and `impossible' respectively.   When rather considering $\oplus$ on ${\bf FdHilb}$ instead of 
$\otimes$ we again have a trivial scalar structure, since there is only
one linear map from the $0$-dimensional Hilbert space to itself.  So
\[
\mathbb{S}_{({\bf FdHilb},\oplus, \mathbb{C})}\simeq \{*\}\,.
\]
So  scalars and scalar multiples are more closely related to the `multiplicative' tensor product structure than to the `additive' direct sum structure.    We also have
\[
\mathbb{S}_{({\bf nCob},+,\emptyset)}\simeq\mathbb{N}\,.  
\]
In general, it is the quantum-like monoidal structures which admit non-trivial scalar structure.  
This might come as a surprise to the reader, given that for vector spaces one typically associates these scalars with linear combinations of vectors, which are very much `additive' in spirit.
\end{example}

The right half of  commutative diagram (\ref{KellyDiagram}) states that 
\begin{diagram}
&s\circ t=\II&\rTo^{\simeq}&\II\otimes\II&\rTo^{s\otimes
t}&\II\otimes\II&\rTo^{\simeq}&\II\,.\quad 
\end{diagram} 
We  generalize this by defining  \em scalar multiples \em of a morphism $A\rTo{f} B$
as 
\begin{diagram} 
&s\bullet f:=A&\rTo^{\simeq}&\II\otimes A&\rTo^{s\otimes f}&\II\otimes B&\rTo^{\simeq}&B\,.\quad
\end{diagram} 
These scalars satisfy the usual properties, namely
\beq\label{eq:tensorlaw1}
(t\bullet g)\circ(s\bullet f)=(t\circ s)\bullet(g\circ f) 
\eeq
and 
\beq\label{eq:tensorlaw2}
(s\bullet f)\otimes (t\bullet g)=(s\circ t)\bullet(f\otimes g)\,, 
\eeq
cf.~in matrix calculus we have 
\[ 
 \left( y\left(\begin{array}{cc}b_{11}&b_{12}\\
b_{21}&b_{22}\end{array}\right) \right) \left( x\left(\begin{array}{cc}a_{11}&a_{12}\\ a_{21}&a_{22}\end{array}\right)
\right)=
yx\left(
\left(\begin{array}{cc}b_{11}&b_{12}\\
b_{21}&b_{22}\end{array}\right)\left(\begin{array}{cc}a_{11}&a_{12}\\
a_{21}&a_{22}\end{array}\right)\right)
\] 
and 
\[ \left(
x\left(\begin{array}{cc}a_{11}&a_{12}\\
a_{21}&a_{22}\end{array}\right)
\right) \otimes \left(
y\left(\begin{array}{cc}b_{11}&b_{12}\\
b_{21}&b_{22}\end{array}\right)
\right) =
xy\left(\left(\begin{array}{cc}a_{11}&a_{12}\\ 
a_{21}&a_{22}\end{array}\right)
\otimes \
\left(\begin{array}{cc}b_{11}&b_{12}\\
b_{21}&b_{22}\end{array}\right)
\right)\,.
\]
Diagrammatically these properties are again implicit and require `artificial' brackets to be made explicit, for example, eq.(\ref{eq:tensorlaw1}) is hidden as:

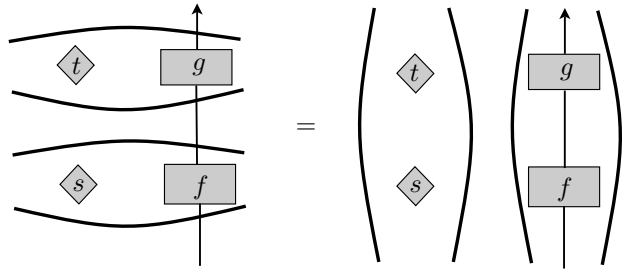
\begin{center} 
\ifx\JPicScale\undefined\def\JPicScale{1}\fi
\psset{unit=\JPicScale mm}
\psset{linewidth=0.3,dotsep=1,hatchwidth=0.3,hatchsep=1.5,shadowsize=1,dimen=middle}
\psset{dotsize=0.7 2.5,dotscale=1 1,fillcolor=black}
\psset{arrowsize=1 2,arrowlength=1,arrowinset=0.25,tbarsize=0.7 5,bracketlength=0.15,rbracketlength=0.15}
\begin{pspicture}(0,0)(85.35,35.17)
\psbezier[linewidth=0.45](71.95,34.4)(69.25,19.4)(69.25,17.21)(72.55,0.61)
\psbezier[linewidth=0.45](51.65,34.56)(48.95,19.56)(48.95,17.37)(52.25,0.77)
\psbezier[linewidth=0.45](81.35,34.2)(85.35,19)(85.15,17)(81.95,0.39)
\psbezier[linewidth=0.45](61.55,34.56)(65.55,19.37)(65.35,17.37)(62.15,0.77)
\rput(42.6,18.5){$=$}
\newrgbcolor{userFillColour}{0.8 0.8 0.8}
\pspolygon[linewidth=0.15,fillcolor=userFillColour,fillstyle=solid](72.15,8.12)(81.6,8.12)(81.6,13.41)(72.15,13.41)
\rput(76.97,10.36){$f$}
\psline[linewidth=0.25](76.87,-0.11)(76.87,8.2)
\newrgbcolor{userFillColour}{0.8 0.8 0.8}
\pspolygon[linewidth=0.15,fillcolor=userFillColour,fillstyle=solid](23.75,8.48)(33.2,8.48)(33.2,13.75)(23.75,13.75)
\rput(28.57,10.71){$f$}
\psline[linewidth=0.25](28.47,0.25)(28.47,8.55)
\newrgbcolor{userFillColour}{0.8 0.8 0.8}
\pspolygon[linewidth=0.15,fillcolor=userFillColour,fillstyle=solid](23.26,24.32)(32.72,24.32)(32.72,29.02)(23.26,29.02)
\rput(28.39,26.67){$g$}
\psline[linewidth=0.25]{->}(28.09,29.07)(28.09,35.17)
\psline[linewidth=0.25](28.08,13.75)(27.98,24.25)
\newrgbcolor{userFillColour}{0.8 0.8 0.8}
\pspolygon[linewidth=0.15,fillcolor=userFillColour,fillstyle=solid](72.15,23.71)(81.61,23.71)(81.61,28.4)(72.15,28.4)
\rput(77.28,26.06){$g$}
\psline[linewidth=0.25]{->}(76.98,28.46)(76.98,34.56)
\psline[linewidth=0.25](76.97,13.26)(76.97,23.56)
\psbezier[linewidth=0.45](34.56,8.19)(20.06,4.69)(17.66,4.69)(3.76,7.89)
\psbezier[linewidth=0.45](34.26,23.69)(19.76,20.19)(17.36,20.19)(3.46,23.39)
\psbezier[linewidth=0.45](34.36,15.29)(20.06,17.39)(18.06,17.49)(3.56,14.99)
\psbezier[linewidth=0.45](33.86,30.39)(19.56,32.49)(17.56,32.59)(3.06,30.09)
\newrgbcolor{userFillColour}{0.8 0.8 0.8}
\pspolygon[linewidth=0.15,fillcolor=userFillColour,fillstyle=solid](59.56,25.84)(57.16,23.26)(54.65,25.87)(57.05,28.44)
\newrgbcolor{userFillColour}{0.8 0.8 0.8}
\pspolygon[linewidth=0.15,fillcolor=userFillColour,fillstyle=solid](59.56,10.84)(57.16,8.26)(54.65,10.86)(57.05,13.44)
\rput(57.05,25.87){$t$}
\rput(53.15,13.76){}
\newrgbcolor{userFillColour}{0.8 0.8 0.8}
\pspolygon[linewidth=0.15,fillcolor=userFillColour,fillstyle=solid](14.4,26.95)(12,24.37)(9.49,26.97)(11.89,29.55)
\newrgbcolor{userFillColour}{0.8 0.8 0.8}
\pspolygon[linewidth=0.15,fillcolor=userFillColour,fillstyle=solid](14.77,11.07)(12.37,8.49)(9.86,11.09)(12.26,13.67)
\rput(12.06,26.89){$t$}
\rput(58.15,8.76){}
\rput(12.36,10.99){$s$}
\rput(57.15,10.86){$s$}
\end{pspicture}
 
\end{center} 

\noindent Of course, we could still prove these properties with commutative diagrams.  For  eq.(\ref{eq:tensorlaw1}) the left-hand-side and the right-hand-side are respectively the top and the bottom path of the following diagram:
\[
\xymatrix@=.28in{ 
& \II\otimes B\ar[dr]^{\rho_\II\otimes 1_B} \ar[rr]^{1_\II\otimes 1_B}&  & \II\otimes B\ar[dr]^{t\otimes g} & & \\ 
A\simeq \II\otimes A\ar[ur]^{s\otimes f}\ar[dr]_{\rho_\II\otimes 1_A} & & (\II\otimes\II)\otimes B\ar[ur]^{\lambda_\II^{-1}\otimes 1_B}\ar[dr]^{(1_\II\otimes t)\otimes g} & & \II\otimes C\simeq C\\ 
& (\II\otimes\II)\otimes A\ar[ur]^{(s\otimes 1_\II)\otimes f}\ar[rr]_{(s\otimes t)\otimes (g\circ f)} & & (\II\otimes\II)\otimes C \ar[ur]_{\lambda_\II^{-1} \otimes 1_C } }
\]
where we use the fact that $t\circ s=\lambda^{-1}_\II\circ(s\otimes t)\circ\rho_\II$. The diamond on the left
commutes by naturality of $\rho_\II$. The top triangle commutes because both paths are equal to $1_{\II\otimes B}$ as $\lambda_\II=\rho_\II$. The bottom triangle commutes by eq.(\ref{eq:SMCsliding}). Finally, the right diamond commutes by naturality of $\lambda_\II$. 

\section{Quantum-like tensors}\label{sec:quantumlike} 

So what makes $\otimes$ so different from $\oplus$ in the category ${\bf FdHilb}$, what makes
$\times$ so different in the categories ${\bf Rel}$ and ${\bf Set}$, and what
makes $\times$ so similar in the category ${\bf Rel}$ to $\otimes$ in the
category ${\bf FdHilb}$?  

\subsection{Compact categories}\label{sec:compact} 

\begin{definition}\label{coclc}\em A {\em compact (closed) category} $\cat$ is
a symmetric monoidal category in which  every object $A\in|\cat|$ comes with
\begin{enumerate} 
\item another  object $A^*$, the  {\em dual}
of $A$, 
\item a pair of morphisms 
\[
\II\rTo^{\eta_A} A^*\otimes A\qquad\mbox{\rm
and}\qquad A\otimes A^*\rTo^{\epsilon_A} \II\,, 
\]
respectively called \em unit \em and \em counit\em, 
\end{enumerate}
which are such that the following two diagrams commute:
\beq\label{diag:compactness1}
\xymatrix@=.6in{
A\ar[d]_{1_A}\ar[r]^{\rho_A\ \   } & A\otimes \II\ar[r]^{1_A\otimes\eta_A\ \ \ \ }& A\otimes (A^*\otimes A)\ar[d]^{\alpha_{A,A^*,A}}\\ 
A & \II\otimes A\ar[l]^{\lambda^{-1}_A} & (A\otimes A^*)\otimes A\ar[l]^{\epsilon_A\otimes 1_A\ \ \ }} 
\eeq
\beq\label{diag:compactness2} 
\xymatrix@=.6in{ A^*\ar[d]_{1_{A^*}}\ar[r]^{\lambda_{A^*}} & \II\otimes A^*\ar[r]^{\eta_A\otimes 1_{A^*}\ \ \ \ } & (A^*\otimes A)\otimes A^*\ar[d]^{\alpha^{-1}_{A^*,A,A^*}}\\ 
A^* & A^*\otimes\II\ar[l]^{\rho^{-1}_{A^*}} & A^*\otimes (A\otimes A^*)\ar[l]^{1_{A^*}\otimes\epsilon_A\ \ \ \ } } 
\eeq 
\end{definition}

In the case that $\cat$ is strict the above diagrams simplify to 
\beq 
\xymatrix{
A \ar[d]_{1_A\otimes\eta_A} \ar[drr]^{1_{A}} 
& & & A^*\ar[rr]^{\eta_A\otimes 1_{A^*}}\ar[drr]_{1_{A^*}} & & A^*\otimes A\otimes A^*\ar[d]^{1_{A^*}\otimes\epsilon_A}\\ 
A\otimes A^*\otimes A\ar[rr]_{\epsilon_A\otimes 1_A} & & A & & & A^*} 
\eeq 
\noindent 
Definition
\ref{coclc} can also be expressed diagrammatically, provided we introduce some new graphical elements:
\bit 
\item As before $A$ will be represented by an upward arrow:
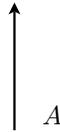
\begin{center} 
\ifx\JPicScale\undefined\def\JPicScale{1}\fi
\psset{unit=\JPicScale mm}
\psset{linewidth=0.3,dotsep=1,hatchwidth=0.3,hatchsep=1.5,shadowsize=1,dimen=middle}
\psset{dotsize=0.7 2.5,dotscale=1 1,fillcolor=black}
\psset{arrowsize=1 2,arrowlength=1,arrowinset=0.25,tbarsize=0.7 5,bracketlength=0.15,rbracketlength=0.15}
\begin{pspicture}(0,0)(6,19)
\psline{->}(1,2)(1,19)
\rput(6,4){$A$}
\end{pspicture}
 \end{center}
\noindent 
On the other hand, we depict $A^*$, the dual object to $A$,  either by an upward arrow labelled by $A^*$, or by a downward arrow labelled 
$A$:
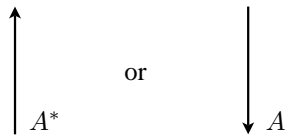
\begin{center} 
\ifx\JPicScale\undefined\def\JPicScale{1}\fi
\psset{unit=\JPicScale mm}
\psset{linewidth=0.3,dotsep=1,hatchwidth=0.3,hatchsep=1.5,shadowsize=1,dimen=middle}
\psset{dotsize=0.7 2.5,dotscale=1 1,fillcolor=black}
\psset{arrowsize=1 2,arrowlength=1,arrowinset=0.25,tbarsize=0.7 5,bracketlength=0.15,rbracketlength=0.15}
\begin{pspicture}(0,0)(35.62,19)
\psline{->}(1,2)(1,19)
\psline{<-}(32,2)(32,19)
\rput(35.62,3.75){$A$}
\rput(5,3.75){$A^*$}
\rput(17,10){or}
\end{pspicture}
 \end{center}
\noindent 
\item 
The unit $\eta_A$ and counit $\epsilon_A$ are respectively
depicted as \begin{center} 
\psset{xunit=1mm,yunit=1mm,runit=1mm}
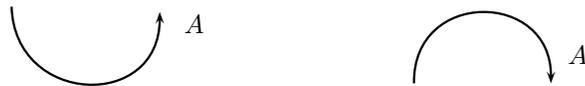
\begin{pspicture}(0,0)(76.70,13.10)
\rput(25.70,8.40){$A$}
\rput(76.70,4.30){$A$}
\psbezier[linewidth=0.30,linecolor=black]{<-}(73.10,0.50)(73.20,12.90)(54.70,13.10)(54.90,0.50)
\psbezier[linewidth=0.30,linecolor=black]{<-}(21.10,10.00)(21.10,-3.20)(1.80,-2.50)(1.40,10.70)
\end{pspicture}
 \end{center} \item Commutation of the
two diagrams now boils down to:
\bigskip\begin{center} 
\ifx\JPicScale\undefined\def\JPicScale{1}\fi
\psset{unit=\JPicScale mm}
\psset{linewidth=0.3,dotsep=1,hatchwidth=0.3,hatchsep=1.5,shadowsize=1,dimen=middle}
\psset{dotsize=0.7 2.5,dotscale=1 1,fillcolor=black}
\psset{arrowsize=1 2,arrowlength=1,arrowinset=0.25,tbarsize=0.7 5,bracketlength=0.15,rbracketlength=0.15}
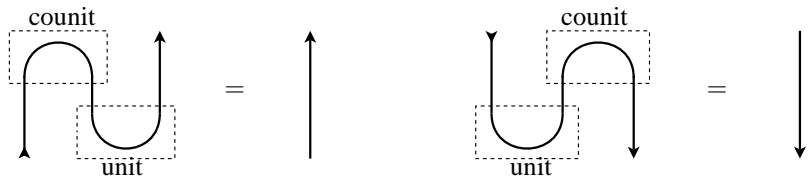
\begin{pspicture}(0,0)(105.1,21)
\psbezier(11,12.78)(11,19)(2,19)(2,12.78)
\psbezier(20,8)(20,1.78)(11,1.78)(11,8)
\psline{-<}(2,13)(2,2)
\psline(11,8)(11,13)
\psline{->}(20,8)(20,19)
\rput(30,11){$=$}
\rput(32,11){}
\psline{->}(40,2)(40,19)
\pspolygon[linewidth=0.15,linestyle=dashed,dash=0.5 0.5](9,9)(22,9)(22,2)(9,2)
\pspolygon[linewidth=0.15,linestyle=dashed,dash=0.5 0.5](0,19)(13,19)(13,12)(0,12)
\rput(15,1){unit}
\rput(7,21){counit}
\rput(32,16){}
\psbezier(73.55,12.78)(73.55,19)(83,19)(83,12.78)
\psbezier(64.1,8)(64.1,1.78)(73.55,1.78)(73.55,8)
\psline{->}(83,13)(83,2)
\psline(73.55,8)(73.55,13)
\psline{-<}(64.1,8)(64.1,19)
\rput(94.1,11){$=$}
\rput(65.5,11){}
\psline{<-}(105.1,2)(105.1,19)
\pspolygon[linewidth=0.15,linestyle=dashed,dash=0.5 0.5](75.65,9)(62,9)(62,2)(75.65,2)
\pspolygon[linewidth=0.15,linestyle=dashed,dash=0.5 0.5](85.1,19)(71.45,19)(71.45,12)(85.1,12)
\rput(69.35,1){unit}
\rput(77.75,21){counit}
\rput(65.5,16){}
\end{pspicture}
 \end{center}\smallskip
\eit 
When expressed diagrammatically, these equational constraints admit the
simple interpretation of `yanking a wire'.  While at first sight compactness of
a category as stated in Definition \ref{coclc} seems to be a somewhat ad hoc notion,
this graphical interpretation establishes it as a very canonical one which
extends the graphical calculus for symmetric monoidal categories with \em cup\em- and \em cap\em-shaped wires. As the following lemma shows, the equational constraints imply that we are allowed
to `slide' morphisms also along these cups and caps.

\begin{lemma}\label{lem:sliding} 
Given a morphism $f:A\rTo B$ define its transpose to be 
\[ 
f^*:= (1_{A^*}\otimes\epsilon_B)\circ(1_{A^*}\otimes f\otimes 1_{B^*})\circ(\eta_A\otimes 1_{B^*}): B^*\rTo A^*\,.
\]
Diagrammatically, when depicting the morphism $f$ as
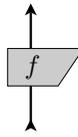
\begin{center} 
\ifx\JPicScale\undefined\def\JPicScale{1}\fi
\psset{unit=\JPicScale mm}
\psset{linewidth=0.3,dotsep=1,hatchwidth=0.3,hatchsep=1.5,shadowsize=1,dimen=middle}
\psset{dotsize=0.7 2.5,dotscale=1 1,fillcolor=black}
\psset{arrowsize=1 2,arrowlength=1,arrowinset=0.25,tbarsize=0.7 5,bracketlength=0.15,rbracketlength=0.15}
\begin{pspicture}(0,0)(10,17)
\psline{>->}(3,0)(3,17)
\newrgbcolor{userFillColour}{0.8 0.8 0.8}
\psline[linewidth=0.1,fillcolor=userFillColour,fillstyle=solid](-0.09,11)
(-0.09,6.02)
(6.64,6.02)
(10,11)(-0.09,11)
\rput(3.28,8.51){$f$}
\end{pspicture}
 \end{center}
\noindent then its transpose is depicted as 
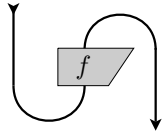
\begin{center} 
\ifx\JPicScale\undefined\def\JPicScale{1}\fi
\psset{unit=\JPicScale mm}
\psset{linewidth=0.3,dotsep=1,hatchwidth=0.3,hatchsep=1.5,shadowsize=1,dimen=middle}
\psset{dotsize=0.7 2.5,dotscale=1 1,fillcolor=black}
\psset{arrowsize=1 2,arrowlength=1,arrowinset=0.25,tbarsize=0.7 5,bracketlength=0.15,rbracketlength=0.15}
\begin{pspicture}(0,0)(19.9,17.22)
\psbezier(10.45,11)(10.45,17.22)(19.9,17.22)(19.9,11)
\psbezier(1,6.22)(1,0)(10.45,0)(10.45,6.22)
\psline{->}(19.9,11.22)(19.9,0.22)
\psline(10.45,6.22)(10.45,11.22)
\psline{-<}(1,6.22)(1,17.22)
\rput(2.4,9.22){}
\rput(2.4,14.22){}
\newrgbcolor{userFillColour}{0.8 0.8 0.8}
\psline[linewidth=0.1,fillcolor=userFillColour,fillstyle=solid](6.9,11.2)
(6.9,6.22)
(13.63,6.22)
(16.99,11.2)(6.9,11.2)
\rput(10.27,8.71){$f$}
\end{pspicture}
 \end{center}
\noindent Anticipating what will follow, we abbreviate this notation for $f^*$ to
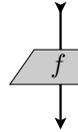
\begin{center} 
\ifx\JPicScale\undefined\def\JPicScale{1}\fi
\psset{unit=\JPicScale mm}
\psset{linewidth=0.3,dotsep=1,hatchwidth=0.3,hatchsep=1.5,shadowsize=1,dimen=middle}
\psset{dotsize=0.7 2.5,dotscale=1 1,fillcolor=black}
\psset{arrowsize=1 2,arrowlength=1,arrowinset=0.25,tbarsize=0.7 5,bracketlength=0.15,rbracketlength=0.15}
\begin{pspicture}(0,0)(9.84,16.99)
\psline{<-<}(6.87,-0.01)(6.87,16.99)
\newrgbcolor{userFillColour}{0.8 0.8 0.8}
\psline[linewidth=0.1,fillcolor=userFillColour,fillstyle=solid](9.84,5.99)
(9.84,10.61)
(3.37,10.61)
(0.14,5.99)(9.84,5.99)
\rput(6.6,8.5){$f$}
\end{pspicture}
 \end{center}
\noindent With this graphical notation we have:
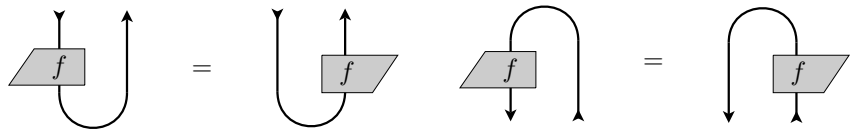
\begin{center} 
\ifx\JPicScale\undefined\def\JPicScale{1}\fi
\psset{unit=\JPicScale mm}
\psset{linewidth=0.3,dotsep=1,hatchwidth=0.3,hatchsep=1.5,shadowsize=1,dimen=middle}
\psset{dotsize=0.7 2.5,dotscale=1 1,fillcolor=black}
\psset{arrowsize=1 2,arrowlength=1,arrowinset=0.25,tbarsize=0.7 5,bracketlength=0.15,rbracketlength=0.15}
\begin{pspicture}(0,0)(116,20.99)
\psbezier(20,8)(20,1.78)(11,1.78)(11,8)
\psline{->}(20,8)(20,19)
\rput(30,11){$=$}
\rput(32,11){}
\rput(32,16){}
\psline{-<}(11,8)(11,19)
\psbezier(49,8.22)(49,2)(40,2)(40,8.22)
\psline{->}(49,8.22)(49,19.22)
\psline{-<}(40,8.22)(40,19.22)
\psbezier(80,14.92)(80,20.99)(71,20.99)(71,14.92)
\psline{-<}(80,14.92)(80,4.2)
\rput(90,12){$=$}
\rput(92,12){}
\rput(92,7.13){}
\psline{->}(71,14.92)(71,4.2)
\psbezier(109,14.71)(109,20.77)(100,20.77)(100,14.71)
\psline{-<}(109,14.71)(109,3.99)
\psline{->}(100,14.71)(100,3.99)
\newrgbcolor{userFillColour}{0.8 0.8 0.8}
\psline[linewidth=0.1,fillcolor=userFillColour,fillstyle=solid](14.39,9.02)
(14.39,13.85)
(7.66,13.85)
(4.3,9.02)(14.39,9.02)
\rput(11.03,11.36){$f$}
\newrgbcolor{userFillColour}{0.8 0.8 0.8}
\psline[linewidth=0.1,fillcolor=userFillColour,fillstyle=solid](74.36,8.66)
(74.36,13.49)
(67.63,13.49)
(64.27,8.66)(74.36,8.66)
\rput(71,11){$f$}
\newrgbcolor{userFillColour}{0.8 0.8 0.8}
\psline[linewidth=0.1,fillcolor=userFillColour,fillstyle=solid](45.91,13)
(45.91,8.02)
(52.64,8.02)
(56,13)(45.91,13)
\rput(49.28,10.51){$f$}
\newrgbcolor{userFillColour}{0.8 0.8 0.8}
\psline[linewidth=0.1,fillcolor=userFillColour,fillstyle=solid](105.91,13)
(105.91,8.02)
(112.64,8.02)
(116,13)(105.91,13)
\rput(109.28,10.51){$f$}
\end{pspicture}
 \end{center}
\noindent that is, we can `slide' morphisms along cup- and cap-shaped wires.
\end{lemma} 

The proof of the first equality simply is
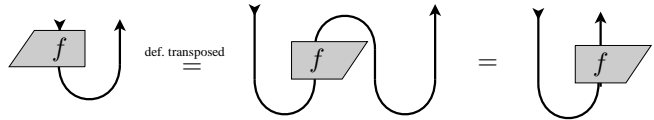
\begin{center} 
\ifx\JPicScale\undefined\def\JPicScale{1}\fi
\psset{unit=\JPicScale mm}
\psset{linewidth=0.3,dotsep=1,hatchwidth=0.3,hatchsep=1.5,shadowsize=1,dimen=middle}
\psset{dotsize=0.7 2.5,dotscale=1 1,fillcolor=black}
\psset{arrowsize=1 2,arrowlength=1,arrowinset=0.25,tbarsize=0.7 5,bracketlength=0.15,rbracketlength=0.15}
\begin{pspicture}(0,0)(86,15.51)
\psline{->}(79.28,4.51)(79.28,14.51)
\psbezier(41.28,5.51)(41.28,-0.71)(33.28,-0.71)(33.28,5.51)
\psbezier(15.37,7.51)(15.37,1.29)(7.25,1.29)(7.25,7.51)
\psbezier(49.28,9.51)(49.28,15.42)(41.28,15.42)(41.28,9.51)
\psline{-<}(7.37,7.51)(7.37,13.51)
\psline{->}(15.37,7.51)(15.37,13.51)
\newrgbcolor{userFillColour}{0.8 0.8 0.8}
\psline[linewidth=0.1,fillcolor=userFillColour,fillstyle=solid](10.73,7.17)
(10.73,12)
(4,12)
(0.64,7.17)(10.73,7.17)
\rput(7.37,9.51){$f$}
\rput(25,7){$=$}
\rput(24,9){{\tiny def. transposed}}
\rput(41.28,5.51){}
\psline{-<}(33.28,5.51)(33.28,15.51)
\psline(41.28,5.51)(41.28,9.51)
\psline(49.28,5.51)(49.28,9.51)
\psbezier(57.28,5.51)(57.28,-0.71)(49.28,-0.71)(49.28,5.51)
\psline{->}(57.28,5.51)(57.28,15.51)
\newrgbcolor{userFillColour}{0.8 0.8 0.8}
\psline[linewidth=0.1,fillcolor=userFillColour,fillstyle=solid](38.19,10.51)
(38.19,5.53)
(44.92,5.53)
(48.28,10.51)(38.19,10.51)
\rput(41.56,8.02){$f$}
\rput(64.28,7.51){$=$}
\psbezier(79,5)(79,-1.22)(71,-1.22)(71,5)
\rput(79,5){}
\psline{-<}(71,5)(71,15)
\psline(79,5)(79,9)
\newrgbcolor{userFillColour}{0.8 0.8 0.8}
\psline[linewidth=0.1,fillcolor=userFillColour,fillstyle=solid](75.91,10)
(75.91,5.02)
(82.64,5.02)
(86,10)(75.91,10)
\rput(79.28,7.51){$f$}
\end{pspicture}
 \end{center} 
\noindent The proof for the second equality proceeds analogously.

\begin{example}\label{excc} The category ${\bf FdVect}_\mathbb{K}$ is compact.
We take the usual linear algebraic dual space $V^*$ to be $V$'s dual
object and the unit to be 
\[ 
\eta_V:\mathbb{K}\rightarrow V^*\otimes V::1\mapsto \sum_{i=1}^{n} f_i\otimes e_i 
\] 
where $\{ e_i\}_{i=1}^n$ is a basis of $V$ and $f_j\in V^*$ is the linear functional
such that $f_j(e_i)=\delta_{i,j}$ for all $1\leq i,j\leq n$. Finally, we
take the counit to be 
\[ 
\epsilon_V:V\otimes V^*\rightarrow\mathbb{K}:: e_i\otimes f_j\mapsto f_j(e_i)\,.
\] 
We leave it to the reader to verify commutation of diagrams \ref{diag:compactness1} and \ref{diag:compactness2}.
Two important points need to be made here: 
\bit 
\item The linear maps $\eta_V$ and  $\epsilon_V$ do not depend on the choice of the basis $\{ e_i\}_{i=1}^n$.  It suffices to verify that there is a canonical isomorphism 
\[ 
{\bf FdVect}_{\mathbb{K}}(V,V)\stackrel{\simeq}{\longrightarrow}{\bf FdVect}_{\mathbb{K}}(\mathbb{K},V^*\otimes V)
\]

\noindent
which does not depend on the choice of basis.  The unit $\eta_V$ is the image of $1_V$ under this isomorphism and since $1_V$ is independent of the choice of basis it follows that $\eta_V$
does not depend on any choice of basis. The argument for $\epsilon_V$ proceeds analogously.  
\noindent \item There are other possible choices for $\eta_V$ and  $\epsilon_V$  which turn
${\bf FdVect}_\mathbb{K}$ into a compact category.  For example, if
$f:V\to V$ is invertible  then 
\[ 
\eta_V':= (1_{V^*}\otimes f)\circ\eta_V\qquad\mbox{\rm and}\qquad\epsilon_V':=\epsilon_V\circ(f^{-1}\otimes 1_{V^*}) 
\] 
make diagrams (\ref{diag:compactness1}) and  (\ref{diag:compactness2}) commute.
Indeed, graphically we have:
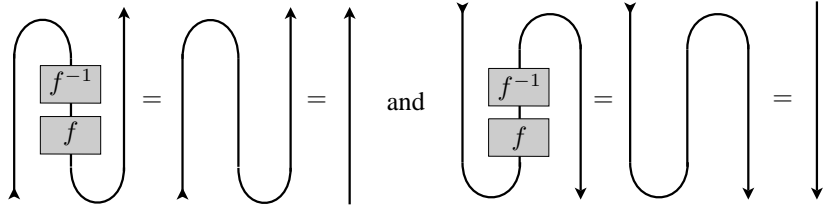
\begin{center} 
\ifx\JPicScale\undefined\def\JPicScale{1}\fi
\psset{unit=\JPicScale mm}
\psset{linewidth=0.3,dotsep=1,hatchwidth=0.3,hatchsep=1.5,shadowsize=1,dimen=middle}
\psset{dotsize=0.7 2.5,dotscale=1 1,fillcolor=black}
\psset{arrowsize=1 2,arrowlength=1,arrowinset=0.25,tbarsize=0.7 5,bracketlength=0.15,rbracketlength=0.15}
\begin{pspicture}(0,0)(107.74,27.62)
\psline(8.57,5.64)(8.57,20.02)
\newrgbcolor{userFillColour}{0.8 0.8 0.8}
\psline[linewidth=0.1,fillcolor=userFillColour,fillstyle=solid](4.5,19.08)
(4.5,14.1)
(12.52,14.08)
(12.52,19.08)(4.5,19.08)
\newrgbcolor{userFillColour}{0.8 0.8 0.8}
\psline[linewidth=0.1,fillcolor=userFillColour,fillstyle=solid](4.5,12.3)
(4.5,7.32)
(12.52,7.3)
(12.52,12.3)(4.5,12.3)
\psline(68.22,5.62)(68.22,21.25)
\newrgbcolor{userFillColour}{0.8 0.8 0.8}
\psline[linewidth=0.1,fillcolor=userFillColour,fillstyle=solid](64.09,12)
(64.09,7.02)
(72.1,7)
(72.1,12)(64.09,12)
\psbezier(8.57,20.02)(8.57,26.89)(1.02,26.89)(1.02,20.02)
\psbezier(15.62,5.52)(15.62,-0.7)(8.57,-0.7)(8.57,5.52)
\psline{-<}(1.02,20.64)(1.02,1.27)
\psline{->}(15.62,5.02)(15.62,26.89)
\rput(19.38,14.38){$=$}
\rput(27.96,8.48){}
\psline{->}(45.62,0.62)(45.62,26.88)
\rput(27.96,23.86){}
\psbezier(68.22,21.28)(68.22,27.5)(76.33,27.47)(76.33,21.25)
\psbezier(60.62,6.22)(60.62,0)(68.28,0)(68.28,6.22)
\psline{->}(76.33,21.88)(76.33,1.25)
\psline{-<}(60.62,6.25)(60.62,27.5)
\rput(79.37,14.38){$=$}
\rput(62.04,9.12){}
\psline{<-}(107.74,1.25)(107.74,27.62)
\rput(62.04,24.5){}
\newrgbcolor{userFillColour}{0.8 0.8 0.8}
\psline[linewidth=0.1,fillcolor=userFillColour,fillstyle=solid](64.1,18.7)
(64.1,13.72)
(72.12,13.7)
(72.12,18.7)(64.1,18.7)
\rput(68.22,9.38){$f$}
\rput(68.22,16.25){$f^{-1}$}
\psbezier(90.51,21.28)(90.51,27.5)(98.62,27.47)(98.62,21.25)
\psbezier(82.91,6.22)(82.91,0)(90.57,0)(90.57,6.22)
\psline{->}(98.62,21.88)(98.62,1.25)
\psline(90.51,5.62)(90.51,21.25)
\psline{-<}(82.91,6.25)(82.91,27.5)
\rput(84.33,9.12){}
\rput(84.33,24.5){}
\rput(103.18,14.5){$=$}
\rput(8.57,10.02){$f$}
\rput(8.57,16.89){$f^{-1}$}
\psbezier(30.83,19.98)(30.83,26.86)(23.4,26.86)(23.4,19.98)
\psbezier(37.77,5.48)(37.77,-0.74)(30.83,-0.74)(30.83,5.48)
\psline{-<}(23.4,20.61)(23.4,1.23)
\psline(30.83,5.61)(30.83,19.98)
\psline{->}(37.77,4.98)(37.77,26.86)
\rput(41.25,14.38){$=$}
\rput(53.12,14.38){and}
\end{pspicture}
 \end{center} 
\eit 
\end{example}

\begin{example} The category ${\bf Rel}$ of sets and relations is also compact
relative to the Cartesian product as we shall see in detail in Section \ref{sect:rel}.
\end{example}

\begin{example} 
The category $\mathbf{QuantOpp}$ is compact. We can pick
Bell-states as the units and the corresponding Bell-effects as counits.
As shown in \cite{AC2004, C2005c}, compactness is exactly what enables
modeling protocols such as quantum teleportation:
\medskip
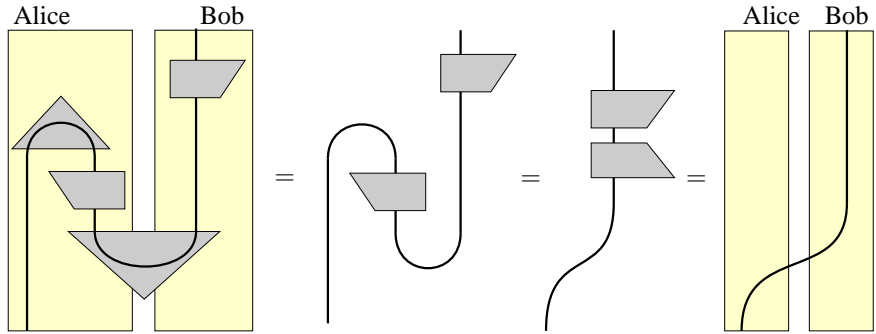
\begin{center} 
\ifx\JPicScale\undefined\def\JPicScale{1}\fi
\psset{unit=\JPicScale mm}
\psset{linewidth=0.3,dotsep=1,hatchwidth=0.3,hatchsep=1.5,shadowsize=1,dimen=middle}
\psset{dotsize=0.7 2.5,dotscale=1 1,fillcolor=black}
\psset{arrowsize=1 2,arrowlength=1,arrowinset=0.25,tbarsize=0.7 5,bracketlength=0.15,rbracketlength=0.15}
\begin{pspicture}(0,0)(115.5,42)
\newrgbcolor{userFillColour}{1 1 0.8}
\pspolygon[linewidth=0.1,fillcolor=userFillColour,fillstyle=solid](95.75,39.9)(104.25,39.9)(104.25,0)(95.75,0)
\newrgbcolor{userFillColour}{1 1 0.8}
\pspolygon[linewidth=0.1,fillcolor=userFillColour,fillstyle=solid](107,39.9)(115.5,39.9)(115.5,0)(107,0)
\rput(102,42){Alice}
\newrgbcolor{userFillColour}{1 1 0.8}
\pspolygon[linewidth=0.1,fillcolor=userFillColour,fillstyle=solid](0.5,40)(17,40)(17,0)(0.5,0)
\newrgbcolor{userFillColour}{1 1 0.8}
\pspolygon[linewidth=0.1,fillcolor=userFillColour,fillstyle=solid](20,40)(33,40)(33,0)(20,0)
\psline(81,17)(81,40)
\newrgbcolor{userFillColour}{0.8 0.8 0.8}
\pscustom[linewidth=0.1,fillcolor=userFillColour,fillstyle=solid]{\psline(28.63,13.22)(8.49,13.22)
\psline(8.49,13.22)(18.58,4.22)
\psline(18.58,4.22)(28.63,13.22)
\closepath}
\newrgbcolor{userFillColour}{0.8 0.8 0.8}
\pscustom[linewidth=0.1,fillcolor=userFillColour,fillstyle=solid]{\psline(14,24.22)(1,24.22)
\psline(1,24.22)(7.51,31.22)
\psline(7.51,31.22)(14,24.22)
\closepath}
\psbezier(25.47,13.22)(25.47,7)(12,7)(12,13.22)
\psline(25.47,13.22)(25.47,40.22)
\rput(37.27,20.22){$=$}
\rput(27.05,11){}
\rput(27.05,16){}
\psbezier(12,23.14)(12,29.21)(3,29.21)(3,23.14)
\psline(12,23.14)(12,13.22)
\rput(20.74,20.22){}
\rput(20.74,15.35){}
\psline(3,23.14)(3,0)
\newrgbcolor{userFillColour}{0.8 0.8 0.8}
\psline[linewidth=0.1,fillcolor=userFillColour,fillstyle=solid](22.03,36)
(22.03,31.02)
(28.68,31.02)
(32,36)(22.03,36)
\newrgbcolor{userFillColour}{0.8 0.8 0.8}
\psline[linewidth=0.1,fillcolor=userFillColour,fillstyle=solid](16,21.2)
(16,16.22)
(9.27,16.22)
(5.91,21.2)(16,21.2)
\psbezier(60.64,13)(60.64,6.78)(52,6.78)(52,13)
\psline(60.64,13)(60.64,40)
\rput(61.6,10.78){}
\rput(61.6,15.78){}
\psbezier(52,22.92)(52,28.99)(43,28.99)(43,22.92)
\psline(52,22.92)(52,13)
\rput(59.88,20){}
\rput(59.88,15.13){}
\psline(43,22.92)(43,1)
\newrgbcolor{userFillColour}{0.8 0.8 0.8}
\psline[linewidth=0.1,fillcolor=userFillColour,fillstyle=solid](58,36.78)
(58,31.8)
(64.67,31.8)
(68,36.78)(58,36.78)
\newrgbcolor{userFillColour}{0.8 0.8 0.8}
\psline[linewidth=0.1,fillcolor=userFillColour,fillstyle=solid](56,20.98)
(56,16)
(49.27,16)
(45.91,20.98)(56,20.98)
\rput(70,20){$=$}
\newrgbcolor{userFillColour}{0.8 0.8 0.8}
\psline[linewidth=0.1,fillcolor=userFillColour,fillstyle=solid](78,20.38)
(78,25)
(85.39,25)
(89.08,20.38)(78,20.38)
\newrgbcolor{userFillColour}{0.8 0.8 0.8}
\psline[linewidth=0.1,fillcolor=userFillColour,fillstyle=solid](78,31.98)
(78,27)
(85.39,27)
(89.08,31.98)(78,31.98)
\psbezier(72,0)(72,12)(81,6)(81,17)
\rput(92,20){$=$}
\psline(112,17)(112,40)
\psbezier(98,0)(98,12)(112,6)(112,17)
\rput(5,42){Alice}
\rput(29,42){Bob}
\rput(112,42){Bob}
\end{pspicture}
 \end{center}
\noindent 
where the trapezoid is assumed to be unitary and hence, its adjoint coincides with its  inverse.
The classical information flow is (implicitly) encoded in the fact that the same trapezoid appears in the left-hand-side picture both at Alice's and Bob's side.
\end{example}

Given a morphism $f:A\rTo B$ in a compact category, its {\em
name} 
\[
\II\rTo^{\name{f}} A^*\otimes B
\]
and its \em coname \em
\[
A\otimes B^*\rTo^{\coname{f}}\II
\]
are defined by: 
\[ \xymatrix{A^*\otimes A\ar[r]^{1_{A^*}\otimes f} & A^*\otimes B &\ar@{}[d]|{\mbox{and}} & & \II\\%
\II\ar[u]^{\eta_A}\ar[ur]_{\name{f}} & & & A\otimes B^*\ar[ur]^{\coname{f}}
\ar[r]_{f\otimes 1_{B^*}} & B\otimes B^*\ar[u]_{\epsilon_B}} 
\] 
Following  \cite{AC2004} we can show that for $f:A\rTo B$ and
$g:B\rTo C$ 
\[ 
\lambda^{-1}_C\circ(\coname{f}\otimes 1_C)\circ (1_A\otimes\name{g})\circ\rho_A=g\circ f 
\] 
always holds.  The graphical proof is again trivial: 

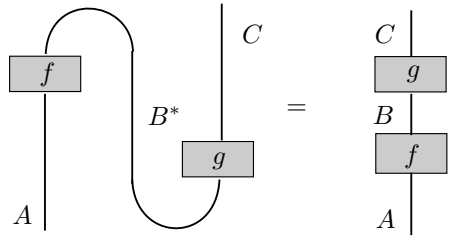
\begin{center} 
\ifx\JPicScale\undefined\def\JPicScale{1}\fi
\psset{unit=\JPicScale mm}
\psset{linewidth=0.3,dotsep=1,hatchwidth=0.3,hatchsep=1.5,shadowsize=1,dimen=middle}
\psset{dotsize=0.7 2.5,dotscale=1 1,fillcolor=black}
\psset{arrowsize=1 2,arrowlength=1,arrowinset=0.25,tbarsize=0.7 5,bracketlength=0.15,rbracketlength=0.15}
\begin{pspicture}(0,0)(59.33,32.9)
\psbezier[linewidth=0.25](29.1,7.7)(28.7,-1.1)(17.8,-0.8)(17.5,7.7)
\psbezier[linewidth=0.25](6,24.4)(6.4,32.9)(17.4,31.9)(17.6,24.7)
\rput(39.4,17){$=$}
\newrgbcolor{userFillColour}{0.8 0.8 0.8}
\pspolygon[linewidth=0.15,fillcolor=userFillColour,fillstyle=solid](1.2,19.4)(10.65,19.4)(10.65,24.1)(1.2,24.1)
\rput(6.2,21.7){$f$}
\newrgbcolor{userFillColour}{0.8 0.8 0.8}
\pspolygon[linewidth=0.15,fillcolor=userFillColour,fillstyle=solid](24.17,8.05)(33.62,8.05)(33.62,12.75)(24.17,12.75)
\rput(29.1,10.2){$g$}
\psline[linewidth=0.25](5.9,0.9)(5.9,19.1)
\psline[linewidth=0.25](17.5,24.8)(17.5,7.1)
\psline[linewidth=0.25](29.4,12.9)(29.4,31.1)
\newrgbcolor{userFillColour}{0.8 0.8 0.8}
\pspolygon[linewidth=0.15,fillcolor=userFillColour,fillstyle=solid](49.77,19.35)(59.23,19.35)(59.23,24.05)(49.77,24.05)
\newrgbcolor{userFillColour}{0.8 0.8 0.8}
\pspolygon[linewidth=0.15,fillcolor=userFillColour,fillstyle=solid](49.88,8.53)(59.33,8.53)(59.33,13.8)(49.88,13.8)
\rput(54.7,10.76){$f$}
\rput(54.9,21.7){$g$}
\psline[linewidth=0.25](54.6,24.1)(54.6,30.2)
\psline[linewidth=0.25](54.6,0.3)(54.6,8.6)
\psline[linewidth=0.25](54.6,13.6)(54.6,19.1)
\rput(2.9,3){$A$}
\rput(51.2,2.4){$A$}
\rput(50.9,16.2){$B$}
\rput(21.8,16.2){$B^*$}
\rput(51.1,27.1){$C$}
\rput(33.5,27){$C$}
\end{pspicture}
 \end{center} 

In contrast a (non-strict) symbolic proof goes as follows: 
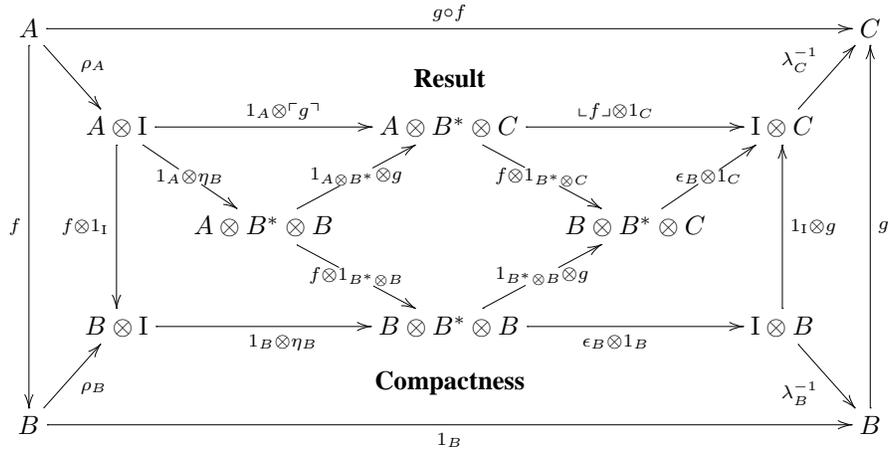
\[ 
\xymatrix@C=.155in{%
A \ar[rrrrrr]^{g\circ f}_{}="A"  \ar[dddd]_{f} \ar[dr]^{\rho_A}& & & & &
& C\\%
\ar@{}[rrrrrr]^{}="B" \ar@{}|{\mbox{\bf Result}} "A";"B" &
A\otimes \II\ar[rr]^{1_A\otimes\name{g}} \ar[dd]_{f\otimes
1_\II}\ar[dr]|{1_A\otimes\eta_B} & &  A\otimes B^*\otimes
C\ar[rr]^{\coname{f}\otimes 1_C} \ar[dr]|{f\otimes 1_{B^*\otimes C}} &
& \II\otimes C \ar[ur]^{\lambda^{-1}_C}& \\%
& & A\otimes B^*\otimes  B
\ar[ur]|{1_{A\otimes B^*}\otimes g} \ar[dr]|{f\otimes 1_{B^*\otimes
B}}& & B\otimes B^*\otimes C \ar[ur]|{\epsilon_B\otimes 1_C}& & \\%
\ar@{}[rrrrrr]^{}="C" & B\otimes \II \ar[rr]_{1_B\otimes\eta_B}& &
B\otimes B^* \otimes B \ar[ur]|{1_{B^*\otimes B}\otimes g}
\ar[rr]_{\epsilon_B\otimes 1_B} & & \II\otimes B \ar[uu]_{1_\II\otimes
g}\ar[dr]_{\lambda^{-1}_B}& \\%
B \ar[rrrrrr]_{1_B}="D"
\ar@{}|{\mbox{\bf Compactness}} "C";"D"\ar[ur]_{\rho_B}& & & & & & B
\ar[uuuu]_{g}} 
\] 
Both paths on the outside are equal to $g\circ f$. We
want to show that the pentagon labelled `Result' commutes.  To do this
we will `unfold' arrows using equations which hold in compact
categories in order to pass from the composite $g\circ f$ at the
left/bottom/right to $\lambda^{-1}_C\circ(\coname{f}\otimes 1_C)\circ
(1_A\otimes\name{g})\circ\rho_A$.  This will transform the tautology
$g\circ f=g\circ f$ into commutation of the pentagon labelled `Result'.
For instance, we use compactness to go from the identity arrow at the
bottom of the diagram  to the composite $\lambda_B^{-1}\circ(\epsilon_B\otimes
1_B)\circ(1_B\otimes\eta_B)\circ\rho_B$. The outer left and right
trapezoids express naturality of $\rho$ and $\lambda$. The remaining
triangles/diamonds express bifunctoriality and the definitions of
name/coname.

The scalar $\epsilon_A\circ\sigma_{A^*\!,A}\circ\eta_A:\mathbb{K}\to\mathbb{K}$
depicts as 

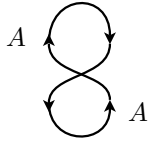
\begin{center} 
\ifx\JPicScale\undefined\def\JPicScale{1}\fi
\psset{unit=\JPicScale mm}
\psset{linewidth=0.3,dotsep=1,hatchwidth=0.3,hatchsep=1.5,shadowsize=1,dimen=middle}
\psset{dotsize=0.7 2.5,dotscale=1 1,fillcolor=black}
\psset{arrowsize=1 2,arrowlength=1,arrowinset=0.25,tbarsize=0.7 5,bracketlength=0.15,rbracketlength=0.15}
\begin{pspicture}(0,0)(18.75,20)
\psbezier{<-<}(15,12.81)(15,20)(6.88,19.75)(6.88,12.57)
\psbezier{<-<}(15,5.41)(15,-0.81)(6.88,-0.81)(6.88,5.41)
\psbezier(6.88,12.88)(6.88,8.52)(15,9.14)(15,5.41)
\psbezier(15,12.88)(15,8.28)(6.88,8.93)(6.88,5)
\rput(2.5,13.72){$A$}
\rput(18.75,3.75){$A$}
\end{pspicture}
 \end{center}

\noindent and when setting 

\begin{center} 
\ifx\JPicScale\undefined\def\JPicScale{1}\fi
\psset{unit=\JPicScale mm}
\psset{linewidth=0.3,dotsep=1,hatchwidth=0.3,hatchsep=1.5,shadowsize=1,dimen=middle}
\psset{dotsize=0.7 2.5,dotscale=1 1,fillcolor=black}
\psset{arrowsize=1 2,arrowlength=1,arrowinset=0.25,tbarsize=0.7 5,bracketlength=0.15,rbracketlength=0.15}
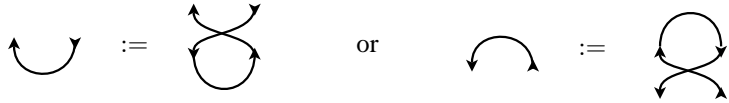
\begin{pspicture}(0,0)(95,12.91)
\psbezier(33,6)(33,-0.22)(24.88,-0.22)(24.88,6)
\psbezier{<-<}(24.88,12.69)(24.88,8.33)(33,8.95)(33,5.22)
\psbezier{>->}(33,12.69)(33,8.33)(24.88,8.95)(24.88,5.22)
\psbezier{>->}(9.12,8.22)(9.12,2)(1,2)(1,8.22)
\rput(17,7){$:=$}
\psbezier(95,7)(95,12.91)(86.88,12.91)(86.88,7)
\psbezier{<-<}(86.88,0)(86.88,4.14)(95,3.55)(95,7.09)
\psbezier{>->}(95,0)(95,4.14)(86.88,3.55)(86.88,7.09)
\psbezier{>->}(70,3.64)(70,9.55)(61.88,9.55)(61.88,3.64)
\rput(77.88,6.55){$:=$}
\rput(48,7){or}
\end{pspicture}
 \end{center} 

\noindent it becomes an  `$A$-labelled circle' 

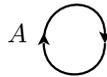
\begin{center} 
\ifx\JPicScale\undefined\def\JPicScale{1}\fi
\psset{unit=\JPicScale mm}
\psset{linewidth=0.3,dotsep=1,hatchwidth=0.3,hatchsep=1.5,shadowsize=1,dimen=middle}
\psset{dotsize=0.7 2.5,dotscale=1 1,fillcolor=black}
\psset{arrowsize=1 2,arrowlength=1,arrowinset=0.25,tbarsize=0.7 5,bracketlength=0.15,rbracketlength=0.15}
\begin{pspicture}(0,0)(14.13,11.24)
\psbezier{<-<}(14.13,4.51)(14.13,11.24)(6,11)(6,4.27)
\psbezier(14.13,4.53)(14.13,-1)(6,-1)(6,4.53)
\rput(2.5,5.81){$A$}
\end{pspicture}
 \end{center} 

\begin{example} 
In ${\bf FdVect}_\mathbb{K}$ the $V$-labelled  circle stands for the dimension of the vector space $V$.  By the definitions of $\eta_V$ and $\epsilon_V$, the previous composite is equal to 
\[ 
\sum_{ij}
f_j(e_i)=\sum_{ij}\delta_{i,j}=\sum_{i}1={\rm dim}(V)\,.
\] 
\end{example}

\begin{definition}\em A \em dagger compact category \em $\cat$ is both a compact
category and a dagger symmetric monoidal category,  such that for all $A\in|\cat|$,
$\epsilon_A=\eta_A^\dagger\circ\sigma_{A, A^*}$.  
\end{definition} 

\begin{example} 
The category ${\bf FdHilb}$ is dagger compact.  
\end{example}

\subsection{The category of relations}\label{sect:rel} 

We now turn our attention to the category ${\bf Rel}$ of sets and relations, a category which
we briefly encountered in previous sections. Perhaps surprisingly, ${\bf Rel}$
possesses more `quantum features' than the category ${\bf Set}$ of sets and
functions.   In particular, just like ${\bf FdHilb}$ it is a dagger compact category. 

A {\em relation} $R:X\rightarrow Y$ between two sets $X$ and $Y$ is a subset of
the set of all their ordered pairs, that is, $R\subseteq X\times Y$.  Thus, given
an element $(x,y)\in R$, we say that $x\in X$  {\em relates} to $y\in Y$,  which we 
denote as $xRy$. The set
\[
R:=\{(x,y)\ |\ xRy\}
\]
is also referred to as the \em graph \em of the relation.

\begin{example} 
For the relation ``strictly less than''
or `$<$' on the natural numbers,  we have that $2$ relates to $5$, which
is denoted as $2<5$ or $(2,5)\in\ <\ \subseteq\mathbb{N}\times\mathbb{N}$. For the relation ``is a divisor of'' or `$|$' on the natural numbers, we have $6|36$ or $(6,36)\in |\subseteq \mathbb{N}\times\mathbb{N}$.  Other examples are general preorders or equivalence relations.
\end{example} 

\begin{definition} 
\em The monoidal category ${\bf Rel}$ is defined as follows:
\begin{itemize} 
	\item The objects are sets.
	\item The morphisms are all relations $R:X\rightarrow Y$.
	\item For $R_1:X\rightarrow Y$ and $R_2:Y\rightarrow Z$ the composite $R_2\circ
R_1\subseteq X\times Z$ is 
\[
R_2\circ R_1:=\{(x,z)\ |\ \mbox{there exists a }y\in Y\ \mbox{such that }xR_1 y\ \mbox{and } yR_2 z\}\,.
\]
Composition is easily seen to be associative. For $X\in|{\bf Rel}|$ we have
\[
1_X:=\{(x,x)\ |\ x\in X\}\,.
\]
	\item The monoidal product of two sets is their Cartesian product, the unit for the monoidal structure is any singleton,  and for two relations $R_1:X_1\rightarrow Y_1$ and $R_2:X_2\to Y_2$ the monoidal product  $R_1\times R_2\subseteq X_1\times X_2\to Y_1\times Y_2$ is  
\[
R_1\times R_2:=\{((x,x'),(y,y'))\ |\ xR_1y\mbox{ and }x'R_2y'\}\subseteq(X_1\times X_2)\times(Y_1\times Y_2)\,.
\]		
\end{itemize}\em 
\end{definition}

We mentioned  before that ${\bf Set}$ was contained in ${\bf Rel}$ as a `sub-monoidal
category'. In ${\bf Rel}$, the left- and right-unit natural isomorphisms respectively  are
\[
\lambda_X:=\{(x,(*,x))\ |\ x\in X\}\ \ \ \ \mbox{and}\ \ \ \ \rho_X:=\{(x,(x,*))\ |\ x\in X\}\,,
\]
and the associativity natural isomorphism is
\[ 
\alpha_{X,Y,Z}:=\{( (x,(y,z)) , ((x,y),z) ) \ |\ x\in X,y\in Y\mbox{ and }z\in Z\}.  
\]
These relations are all single-valued, so they are also functions, and 
they are the same functions as the natural isomorphisms for the Cartesian product in ${\bf Set}$.  Let us verify the coherence conditions for them:
\begin{itemize} 
\item[](i) The pentagon
\[
\xymatrix{ 
W\times (X\times (Y\times Z)) \ar[r]^{\alpha_{-}}
\ar[d]_{1\times \alpha_{-}}& (W\times X)\times (Y\times
Z)\ar[r]^{\alpha_{-}}& ((W\times X)\times Y)\times Z\\ W\times
((X\times Y)\times Z) \ar[rr]_{\alpha_{-}}& & (W\times
(X\times Y))\times Z \ar[u]_{\alpha_{-}\times 1} }
\] 
indeed commutes. The top part 
\[
\alpha_{-}\circ\alpha_{-}:W\times (X\times (Y\times Z))\rightarrow ((W\times X)\times Y)\times Z
\] 
is by definition  a subset of  
\[
(W\times (X\times (Y\times Z)))\times (((W\times X)\times Y)\times Z)\,.
\]
Unfolding the definition of relational composition we obtain
\[
\alpha_{-}\circ\alpha_{-}=\Bigl\{((w,(x,(y,z))),(((w'',x''),y''),z''))\
\Bigm|\ \exists((w',x'),(y',z'))\mbox{ s.t.}
\]
\[
(w,(x,(y,z)))\alpha((w',x'),(y',z'))\mbox{ and }((w',x'),(y',z'))\alpha(((w'',x''),y''),z'')\Bigr\}\,,
\] 
which by the definition of $\alpha$ simplifies to
\[
\alpha_{-}\circ\alpha_{-}=\{((w,(x,(y,z))),(((w,x),y),z))\
|\ w\in W,x\in X, y\in Y,z\in Z\}\,.
\] 
The bottom path  yields the same result, hence making the pentagon commute. For the remaining diagrams we leave the details to the reader. 
\item[] (ii) The triangle 
\[
\xymatrix{ X\times Y\ar[r]^{1_A\times\lambda_Y\ \ \ }\ar[dr]_{\rho_X\times 1_Y} & X\times (\{*\}\times
Y)\ar[d]^{\alpha_{X,\{*\},Y}}\\ & (X\times \{*\})\times Y }
\] 
commutes as both paths are now equal to 
\[
\{((x,y),((x,*),y))\ |\ x\in X\mbox{ and }y\in Y\}\,.
\]
\end{itemize} 
As $\times$ is symmetric in ${\bf Set}$ we also expect
${\bf Rel}$ to be symmetric monoidal. For any $X$ and $Y\in |{\bf
Rel}|$, the natural isomorphism 
\[
\sigma_{X,Y}:=\{((x,y),(y,x))\ |\ x\in X\mbox{ and }y\in Y\}
\]
also obeys the coherence conditions:
\begin{itemize} 
\item[] (i) The two triangles
\[
\xymatrix{X\times Y\ar[r]^{\sigma_{X,Y}} \ar@{=}[dr]
& Y\times X \ar[d]^{\sigma_{Y,X}} & \ar@{}[d]|{,} &
X\ar[r]^{\lambda_X} \ar[dr]_{\rho_X} & \{*\}\times
X\ar[d]^{\sigma_{\{*\},X}} \\%
& X\times Y & & &
X\times \{*\} }
\] 
commute since both paths of the left triangle are equal to 
\[
\{((x,y),(x,y))\ |\ x\in X\mbox{ and }y\in Y\}\,,
\] 
while the paths of the right triangle are equal to 
\[
\{(x,(x,*))\ |\ x\in X\}\,.
\]
\item[] (ii) The hexagon
\[
\xymatrix@C=.5in{X\times (Y\times
Z)\ar[r]^{\alpha_{-}}\ar[d]_{1_X\times\sigma_{Y,Z}} &
(X\times Y)\times Z\ar[r]^{\sigma_{(X\times Y),Z}} &
Z\times (X\times Y) \ar[d]^{\alpha_{-}}\\ 
X\times (Z\times Y)\ar[r]_{\alpha_{-}} & (X\times Z)\times
Y\ar[r]_{\sigma_{X,Y}\times 1_Z} & (Z\times X)\times Y
}
\] 
commutes since both paths are equal to
\[
\{((x,(y,z)),((z,x),y))\ |\ x\in X, y\in Y\mbox{ and }z\in Z\}\,.
 \]
\end{itemize} 
So ${\bf Rel}$ is indeed a symmetric monoidal category as expected.  ${\bf Rel}$ shares many common characteristics with ${\bf FdHilb}$, one of them being a $\dagger$-compact structure. Firstly, ${\bf Rel}$ is compact closed with self-dual objects that is, $X^*=X$ for any $X\in|{\bf Rel}|$. Moreover, for any $X\in|{\bf Rel}|$ let 
\[
 \eta_X:\{*\}\rightarrow X\times X:=\{( *,( x,x))\ |\ x\in X\} 
\]
and 
\[
\ \epsilon_X:X\times X\rightarrow\{*\} :=\{(( x,x),*)\ |\ x\in X\}\,.   
\]
These morphisms make
\[
\xymatrix@=.4in{X\ar[d]_{1_X}\ar[r]^{\rho_X\ \   } & X\times
\{*\}\ar[r]^{1_X\times\eta_X\ \ \ \ }& X\times (X\times
X)\ar[d]^{\alpha_{-}}\\ X & \{*\}\times X\ar[l]^{\lambda^{-1}_X} & (X\times
X)\times X\ar[l]^{\epsilon_X\times 1_X\ \ \ } }
\] 
and its dual both commute. Indeed:
\begin{itemize} 
\item[] (a) The composite 
\[(1_X\times\eta_X)\circ
\rho_X:X\rightarrow X\times (X\times X)
\]
is the set of tuples
\[
\{(x,(x',(x'',x''')))\}\subseteq X\times (X\times (X\times X))
\] 
such that there exists an $(x'''',*)\in X\times \{*\}$ with 
\[
x\ \rho_X\  (x'''',*)\quad\mbox{ and }\quad(x'''',*)\ (1_X\times\eta_X)\ (x',(x'',x'''))\,.
\] 
By definition of $\rho$ and $1_X$, and of the product of relations, this entails that
$x,x''''$ and $x'$ are all equal. Moreover, by definition of
$\eta_X$, and of the product of relations, we have that $x''$ and
$x'''$ are also equal. Thus,
\[
(1_X\times\eta_X)\circ\rho_X:=\{(x,(x,(x',x')))\ |\ x,x'\in X\}\,.
\]
\item[] (b) Hence the composite 
\[
\alpha\circ ((1_X\times\eta_X)\circ \rho):X\rightarrow (X\times X)\times X
\]
is 
\[
\alpha\circ ((1_X\times\eta_X)\circ \rho)=\{(x,((x,x'),x')\ |\ x,x'\in X\}\,.
 \]
\item[] (c) The composite 
\[
(\epsilon_X\times 1_X)\circ(\alpha\circ (1_X\times\eta_X)\circ \rho):X\rightarrow\{*\}\times X 
\]
is a set of tuples 
\[
\{(x,(*,x'))\}\subseteq X\times(\{*\}\times X)
\]
such that there exists an $((x'',x'''),x'''')\in (X\times X)\times X$
with 
\[
x\ (\alpha\circ (1_X\times\eta_X)\circ \rho)\ ((x'',x'''),x'''')
\ \ \mbox{ and }\ \ 
((x'',x'''),x'''')\ (\epsilon_X\times 1_X)\ (*,x')\,.
\] 
By the
computation in (b) we have that $x=x''$ and $x'''=x''''$. By definition of
$\epsilon_X$, $1_X$ and the product of relations we have $x''=x'''$
and $x''''=x'$. All this together yields $x=x''=x'''=x''''=x'$
and hence 
\[
(\epsilon_X\otimes 1_X)\circ(\alpha\circ (1_X\otimes\eta_X)\circ \rho)=\{(x,(*,x))\ |\ x\in X\}\,.
\]
\item[] (d) Post-composing the previous composite with the natural isomorphism $\lambda^{-1}_X$ yields a morphism of type $X\rightarrow X$, namely
\[
\lambda_X^{-1}\circ(\epsilon_X\otimes 1_X)\circ\alpha\circ (1_X\otimes\eta_X)\circ \rho=\{(x,x)\ |\ x\in X\}
\]
which is the identity relation as required.  
\end{itemize}
Commutation of the dual diagram is done analogously. From this,
we conclude that ${\bf Rel}$ is compact closed.
The obvious candidate for the dagger 
\[
\dagger:{\bf Rel}^{\scriptsize \mbox{\em op}}\longrightarrow {\bf Rel}
\] 
is the relational converse.  For  relation  $R:X\rightarrow Y$ its converse $R^{\cup}:Y\rightarrow X$ is 
\[
R^{\cup}:=\{(y,x)\ |\ xRy\}\,.
\]
We define the contravariant identity-on-objects involutive functor 
\[
\dagger:{\bf Rel}\longrightarrow{\bf Rel}::R\mapsto R^{\cup}\,.
\]
Note that the adjoint and the transpose coincide, that is, 
\[
R^*=(1_X\times \epsilon_Y)\circ(1_X\times R\times 1_Y)\circ(\eta_X\times 1_Y)=R^\dagger
\]
which the reader may easily check. 
Finally, we verify that ${\bf Rel}$ is dagger compact: 
\begin{itemize} 
\item The category ${\bf Rel}$ is dagger monoidal: 
\begin{itemize}
\item[] (i) From the definition of the monoidal
product of two relations
\[
R_1:=\{(x,y)\ |\ xRy\}\quad \mbox{ and } \quad R_2:=\{(x',y')\ |\ x'Ry'\}
\]
we have that 
\[
(R_1\times R_2)^\dagger=\{((y,y'),(x,x'))\ |\ xR_1y\mbox{ and }x'R_2y'\}=R_1^\dagger\times R_2^\dagger.
\]
\item[] (ii) The fact that $\alpha^\dagger=\alpha^{-1}$, $\lambda^\dagger=\lambda^{-1}$, $\rho^\dagger=\rho^{-1}$ and $\sigma^\dagger=\sigma^{-1}$ is trivial as the inverse of all these morphisms is the relational
converse.  
\end{itemize} 
\item  The diagram
\[
\xymatrix{\{*\}\ar[r]^{\epsilon^\dagger_X} \ar[dr]_{\eta_X} & X\times X\ar[d]^{\sigma_{X,X}}\\ 
& X\times X}
\] 
commutes since from 
\[
\epsilon_X:=\{((x,x),*)\ |\ x\in X\}
\]
follows
\[
\epsilon^\dagger_X:=\{(*,(x,x))\ |\ x\in X\}
\] 
and hence $\sigma\circ\epsilon^\dagger_X=\epsilon^\dagger_X=\eta_X$.
\end{itemize} 
So  ${\bf Rel}$ is indeed a dagger compact category. 

\subsection{The category of 2D cobordisms}\label{sect:2cob} 


The category ${\bf 2Cob}$ can be informally described as a category whose
morphisms, so-called {\em cobordisms}, describe the `topological evolution' of manifolds of
dimension $2-1=1$ through time. 
For instance, consider some snapshots of two circles which merge into a single
circle, with time going upwards: 
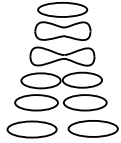
\begin{center} 
\ifx\JPicScale\undefined\def\JPicScale{1}\fi
\psset{unit=\JPicScale mm}
\psset{linewidth=0.3,dotsep=1,hatchwidth=0.3,hatchsep=1.5,shadowsize=1,dimen=middle}
\psset{dotsize=0.7 2.5,dotscale=1 1,fillcolor=black}
\psset{arrowsize=1 2,arrowlength=1,arrowinset=0.25,tbarsize=0.7 5,bracketlength=0.15,rbracketlength=0.15}
\begin{pspicture}(0,0)(15.41,18.74)
\rput{0}(12.14,1.99){\psellipse[linewidth=0.25](0,0)(3.28,1.07)}
\rput{0}(4.05,1.91){\psellipse[linewidth=0.25](0,0)(3.27,1.07)}
\rput{0}(11.08,5.47){\psellipse[linewidth=0.25](0,0)(2.85,1.03)}
\rput{0}(8.08,17.8){\psellipse[linewidth=0.25](0,0)(3.27,0.94)}
\rput{0}(4.63,5.47){\psellipse[linewidth=0.25](0,0)(2.85,1.03)}
\rput{0}(5.21,8.37){\psellipse[linewidth=0.25](0,0)(2.61,1)}
\rput{0}(10.72,8.46){\psellipse[linewidth=0.25](0,0)(2.61,1)}
\psbezier[linewidth=0.25](3.8,11.6)(3.86,9.01)(7.2,11.52)(8.54,11.3)
\psbezier[linewidth=0.25](3.8,11.6)(3.86,14.26)(7.13,11.74)(8.29,11.82)
\psbezier[linewidth=0.25](12.3,11.6)(12.26,13.96)(9.5,11.82)(8.32,11.82)
\psbezier[linewidth=0.25](12.33,11.6)(12.26,9.08)(9.31,11.3)(8.35,11.3)
\psbezier[linewidth=0.25](4.44,14.93)(4.49,12.79)(7.52,14.44)(8.48,14.32)
\psbezier[linewidth=0.25](4.44,14.93)(4.6,17.38)(7.26,15.48)(8.29,15.48)
\psbezier[linewidth=0.25](11.81,14.93)(11.66,17.38)(9.25,15.42)(8.22,15.48)
\psbezier[linewidth=0.25](11.77,14.87)(11.77,12.85)(9.18,14.32)(8.41,14.26)
\end{pspicture}
 \end{center} 
Passing to the continuum, the same process can be described by the cobordism 
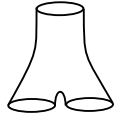
\begin{center} 
\ifx\JPicScale\undefined\def\JPicScale{1}\fi
\psset{unit=\JPicScale mm}
\psset{linewidth=0.3,dotsep=1,hatchwidth=0.3,hatchsep=1.5,shadowsize=1,dimen=middle}
\psset{dotsize=0.7 2.5,dotscale=1 1,fillcolor=black}
\psset{arrowsize=1 2,arrowlength=1,arrowinset=0.25,tbarsize=0.7 5,bracketlength=0.15,rbracketlength=0.15}
\begin{pspicture}(0,0)(14.94,15.62)
\rput{0.33}(11.83,1.99){\psellipse[linewidth=0.25](0,0)(3.11,0.86)}
\rput{0.33}(4.15,1.89){\psellipse[linewidth=0.25](0,0)(3.11,0.86)}
\rput{0.33}(7.91,14.74){\psellipse[linewidth=0.25](0,0)(3.11,0.86)}
\psbezier[linewidth=0.25](7.26,1.85)(7.24,4.28)(8.64,4.29)(8.66,1.86)
\psbezier[linewidth=0.25](1.04,1.93)(5.32,9.54)(4.72,8.94)(4.8,14.87)
\psbezier[linewidth=0.25](14.93,2.19)(10.94,9.1)(11.12,7.8)(11.08,14.73)
\end{pspicture}
 \end{center} 
Thus, we take a cobordism to be a (compact) 2-dimensional manifold whose boundary is partitioned in two.  We take these closed one-dimensional manifolds to be the domain and the codomain of the cobordism. 
Since we are only interested in the topology of the manifolds, each (co)domain consists of a finite number of closed strings. 

\begin{definition}\em 
The category ${\bf 2Cob}$ is defined as follows:
\begin{itemize} 
\item Each object is a finite number of closed strings. Hence each object can be  
equivalently represented by a natural number $n\in\mathbb{N}$:
\medskip
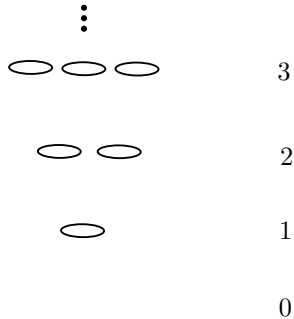
\begin{center} 
\ifx\JPicScale\undefined\def\JPicScale{1}\fi
\psset{unit=\JPicScale mm}
\psset{linewidth=0.3,dotsep=1,hatchwidth=0.3,hatchsep=1.5,shadowsize=1,dimen=middle}
\psset{dotsize=0.7 2.5,dotscale=1 1,fillcolor=black}
\psset{arrowsize=1 2,arrowlength=1,arrowinset=0.25,tbarsize=0.7 5,bracketlength=0.15,rbracketlength=0.15}
\begin{pspicture}(0,0)(39.21,40.93)
\rput{-0.7}(12.04,11.61){\psellipse[linewidth=0.25](0,0)(2.86,0.86)}
\rput(39.21,21.54){$2$}
\rput(38.81,32.84){$3$}
\rput(39.01,11.64){$1$}
\rput(39,1.4){$0$}
\rput(12.2,40.93){{\huge $\vdots$}}
\rput{-0.7}(8.92,22.14){\psellipse[linewidth=0.25](0,0)(2.85,0.86)}
\rput{-0.7}(16.88,22.11){\psellipse[linewidth=0.25](0,0)(2.86,0.85)}
\rput{-0.7}(19.25,33.13){\psellipse[linewidth=0.25](0,0)(2.86,0.86)}
\rput{-0.7}(12.2,33.16){\psellipse[linewidth=0.25](0,0)(2.86,0.86)}
\rput{-0.7}(5.14,33.31){\psellipse[linewidth=0.25](0,0)(2.86,0.86)}
\end{pspicture}
 \end{center} 
\item Morphisms are cobordisms $M:n\rightarrow m$ taking $n\in\mathbb{N}$
(strings) to $m\in\mathbb{N}$ (strings), which are defined up to homeomorphic  equivalence. Hence, if a cobordism can be continuously deformed into another cobordism, then these two cobordisms correspond to the same morphisms.  
\item For each object $n$, the identity $1_n:n\rightarrow n$ which is
given by $n$ parallel cylinders: 
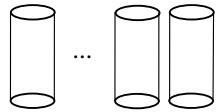
\begin{center} 
\ifx\JPicScale\undefined\def\JPicScale{1}\fi
\psset{unit=\JPicScale mm}
\psset{linewidth=0.3,dotsep=1,hatchwidth=0.3,hatchsep=1.5,shadowsize=1,dimen=middle}
\psset{dotsize=0.7 2.5,dotscale=1 1,fillcolor=black}
\psset{arrowsize=1 2,arrowlength=1,arrowinset=0.25,tbarsize=0.7 5,bracketlength=0.15,rbracketlength=0.15}
\begin{pspicture}(0,0)(27.41,14.07)
\rput{-0.04}(24.49,13.1){\psellipse[linewidth=0.25](0,0)(2.92,0.96)}
\rput{-0.04}(17.28,13.03){\psellipse[linewidth=0.25](0,0)(2.92,0.96)}
\rput{-0.04}(24.42,1.34){\psellipse[linewidth=0.25](0,0)(2.91,0.96)}
\rput{-0.04}(17.21,1.27){\psellipse[linewidth=0.25](0,0)(2.91,0.96)}
\psline[linewidth=0.15](27.41,13.06)(27.34,1.33)
\psline[linewidth=0.15](21.57,1.34)(21.58,13.13)
\psline[linewidth=0.15](20.14,13)(20.13,1.2)
\psline[linewidth=0.15](14.3,1.27)(14.31,13.13)
\rput{-0.04}(3.39,13.11){\psellipse[linewidth=0.25](0,0)(2.91,0.96)}
\rput{-0.04}(3.32,1.35){\psellipse[linewidth=0.25](0,0)(2.91,0.96)}
\psline[linewidth=0.15](6.25,13.08)(6.24,1.28)
\psline[linewidth=0.15](0.41,1.35)(0.42,13.21)
\rput(9.92,7.19){...}
\end{pspicture}
 \end{center}
\item
Composition is given by ``gluing'' manifolds
together, e.g.
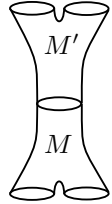
\begin{center} 
\ifx\JPicScale\undefined\def\JPicScale{1}\fi
\psset{unit=\JPicScale mm}
\psset{linewidth=0.3,dotsep=1,hatchwidth=0.3,hatchsep=1.5,shadowsize=1,dimen=middle}
\psset{dotsize=0.7 2.5,dotscale=1 1,fillcolor=black}
\psset{arrowsize=1 2,arrowlength=1,arrowinset=0.25,tbarsize=0.7 5,bracketlength=0.15,rbracketlength=0.15}
\begin{pspicture}(0,0)(13.57,26.69)
\rput{0}(10.64,1.2){\psellipse[linewidth=0.25](0,0)(2.94,0.8)}
\rput{0}(3.4,1.14){\psellipse[linewidth=0.25](0,0)(2.93,0.81)}
\psbezier[linewidth=0.25](6.33,1.09)(6.33,3.36)(7.65,3.36)(7.65,1.09)
\psbezier[linewidth=0.25](0.47,1.2)(4.54,8.29)(3.97,7.73)(4.08,13.27)
\psbezier[linewidth=0.25](13.57,1.37)(9.83,7.84)(10,6.63)(10,13.1)
\rput{0}(10.63,25.78){\psellipse[linewidth=0.25](0,0)(2.93,0.85)}
\rput{0}(3.4,25.84){\psellipse[linewidth=0.25](0,0)(2.93,0.85)}
\rput{0}(7.01,13.14){\psellipse[linewidth=0.25](0,0)(2.93,0.86)}
\psbezier[linewidth=0.25](6.33,25.9)(6.33,23.49)(7.65,23.49)(7.65,25.9)
\psbezier[linewidth=0.25](0.47,25.78)(4.54,18.27)(3.97,18.86)(4.08,13)
\psbezier[linewidth=0.25](13.57,25.6)(9.83,18.74)(10,20.03)(10,13.17)
\rput(6.9,7.8){$M$}
\rput(7.36,21.07){$M'$}
\end{pspicture}
 \end{center}
is a composite 
\[
M'\circ M:2\rightarrow 2
\]
where the cobordism $M':1\rightarrow 2$ is glued to
$M:2\rightarrow 1$ along the object $1$.  
\item The disjoint union of manifolds provides this category with a monoidal structure.  For example, if $M:1\rightarrow 0$ and $M':2\rightarrow 1$ are
cobordisms, then the cobordism  $M + M':1 + 2\rightarrow 0 + 1$ depicts as:
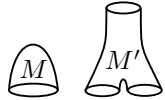
\begin{center} 
\ifx\JPicScale\undefined\def\JPicScale{1}\fi
\psset{unit=\JPicScale mm}
\psset{linewidth=0.3,dotsep=1,hatchwidth=0.3,hatchsep=1.5,shadowsize=1,dimen=middle}
\psset{dotsize=0.7 2.5,dotscale=1 1,fillcolor=black}
\psset{arrowsize=1 2,arrowlength=1,arrowinset=0.25,tbarsize=0.7 5,bracketlength=0.15,rbracketlength=0.15}
\begin{pspicture}(0,0)(22.09,14.18)
\rput{0}(5.05,2.52){\psellipse[linewidth=0.25](0,0)(3.25,0.85)}
\rput{0}(19.83,2.5){\psellipse[linewidth=0.25](0,0)(2.26,0.74)}
\rput{0}(14.23,2.44){\psellipse[linewidth=0.25](0,0)(2.26,0.74)}
\rput{0}(17.03,13.44){\psellipse[linewidth=0.25](0,0)(2.26,0.74)}
\psbezier[linewidth=0.25](16.49,2.4)(16.49,4.48)(17.51,4.48)(17.51,2.4)
\psbezier[linewidth=0.25](11.96,2.5)(15.12,9)(14.67,8.49)(14.76,13.58)
\psbezier[linewidth=0.25](22.09,2.65)(19.2,8.6)(19.33,7.48)(19.33,13.42)
\psbezier[linewidth=0.25](1.8,2.5)(2.32,10.45)(7.54,10.8)(8.3,2.5)
\rput(5.1,5.4){$M$}
\rput(17,6.6){$M'$}
\end{pspicture}
 \end{center}
\item 
The empty manifold 0 is the identity for the disjoint union. 
\item
The  {\em twist}
cobordism provides symmetry. For example, the twist 
\[
T_{1,1}:1 + 1\rightarrow 1 + 1
\]
is depicted as 
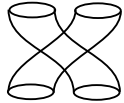
\begin{center} 
\ifx\JPicScale\undefined\def\JPicScale{1}\fi
\psset{unit=\JPicScale mm}
\psset{linewidth=0.3,dotsep=1,hatchwidth=0.3,hatchsep=1.5,shadowsize=1,dimen=middle}
\psset{dotsize=0.7 2.5,dotscale=1 1,fillcolor=black}
\psset{arrowsize=1 2,arrowlength=1,arrowinset=0.25,tbarsize=0.7 5,bracketlength=0.15,rbracketlength=0.15}
\begin{pspicture}(0,0)(15.24,13.15)
\rput{-0.03}(12.02,1.22){\psellipse[linewidth=0.25](0,0)(3.04,0.98)}
\rput{-0.03}(3.27,1.29){\psellipse[linewidth=0.25](0,0)(3.04,0.98)}
\rput{-0.03}(12.21,12.1){\psellipse[linewidth=0.25](0,0)(3.04,0.98)}
\rput{-0.03}(3.46,12.17){\psellipse[linewidth=0.25](0,0)(3.04,0.98)}
\psbezier[linewidth=0.25](0.3,1.39)(0.3,5.67)(9.23,8.13)(9.23,12.23)
\psbezier[linewidth=0.25](6.31,1.2)(6.31,5.49)(15.24,7.94)(15.24,12.04)
\psbezier[linewidth=0.25](8.98,1.26)(8.86,6.37)(0.3,7.06)(0.3,12.24)
\psbezier[linewidth=0.25](15.12,1.31)(15,6.42)(6.43,7.13)(6.43,12.3)
\end{pspicture}
 \end{center} 
The generalisation  to 
\[
T_{n,m}:m + n\rightarrow n + m
\]
for any $m,n\in\mathbb{N}$ should be obvious.  
\item The unit and counit 
\[
\eta_1:0\rightarrow 1+1 \quad\qquad  \mbox{and} \quad\qquad \epsilon_1:1+1\rightarrow 0
\]
of the   compact structure on $1$ are the cobordisms  
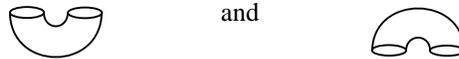
\begin{center} 
\ifx\JPicScale\undefined\def\JPicScale{1}\fi
\psset{unit=\JPicScale mm}
\psset{linewidth=0.3,dotsep=1,hatchwidth=0.3,hatchsep=1.5,shadowsize=1,dimen=middle}
\psset{dotsize=0.7 2.5,dotscale=1 1,fillcolor=black}
\psset{arrowsize=1 2,arrowlength=1,arrowinset=0.25,tbarsize=0.7 5,bracketlength=0.15,rbracketlength=0.15}
\begin{pspicture}(0,0)(12.63,6.87)
\rput{0.08}(10.37,6.12){\psellipse[linewidth=0.25](0,0)(2.27,0.71)}
\rput{0.08}(2.67,6.13){\psellipse[linewidth=0.25](0,0)(2.26,0.73)}
\psbezier[linewidth=0.25](4.9,6.2)(4.9,3.68)(8.1,3.68)(8.1,6.2)
\psbezier[linewidth=0.25](0.4,6.21)(0.41,-1.81)(12.61,-1.79)(12.6,6.23)
\end{pspicture}
 \qquad\qquad\raisebox{5mm}{\mbox{and}}\qquad\qquad 
\ifx\JPicScale\undefined\def\JPicScale{1}\fi
\psset{unit=\JPicScale mm}
\psset{linewidth=0.3,dotsep=1,hatchwidth=0.3,hatchsep=1.5,shadowsize=1,dimen=middle}
\psset{dotsize=0.7 2.5,dotscale=1 1,fillcolor=black}
\psset{arrowsize=1 2,arrowlength=1,arrowinset=0.25,tbarsize=0.7 5,bracketlength=0.15,rbracketlength=0.15}
\begin{pspicture}(0,0)(13.24,8.63)
\rput{-0.41}(10.95,1.05){\psellipse[linewidth=0.25](0,0)(2.27,-0.68)}
\rput{-0.41}(3.25,1.09){\psellipse[linewidth=0.25](0,0)(2.27,-0.7)}
\psbezier[linewidth=0.25](5.48,1.01)(5.5,3.4)(8.7,3.38)(8.68,0.99)
\psbezier[linewidth=0.25](0.98,1.02)(1.04,8.63)(13.24,8.54)(13.18,0.93)
\end{pspicture}
 \end{center}
We recover the equations of compactness as 
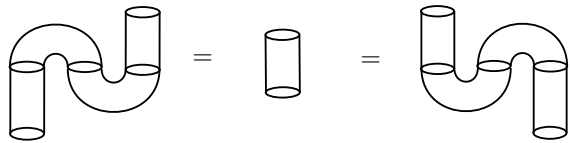
\begin{center} 
\ifx\JPicScale\undefined\def\JPicScale{1}\fi
\psset{unit=\JPicScale mm}
\psset{linewidth=0.3,dotsep=1,hatchwidth=0.3,hatchsep=1.5,shadowsize=1,dimen=middle}
\psset{dotsize=0.7 2.5,dotscale=1 1,fillcolor=black}
\psset{arrowsize=1 2,arrowlength=1,arrowinset=0.25,tbarsize=0.7 5,bracketlength=0.15,rbracketlength=0.15}
\begin{pspicture}(0,0)(74.3,18.31)
\rput{-0.07}(10.17,10.24){\psellipse[linewidth=0.25](0,0)(2.26,-0.68)}
\rput{-0.07}(2.47,10.23){\psellipse[linewidth=0.25](0,0)(2.27,-0.7)}
\psbezier[linewidth=0.25](4.71,10.16)(4.71,12.55)(7.91,12.55)(7.91,10.16)
\psbezier[linewidth=0.25](0.21,10.15)(0.22,17.76)(12.42,17.74)(12.41,10.13)
\rput{-0.07}(17.87,9.85){\psellipse[linewidth=0.25](0,0)(2.26,0.68)}
\psbezier[linewidth=0.25](12.41,9.93)(12.41,7.52)(15.61,7.52)(15.61,9.93)
\psbezier[linewidth=0.25](7.91,9.96)(7.9,2.3)(20.1,2.29)(20.11,9.95)
\psline[linewidth=0.25](15.61,9.83)(15.62,17.53)
\psline[linewidth=0.25](20.11,9.83)(20.12,17.53)
\rput{-0.07}(17.88,17.51){\psellipse[linewidth=0.25](0,0)(2.27,0.69)}
\psline[linewidth=0.25](0.28,10.28)(0.27,1.33)
\psline[linewidth=0.25](4.78,10.27)(4.77,1.32)
\rput{-0.07}(2.53,1.34){\psellipse[linewidth=0.25](0,0)(2.26,-0.79)}
\rput{0.42}(36.53,6.83){\psellipse[linewidth=0.25](0,0)(2.27,0.69)}
\psline[linewidth=0.25](34.26,6.79)(34.2,14.49)
\psline[linewidth=0.25](38.76,6.82)(38.7,14.52)
\rput{0.42}(36.47,14.49){\psellipse[linewidth=0.25](0,0)(2.26,0.69)}
\rput{-0}(64.6,10.34){\psellipse[linewidth=0.25](0,0)(2.21,0.68)}
\rput{-0}(72.1,10.32){\psellipse[linewidth=0.25](0,0)(2.21,0.7)}
\psbezier[linewidth=0.25](69.91,10.26)(69.91,12.65)(66.8,12.65)(66.8,10.26)
\psbezier[linewidth=0.25](74.3,10.24)(74.3,17.86)(62.42,17.86)(62.42,10.24)
\rput{-0}(57.1,9.97){\psellipse[linewidth=0.25](0,0)(2.2,-0.69)}
\psbezier[linewidth=0.25](62.42,10.04)(62.42,7.64)(59.3,7.64)(59.3,10.04)
\psbezier[linewidth=0.25](66.8,10.06)(66.8,2.4)(54.92,2.4)(54.92,10.06)
\psline[linewidth=0.25](59.3,9.94)(59.3,17.64)
\psline[linewidth=0.25](54.92,9.94)(54.92,17.64)
\rput{-0}(57.1,17.63){\psellipse[linewidth=0.25](0,0)(2.2,-0.69)}
\psline[linewidth=0.25](74.23,10.37)(74.23,1.42)
\psline[linewidth=0.25](69.85,10.37)(69.85,1.42)
\rput{-0}(72.02,1.44){\psellipse[linewidth=0.25](0,0)(2.21,0.8)}
\rput(25.9,11.2){$=$}
\rput(48.2,11){$=$}
\end{pspicture}
 \end{center}
which hold since all cobordisms involved are homeomorphically equivalent. The generalisation of the units to arbitrary $n$ is again obvious:
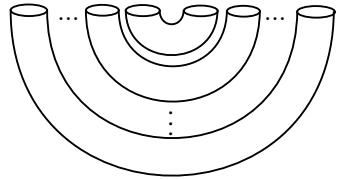
\begin{center} 
\ifx\JPicScale\undefined\def\JPicScale{1}\fi
\psset{unit=\JPicScale mm}
\psset{linewidth=0.3,dotsep=1,hatchwidth=0.3,hatchsep=1.5,shadowsize=1,dimen=middle}
\psset{dotsize=0.7 2.5,dotscale=1 1,fillcolor=black}
\psset{arrowsize=1 2,arrowlength=1,arrowinset=0.25,tbarsize=0.7 5,bracketlength=0.15,rbracketlength=0.15}
\begin{pspicture}(0,0)(43.76,26.31)
\rput{-0.37}(25.89,22.98){\psellipse[linewidth=0.25](0,0)(2.26,0.71)}
\rput{-0.37}(18.19,23.06){\psellipse[linewidth=0.25](0,0)(2.27,0.74)}
\psbezier[linewidth=0.25](20.43,23.11)(20.41,20.59)(23.61,20.57)(23.63,23.09)
\psbezier[linewidth=0.25](15.93,23.16)(15.88,15.14)(28.08,15.06)(28.13,23.08)
\rput{-0.37}(31.57,22.96){\psellipse[linewidth=0.25](0,0)(2.19,0.72)}
\rput{-0.37}(41.24,22.85){\psellipse[linewidth=0.25](0,0)(2.52,0.77)}
\rput{-0.37}(3.05,23.09){\psellipse[linewidth=0.25](0,0)(2.42,0.77)}
\rput{-0.37}(12.81,23.08){\psellipse[linewidth=0.25](0,0)(2.18,0.72)}
\psbezier[linewidth=0.25](10.63,23.19)(10.52,6.93)(33.65,6.78)(33.76,23.04)
\psbezier[linewidth=0.25](15,23.16)(14.93,13.16)(29.31,13.07)(29.38,23.07)
\psbezier[linewidth=0.25](5.48,23.18)(5.33,0.5)(38.58,0.29)(38.73,22.97)
\psbezier[linewidth=0.25](0.63,23.21)(0.44,-6.13)(43.4,-6.4)(43.59,22.94)
\rput(8.3,22){$...$}
\rput(35.8,21.9){$...$}
\rput(10.02,26.31){}
\rput(21.9,9.1){$\vdots$}
\end{pspicture}
 \end{center}
These together with corresponding counits  are easily seen to always satisfy the equations of compactness.    
\item The dagger consists in `flipping' the cobordisms, e.g.~if $M:2\rightarrow 1$ is 
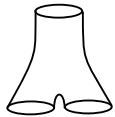
\begin{center}  \end{center} 
then $M^{\dagger}:1\rightarrow 2$ is 
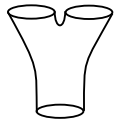
\begin{center} 
\ifx\JPicScale\undefined\def\JPicScale{1}\fi
\psset{unit=\JPicScale mm}
\psset{linewidth=0.3,dotsep=1,hatchwidth=0.3,hatchsep=1.5,shadowsize=1,dimen=middle}
\psset{dotsize=0.7 2.5,dotscale=1 1,fillcolor=black}
\psset{arrowsize=1 2,arrowlength=1,arrowinset=0.25,tbarsize=0.7 5,bracketlength=0.15,rbracketlength=0.15}
\begin{pspicture}(0,0)(15.1,16.04)
\rput{0}(11.99,15.09){\psellipse[linewidth=0.25](0,0)(3.11,-0.89)}
\rput{0}(4.31,15.15){\psellipse[linewidth=0.25](0,0)(3.11,-0.89)}
\rput{0}(8.15,1.92){\psellipse[linewidth=0.25](0,0)(3.11,-0.88)}
\psbezier[linewidth=0.25](7.42,15.21)(7.42,12.71)(8.82,12.71)(8.82,15.21)
\psbezier[linewidth=0.25](1.2,15.09)(5.53,7.28)(4.92,7.88)(5.04,1.77)
\psbezier[linewidth=0.25](15.1,14.9)(11.14,7.76)(11.32,9.1)(11.32,1.96)
\end{pspicture}
 \end{center} 
Clearly the dagger is compatible with the disjoint union which makes ${\bf 2Cob}$ a  dagger monoidal category.  It is also dagger compact  since $\sigma_{1,1}\circ\epsilon_1^\dagger$ is
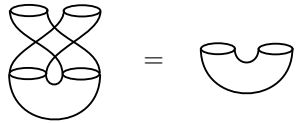
\begin{center} 
\ifx\JPicScale\undefined\def\JPicScale{1}\fi
\psset{unit=\JPicScale mm}
\psset{linewidth=0.3,dotsep=1,hatchwidth=0.3,hatchsep=1.5,shadowsize=1,dimen=middle}
\psset{dotsize=0.7 2.5,dotscale=1 1,fillcolor=black}
\psset{arrowsize=1 2,arrowlength=1,arrowinset=0.25,tbarsize=0.7 5,bracketlength=0.15,rbracketlength=0.15}
\begin{pspicture}(0,0)(38.34,15.87)
\psbezier[linewidth=0.25](5.54,6.64)(5.54,4.82)(7.74,4.82)(7.74,6.64)
\psbezier[linewidth=0.25](0.64,6.63)(0.65,-1.39)(12.75,-1.38)(12.74,6.64)
\rput{0.03}(10.2,6.69){\psellipse[linewidth=0.25](0,0)(2.45,0.76)}
\rput{0.03}(3.13,6.74){\psellipse[linewidth=0.25](0,0)(2.45,0.75)}
\rput{0.03}(10.34,15.07){\psellipse[linewidth=0.25](0,0)(2.45,0.75)}
\rput{0.03}(3.28,15.11){\psellipse[linewidth=0.25](0,0)(2.46,0.75)}
\psbezier[linewidth=0.25](0.73,6.81)(0.73,10.1)(7.94,12.01)(7.94,15.17)
\psbezier[linewidth=0.25](5.58,6.67)(5.58,9.97)(12.79,11.87)(12.79,15.02)
\psbezier[linewidth=0.25](7.74,6.72)(7.64,10.65)(0.73,11.17)(0.73,15.16)
\psbezier[linewidth=0.25](12.7,6.76)(12.6,10.7)(5.68,11.24)(5.68,15.22)
\rput{0.09}(36.07,10){\psellipse[linewidth=0.25](0,0)(2.27,0.71)}
\rput{0.09}(28.38,10.02){\psellipse[linewidth=0.25](0,0)(2.26,0.73)}
\psbezier[linewidth=0.25](30.62,10.09)(30.62,7.57)(33.81,7.57)(33.81,10.09)
\psbezier[linewidth=0.25](26.12,10.1)(26.13,2.08)(38.32,2.1)(38.31,10.12)
\rput(19.84,8.22){$=$}
\end{pspicture}
 \end{center}
which is again easily seen to be true for arbitrary $n$. 
\end{itemize}\em
\end{definition}

Obviously, we have been very informal here.  For a more elaborated discussion and technical details we refer the reader to \cite{B2004, BaezDolan, Kock, Turaev}.  The key thing to remember is that there are important `concrete' categories in which the morphisms are nothing like maps from the domain to the codomain.  

Note also that we can conceive --- again somewhat informally ---   the diagrammatic calculus of the previous sections as the result of contracting the diameter of the strings in ${\bf 2Cob}$ to  zero.  These categories of cobordisms play a key role in \em topological quantum field theory \em (TQFT). We discuss this topic in Section \ref{sec:TQFT}. 


\section{Classical-like tensors}\label{sec:classicallike} 

The tensors to which we referred as classical-like are not compact.  Instead
they do come with some other structure which, in all non-trivial cases, turns
out to be incompatible with compactness \cite{AbrClone}. In fact, this
incompatibility is the abstract incarnation of the No-Cloning theorem which
plays a key role in quantum information \cite{Dieks,WZ}.


\subsection{Cartesian categories}\label{sec:cart_cat}

Consider the category ${\bf Set}$ with the Cartesian product as the monoidal
tensor, as defined in Example \ref{Set_cart_natural_isos}. Given sets
$A_1,A_2\in|{\bf Set}|$,  their Cartesian product $A_1\times A_2$ consists of
all pairs $(x_1,x_2)$ with $x_1\in A_1$ and $x_2\in A_2$.  The fact that
Cartesian products consist of pairs is witnessed by the  \em projection \em
maps 
\[ 
\pi_1:A_1\times A_2\to A_1::(x_1,x_2)\mapsto x_1\quad\mbox{\rm
and}\quad\pi_2:A_1\times A_2\to A_2::(x_1,x_2)\mapsto x_2\,, 
\] 
which identify the respective components, together with the fact that, in turn, we can \em
pair \em $x_1=\pi_1(x_1,x_2)\in A_1$ and $x_2=\pi_2(x_1,x_2)\in A_2$ back
together into $(x_1,x_2)\in A_1\times A_2$, merely by putting brackets around
them.  We would like to express this fact purely in category-theoretic terms.
But both the projections and the pairing operation are expressed in terms of their 
action on elements, while categorical structure only recognises hom-sets, and 
not the internal structure of the underlying objects.
Therefore, we consider the action of projections on hom-sets, namely 
\[
\pi_1\circ -: {\bf Set}(C, A_1\times A_2)\to{\bf Set}(C, A_1) ::f\mapsto \pi_1\circ f\ 
\] 
and
\[ 
\pi_2\circ -: {\bf Set}(C, A_1\times A_2)\to{\bf Set}(C, A_2) ::f\mapsto \pi_2\circ f\,, 
\] 
which we can combine into a single operation `decompose'
\[ 
dec_C^{A_1,A_2}:{\bf Set}(C, A_1\times A_2)\to{\bf Set}(C, A_1)\times{\bf Set}(C, A_2):: f\mapsto (\pi_1\circ f, \pi_2\circ f)\,,
\]
 together with an  operation `recombine'
\[ rec_C^{A_1,A_2}:{\bf Set}(C, A_1)\times{\bf Set}(C, A_2)\to{\bf Set}(C, A_1\times A_2):: (f_1,f_2)\mapsto \langle f_1,f_2\rangle 
\] 
where 
\[ 
\langle f_1,f_2\rangle:C\to A_1\times A_2:: c\mapsto (f_1(c),f_2(c))\,. 
\] 
In this form we have 
\[ 
dec_C^{A_1,A_2}\,\circ\, rec_C^{A_1,A_2}=1_{{\bf Set}(C, A_1)\times{\bf Set}(C, A_2)} 
\] 
and 
\[ 
\ \,rec_C^{A_1,A_2}\,\circ\, dec_C^{A_1,A_2}= 1_{{\bf Set}(C, A_1\times A_2)}\,,
\] 
so $dec_C^{A_1,A_2}$ and $rec_C^{A_1,A_2}$ are now effectively each others inverses.  In the light
of Example \ref{ex:elements}, setting $C:=\{*\}$, we obtain  
\[
\xymatrix{
{\bf Set}({\{*\}}, A_1\times A_2) \ar@/^1pc/[rr]^{dec_{\{*\}}^{A_1,A_2}}
&&\ar@/^1pc/[ll]^{rec_{\{*\}}^{A_1,A_2}}\qquad\qquad\ \ {\bf Set}({\{*\}},
A_1)\times{\bf Set}({\{*\}}, A_2)\,, }
\] 
which corresponds to  projecting and pairing elements exactly as in the  discussion at the beginning of this section.   All of this extends in abstract generality. 

\begin{definition}\label{prod}\em 
A \em product \em of $A_1$ and $A_2\in |\cat|$ is a triple which consists of another object
$A_1\times A_2\in|\cat|$ together with two morphisms 
\[ 
\pi_1:A_1\times A_2\rTo A_1\ \ \ \ \mbox{ and }\ \ \ \ \pi_2:A_1\times
A_2\rTo A_2\,,
\] 
and which is such that for all $C\in|\cat|$ the mapping
\begin{equation}\label{isomapcart} 
(\pi_1\circ-,\pi_2\circ-): {\bf C}(C, A_1\times A_2)\to {\bf C}(C, A_1)\times{\bf C}(C, A_2)
\end{equation} 
admits an inverse $\langle -, -\rangle_{C, A_1, A_2}$.
\end{definition}

Below we omit the indices $C, A_1, A_2$ in $\langle -, -\rangle_{C, A_1, A_2}$.

\begin{definition}[Cartesian category] \em 
A category $\cat$  is  \em Cartesian
\em if any pair of objects $A,B\in|\cat|$ admits  a (not necessarily
unique) product. 
\end{definition}

\begin{proposition} 
If a pair of objects admits two distinct products then the
carrier objects are isomorphic in the category-theoretic sense of
Definition \ref{def:isomorphic}.  \end{proposition}

Indeed, suppose that  $A_1$ and $A_2\in |\cat|$ have two products $A_1 \times A_2$ and $A_1 \boxtimes A_2$ with respective projections 
\[
\pi_i:A_1\times A_2\rTo A_i\quad\mbox{ and }\quad
\pi'_j:A_1\boxtimes A_2\rTo A_j\,. 
\]
Consider the pairs of morphisms
\[
(\pi'_1,\pi'_2)\in\cat(A_1\boxtimes A_2,A_1)\times \cat(A_1\boxtimes A_2,A_2)
\]
and
\[
\ (\pi_1,\pi_2)\in\cat(A_1\times A_2,A_1)\times\cat(A_1\times A_2,A_2)\,.
\] 
By Definition~\ref{prod} we can apply the respective inverses of 
\[
(\pi_1\circ-,\pi_2\circ-)\qquad\mbox{and}\qquad
(\pi_1'\circ-,\pi_2'\circ-) 
\]
to these pairs, yielding morphisms in 
\[
\cat(A_1\boxtimes A_2,A_1\times A_2)\quad\mbox{ and }\quad\cat(A_1\times A_2,A_1\boxtimes A_2)\,,
\] 
 say $f$ and $g$ respectively, for which we have 
\[ 
\pi'_1=\pi_1\circ f,\ \ \ \pi_2'=\pi_2 \circ f,\ \ \ \pi_1=\pi'_1\circ g\ \ \ \mbox{and}\ \ \ \pi_2=\pi'_2\circ g.
\]
Then, it follows that 
\begin{eqnarray*}
(\pi'_1\circ 1_{A_1\boxtimes A_2},\pi'_2\circ 1_{A_1\boxtimes A_2}) =
(\pi_1\circ f,\pi_2\circ f) = (\pi'_1\circ g\circ f,\pi'_2\circ
g\circ f)\,,
\end{eqnarray*}	
and applying the inverse to $(\pi_1'\circ-,\pi_2'\circ-)$ now gives $1_{A_1\boxtimes A_2}=g\circ f$. An analogue argument gives $f\circ g=1_{A_1\times A_2}$ so $f$ is an isomorphism between the two objects $A_1 \times A_2$ and $A_1 \boxtimes A_2$ with $g$ as its inverse.
%

The above definition of products in terms of `decomposing and recombining
compound objects' is not the one that one usually finds in the literature. 

\begin{definition}\label{prodBIS}\em 
A \em product \em of two objects $A_1$ and
$A_2$ in a category $\cat$ is a triple consisting of another object
$A_1\times A_2\in|\cat|$ together with two morphisms 
\[ 
\pi_1:A_1\times A_2\rTo A_1\quad\mbox{\rm and}\quad\pi_2:A_1\times A_2\rTo A_2\,,
\] 
and which is such that for any object $C\in|\cat|$, and any pair of morphisms
$C\rTo^{f_1} A_1$ and  $C\rTo^{f_2} A_2$ in ${\bf C}$, there exists a
\underline{unique} morphism $C\rTo^{f} A_1\times A_2$ such that 
\[
f_1=\pi_1\circ f\qquad\mbox{\rm and}\qquad f_2=\pi_2\circ f\,.
\]
\end{definition}

We can concisely summarise this \em universal \em property by the commutative diagram 
\[ 
\xymatrix@=.6in{& \forall C\ar[dl]_{\forall
f_1}\ar[dr]^{\forall f_2}\ar@{..>}[d]|{\exists ! f}& \\ A_1 & A_1\times A_2
\ar[l]^{\pi_1}\ar[r]_{\pi_2} & A_2\,.} 
\]

It is easy to see that this definition is equivalent to the
previous one: the inverse $\langle -,-\rangle$ to $(\pi_1\circ-,\pi_2\circ-)$ provides for any pair $(f_1, f_2)$ a unique morphism $f:=\langle f_1, f_2\rangle$ which is such that $(\pi_1\circ f,\pi_2\circ f)=(f_1, f_2)$. Conversely, uniqueness of $C\rTo^{f} A_1\times A_2$ guarantees $(\pi_1\circ-,\pi_2\circ-)$ to have an inverse $\langle -,-\rangle$, which is obtained by setting $\langle f_1,f_2\rangle:= f$.  

For  more details on  this definition, and the reason for its prominence in the literature, we refer to \cite{AbramskyT} and standard textbooks such as \cite{Adamek,SML}.  

\begin{proposition}\label{prop:Cart_monoid} 
If a category $\cat$ is Cartesian, then each choice of a product for each pair of objects always defines a
symmetric monoidal structure on $\cat$ with $A\otimes B:=A\times B$, and with the terminal object as the monoidal unit.  
\end{proposition}

Proving this requires work. First, for $f:A_1\rTo B_1$ and $g:A_2\rTo B_2$ 
let 
\[
f\times g: A_1\times A_2\rTo B_1\times B_2
\]
be the unique morphism defined in terms of Definition \ref{prodBIS} within
\[
\xymatrix@=.6in{ &
A_1\times A_2\ar[dl]_{f\circ\pi_1}\ar[dr]^{g\circ \pi_2}
\ar@{..>}[d]|{f\times g}\\ B_1 & B_1\times
B_2\ar[l]^{\pi'_1}\ar[r]_{\pi'_2} & B_2 } 
\] 
Then it immediately follows that the diagrams
\beq\label{diag:proofcartmonoid}
\xymatrix@=.6in{ 
A_1\ar[d]_{f} & A_1\times A_2\ar[d]|{f\times g}\ar[r]^{\pi_2}\ar[l]_{\pi_1} & A_2\ar[d]^{g}\\ 
B_1 & B_1\times B_2\ar[r]_{\pi'_2}\ar[l]^{\pi'_1} & B_2
}
\eeq
commute.  From Definition \ref{prod} we know that for any $h$,
\begin{equation}\label{pro_form}
\langle\pi_1\circ h,\pi_2\circ h\rangle = h\,,
\end{equation}
and,  in particular, this  entails
\begin{equation}\label{projid}
\langle\pi_1,\pi_2\rangle=\langle\pi_1\circ 1_{A_1\times A_2},\pi_2\circ 1_{A_1\times A_2}\rangle=1_{A_1\times A_2}. 
\end{equation}
Using eq.(\ref{pro_form}) for  $A\rTo^{f} B$, $B\rTo^{g} C$ and $B\rTo^{h} D$
we have 
\begin{eqnarray*} \langle g,h\rangle\circ f&=& \langle
\pi_1\circ(\langle g,h\rangle\circ f),\pi_2\circ(\langle
g,h\rangle\circ f)\rangle\\ &=& \langle (\pi_1\circ\langle
g,h\rangle)\circ f,(\pi_2\circ\langle g,h\rangle)\circ
f\rangle\\ &=& \langle g\circ f, h\circ f\rangle\,.
\end{eqnarray*} 
Using this, for $A\rTo^f B$,
$A\rTo^g C$, $B\rTo^h D$ and $C\rTo^k E$\,, we have
\begin{eqnarray*}
(h\times k)\circ \langle f,g\rangle
&=& \langle h\circ \pi_1, k\circ \pi_2\rangle'\circ \langle f,g\rangle\\
&=& \langle h\circ \pi_1\circ \langle f,g\rangle, k\circ \pi_2\circ \langle f,g\rangle\rangle'\\
&=& \langle h\circ f, k\circ g\rangle'\,,
\end{eqnarray*} 
where $\langle -,-\rangle'$ is the pairing operation relative to $(\pi_1'\circ-, \pi_2'\circ-)$. In a similar manner the reader can verify that  $-\times -$ is bifunctorial.

To support the claim in Proposition \ref{prop:Cart_monoid} we will now also construct the required natural isomorphisms, and leave verification of the  coherence diagrams to the reader. Let $!_A$ be the unique morphism of type $A\rTo\top$.  Setting
\begin{equation}\nonumber
\lambda_A:=\langle !_A,1_A\rangle:A\rTo \top\times A
\end{equation} 
we have
\begin{equation}\nonumber
\langle !_B,1_B\rangle\circ f=\langle !_B\circ f, 1_B\circ f\rangle=\langle !_A,f\circ 1_A\rangle=(!_\top\times f)\circ\langle !_A,1_A\rangle\,,
\end{equation}
so we have established commutation of 
\begin{equation}\nonumber
\xymatrix{
A\ar[r]^{\lambda_A\ \ \ }\ar[d]_f & \top\times A\ar[d]^{1_\top\times f}\\
B\ar[r]_{\lambda_B\ \ \ } & \top\times B
}
\end{equation} 
that is, $\lambda$ is natural. The components are moreover isomorphisms with $\pi_2$ as inverse.  The fact that $\pi_2\circ\lambda_A=1_A$ holds by definition,  and from 
\[
\xymatrix @=.6in{
\top\ar[d]_{!_\top} & \top\times A\ar[r]^{\pi_2}\ar[l]_{\pi_1}\ar[d]|{!_\top\times 1_A} & A\ar[d]^{1_A}\\
\top & \top\times A\ar[l]^{\pi'_1}\ar[r]_{\pi'_2} & A 
}
\]
and  the fact that by the terminality of $\top$ we have
\begin{equation}\nonumber
!_{\top\times A}=!_\top\circ\pi_1=!_{A}\circ\pi_2
\end{equation}
it follows that
\[
\xymatrix @=.6in{
& \top\times A\ar[dr]^{1_{A}\circ\pi_2}\ar[dl]_{!_A\circ\pi_2}\ar@{..>}[d]|{\langle !_A\circ\pi_2,1_A\circ\pi_2\rangle}\\
\top & \top\times A\ar[r]_{\pi'_2}\ar[l]^{\pi'_1} & A 
}
\]
commutes, so by uniqueness, it follows that $\langle !_A\circ\pi_2,1_A\circ\pi_2\rangle=!_\top\times 1_A$, and hence
\begin{equation}\nonumber
\langle !_A,1_A\rangle\circ\pi_2=\langle !_A\circ\pi_2,1_A\circ\pi_2\rangle=!_\top\times 1_A=1_\top\times 1_A=1_{\top\times A}\,.
\end{equation}
Similarly the components $\rho_A:=\langle 1_A,!_A\rangle$ also define a natural isomorphism.

For associativity, 
let us fix some notation for the projections:
\[
A \lTo^{\pi_1} A\times (B\times C)\rTo^{\pi_2} B\times C \quad\mbox{and}\quad B \lTo^{\pi'_1} B\times C\rTo^{\pi'_2}  C\,.
\]
We define a morphism of type $A\times (B\times C)\rightarrow A\times B$ within
\begin{equation}\nonumber
\xymatrix@=.6in {
& A\times (B\times C)\ar[dl]_{\pi_1}\ar[dr]^{\pi'_1\circ\pi_2}\ar@{..>}[d]|{\langle \pi_1,\pi'_1\circ\pi_2\rangle} &\\
A & A\times B\ar[l]^{\pi''_1}\ar[r]_{\pi''_2} & B }
\end{equation}
and we define $\alpha_{A,B,C}$ within
\begin{equation}\nonumber
\xymatrix@=.6in {
& A\times (B\times C)\ar[dl]_{\langle\pi_1,\pi'_1\circ\pi_2\rangle}\ar[dr]^{\pi'_2\circ\pi_2}\ar@{..>}[d]|{\langle\langle\pi_1,\pi'_1\circ\pi_2\rangle,\pi'_2\circ\pi_2\rangle} & \\
A\times B & (A\times B)\times C\ar[l]^{\pi'''_1} \ar[r]_{\pi'''_2} & C}
\end{equation}
Naturality as well as the fact that the components are isomorphisms relies on uniqueness of the morphisms as defined above and is left to the reader.  

For symmetry, the components $\sigma_{A,B}:A\times B\rTo B\times A$ are defined within
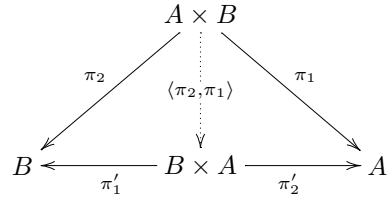
\begin{equation}\nonumber
\xymatrix@=.6in{	& A\times B\ar[dl]_{\pi_2}\ar[dr]^{\pi_1}\ar@{..>}[d]|{\langle\pi_2,\pi_1\rangle} & \\
B & B\times A\ar[l]^{\pi'_1}\ar[r]_{\pi'_2} & A}
\end{equation}
where again we leave verifications to the reader.

\subsection{Copy-ability and delete-ability}   

So how does all this translate in term of morphisms as physical processes?  By
a \em uniform copying operation \em or \em diagonal \em in a monoidal category
$\cat$ we mean a natural transformation 
\[ 
\diag=\left\{A\rTo^{\scdiag _A} A\otimes A\bigm|A\in|\cat|\right\}\,.  
\] 
The corresponding commutativity requirement
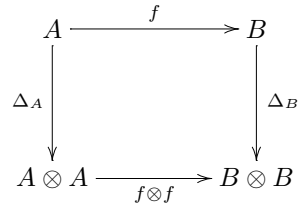
\[ 
\xymatrix@=0.6in{ A\ar[d]_{\scdiag_A}\ar[r]^{f} & B\ar[d]^{\scdiag_B}\\%
A\otimes A\ar[r]_{f\otimes f} & B\otimes B} 
\] 
expresses that `when performing operation $f$ on a system $A$ and then copying it', is the same as
`copying system $A$ and then performing operation $f$ on each copy'.   For
example, correcting typos on a sheet of written paper and then Xeroxing it is
the same as first Xeroxing it and then correcting the typos on each of the copies.  

The category ${\bf Set}$ has
\[ 
\left\{\scdiag_X: X\to X\times X::x\mapsto (x,x)\bigm|X\in|{\bf Set}|\right\} 
\] 
as a uniform copying operation since we have commutation of
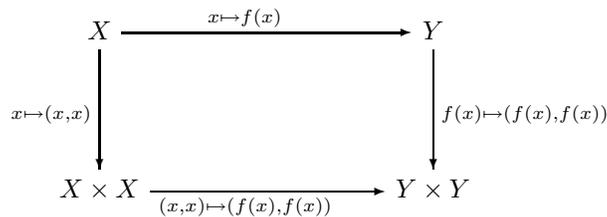
\begin{diagram} X&\rTo^{x\mapsto f(x)}&Y\\
\dTo^{x\mapsto(x,x)}&&\dTo_{f(x)\mapsto(f(x),f(x))}\\ 
X\times X&\rTo_{(x,x)\mapsto (f(x),f(x))}&Y\times Y 
\end{diagram}

\begin{example} 
Is there a uniform copying operation in ${\bf FdHilb}$?
We cannot just set 
\[ 
\scdiag_{\cal H}: {\cal H}\to {\cal H}\otimes
{\cal H}::\psi\mapsto \psi\otimes\psi 
\] 
since this map is not even linear.  On
the other hand, when for each Hilbert space ${\cal H}$ a basis
$\{|i\rangle\}_i$ is specified, we can consider 
\[ 
\left\{\scdiag_{\cal H}:
{\cal H}\to {\cal H}\otimes {\cal H}::|i\rangle\mapsto |i\rangle\otimes
|i\rangle\bigm|{\cal H}\in|{\bf FdHilb}|\right\}\,.
\]

\noindent But now the diagram 
\begin{diagram} \mathbb{C}&\rTo^{1\mapsto |0\rangle+|1\rangle}&\mathbb{C}\oplus\mathbb{C}\\ \dTo^{1\mapsto
1\otimes 1}&&\dTo_{\begin{array}{c} |0\rangle\mapsto
|0\rangle\otimes|0\rangle\vspace{1mm}\\ |1\rangle\mapsto
|1\rangle\otimes|1\rangle \end{array}}\\
\hspace{-0.7cm}\mathbb{C}\simeq\mathbb{C}\otimes\mathbb{C}&\rTo_{1\otimes
1\mapsto
(|0\rangle+|1\rangle)\otimes(|0\rangle+|1\rangle)}&(\mathbb{C}\oplus\mathbb{C})\otimes(\mathbb{C}\oplus\mathbb{C})
\end{diagram} 
fails to commute, since via one path we obtain the (unnormalized) \em Bell-state \em 
\[ 
1\mapsto |0\rangle\otimes|0\rangle+ |1\rangle\otimes|1\rangle\,, 
\] 
while via the other path we obtain an (unnormalized) \em disentangled state\em 
\[ 
1\mapsto (|0\rangle+|1\rangle)\otimes(|0\rangle+|1\rangle)\,.
\] 
This inability to define a uniform copying operation reflects the fact that we cannot
copy (unknown) quantum states.
\end{example} 

\begin{example} 
Let us now  turn our attention to ${\bf Rel}$ and, given that
every function is also a relation, consider the family of
functions which provided a uniform copying operation for ${\bf Set}$.  In more typical relational notation we have
\[ 
\scdiag_X:=\{(x,(x,x))\mid x\in X\}\subseteq X\times (X\times X)\,.
\]
However, the diagram 
\begin{diagram} 
\{*\}&\rTo^{\{(*,0),(*,1)\}}&\{0,1\}\\
\dTo^{\{(*,(*,*))\}}&&\dTo_{\{(0,(0,0)),(1,(1,1))\}}\\
\hspace{-1.6cm}\{(*,*)\}=\{*\}\times\{*\}&\rTo_{\{(*,0),(*,1)\}\times\{(*,0),(*,1)\}}&\{0,1\}\times\{0,1\}
\end{diagram} 
fails to commute, since via one path we have 
\[
\{(*,(0,0)),(*,(1,1))\}=\{*\}\times\{(0,0),(1,1)\}\,, 
\] 
while the other path yields 
\[ 
\{(*,(0,0)),(*,(0,1)),(*,(1,0)),(*,(1,1))\} =\{*\}\times(\{0,1\}\times\{0,1\})\,.  
\] 
Note here in particular the similarity with the counterexample that we provided for the case of ${\bf
FdHilb}$, by identifying 
\beqa 
|0\rangle\otimes|0\rangle+ |1\rangle\otimes|1\rangle \ &\stackrel{\sim}{\longleftrightarrow}&\ \
\{(0,0),(1,1)\}\vspace{4mm}\\ 
(|0\rangle+|1\rangle)\otimes(|0\rangle+|1\rangle)\ &\stackrel{\sim}{\longleftrightarrow}&\ \ \{0,1\}\times\{0,1\}\,. 
\eeqa
\end{example} 

\begin{example} 
Similarly, the cobordism 
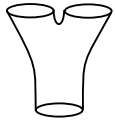
\begin{center}  \end{center} 
is not a component of a uniform copying relation
\[
\{\diag_n:n\rightarrow n+n\mid n\in \mathbb{N}\}\,,
\]
since in
\[
\xymatrix@=0.6in{ 0\ar[r]^{\scdiag_0}\ar[d]_M & 0+0\ar[d]^{M+M}\\
1\ar[r]_{\scdiag_1} & 1+1 } 
\]
where $M:0\rightarrow 1$ is 
\begin{center}
\ifx\JPicScale\undefined\def\JPicScale{1}\fi
\psset{unit=\JPicScale mm}
\psset{linewidth=0.3,dotsep=1,hatchwidth=0.3,hatchsep=1.5,shadowsize=1,dimen=middle}
\psset{dotsize=0.7 2.5,dotscale=1 1,fillcolor=black}
\psset{arrowsize=1 2,arrowlength=1,arrowinset=0.25,tbarsize=0.7 5,bracketlength=0.15,rbracketlength=0.15}
\begin{pspicture}(0,0)(6.15,6.46)
\rput{-0.44}(3.88,5.85){\psellipse[linewidth=0.25](0,0)(2.27,-0.6)}
\psbezier[linewidth=0.25](1.62,5.87)(1.94,0.25)(5.57,-0.02)(6.15,5.84)
\end{pspicture}
 \end{center} 
the upper path gives 
\begin{center} 
\ifx\JPicScale\undefined\def\JPicScale{1}\fi
\psset{unit=\JPicScale mm}
\psset{linewidth=0.3,dotsep=1,hatchwidth=0.3,hatchsep=1.5,shadowsize=1,dimen=middle}
\psset{dotsize=0.7 2.5,dotscale=1 1,fillcolor=black}
\psset{arrowsize=1 2,arrowlength=1,arrowinset=0.25,tbarsize=0.7 5,bracketlength=0.15,rbracketlength=0.15}
\begin{pspicture}(0,0)(11.32,5.51)
\rput{-0.81}(9.05,4.79){\psellipse[linewidth=0.25](0,0)(2.26,-0.6)}
\psbezier[linewidth=0.25](6.78,4.83)(7.06,-0.79)(10.7,-1.08)(11.31,4.77)
\rput{-0.81}(3.05,4.87){\psellipse[linewidth=0.25](0,0)(2.26,-0.6)}
\psbezier[linewidth=0.25](0.78,4.92)(1.06,-0.71)(4.7,-1)(5.31,4.85)
\end{pspicture}
 \end{center} 
while the lower path gives 
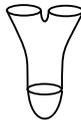
\begin{center} 
\ifx\JPicScale\undefined\def\JPicScale{1}\fi
\psset{unit=\JPicScale mm}
\psset{linewidth=0.3,dotsep=1,hatchwidth=0.3,hatchsep=1.5,shadowsize=1,dimen=middle}
\psset{dotsize=0.7 2.5,dotscale=1 1,fillcolor=black}
\psset{arrowsize=1 2,arrowlength=1,arrowinset=0.25,tbarsize=0.7 5,bracketlength=0.15,rbracketlength=0.15}
\begin{pspicture}(0,0)(11.77,17.45)
\rput{-0.22}(9.5,16.67){\psellipse[linewidth=0.25](0,0)(2.27,-0.7)}
\rput{-0.22}(3.9,16.74){\psellipse[linewidth=0.25](0,0)(2.26,-0.7)}
\rput{-0.22}(6.66,6.25){\psellipse[linewidth=0.25](0,0)(2.26,-0.7)}
\psbezier[linewidth=0.25](6.17,16.77)(6.16,14.79)(7.18,14.79)(7.19,16.77)
\psbezier[linewidth=0.25](1.64,16.7)(4.78,10.49)(4.33,10.98)(4.4,6.13)
\psbezier[linewidth=0.25](11.77,16.51)(8.86,10.86)(8.99,11.92)(8.97,6.26)
\psbezier[linewidth=0.25](4.46,6.1)(4.8,1.2)(8.44,0.98)(8.99,6.08)
\end{pspicture}
 \end{center}
\end{example} 


The category ${\bf Set}$ admits a uniform copying operation as  a consequence of being Cartesian. We indeed have the following general result.

\begin{proposition} 
Each Cartesian category admits a uniform copying operation.
\end{proposition}

Indeed, let 
\[
\diag_A:=\langle 1_A,1_A\rangle
\]
and let $A\rTo^f B$ be arbitrary. Then we have 
\[
\langle 1_B,1_B\rangle\circ f=\langle 1_B\circ f,1_B\circ f\rangle =\langle
f\circ 1_A,f\circ 1_A\rangle=(f\times f)\circ \langle 1_A,1_A\rangle\,,
\] 
so $\diag$ is a natural transformation, and hence a uniform copying operation. 

In fact, one can define Cartesian categories in terms of the existence of a
uniform copying operation and  a corresponding uniform deleting operation 
\[
{\cal E}= \left\{A\rTo^{{\cal E}_A} \II\bigm|A\in|\cat|\right\}\,,  
\] 
for which the naturality constraint now means that 
\[ \xymatrix@=0.6in{
A\ar[d]_{{\cal E}_A}\ar[r]^{f} & B\ar[ld]^{{\cal E}_B}\\%
\II} 
\] 
commutes. There are some additional constraints such as `first copying and then deleting results
in the same as doing nothing', and similar ones, which all together
formally boil down to saying that for each object $A$ in the category the
triple $(A,\scdiag_A,{\cal E}_A)$ has to be an \em internal commutative
comonoid\em.  We will define the concept of internal commutative
comonoid below in Section \ref{sec:comonoids}.

\begin{example} 
The fact that the diagonal in ${\bf Set}$ fails to be a diagonal in ${\bf Rel}$ seems to indicate that in ${\bf Rel}$ the Cartesian product does not provide a product in the sense of Definition \ref{prod}.  Consider 
\[
\xymatrix@=.6in{&\{*\}\ar[dl]_{\emptyset}\ar[dr]^{1_{\{*\}}}\ar@{..>}[d]|{\exists
! f}& \\ \{*\} & \{*\}\times \{*\} \ar[l]^{\pi_1}\ar[r]_{\pi_2} &
\{*\}} 
\] 
where $\emptyset$ stands for the empty relation.  Since
$\{*\}\times \{*\}=\{(*,*)\}$ is a singleton there are only two
possible choices for $\pi_1$ and $\pi_2$, namely the empty relation and
the singleton relation $\{((*,*), *)\}\subseteq\{(*,*)\}\times\{*\}$.
Similarly there are also only two candidate relations to play the
role of $f$.  So since $\pi_1\circ f=\emptyset$ either $\pi_1$ or $f$
has to be $\emptyset$ and since  $\pi_2\circ f=1_{\{*\}}$ neither
$\pi_2$ nor $f$ can be $\emptyset$.  Thus $\pi_1$ has to be the empty
relation and $\pi_2$ has to be the singleton relation.  However, when
considering 
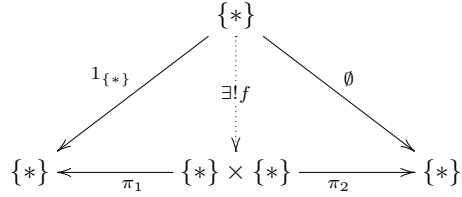
\[
\xymatrix@=.6in{&\{*\}\ar[dl]_{1_{\{*\}}}\ar[dr]^{\emptyset}\ar@{..>}[d]|{\exists
! f}& \\ \{*\} & \{*\}\times \{*\} \ar[l]^{\pi_1}\ar[r]_{\pi_2} &
\{*\}} 
\] 
$\pi_2$ has to be the empty relation and $\pi_1$ has to be
the singleton relation, so we have a contradiction.  Key to all this is
the fact that the empty relation is a relation, while it is not a
function, or more generally, that relations need not be \em total \em
(total = each argument is assigned to a value).  On the other hand, when
showing that the diagonal in ${\bf Set}$ was not a diagonal in ${\bf
Rel}$ we relied on the multi-valuedness of the relation  $\{(*,
0),(*,1)\}\subseteq\{*\}\times\{0,1\}$.  Hence multi-valuedness of
certain relations obstructs the existence of a natural diagonal in
${\bf Rel}$, while the lack of totality of certain relations obstructs
the existence of faithful projections in ${\bf Rel}$, causing a
break-down of the Cartesian structure of $\times$ in ${\bf Rel}$ as
compared to the role it plays in ${\bf Set}$.  \end{example}

\subsection{Disjunction vs.~conjunction}\label{sec:coprod}

As we saw in Section \ref{sec:cart_cat}, the fact that in ${\bf Set}$
Cartesian products $X\times Y$ consist of pairs $(x,y)$ of elements $x\in X$
and $y\in Y$ can be expressed in terms of a bijective correspondence 
\[ 
{\bf Set}(C, A_1\times A_2)\simeq {\bf Set}(C, A_1)\times{\bf Set}(C, A_2)\,.  
\]
One can then naturally ask whether we also have that 
\[ {\bf Set}(A_1\times A_2, C)\stackrel{?}{\simeq}{\bf Set}(A_1, C)\times{\bf Set}(A_2, C)\,. 
\] 
The answer is no.  But we do have 
\[ {\bf Set}(A_1+ A_2, C)\simeq {\bf Set}(A_1, C)\times{\bf Set}(A_2, C)\,. 
\] 
where $A_1+ A_2$ is the disjoint union of two sets $A_1$ and $A_2$, that is, we repeat, 
\[ A_1+ A_2:=\{(x_1, 1)\mid x_1\in A_1\}\cup\{(x_2, 2)\mid x_2\in A_2\}\,. 
\] 
This isomorphism now involves \em
injection \em maps 
\[ 
\iota_1:A_1\to A_1 + A_2::x_1\mapsto (x_1, 1)\quad\mbox{\rm and}\quad\iota_2:A_2\to A_1 + A_2::x_2\mapsto (x_2, 2)\,
\]
They embed the elements of $A_1$ and $A_2$ within $A_1 + A_2$.
Their action on hom-sets is 
\[ -\circ\iota_1: {\bf Set}(A_1+ A_2, C)\to{\bf Set}(A_1, C) ::f\mapsto f\circ\iota_1\ \] \[ -\circ\iota_2: {\bf Set}(A_1+ A_2,
C)\to{\bf Set}(A_2, C) ::f\mapsto f\circ\iota_2\,,
\] 
which converts a function that takes values on all elements that either live in $A_1$ \em or \em $A_2$,   into two functions, one that takes values in $A_1$, and one that takes values in $A_2$.  We can again recombine
these two operations in a single one 
\[ codec_C^{A_1,A_2}:{\bf Set}(A_1+ A_2, C)\to{\bf Set}(A_1, C)\times{\bf Set}(A_2, C):: f\mapsto
(f\circ\iota_1,f\circ\iota_2) 
\] 
which has an inverse, namely 
\[
corec_C^{A_1,A_2}:{\bf Set}(A_1, C)\times{\bf Set}(A_2, C)\to{\bf Set}(A_1+
A_2, C):: (f_1,f_2)\mapsto [f_1,f_2] \] where \[ [ f_1,f_2]:A_1+ A_2\to C::
\left\{\begin{array}{l c c} x\mapsto f_1(x) & \mbox{\it iff} & x\in A_1\\
x\mapsto f_2(x) & \mbox{\it iff} & x\in A_2 \end{array}\right.\,.
\] 
The binary operation $[-,-]$ on functions now recombines two functions $f_1$ and $f_2$ into a single one.  We have an isomorphism 
\[\xymatrix{ {\bf Set}(A_1+ A_2, C) \ar@/^1pc/[rr]^{codec_{\{*\}}^{A_1,A_2}}
&&\ar@/^1pc/[ll]^{corec_{\{*\}}^{A_1,A_2}}\qquad\qquad\ \ {\bf
Set}(A_1, C)\times{\bf Set}(A_2, C)\,.  }
\] 
Note that while $[f_1,f_2]$ produces an image either for the function $f_1$ \em
\underline{or} \em the function $f_2$, in contrast $\langle f_1,f_2\rangle$ produces
an image both for the function $f_1$ \em \underline{and} \em the
function $f_2$.  In operational terms, while the product allows to  describe a
pair of (classical) systems, the disjoint union allows to describe a situation where we have either of two systems.  For example, it allows to describe the \em branching structure \em that arises as a consequence of  non-determinism.  

\begin{definition}\label{coprod}\em 
A \em coproduct \em of two objects $A_1$
and $A_2$ in a category $\cat$ is a triple consisting of another object
$A_1+A_2\in|\cat|$ together with two morphisms 
\[ \iota_1:A_1\rTo A_1 + A_2\ \ \ \ \mbox{ and }\ \ \ \ \iota_2:A_2\rTo A_1 + A_2\,,
\] 
and which is such that  for all $C\in|\cat|$ the mapping 
\[ 
(-\circ\iota_1,-\circ\iota_2):{\bf C}(A_1+ A_2, C)\to{\bf C}(A_1, C)\times{\bf C}(A_2, C) 
\] 
admits an inverse.  A category $\cat$ \em  is co-Cartesian \em if any
pair of objects $A,B\in|\cat|$ admits  a (not necessarily unique)
coproduct.  
\end{definition}

As in the case of products, we also have the following variant:

\begin{definition}\label{coprodBIS}\em A \em coproduct \em of two objects $A_1$
and $A_2$ in a category $\cat$ is a triple consisting of another object
$A_1+A_2\in|\cat|$ together with two morphisms 
\[ 
\iota_1:A_1\rTo A_1 + A_2\ \ \ \ \mbox{ and }\ \ \ \ \iota_2:A_2\rTo A_1 + A_2\,,
\] 
and which is such that  for any object $C\in|\cat|$, and any pair of morphisms $A_1\rTo^{f_1}
C$ and  $A_2\rTo^{f_2} C$ in ${\bf C}$, there exists a
\underline{unique} morphism $A_1+ A_2\rTo^{f} C$ such that 
\[ 
f_1=f\circ\iota_1\qquad\mbox{\rm and}\qquad f_2=f\circ\iota_2\,.
\]
\end{definition}

We can again represent this in a commutative diagram:
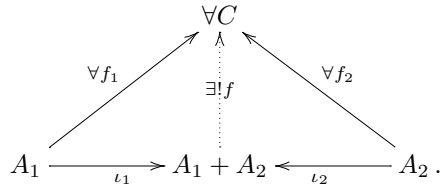
\[ 
\xymatrix@=.6in{& \forall C& \\ A_1\ar[ur]^{\forall f_1}\ar[r]_{\iota_1} &
A_1+A_2\ar@{..>}[u]|{\exists ! f} & A_2\ar[l]^{\iota_2}\ar[ul]_{\forall
f_2}\,.} 
\] 

As a counterpart to the diagonal which we have in Cartesian categories we now
have a \em codiagonal\em, with components \[ \nabla_A:=[1_A,1_A]:A+A\rTo A\,.
\]

\begin{example}\label{ex:lattice} 
As explained in Example \ref{excat6}, we can
think of a partially ordered set $P$ as a category ${\bf P}$.  In such
a category  products turn out to be  \em greatest lower bounds \em or
\em meets\em, and coproducts turn out to be \em least upper bounds \em
or \em joins\em.  The existence of an isomorphism 
\[\xymatrix{ {\bf P}(a_1+ a_2, c) \ar@/^1pc/[rr]^{codec_{c}^{a_1,a_2}}
&&\ar@/^1pc/[ll]^{corec_{c}^{a_1,a_2}}\qquad\qquad\ \ {\bf P}(a_1,
c)\times{\bf P}(a_2, c)\,, }
\] 
given that ${\bf P}(a_1+ a_2, c)$, ${\bf P}(a_1, c)$ and ${\bf P}(a_2, c)$ and hence also ${\bf P}(a_1,
c)\times{\bf P}(a_2, c)$ are all either singletons or empty, means that
${\bf P}(a_1+ a_2, c)$ is non-empty if and only if ${\bf P}(a_1,
c)\times{\bf P}(a_2, c)$ is non-empty, that is, if and only if both ${\bf P}(a_1,
c)$ and ${\bf P}(a_2, c)$ are non-empty.  Since non-emptiness of ${\bf
P}(a, b)$ means that $a\leq b$, we indeed have
\[ 
a_1+ a_2 \leq c\  \ \Longleftrightarrow\  \ a_1 \leq c\  \&\ a_2\leq c 
\] 
so $a_1+ a_2$ is indeed the least upper bounds of $a_1$ and $a_2$.  So Definition
\ref{coprodBIS} provides us with a complementary but equivalent
definition of  least upper bounds.  In 
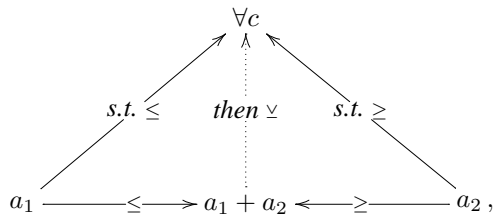
\[ \xymatrix@=.8in{& \forall c& \\ a_1\ar[ur]|{\mbox{\it s.t.}\ \leq}\ar[r]|{\leq} &
a_1+a_2\ar@{..>}[u]|{\mbox{\it then}\ \veebar} & a_2\ar[l]|{\geq}\ar[ul]|{\mbox{\it s.t.}\ \geq}\,,} \] 
we now have that the existence of $\iota_1$ and $\iota_2$ assert that
$a_1\leq a_1+ a_2$ and $a_2\leq a_1+ a_2$, so $a_1+ a_2$ is an upper bound for
$a_1$ and $a_2$, and whenever there exists an element $c\in P$ which is such
that both $a_1\leq c$ and $a_2\leq c$ hold,  then we have that $a_1+ a_2\leq
c$, so $a_1+ a_2$ is indeed the least upper bound for $a_1$ and $a_2$.  
\end{example}

Dually to what we did in a category with products, in a category with coproducts we can define sum morphisms $f+g$ in terms of commutation of
\[
\xymatrix @=.6in{
A_1\ar[d]_{f}\ar[r]^{\iota_1} & A_1+A_2\ar[d]|{\mbox{\normalsize $f+g$}} & A_2\ar[l]_{\iota_2}\ar[d]^{g}\\
B_1  \ar[r]_{\iota'_1}& B_1+B_2 & B_2\ar[l]^{\iota'_2}
}
\]
and we have 
\begin{equation}\nonumber
	h\circ [f,g]=[h\circ f,h\circ g] \quad\mbox{ and }\quad [f,g]\circ (h+k)=[f\circ h,g\circ k]\,.
\end{equation}
From this we can derive that coproducts provide a monoidal structure.

We already hinted at the fact that while a product can be interpreted as a
conjunction, the coproduct can be interpreted as a disjunction.  The distributive law
\[ 
A\ and\ (B\ or\ C)=(A\ and\ B)\ or\ (A\ and\ C)
\]
of classical logic incarnates in \em categorical logic \em as the existence of a natural isomorphism wich effectively  `distributes', namely
\[ \begin{diagram}
\{A\times (B + C)&\rTo^{dist_{A,B,C}}&(A\times B) + (A\times C)\mid
A,B,C\in|\cat|\}\,. \end{diagram} \] 
Shorter, we can write 
\[ 
A\times (B + C)\simeq(A\times B) + (A\times C)\,.
\]
This of course requires the category to be both Cartesian and
co-Cartesian. 

Such an isomorphism does not always exist, as the following example illustrates.

\begin{example}\label{Quant_log}
Let ${\cal H}$ be a Hilbert space and let $L({\cal H})$ be the
set of all of its (closed, in the infinite-dimensional case) subspaces,
ordered by inclusion.  Again this can be thought of as a category ${\bf
L}$.   It has an initial object, namely the zero-dimensional subspace,
and it has a terminal object, namely the whole Hilbert space itself.
This category is Cartesian with intersection as product, and it is also
co-Cartesian for 
\[ 
V + W:=\bigcap\{X\in L({\cal H})\mid V,W\subseteq X\}\,, 
\] that is, the (closed) linear span of $V$ and $W$.  However,
as observed by Birkhoff and von Neumann in \cite{BvN}, this lattice does not satisfy the
distributive law.  Take for example two vectors $\psi,\phi\in{\cal H}$
with $\phi\perp\psi$.  Then we have
\[ 
{\rm span}(\psi+\phi)\cap ({\rm
span}(\psi) + {\rm span}(\phi))={\rm span}(\psi+\phi)\cap {\rm
span}(\psi,\phi)={\rm span}(\psi+\phi)\,,
\] 
while since
\[ ({\rm
span}(\psi+\phi)\cap {\rm span}(\psi)) \quad\mbox{\rm and}\quad ({\rm
span}(\psi+\phi)\cap {\rm span}(\phi)) 
\] 
only include the zero-vector $0$, we have 
\[ 
({\rm span}(\psi+\phi)\cap {\rm span}(\psi)) + ({\rm span}(\psi+\phi)\cap {\rm span}(\phi))=0\,, 
\] 
and as a consequence
\beqa 
&{\rm span}(\psi+\phi)\cap ({\rm span}(\psi) + {\rm
span}(\phi))&\\ &\nparallel&\\ &({\rm span}(\psi+\phi)\cap {\rm
span}(\psi)) + ({\rm span}(\psi+\phi)\cap {\rm span}(\phi))&\,.
\eeqa
\end{example} 

Recall that an isomorphism consists of a pair of morphisms that are mutually inverse. So a natural isomorphism consists of a pair of  natural transformations. In a category which is both Cartesian and co-Cartesian one of the two components of the distributivity natural isomorphism always exists, namely the natural transformation 
\[ 
\begin{diagram} \{(A\times B) + (A\times
C)&\rTo^{\theta_{A,B,C}}&A\times (B + C)\mid A,B,C\in|\cat|\}\,,
\end{diagram} 
\] 
which we conveniently denote by 
\[ 
(A\times B) + (A\times
C)\leadsto A\times (B + C)\,.
\]
Indeed, by the assumption that the category is both
Cartesian and co-Cartesian there exist unique morphisms $f$ and $g$ such that
\[ 
\xymatrix{ & A & \\ A\times B \ar[ur]^{\pi_1}\ar[r]_{\iota_1\ \ \ \ \ \ \ \
\ } & (A\times B)+(A\times C)\ar@{..>}[u]|{f} & A\times C\ar[l]^{\ \ \ \ \ \ \
\ \ \iota_2}\ar[ul]_{\pi_1}}
\] 
and 
\[ 
\xymatrix{ & B+C & \\ A\times B
\ar[ur]^{\iota_1\circ\pi_2}\ar[r]_{\iota_1\ \ \ \ \ \ \ \ \ } & (A\times
B)+(A\times C)\ar@{..>}[u]|{g} & A\times C\ar[l]^{\ \ \ \ \ \ \ \ \
\iota_2}\ar[ul]_{\iota_2\circ\pi_2}} 
\] 
commute, namely $f:=[\pi_1, \pi_1]$ and $g:=[\iota_1\circ\pi_2, \iota_2\circ \pi_2]$, and hence there also exists a unique morphism $\theta_{A,B,C}$ such that 
\[ 
\xymatrix{ & (A\times B)+(A\times 
C)\ar[dl]_{f}\ar[dr]^{g}\ar@{..>}[d]|{\ \ \ \theta_{A,B,C}}& \\ A & A\times
(B+C) \ar[l]^{\pi_1}\ar[r]_{\pi_2} & B+C} 
\]
commutes, namely 
\[
\theta_{A,B,C}=\langle f,g\rangle=\langle[\pi_1, \pi_1],[\iota_1\circ\pi_2, \iota_2\circ \pi_2] \rangle\,.
\]
The collection 
\[
\theta= \{\theta_{A,B,C}\mid A,B,C\in|\cat|\}
\]
is moreover a natural transformation since given
\begin{equation}\nonumber (f\times g)+(f\times h):(A\times B)+(A\times
C)\rTo (A'\times B')+(A'\times C')\,,
\end{equation} 
using the various lemmas for products and coproducts, we have that 
\begin{eqnarray*} && \langle
[\pi_1',\pi_1'],[\iota_1'\circ\pi_2',\iota_2'\circ\pi_2']\rangle
\circ ((f\times g)+(f\times h)) \\ && \ \ \ \ \ \ \ \ \ = \langle
[\pi_1',\pi_1']\circ
((f\times g)+(f\times h)),[\iota_1'\circ\pi_2',\iota_2'\circ\pi_2'] \circ
((f\times g)+(f\times h))\rangle \\ && \ \ \ \ \ \ \ \ \ = \langle [\pi_1'
\circ (f\times g),\pi_1'\circ (f\times h)],[\iota_1'\circ\pi_2'\circ (f\times g),\iota_2'\circ\pi_2'\circ
(f\times h)]\rangle \\ && \ \ \ \ \ \ \ \ \ = \langle [f\circ
\pi_1,f\circ\pi_1],[\iota_1'\circ g\circ\pi_2,\iota_2'\circ
h\circ\pi_2]\rangle \\ && \ \ \ \ \ \ \ \ \ = \langle f\circ
[\pi_1,\pi_1],(g+h)\circ[\iota_1\circ\pi_2,\iota_2\circ\pi_2]\rangle\\
&& \ \ \ \ \ \ \ \ \ = (f\times (g+h)) \circ\langle
[\pi_1,\pi_1],[\iota_1\circ\pi_2,\iota_2\circ\pi_2]\rangle\,,
\end{eqnarray*} 
which results in commutation of 
\[
\xymatrix @=.8in{
(A\times B)+(A\times C)\ar[r]^{(f\times g)+(f\times h)\ \ }\ar[d]_{\theta_{A,B,C}} & (A'\times B')+(A'\times C')\ar[d]^{\theta_{A',B',C'}}\\
A\times (B+C)\ar[r]_{f\times (g+h)\ \ } & A'\times (B'+C')
}
\]
If this natural transformation is an isomorphism we have a \em
distributive category\em.  

\begin{example}
From the above it follows that in any lattice we have 
\[
(a\wedge b) +(a\wedge c)\leq a\wedge(b+c)\,. 
\] 
Below we will see that also the so-called orthomodular law can be given a purely category-theoretic form, 
so Birkhoff-von Neumann style quantum logic can be entirely casted in purely category-theoretic terms.  
\end{example}

\subsection{Direct sums}\label{sec:biprod}

\begin{example}\label{fdvbiprod} 
The {\em direct sum} $V\oplus
V'$ of two vector spaces $V$ and $V'$ is both a product and a coproduct in ${\bf FdVect}_{\mathbb{K}}$. Indeed, consider matrices
\[ 
\pi_1:=(1_W|0_{W,W'}) \qquad\mbox{and}\qquad \ \ \ \pi_2:=(0_{W',W}|1_{W'})\,,
\]
where $1_{U}$ denotes the identity on $U$  and $0_{U,U'}$ is a matrix of 0's of dimension $\mbox{dim}(U)\times\mbox{dim}(U')$. Let $M:V\rightarrow W$ and $N:V\rightarrow W'$ also be represented as matrices.
The unique matrix $P$ which makes
\[
\xymatrix@=.6in{ &
V\ar[dl]_{M}\ar[dr]^{N}\ar@{..>}[d]^P & \\
W & W\oplus W'\ar[l]^{\pi_1}\ar[r]_{\pi_2} & W' } 
\]
commute is  
\begin{equation}\nonumber
\left(\begin{array}{c} M\\ \hline N\end{array}\right)\,.
\end{equation}	
Therefore $\oplus$ is a product. The dual is obtained by transposing the matrices in this diagram.
Setting $\iota_i$ for the transpose of $\pi_i$ the diagram
\[
\xymatrix@=.6in{
W\ar[r]^{\iota_1}\ar[dr]_{M} & W\oplus W'\ar@{..>}[d]|{(M|N)} & W'\ar[l]_{\iota_2}\ar[dl]^{N}\\
& V & } 
\] 
commutes. This shows that $W\oplus W'$ is  indeed also a coproduct. Moreover, the zero-dimensional space is both initial and terminal. 
\end{example}

\begin{example}\label{relbiprod} 
In the category ${\bf Rel}$ we can extend the disjoint union to morphisms.   For any  two relations $R_1:X\rightarrow X'$ and $R_2:Y\rightarrow Y'$ we set
\[
R_1+R_2:=\{(( x,1),( x',1))\ |\ xR_1x'\}\cup \{(( y,2),( y',2))\ |\ yR_2y'\}\,.
\] 
We define injection relations $\iota_1:X\rightarrow X+Y$ and $\iota_2:Y\rightarrow X+Y$ to be  
\[
\iota_1:=\{ ( x,( x,1))\ |\ x\in X \}\ \ \ \ \mbox{and}\ \ \ \ \iota_2:=\{( y,( y,2))\ |\ y \in Y \}
\]
and the copairing relation $[R_1,R_2]:X+Y\rightarrow Z$ to be 
\[
[R_1,R_2]:=\{(( x,1),z)\ |\ xR_1z\}\cup\{(( y,2),z)\ |\ yR_2z\}\,. 
\]
One easily verifies that all these data define a coproduct.  We define projection relations as the relational converse of the injection relations, that is,
\begin{equation}\nonumber
\pi_1:=\{( (x,1),x)\ |\ x\in X\}\ \ \ \ \mbox{and}\ \ \ \ \pi_2:=\{( (y,2),y)\ |\ y\in Y\}.
\end{equation}
One easily verifies that this defines a product.  So the diagrams expressing the product properties are converted into the diagrams expressing the coproduct properties by the relational converse.  Since for any $X\in |{\bf Rel}|$ there is only one relation of type
\[
\emptyset\rightarrow X\quad \ \ \ \ \
\mbox{and}\quad \ \ \ \ X\rightarrow\emptyset
\] 
it follows that the empty set is both initial and terminal.  All of this makes the disjoint union within ${\bf Rel}$ very similar to the direct sum in ${\bf FdVect}_\mathbb{K}$. 
\end{example}

\begin{definition}\em 
A category $\cat$ is \em enriched in commutative monoids \em if each hom-set 
$\cat(A,B)$ is a commutative monoid
\[
(\cat(A,B)\,, +\,, 0_{A,B})\,, 
\]
and if  for all $f\in\cat(A,B)$, all $g_1, g_2\in\cat(B,C)$ and 
all $h\in\cat(C,D)$ we have
\[
(g_1+g_2)\circ f=(g_1\circ f) + (g_2\circ f)  \qquad\quad  0_{B,C}\circ f= 0_{A,C}\ 
\]
\[
h\circ (g_1+g_2)=(h\circ g_1) + (h\circ g_2)  \qquad\quad  h\circ0_{B,C}= 0_{B,D}\,.
\]
\end{definition}

\begin{example}
The category ${\bf FdVect}_\mathbb{K}$ is enriched in commutative monoids.  The monoid operation is addition of linear maps and the unit is the zero linear map.  Also the category ${\bf Rel}$ is enriched in commutative monoids. The monoid operation is the union of relations and the unit is the empty relation. 
\end{example}

\begin{definition}\label{biprod}\em 
The \em direct sum \em or \em biproduct \em of two
objects $A_1,A_2\in|\cat|$ is a quintuple consisting of another object
$A_1\oplus A_2\in |\cat|$ together with four morphisms 
\[\xymatrix{
A_1\ar@/^1pc/[r]^{\iota_1} & \ar@/^1pc/[l]^{\pi_1}\qquad\quad A_1\oplus
A_2\ar@/_1pc/[r]_{\pi_2}\qquad\quad & \ar@/_1pc/[l]_{\iota_2}A_2 }
\]
satisfying 
\beq\label{eqs:niprodinnerp}
\pi_1\circ\iota_1=1_{A_1}\ \ \quad\pi_2\circ\iota_2=1_{A_2}\ \ \quad\pi_2\circ\iota_1=0_{A_1,A_2}\ \ \quad
\pi_1\circ\iota_2=0_{A_2,A_1}
\eeq
and 
\[
\iota_1\circ \pi_1 + \iota_2 \circ \pi_2= 1_{A_1\oplus A_2}\,.
\]
\end{definition}

When setting 
\[ 
\delta_{ij}:= \left\{\begin{array}{ll} 1_{A_i} & i=j\\ 
0_{A_j,A_i}\qquad & i\not= j 
\end{array}\right.  
\] 
then eqs.(\ref{eqs:niprodinnerp}) can be rewritten as 
\[
\pi_i\circ\iota_j=\delta_{ij}\,. 
\] 

Note that Definition \ref{eqs:niprodinnerp} does not explicitly require that $A_1\oplus A_2$ is both a product and a coproduct.  In particular, it does not make any reference to other objects $C$ as the definitions of product and  coproduct do. 

\begin{definition}\em
A \em zero object \em  is an object which is both initial and terminal.  
\end{definition}

If a category $\cat$ has a zero object, then for each pair of objects
$A,B\in|\cat|$ we can construct a canonical \em zero map \em by relying on the
uniqueness of morphism from the initial object to $B$ and from $A$ to the terminal object:
\[\xymatrix{ A\ar@/^1pc/[rr]^{0_{A,B}}\ar[r]_{\exists!} & 0\ar[r]_{\exists!} & B\,.  }\]
One can show that if a category with a zero object is enriched in commutative monoids, that 
these unique morphisms must be the units for the monoids.

\begin{definition}
A \em biproduct category \em is a category with a zero object in which for
any two objects $A_1$ and $A_2$ a biproduct $(A_1\oplus A_2,
\pi_1,\pi_2,\iota_1,\iota_2)$ is specified.\footnote{There is no
particular reason why we ask for biproducts to be specified while in
the case of Cartesian categories we only required existence.  This is a
matter of taste, whether one prefers `being Cartesian' or `being a
biproduct category' to be conceived as a `property a category
possesses' or `some extra structure it comes with'.  There are
different `schools' of category theory which have strong arguments for
either of these.  Each of these have their virtues and therefore we decided to give an example of both.}
\end{definition}

One can show that the above definition is equivalent to the following one,  which does make explicit reference to products and coproducts \cite{Houston}.

\begin{definition}\label{biprodBIS}\em 
Let $\cat$ be both Cartesian and
co-Cartesian with specified products and coproducts, and let $\bot$ and $\top$ respectively  denote an initial and a terminal object of $\cat$.
Then $\cat$ is a \em biproduct category \em if:
\ben 
\item
the (unique) morphism $\bot\rTo \top$ is an isomorphism\,,
\item setting 
\[\xymatrix{ 
A_1\ar@/^1.2pc/[rrr]^{0_{A_1,A_2}}\ar[r] &1\ar@/^0.4pc/[r]_{\simeq} &\ar@/^0.4pc/[l] 0\ar[r] & A_2  }
\] 
the morphism 
\[ 
[\langle 1_{A_1}, 0_{A_1,A_2}\rangle,\langle 0_{A_2,A_1},
1_{A_2}\rangle]:A_1+ A_2\rTo A_1\times A_2 
\] 
is an isomorphism for all objects $A_1,A_2\in|\cat|$. 
\een
\end{definition}

In fact, any morphism
\[
A_1+ A_2\rTo^f B_1\times B_2
\]
is fully characterised by
four `component' morphisms, namely 
\[
f_{ij}:=\pi_i\circ f\circ \iota_j\qquad \mbox{for}\qquad i=1,2\,,
\] 
since
\[ 
f=[\langle f_{1,1},f_{2,1}\rangle,\langle f_{1,2}, f_{2,2}\rangle]\,.  
\] 
Indeed,
\begin{eqnarray*} 
&&\hspace{-1.9cm}[\langle f_{1,1}, f_{2,1}\rangle,\langle f_{1,2}, f_{2,2}\rangle]\\
&=&[\langle\pi_1\circ (f\circ\iota_1),\pi_2\circ (f\circ\iota_1)\rangle,\langle\pi_1\circ (f\circ\iota_2),\pi_2\circ 	(f\circ\iota_2)\rangle]\hspace{-1.9cm}\\ 
&=& [f\circ\iota_1,f\circ\iota_2]\\
&=& f\circ [\iota_1,\iota_2]\\
&=&f\,.
\end{eqnarray*}	
\noindent Therefore it makes sense to think of $f$ as the
matrix
\[ 
f= \left(\begin{array}{cc} f_{1,1} & f_{1,2}\\ f_{2,1} & f_{2,2}
\end{array}\right)\,. 
\] 
Using this, condition 2 in Definition \ref{biprodBIS} can now be
stated as the requirement that
\[ 
\left(\begin{array}{cc} 1_{A_1} & 0_{A_2,A_1}\\
0_{A_1,A_2} & 1_{A_2} \end{array}\right) 
\] 
is an isomorphism.


\begin{example} 
In ${\bf FdVect}_{\mathbb{K}}$ the direct sum $\oplus$ is a biproduct.   We have
\begin{equation}\nonumber
\pi_1\circ\iota_1=\pi_1\circ\pi_1^T=(1_W|0_{W,W'})\left(\begin{array}{c}1_W\\ \hline 0_{W',W} \end{array}\right)=1_W.
\end{equation}
We also have
\begin{equation}\nonumber 
\pi_1\circ \iota_2=\pi_1\circ\pi_2^T=(1_W|0_{W,W'})\left(\begin{array}{c}0_{W,W'}\\ \hline 1_{W'} \end{array}\right)=0_{W',W}.
\end{equation}
The two remaining equations are obtained in the same manner. 
\end{example}

\begin{example} In ${\bf Rel}$ the disjoint union $+$ is a biproduct. The morphism 
\begin{equation}\nonumber
\pi_1\circ\iota_1:X\rightarrow X+Y\rightarrow X
\end{equation}
is a subset of $X\times X$. The composite of 
\[
\iota_1=\{(x,(x,1))\ |\ x\in X\}\quad\mbox{ and }\quad\pi_1=\{((x,1),x)\ |\ x\in X\}
\]
is $\{(x,x)\ |\ x\in X\}=1_X$. The morphism
\begin{equation}\nonumber
\pi_1\circ\iota_2:Y\rightarrow X+Y\rightarrow X
\end{equation}
is a subset of $X\times Y$, namely  the set of pairs $(x,y)$ such that there exists a $(x,z)\in\iota_2$ and $(z,x)\in\pi_1$. But there are no such elements $z$ since the elements of $X$ are labeled by $1$ and those of $Y$ by $2$ within $X+Y$. Thus, we obtain the empty relation $0_{Y,X}$. 
\end{example}

\subsection{Categorical matrix calculus}

By Definition \ref{biprodBIS} each biproduct category is Cartesian, hence by Proposition \ref{prop:Cart_monoid} it carries monoidal structure.  We show now that from Definition \ref{biprodBIS} it indeed follows that 
each hom-set $\cat(A,B)$ in a biproduct category $\cat$ is a monoid, with  
\[
f+g:=A\rTo^{\scdiag_A}A\oplus A\rTo^{f\oplus g}B\oplus B\rTo^{\nabla_B} B \]
and with $0_{A,B}$ as  the unit.
Indeed, let $f:A\rightarrow B$ and consider
\begin{equation}\nonumber
f+0_{A,B}=A\rTo^{\diag_A}A\oplus A\rTo^{f\oplus 0_{A,B}} B\oplus B\rTo{\nabla_A} B\,.
\end{equation}
The equality $f+0_{A,B}=f$ can be shown via the commutation of
\begin{equation}\label{firstdiag}
\xymatrix@=.7in { A\ar[r]^{\diag_A}\ar[dr]_{\langle 1_A,0_{A,0}\rangle}\ar@/_2.7em/[ddr]_{1_A} & A\oplus A\ar[r]^{f\oplus 0_{A,B}}\ar[d]_{1_A\oplus 0_{A,0}} & B\oplus B\ar[r]^{\nabla_B} & B\\
& A\oplus 0\ar[d]_{\pi_1}\ar[r]_{f\oplus 0_{0,0}} & B\oplus 0\ar[ur]_{[1_B,0_{0,B}]}\ar[u]_{1_B\oplus 0_{0,B}}\\
& A\ar[r]_f & B\ar[u]_{\iota'_1}\ar@/_2.7em/[uur]_{1_B}}
\end{equation}
In the above diagram, all subdiagrams correspond to  definitions, except for the square at the bottom. To show that it commutes, consider 
\begin{equation}\label{seconddiag}
\xymatrix@=.6in { A\ar[d]_f & A\oplus 0\ar[l]_{\pi_1}\ar[r]^{\pi_2}\ar@{..>}[d]|{f\oplus 0_{0,0}} & 0\ar[d]^{0_{0,0}}\\
B & B\oplus 0\ar[l]^{\pi'_1}\ar[r]_{\pi'_2} & 0}
\end{equation}
Since this is a product diagram, $f\oplus 0_{0,0}$ is the unique morphism making it commute. Moreover,  the diagram
\begin{equation}\nonumber
\xymatrix {A\ar@{=}[dr]\ar[ddd]_f & A\oplus 0\ar[dddrr]^{0_{A\oplus 0,0}}\ar[d]^{\pi_1}\ar[l]_{\pi_1}\ar[rr]^{\pi_2} & & 0\ar[ddd]^{0_{0,0}}\\
& A\ar[d]^f & \\
& B \ar[d]^{\iota'_1}\ar@{=}[dl] &\\
B & B\oplus 0\ar[l]^{\pi'_1}\ar[rr]_{\pi_2'} & & 0
}
\end{equation} 
commutes, so it follows that $\iota_1'\circ f\circ\pi_1$ also makes diagram~(\ref{seconddiag}) commute. Thus,  
\[
f\oplus 0_{0,0}=\iota'_1\circ f\circ\pi_1
\]
by uniqueness, that is, 
the square at the bottom of diagram~(\ref{firstdiag}) also commutes. 
To establish  $0_{A,B}+f$ one proceeds similarly. 

We also have to show that 
\[
(f + g)+ h=f+(g+h)\,. 
\]
This is established in terms of  commutation of the diagram 
\begin{equation}\nonumber
\xymatrix {	A\ar[r]^{\diag_A}\ar[dr]_{\diag_A} & A\oplus A\ar[r]^{1_A\oplus\diag_A\ \ \ \ } & A\oplus (A\oplus A)\ar[r]^{f\oplus (g\oplus h)}\ar[dd]_{\alpha_{A,A,A}} & B\oplus (B\oplus B)\ar[dd]^{\alpha_{B,B,B}}\ar[r]^{\ \ \ \ 1_B\oplus\nabla_B} & B\oplus B\ar[r]^{\nabla_B} & B\\
& A\oplus A\ar[dr]_{\diag_A\oplus 1_A} & & & B\oplus B\ar[ur]_{\nabla_B} & \\
& & (A\oplus A)\oplus A\ar[r]_{(f\oplus g)\oplus h} & (B\oplus B)\oplus B\ar[ur]_{\nabla_B\oplus 1_B} & &} 
\end{equation}
where $\alpha_{A,A,A}$ is defined as in Proposition~\ref{prop:Cart_monoid}. The central square commutes by definition.   We now show that the left triangle also commutes. We have
\begin{eqnarray*}
&& \langle\langle \pi_1,\pi'_1\circ\pi_2\rangle,\pi'_2\circ\pi_2\rangle\circ  (1_A\oplus\diag_A)\circ\diag_A\\
&&\ \ \ \ = \langle\langle \pi_1,\pi'_1\circ\pi_2\rangle,\pi'_2\circ\pi_2\rangle \circ \langle 1_A,\langle 1_A,1_A\rangle\rangle \\
&&\ \ \ \ = \langle\langle \pi_1,\pi'_1\circ\pi_2\rangle\circ\langle 1_A,\langle 1_A,1_A\rangle\rangle,\pi'_2\circ\pi_2\circ \langle 1_A,\langle 1_A,1_A\rangle\rangle\rangle\\
&&\ \ \ \ = \langle\langle\pi_1\circ\langle 1_A,\langle 1_A,1_A\rangle\rangle,\pi'_1\circ\pi_2\circ\langle 1_A,\langle 1_A,1_A\rangle\rangle\rangle ,\pi'_2\circ\pi_2\circ \langle 1_A,\langle 1_A,1_A\rangle\rangle\rangle\\
&&\ \ \ \ = \langle\langle 1_A,1_A\rangle,1_A\rangle\\
&&\ \ \ \ = (\diag_A\oplus 1_A)\circ\diag_A\,.
\end{eqnarray*}
The right triangle is also easily seen to commute.

This addition moreover satisfies a distributive law, namely 
\beq\label{eq:distributivity}
(f+g)\circ h=(f\circ h)+(g\circ h)\quad\mbox{\rm and}\quad
h\circ(f+g)=(h\circ f)+(h\circ g)\,.
\eeq
One usually refers to this additive structure on morphisms as \em enrichment in monoids\em.  We leave it up to the reader to verify these distributive laws.  A physicist-friendly introduction to \em enriched category theory \em suitable for the readers of this chapter is \cite{BorceuxStubbe}. An inspiring paper which introduced the concept is \cite{Lawvere}.

We now show that from Definition \ref{biprodBIS} it also follows that for 
\[ {\rm Q}_i:=\iota_i\circ\pi_i:A_1\oplus A_2\rTo
A_1\oplus A_2 \] with $i=1,2$ we have 
\beq\label{eq:sumproj} 
\sum_{i=1,2}{\rm
Q}_i=1_{A_1\oplus A_2}\,. 
\eeq
Indeed, unfolding the definitions we have 
\begin{eqnarray*} 
\sum_{i=1,2}{\rm Q}_i &=& \nabla_{A_1\oplus A_2}\circ ((\iota_1\circ\pi_1)\oplus(\iota_2\circ\pi_2))\circ\diag_{A_1\oplus A_2}\\ 
&=& \nabla_{A_1\oplus A_2}\circ((\iota_1\oplus\iota_2)\circ(\pi_1\oplus\pi_2))
\circ\diag_{A_1\oplus A_2}\\ 
&=& (\nabla_{A_1\oplus A_2}\circ(\iota_1\oplus\iota_2))\circ((\pi_1\oplus\pi_2)\circ\diag_{A_1\oplus A_2})
\end{eqnarray*} 
and using the fact that a biproduct of morphisms
is at the same time  a product of morphisms we obtain 
\[
(\pi_1\oplus \pi_2)\circ
\diag_A=\langle \pi_1\circ 1_{A_1\oplus A_2},\pi_2\circ
1_{A_1\oplus A_2}\rangle=\langle\pi_1,\pi_2\rangle=1_{A_1\oplus A_2}\,.
\]
Analogously, one obtains that
\[
\nabla_A\circ(\iota_1\oplus\iota_2)=1_{A_1\oplus A_2}\,, 
\]
and the composite of
identities being again the identity, we proved the claim.

\begin{definition}\em A 
\em dagger biproduct category \em is a category which
is both a dagger symmetric monoidal category and a biproduct  category for which the monoidal tensor and the biproduct coincide, and with $\iota_i=\pi_i^\dagger$
for all projections and injections.  
\end{definition}

These dagger biproduct categories were introduced in \cite{AC2004,deLL,SelingerPre} in order to enable one to talk about quantum spectra in purely category-theoretic language.  Let 
\[
A_1\oplus A_2\rTo^U B
\]
be unitary in a dagger biproduct category.  By the corresponding \em projector spectrum \em we mean the family $\{{\rm P}_i\}_i$
of \em projectors \em 
\[ 
{\rm P}_i^U:= U\circ{\rm Q}_i\circ U^\dagger : B\rTo B\,.
\]

\begin{proposition}\label{spectra} Binary projector spectra satisfy 
\[
\sum_{i=1,2}{\rm P}_i^U=1_{B}\,. 
\] 
\end{proposition}

This result easily extends to more general biproducts $A_1\oplus\ldots\oplus
A_n$, which can be defined in the obvious manner, and which allow us in addition to
define $n$-ary projector spectra too.  In ${\bf FdHilb}$, this $n$-ary
generalisation of Proposition \ref{spectra} corresponds to the fact that
\[
\sum_{i=1}^{i=n}{\rm P}_i=1_{\cal H}\qquad \mbox{where}\qquad \{{\rm P}_i\}_{i=1}^{i=n}
\]
is the
projector spectrum of an arbitrary  self-adjoint operator. More details on
this abstract view of quantum spectra are in \cite{AC2004,deLL,SelingerPre}.  

Now, consider two biproducts $A_1\oplus\ldots\oplus A_n$ and
$B_1\oplus\ldots\oplus B_m$ each with their respective injections and
projections.  As already indicated in the previous section, with each morphism
\[ 
A_1\oplus\ldots\oplus A_n\rTo^f B_1\oplus\ldots\oplus B_m
\] 
we can associate a matrix 
\[ 
\left(\begin{array}{ccc} \pi_1\circ f\circ \iota_1 &	\hdots & \pi_1\circ f\circ \iota_n\\ \vdots & \ddots & \vdots\\
\pi_m\circ f\circ \iota_1 & \hdots & \pi_m\circ f\circ \iota_n
\end{array}\right)\,.
\] 
Moreover, these matrices obey the usual matrix rules
with respect to composition and the above defined summation.  Indeed, for composition, the composite $g\circ f=h$ also has an associated matrix with entries
\[
h_{ij}=\pi_i\circ (f\circ g)\circ\iota_j\,.
\]
By eq.(\ref{eq:sumproj}) we have
\begin{eqnarray*}
h_{ij}&=&\pi_i\circ (f\circ g)\circ\iota_j\\
&=&\pi_i\circ (f\circ 1\circ g)\circ\iota_j\\
&=&\pi_i\circ \left(f\circ \left(\sum_r\iota'_r\circ\pi'_r\right)\circ g\right)\circ\iota_j\\
&=&\sum_r\pi_i\circ f\circ\iota'_r\circ\pi'_r\circ g\circ\iota_j\\
&=& \sum_r(\pi_i\circ f\circ\iota'_r)\circ(\pi'_r\circ g\circ\iota_j)\\
&=& \sum_r f_{ir}\circ g_{rj}
\end{eqnarray*}
from which we recover matrix multiplication. For the sum, using the distributivity of the composition over the sum, one finds that for individual entries in $f+g$ we have
\begin{eqnarray*}
\pi_i\circ (f+g)\circ\iota_j&=& (\pi_i\circ f+\pi_i\circ g)\circ\iota_j\\
	    &=& \pi_i\circ f\circ\iota_j+\pi_i\circ g\circ\iota_j\\
            &=& f_{ij}+g_{ij}
\end{eqnarray*}
which indeed is the sum of matrices. 

\begin{example} 
We illustrate the concepts of this section for the category
${\bf Rel}$.  Somewhat unfortunately, the disjoint union bifunctor and the monoidal enrichment 
operation share the same notation $+$.  But since their type are essentially different, i.e.
\[
\mbox{tensor }+ : {\bf Rel}(X,Y)\times {\bf Rel}(X',Y')\to {\bf Rel}(X+X',Y+Y')
\]
and
\[
\ \mbox{monoid }+ : {\bf Rel}(X,Y)\times {\bf Rel}(X,Y)\to {\bf Rel}(X,Y)
\]
respectively, this should not confuse the reader.
\begin{itemize} 
\item 
The sum $R_1+R_2:X\rightarrow Y$ of
two relations is, by definition, the composite
\[
X\stackrel{\diag_X}{\longrightarrow}X+X\stackrel{R_1+R_2}{\longrightarrow}Y+Y\stackrel{\nabla_Y}{\longrightarrow} Y\,.
\]
The relation $\diag_X$ consists of all ordered pairs 
\[
\{(x,(x,1))\mid x\in X\}\cup\{(x,(x,2))\mid x\in X\}\,. 
\]
Thus the composite $(R_1+R_2)\circ\diag_X$ is then, by definition, the set 
\[
\{(x,(y,1))\ |\ xR_1y\}\cup\{(x',(y',2))\ |\ x'R_2y'\}\,.
\]
Using the definition of copairing $\nabla_Y:=[1_Y,1_Y]$ we obtain 
\[
\{(x,y)\ |\ xR_1y\}\cup\{(x',y')\ |\ x'R_2y'\}\,,
\]
that is,
\[
R_1+R_2=\{(x,y)\ |\ xR_1y \mbox{ \underline{or} } x R_2y\}\,.
\]
\item 
Relations
\[
Q_X:X + Y\rightarrow X\rightarrow X+ Y
\quad\mbox{ and }\quad
Q_Y:X + Y\rightarrow Y\rightarrow X+ Y
\]
are  defined as $\iota_X\circ\pi_X$ and $\iota_Y\circ\pi_Y$ respectively, that is,   
\[
Q_X=\{((x,1),(x,1))\  |\ x\in X\}
\quad\mbox{ and }\quad
Q_Y=\{((y,2),(y,2))\ | \ y\in Y\}\,.
\]
Using the definition	of the sum we obtain 
\beqa
Q_X+Q_Y
&=&\{((x,1),(x,1))\ \mid\ x\in X\}\cup \{((y,2),(y,2))\ |\ y\in Y\}\\
&=&\{((z,i),(z,i))\  \mid \ (z,i)\in X+Y\}\\
&=&1_{X+Y}
\eeqa
as required. It is easily seen that this generalises to an arbitrary number of terms in the biproduct.  
\item 
The matrix calculus in ${\bf Rel}$ is done over the  semiring (= rig = ring without inverses) $\mathbb{B}$ of Booleans. 
Indeed, there are two relations between $\{*\}$ and itself, namely the empty relation and the identity relation.  These will respectively be denoted by  0 and 1. The semiring operations arise from composing and adding these relations, which amounts to the semiring multiplication and the semiring addition respectively.
By eqs.(\ref{eq:distributivity}), we have distributivity, and we then easily see that we indeed get the Boolean semiring:
\[
0\cdot 0=0\quad 0\cdot 1=0 \quad 1\cdot 1=1 \qquad 0+0=0\quad 0+1=1\quad 1+1=1
\]
--- contra the two-element field where we have $1 + 1=0$ --- so the operations $-\cdot-$ and $-+-$ coincide
with the Boolean logic operations:
\[
\cdot\ \sim \ \wedge\qquad\qquad\mbox{ and }\qquad\qquad +\ \sim \ \vee\,.
\]
A relation $R:\{a,b\}\rightarrow \{c,d\}$ can now be represented by a $2\times 2$ matrix, e.g.
\[
R=\left(\begin{array}{cc} 1 & 1\\ 1 &	0\end{array}\right)
\]	
when $aRc$, $bRc$ and  $aRd$ (and not $b R d$). Similarly, $R':\{c,d\}\rightarrow \{e,f,g\}$ is 
\[
R'=\left(\begin{array}{cc} 
1 & 0 \\ 
1 & 1 \\
0 & 1
\end{array}\right)
\]
when $cRe$, $cRf$, $dRf$ and $dRg$. Their composite 
\[
R'\circ R=\{(a,e),(a,f),(b,e),(b,f),(a,g)\}
\]
can be computed by matrix multiplication:
\[
\left(\begin{array}{cc} 1 & 0 \\ 1 & 1\\ 0 & 1\end{array}\right)\left(\begin{array}{cc} 1 & 1\\ 1 &	0\end{array}\right)= \left(\begin{array}{cc}1 & 1 \\ 1 & 1 \\ 1 & 0\end{array}\right)\,.
\] 
For a relation $R'':\{a,b\}\rightarrow\{c,d\}$ represented by the matrix 
\[
\left(\begin{array}{cc} 0 & 1 \\ 0 & 1\end{array}\right)\,,
\]	
that is, $R''=\{(b,c),(b,d)\}$, the sum $R+R''$ is given by 
\[
\{(a,c),(b,c),(a,d)\}\cup\{(b,c),(b,d)\}=\{(a,b),(a,c),(b,c),(b,d)\}\,,
\]
which indeed corresponds to the matrix sum
\[
\left(\begin{array}{cc} 1 & 1 \\ 1 & 0\end{array}\right)+
\left(\begin{array}{cc} 0 & 1 \\ 0 & 1\end{array}\right)=
\left(\begin{array}{cc} 1 & 1\\ 1 & 1\end{array}\right)\,.
\]
\end{itemize}
\end{example}	

\subsection{Quantum tensors from classical tensors}\label{sec:classquant}

Interesting categories such as ${\bf FdHilb}$ and ${\bf Rel}$ have both
a classical-like and a quantum-like tensor.  Obviously these two structures
interact.  For example, due to very general reasons we have distributivity
natural isomorphisms 
\[ A\otimes (B\oplus C)\simeq (A\otimes B)\oplus (A\otimes
C)\qquad\mbox{\rm and}\qquad A\otimes 0\simeq 0 
\] 
both  in the case of ${\bf FdHilb}$ and ${\bf Rel}$.  We can rely on  so-called \em closedness \em of the $\otimes$-structure to prove this, something for which we refer to other sources.  Another manner to establish this fact for the cases of ${\bf FdHilb}$ and ${\bf Rel}$, is to observe that the $\otimes$-structure arises from the $\oplus$-structure.  

Let ${\bf C}$ be a biproduct category and let $X\in{\bf C}$ be such that
composition commutes in ${\bf C}(X,X)$.  Define a new category ${\bf C}|X$ as
follows: 
\bit 
\item The objects of ${\bf C}|X$ are those objects of ${\bf C}$
which are of the form $X\oplus\ldots\oplus X$.  We denote such an object
consisting of $n$ terms by $[n]$.  
\item For all $n,m\in\mathbb{N}$ we set
${\bf C}|X\!([n],[m]):={\bf C}([n],[m])$. 
\eit 
Note that we can represent all morphisms in ${\bf C}([n],[m])$ by matrices, and hence also those in ${\bf C}|X\!([n],[m])$.  Now we define a monoidal structure: 
\bit \item $\II:=X$ \item $[n]\otimes [m]:=[n\times m]$ 
\item For all $f\in{\bf C}([n],[m])$ and $g\in{\bf C}([n'],[m'])$ we
define 
\[
f\otimes g\in{\bf C}|X\!([n]\otimes [n'],[m]\otimes [m'])
\]
to be the morphism with matrix entries
\[ 
(f\otimes g)_{(i,i'),(j,j')}:=f_{i,j}\circ g_{i'\!,j'}\,.
\] 
\eit 
We leave it to the reader to verify that this provides ${\bf C}|X$ with a symmetric monoidal
structure.  Note that commutativity of ${\bf C}(X,X)$ is necessary, since otherwise we would be in contradiction with  the fact that the scalar monoid in a monoidal category is always
commutative --- cf.~Section \ref{sec:scalars}.  With these
definitions we have:
\[ 
[n]\otimes ([m]\oplus [k])\simeq ([n]\otimes [m])\oplus ([n]\otimes [k])\qquad\mbox{\rm and}\qquad [n]\otimes [0]\simeq
[0]\,.  
\] 
Indeed, note first that since $[n]=\underbrace{\II\oplus \dots\oplus \II}_n$ we have
\begin{equation}\nonumber
[n]\oplus [m]\simeq [n + m]
\end{equation}
where $[n + m]=\underbrace{\II\oplus \dots\oplus \II}_{n+m}$. Therefore, 
\begin{eqnarray*}
[n]\otimes ([m]\oplus [k]) &\simeq & [n]\otimes [m+k]\\
&\simeq & [n\times (m+k)]\\
&=& [(n\times m)+ (n\times k)]\\
&\simeq & [n\times m]\oplus [n\times k]\\
&\simeq & ([n]\otimes [m])\oplus([n]\otimes [k]). 
\end{eqnarray*}
Moreover, 
\begin{eqnarray*}
[n]\otimes [0] &\simeq & [n\times 0]\\
&=& [0].
\end{eqnarray*}

\begin{example}
In ${\bf FdHilb}$ there is one non-trivial object ${\cal H}$ such that ${\bf FdHilb}({\cal H}, {\cal H})$ is commutative, namely $\mathbb{C}$.  The category ${\bf FdHilb}|\mathbb{C}$ has Hilbert spaces of the form $\mathbb{C}^{\oplus n}$ with $n\in\mathbb{N}$ as objects,  linear
maps between these as morphisms,  and the tensor product as the
monoidal structure.  This category is said to be \em categorically equivalent \em (a notion which we define later) to ${\bf FdHilb}$.  The only difference is that ${\bf FdHilb}$ contains for each $n\in\mathbb{N}$ many isomorphic Hilbert spaces of dimension $n$, while in ${\bf FdHilb}|\mathbb{C}$ there is exactly one Hilbert space of dimension $n$.
\end{example}

\begin{example}
In ${\bf Rel}$  it is the non-trivial object $\{*\}$ for which ${\bf Rel}(\{*\},\{*\})\simeq\mathbb{B}$ is commutative. We obtain a category with objects  of the form 
\[
\{*\}+\ldots+\{*\}\,, 
\]
that is, a
$n$-element set for each $n\in\mathbb{N}$, with relations between these as
morphisms, and with the Cartesian product as the monoidal structure.  Again we have that ${\bf Rel}|\mathbb{B}$ is categorically equivalent to ${\bf Rel}$.
\end{example}

We can endow ${\bf C}|X$ with compact structure.  Set: 
\bit 
\item 
$[n]^*:=[n]$ 
\item 
Let $\eta_{[n]}\in {\bf C}|X\!(\II,
[n]^*\otimes [n])$  be the morphism with matrix entries
\[ 
(\eta_{[n]})_{(i,i),1}:=1_\II \qquad\mbox{\rm and}\qquad (\eta_{[n]})_{(i,j\not= i),1}:=0_{\II,\II}\,. 
\] 
\item 
Let $\epsilon_{[n]}\in {\bf C}|X\!([n]\otimes [n]^*,\II)$  to be the morphism with matrix entries
\[ 
(\epsilon_{[n]})_{1,(i,i)}:=1_\II \qquad\mbox{\rm and}\qquad (\epsilon_{[n]})_{1,(i,j\not= i)}:=0_{\II,\II}\,. 
\] 
\eit 
To see that this indeed defines a compact structure, observe that the identity of $[n]$ is
\begin{equation}\nonumber
1_{[n]}=\delta_{i,j}:=\left\{\begin{array}{l} 1_{\II}\ \ \mbox{if}\ \ i=j\\
0_{\II,\II}\ \ \mbox{otherwise}\end{array}\right.\,.
\end{equation}
Using this, we find that
\begin{equation}\nonumber
(1_{[n]}\otimes \eta_{[n]})_{(i,(j,k)),(l,1)}=\delta_{i,l}\circ\eta_{(j,k),1} 
\end{equation}
and
\begin{equation}\nonumber
(\epsilon_{[n]}\otimes 1_{[n]})_{(1,i),((j,k),l)}=\epsilon_{1,(j,k)}\circ\delta_{i,l}.
\end{equation}
We can now verify the equations of compactness by computing the composite --- say $e$ --- of the two preceding morphisms using matrix calculus, i.e.
\[
e_{(m,n)}=\sum_{j,k,l}  (\epsilon_{[n]}\otimes 1_{[n]})_{(1,m),((j,k),l)} (1_{[n]}\otimes \eta_{[n]})_{(j,(k,l)),(n,1)}\,.
\]
Note that the indexation over $j$, $k$ and $l$ has two different bracketings in the above sum. By definition of the identity, unit and counit, the term $e_{(m,n)}$ will be $1_I$ only if $j=k=l$, which entails that $e_{(m,n)}=\delta_{i,j}$, the identity. Since the objects are self-dual the other equation holds too.

Robin Houston proved a surprising result in \cite{Houston} which to some extent
is a converse to the above.  It states that when a compact category is
Cartesian (or co-Cartesian) then it also has direct sums.

\subsection{Internal classical structures}\label{sec:comonoids}

In \cite{AC2004}  unitary biproduct decompositions of the form 
\[
U:A\rTo \underbrace{\II\oplus\ldots\oplus\II}_n
\]
were used to encode the flow of classical data in quantum informatic protocols.  In ${\bf FdHilb}$ such a map indeed singles out a basis.  Explicitly, via the correspondence between vectors in Hilbert space ${\cal H}$ and linear maps  of type $\mathbb{C}\to{\cal H}$, the linear maps
\[
\{U^\dagger\circ \iota_i:\mathbb{C}\to {\cal H}\mid i= 1,\ldots, n \}
\]
define a basis for ${\cal H}$, namely
 \[
\{|i\rangle:= (U^\dagger\circ \iota_i)(1)\mid i = 1,\ldots, n \}\,.
\]
These basis vectors are then identified with outcomes of measurements.  

But there is another way to encode  bases as morphisms in a category, one for which we only need to rely on the tensor structure, and hence we can stay in the diagrammatic realm of Section \ref{sec:graphCalc}.  If we have a basis 
\[
{\cal B}:=\{|i\rangle\mid i = 1,\ldots, n \}
\]
of a Hilbert space ${\cal H}$ then we can consider the linear maps
\[
\delta: {\cal H}\to {\cal H}\otimes  {\cal H}::|i\rangle\mapsto |ii\rangle\quad
\mbox{ and }\quad \epsilon:{\cal H}\to\mathbb{C}::|i\rangle\mapsto1\,.
\]
These two maps indeed faithfully encode the basis ${\cal B}$ since we can extract it back from them.  
It suffices to solve the equation 
\[
\delta(|\psi\rangle)=|\psi\rangle\otimes |\psi\rangle
\]
in the unknown $|\psi\rangle$.  Indeed, the only $|\psi\rangle$'s for which the right-hand-side  is of the form  $|\phi\rangle\otimes|\phi'\rangle$ are the basis vectors. For any other 
$\psi=\sum_i \alpha_i\,|i\rangle$ we have that 
\[
\delta(|\psi\rangle)=\sum_i \alpha_i\,|i\rangle\otimes |i\rangle\,, 
\]
that is, we obtain a genuinely \em entangled \em state.

The pair of maps $(\delta, \epsilon)$ satisfies several properties e.g.
\[
(\delta\otimes 1_{\cal H})\circ\delta=(1_{\cal H}\otimes\delta)\circ\delta:{\cal H}\to {\cal H}\otimes {\cal H}\otimes  {\cal H}::|i\rangle\mapsto|iii\rangle
\]
and
\[
\ (\epsilon\otimes 1_{\cal H})\circ\delta=(1_{\cal H}\otimes\epsilon)\circ\delta= 1_{\cal H}::|i\rangle\mapsto|i\rangle
\]
establishing it as an instance of the following concept in ${\bf FdHilb}$:

\begin{definition} \em
Let $(\cat,\otimes,\II)$ be a monoidal category. An {\em internal comonoid} is an object $C\in|\cat|$ together with a pair of morphims
\[
C\otimes C \lTo^{\delta} C\rTo^{\epsilon} \II\,,
\]
where $\delta$ is the {\em comultiplication} and $\epsilon$ the {\em comultiplicative unit}, which are such that 
\[
\xymatrix@=.44in{C\ar[r]^{\delta}\ar[d]_{\delta} & C\!\otimes\! C\ar[d]^{1_C \otimes \delta} & \ar@{}[d]|{\hspace{-8mm}\mbox{and}\hspace{-8mm}} & & C \ar[d]_{\delta} \ar[dr]^{\simeq}\ar[dl]_{\simeq}& \\%
C\!\otimes\!C\ar[r]_{\delta\otimes 1_C} & C\!\otimes\!C\!\otimes\!C & \hspace{-12mm}& \II\!\otimes\!C & C\!\otimes\!C\ar[l]^{\epsilon \otimes 1_C} \ar[r]_{1_C \otimes \epsilon} & C\!\otimes\!\II
}
\]
commute.
\end{definition}

\begin{example}
The relations 
\[
\delta=   \{(x,(x,x))\mid x\in X\}  \subseteq  X\times (X\times X)
\]
and 
\[
\epsilon = \{(x,*)\mid x\in X\}  \subseteq  X\times \{*\}
\]
define an internal comonoid on $X$ in ${\bf Rel}$ as the reader may verify. We could refer to these as the \em copying \em and \em deleting \em relations.
\end{example}

The notion of internal comonoid is dual to the notion of \em internal monoid\em.  

\begin{definition}\em
Let $(\cat,\otimes,\II)$ be a monoidal category. An {\em internal monoid} is an object $M\in|\cat|$ together with a pair of morphisms
\[
M\otimes M\rTo^\mu M \lTo^{e}\II\,,
\]
where $\mu$ is the {\em multiplication} and $e$ the {\em multiplicative unit}, which are  such that
\[
\hspace{-1.5mm}\xymatrix@=.42in{
M & M\otimes M\ar[l]_{\mu} & \ar@{}[d]|
{\hspace{-8mm}\mbox{and}\hspace{-8mm}} 
& & 
M & \\%
M\!\otimes\! M\ar[u]^{\mu} & M\!\otimes\!M\!\otimes\!M\ar[l]^{\mu \otimes 1_M}\ar[u]_{1_M \otimes \mu} & \hspace{-12mm}& \II\!\otimes\!M\ar[ur]^{\simeq}\ar[r]_{e \otimes 1_C} & M\!\otimes\!M\ar[u]^{\mu}  & M\!\otimes\!\II\ar[ul]_{\simeq}\ar[l]^{1_M\otimes e}
}
\]
commute.  
\end{definition}

The origin of this name is the fact that monoids can equivalently be defined as internal monoids in ${\bf Set}$.  Since the notion of internal monoid applies to arbitrary monoidal categories, it generalises the usual notion of a monoid.

\begin{example}
A strict monoidal category can also be  defined as an internal monoid in the category ${\bf Cat}$, which has categories as objects, functors as morphisms and the product of categories as tensor --- see Section \ref{sec:BifunctAsFunct} below.  Proving this is slightly beyond the scope of this chapter but we invite the interested reader to do so.
\end{example}

%

We now show that internal monoids in ${\bf Set}$ are indeed ordinary monoids.  Given such an internal monoid $(X,\mu,e)$ in  ${\bf Set}$ with functions 
\[
\mu:X\times X\to X\quad\mbox{ and }\quad e:\{*\}\to X,
\]
we take the elements of the monoid to be those of $X$, the monoid operation  to be 
\[
-\bullet-: X\times X\to X::(x,y)\mapsto \mu(x,y)\,,
\]
and the unit of the monoid to be $1:=e(*)\in X$.  The condition 
\[
\xymatrix@=.44in{X\times X\times X\ar[r]^{1_X\times \mu}\ar[d]_{\ \ \ \ \ \mu\times 1_X} & X\times X\ar[d]^{\mu}\\ X\times X\ar[r]_{\ \ \ \ \ \mu} & X}
\]
boils down to the fact that for all $x,y,z\in X$ we have 
\[
x\bullet (y\bullet z)=(x\bullet y)\bullet z\,,
\] 
that is, associativity of the monoid operation, 
and the condition
\[
\xymatrix@=.44in{ & X & \\%
\{*\} \times X\ar[r]_{e\times 1_X}\ar[ur]^{\simeq} & X\times X\ar[u]_{\mu} & X\times\{*\} \ar[l]^{1_X\times e}\ar[ul]_{\simeq}}
\]
boils down to the fact that for all $x\in X$ we have 
\[
x\bullet 1=1\bullet x=x\,, 
\]
that is, the element $1$ is the unit of the monoid.

An internal definition of a group requires a bit more work.

\begin{definition}\label{internal_group} \em
Let $\cat$ be a category with finite products and let $\top$ be the terminal object in $\cat$. An {\em internal group} is an internal monoid $(G,\mu,e)$ together with a morphism ${\rm inv}:G\rTo G$ such that 
we have commutation of
\[
\xymatrix{G\ar[r]^{!}\ar[d]_{\langle 1_G,{\rm inv}\rangle} & \top\ar[d]^e \\ G\times G\ar[r]^{\ \ \ \mu} & G\\ G\ar[u]^{\langle{\rm inv},1_G\rangle} \ar[r]_{!} & \top\ar[u]_{e}}
\]
\end{definition}

The additional operation ${\rm inv}:G\rTo G$ assigns the inverses to the elements of the group.  We leave it to the reader to verify that internal groups in ${\bf Set}$ are indeed ordinary groups.  When we rather consider groups in other categories, in particular those in categories of vector spaces, then one typically speaks about \em Hopf algebras\em, of which \em quantum groups \em are a special case. 
An excellent textbook on this topic is \cite{StreetBook}.  There are also  lectures on this topic  available on-line \cite{CatstersHopf}.
Also the notion of group homomorphism can be `internalized' in a category.  We define a {\em group homomorphism} between two group objects $(G,\mu,e,{\rm inv})$ and $( G',\mu',e',{\rm inv}')$ to be a morphism $\phi:G\rTo G'$ which commutes with all three structural morphisms,  that is, the diagrams 
\[
\xymatrix{G\times G\ar[r]^\mu\ar[d]_{\phi\times\phi} & G\ar[d]^\phi &\ar@{}[d]|{,} &\top\ar[dr]_{e'}\ar[r]^e & G \ar[d]^{\phi} & \ar@{}[d]|{\mbox{and}} & G\ar[r]^{{\rm inv}} \ar[d]_{\phi} & G\ar[d]^{\phi}\\ %
G'\times G'\ar[r]_{\mu'} & G' & & & G' & & G'\ar[r]_{{\rm inv}'} & G'
}
\]
all commute. Again, these diagrams generalise what we know about group homomorphisms, namely that they preserve multiplication, unit and inverses. The notion of (co)monoid homomorphism is defined analogously. 

\subsection{Diagrammatic classicality}\label{sec:comonoidsBis}

In a dagger monoidal category every internal comonoid 
\[
\Bigl(X\,,\,X\rTo^{\delta} X\otimes X\,,\,X\rTo^{\epsilon} \II\Bigr)
\]
defines an internal monoid 
\[
\ \Bigl(X\,,\,X\otimes X\rTo^{\delta^\dagger} X\,,\,\II\rTo^{\epsilon^\dagger} X\Bigr)\,.
\]
This merely involves reversal of the arrows.  We can easily see this in diagrammatic terms.
We represent the comultiplication and its unit as follows:
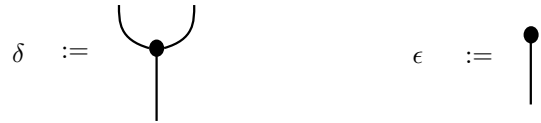
\begin{center} 
\ifx\JPicScale\undefined\def\JPicScale{1}\fi
\psset{unit=\JPicScale mm}
\psset{linewidth=0.3,dotsep=1,hatchwidth=0.3,hatchsep=1.5,shadowsize=1,dimen=middle}
\psset{dotsize=0.7 2.5,dotscale=1 1,fillcolor=black}
\psset{arrowsize=1 2,arrowlength=1,arrowinset=0.25,tbarsize=0.7 5,bracketlength=0.15,rbracketlength=0.15}
\begin{pspicture}(0,0)(72.25,15.73)
\rput{90}(21.5,10){\psellipse[linestyle=none,fillstyle=solid](0,0)(1.15,1)}
\psbezier(22.5,10)(26.5,11.15)(26.5,13.44)(26.5,15.73)
\psbezier(20.5,10)(16.5,11.15)(16.5,13.44)(16.5,15.73)
\rput(3.12,9.38){$\delta$}
\rput(64.38,8.75){$:=$}
\rput(10.62,9.38){$:=$}
\rput(56.25,8.75){$\epsilon$}
\psline(21.5,0.31)(21.5,9.69)
\psline(71.25,2.5)(71.25,11.25)
\rput{90}(71.25,11.77){\psellipse[linestyle=none,fillstyle=solid](0,0)(1.15,1)}
\end{pspicture}
 \end{center}
Then, the corresponding requirements are:
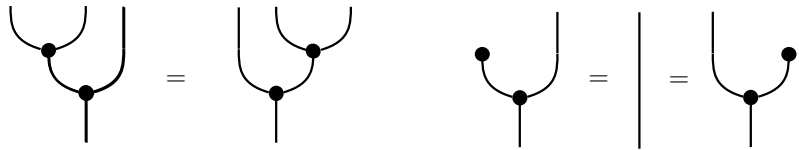
\begin{center} 
\ifx\JPicScale\undefined\def\JPicScale{1}\fi
\psset{unit=\JPicScale mm}
\psset{linewidth=0.3,dotsep=1,hatchwidth=0.3,hatchsep=1.5,shadowsize=1,dimen=middle}
\psset{dotsize=0.7 2.5,dotscale=1 1,fillcolor=black}
\psset{arrowsize=1 2,arrowlength=1,arrowinset=0.25,tbarsize=0.7 5,bracketlength=0.15,rbracketlength=0.15}
\begin{pspicture}(0,0)(105.47,19.66)
\rput{90}(11,8.15){\psellipse[linestyle=none,fillstyle=solid](0,0)(1.02,1)}
\psbezier(12,8.28)(16,9.42)(16,11.71)(16,14.01)
\psbezier(10,8.28)(6,9.42)(6,11.71)(6,14.01)
\psline(11,1.58)(11,7.96)
\psline(84.6,0.68)(84.6,18.83)
\rput{0}(63.7,13.25){\psellipse[linestyle=none,fillstyle=solid](0,0)(1,0.97)}
\rput{41.9}(6.05,13.78){\psellipse[linestyle=none,fillstyle=solid](0,0)(1.02,0.97)}
\psbezier(7,13.93)(11,15.08)(11,17.37)(11,19.66)
\psbezier(5,13.93)(1,15.08)(1,17.37)(1,19.66)
\psline(16,14.08)(16,19.58)
\rput{90}(36.31,8.05){\psellipse[linestyle=none,fillstyle=solid](0,0)(1.02,-0.99)}
\psbezier(35.31,8.18)(31.34,9.32)(31.34,11.61)(31.34,13.91)
\psbezier(37.3,8.18)(41.27,9.32)(41.27,11.61)(41.27,13.91)
\psline(36.31,1.48)(36.31,7.86)
\rput{134.27}(41.22,13.68){\psellipse[linestyle=none,fillstyle=solid](0,0)(1.02,-0.97)}
\psbezier(40.28,13.83)(36.31,14.98)(36.31,17.27)(36.31,19.56)
\psbezier(42.27,13.83)(46.24,14.98)(46.24,17.27)(46.24,19.56)
\psline(31.34,13.98)(31.34,19.48)
\rput(23,9.78){$=$}
\rput{90}(11.1,8.05){\psellipse[linestyle=none,fillstyle=solid](0,0)(1.02,1)}
\psbezier(12.1,8.18)(16.1,9.32)(16.1,11.61)(16.1,13.91)
\psbezier(10.1,8.18)(6.1,9.32)(6.1,11.61)(6.1,13.91)
\psline(11.1,1.48)(11.1,7.86)
\psline(16.1,13.98)(16.1,19.48)
\rput{90}(68.7,7.45){\psellipse[linestyle=none,fillstyle=solid](0,0)(1.02,1)}
\psbezier(69.7,7.58)(73.7,8.72)(73.7,11.01)(73.7,13.31)
\psbezier(67.7,7.58)(63.7,8.72)(63.7,11.01)(63.7,13.31)
\psline(68.7,0.88)(68.7,7.26)
\psline(73.7,13.38)(73.7,18.88)
\rput{0}(104.46,13.27){\psellipse[linestyle=none,fillstyle=solid](0,0)(1.01,-0.97)}
\rput{90}(99.41,7.48){\psellipse[linestyle=none,fillstyle=solid](0,0)(1.02,-1.01)}
\psbezier(98.4,7.6)(94.36,8.75)(94.36,11.04)(94.36,13.33)
\psbezier(100.42,7.6)(104.46,8.75)(104.46,11.04)(104.46,13.33)
\psline(99.41,0.9)(99.41,7.29)
\psline(94.36,13.4)(94.36,18.9)
\rput(79.2,9.78){$=$}
\rput(89.9,9.68){$=$}
\end{pspicture}
 \end{center}
Now, if we flip all of these upside-down we obtain a monoid:
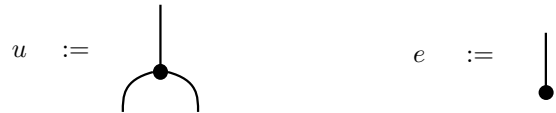
\begin{center} 
\ifx\JPicScale\undefined\def\JPicScale{1}\fi
\psset{unit=\JPicScale mm}
\psset{linewidth=0.3,dotsep=1,hatchwidth=0.3,hatchsep=1.5,shadowsize=1,dimen=middle}
\psset{dotsize=0.7 2.5,dotscale=1 1,fillcolor=black}
\psset{arrowsize=1 2,arrowlength=1,arrowinset=0.25,tbarsize=0.7 5,bracketlength=0.15,rbracketlength=0.15}
\begin{pspicture}(0,0)(74.12,15.62)
\rput{90}(21.88,6.65){\psellipse[linestyle=none,fillstyle=solid](0,0)(1.06,-1)}
\psbezier(22.88,6.66)(26.88,5.59)(26.88,3.47)(26.88,1.36)
\psbezier(20.88,6.66)(16.88,5.59)(16.88,3.47)(16.88,1.36)
\rput(3.12,9.38){$u$}
\rput(64.38,8.75){$:=$}
\rput(10.62,9.38){$:=$}
\rput(56.25,8.75){$e$}
\psline(21.88,15.62)(21.88,6.94)
\psline(73.12,11.89)(73.12,4.38)
\rput{0}(73.12,3.93){\psellipse[linestyle=none,fillstyle=solid](0,0)(1,-0.99)}
\end{pspicture}
 \end{center}
with corresponding requirements:
\begin{center} 
\ifx\JPicScale\undefined\def\JPicScale{1}\fi
\psset{unit=\JPicScale mm}
\psset{linewidth=0.3,dotsep=1,hatchwidth=0.3,hatchsep=1.5,shadowsize=1,dimen=middle}
\psset{dotsize=0.7 2.5,dotscale=1 1,fillcolor=black}
\psset{arrowsize=1 2,arrowlength=1,arrowinset=0.25,tbarsize=0.7 5,bracketlength=0.15,rbracketlength=0.15}
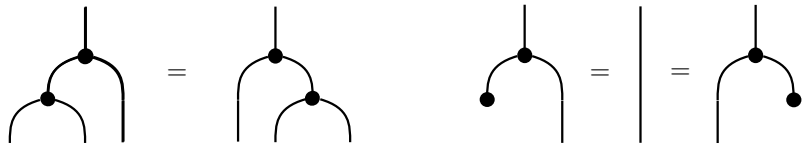
\begin{pspicture}(0,0)(105.97,19.05)
\rput{90}(10.76,12.14){\psellipse[linestyle=none,fillstyle=solid](0,0)(1.02,-1)}
\psbezier(11.76,12.02)(15.76,10.87)(15.76,8.59)(15.76,6.3)
\psbezier(9.76,12.02)(5.76,10.87)(5.76,8.59)(5.76,6.3)
\psline(10.76,18.7)(10.76,12.33)
\psline(84.6,0.68)(84.6,18.83)
\rput{0}(64.2,6.42){\psellipse[linestyle=none,fillstyle=solid](0,0)(1,-0.99)}
\rput{-40.54}(5.81,6.52){\psellipse[linestyle=none,fillstyle=solid](0,0)(1.02,-0.97)}
\psbezier(6.76,6.38)(10.76,5.23)(10.76,2.95)(10.76,0.66)
\psbezier(4.76,6.38)(0.76,5.23)(0.76,2.95)(0.76,0.66)
\psline(15.76,6.23)(15.76,0.74)
\rput{90}(36.07,12.24){\psellipse[linestyle=none,fillstyle=solid](0,0)(1.02,0.99)}
\psbezier(35.07,12.12)(31.1,10.97)(31.1,8.69)(31.1,6.4)
\psbezier(37.06,12.12)(41.03,10.97)(41.03,8.69)(41.03,6.4)
\psline(36.07,18.8)(36.07,12.43)
\rput{44.34}(40.98,6.62){\psellipse[linestyle=none,fillstyle=solid](0,0)(1.02,0.97)}
\psbezier(40.04,6.47)(36.07,5.33)(36.07,3.05)(36.07,0.76)
\psbezier(42.03,6.47)(46,5.33)(46,3.05)(46,0.76)
\psline(31.1,6.33)(31.1,0.84)
\rput(23,9.78){$=$}
\rput{90}(10.86,12.24){\psellipse[linestyle=none,fillstyle=solid](0,0)(1.02,-1)}
\psbezier(11.86,12.12)(15.86,10.97)(15.86,8.69)(15.86,6.4)
\psbezier(9.86,12.12)(5.86,10.97)(5.86,8.69)(5.86,6.4)
\psline(10.86,18.8)(10.86,12.43)
\psline(15.86,6.33)(15.86,0.84)
\rput{90}(69.2,12.33){\psellipse[linestyle=none,fillstyle=solid](0,0)(1.04,-1)}
\psbezier(70.2,12.21)(74.2,11.04)(74.2,8.7)(74.2,6.36)
\psbezier(68.2,12.21)(64.2,11.04)(64.2,8.7)(64.2,6.36)
\psline(69.2,19.05)(69.2,12.53)
\psline(74.2,6.29)(74.2,0.67)
\rput{-0}(104.96,6.39){\psellipse[linestyle=none,fillstyle=solid](0,0)(1.01,0.99)}
\rput{90}(99.91,12.31){\psellipse[linestyle=none,fillstyle=solid](0,0)(1.04,1.01)}
\psbezier(98.9,12.19)(94.86,11.02)(94.86,8.68)(94.86,6.34)
\psbezier(100.92,12.19)(104.96,11.02)(104.96,8.68)(104.96,6.34)
\psline(99.91,19.03)(99.91,12.51)
\psline(94.86,6.26)(94.86,0.65)
\rput(79.2,9.78){$=$}
\rput(89.9,9.9){$=$}
\end{pspicture}
 \end{center}
A {\em dagger (co)monoid} is a (co)monoid satisfying all the preceding requirements.

The dagger comonoids in ${\bf FdHilb}$ and ${\bf Rel}$ which we have seen above both have some additional properties.  For example, they are \em commutative\em: 
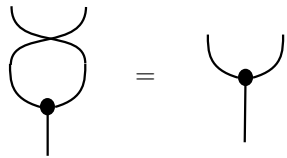
\begin{center} 
\ifx\JPicScale\undefined\def\JPicScale{1}\fi
\psset{unit=\JPicScale mm}
\psset{linewidth=0.3,dotsep=1,hatchwidth=0.3,hatchsep=1.5,shadowsize=1,dimen=middle}
\psset{dotsize=0.7 2.5,dotscale=1 1,fillcolor=black}
\psset{arrowsize=1 2,arrowlength=1,arrowinset=0.25,tbarsize=0.7 5,bracketlength=0.15,rbracketlength=0.15}
\begin{pspicture}(0,0)(38.1,20.9)
\rput{90}(6.8,7.5){\psellipse[linestyle=none,fillstyle=solid](0,0)(1.15,1)}
\psbezier(7.8,7.5)(11.8,8.65)(11.8,10.94)(11.8,13.23)
\psbezier(5.8,7.5)(1.8,8.65)(1.8,10.94)(1.8,13.23)
\psline(6.8,1)(6.8,7.19)
\psbezier(1.9,13.1)(1.9,18)(11.8,14.7)(11.9,20.8)
\psbezier(11.8,13.3)(11.8,18.2)(1.9,14.8)(2,20.9)
\rput(19.8,11.3){$=$}
\rput{90}(33.1,11.4){\psellipse[linestyle=none,fillstyle=solid](0,0)(1.15,1)}
\psbezier(34.1,11.4)(38.1,12.55)(38.1,14.84)(38.1,17.13)
\psbezier(32.1,11.4)(28.1,12.55)(28.1,14.84)(28.1,17.13)
\psline(33,2.8)(33.1,11.09)
\end{pspicture}
 \end{center}
that is, symbolically, 
\[
\sigma_{X,X}\circ\delta=\delta\,.  
\]
The comultiplication is \em isometric \em or \em special\em:
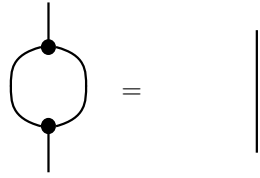
\begin{center} 
\ifx\JPicScale\undefined\def\JPicScale{1}\fi
\psset{unit=\JPicScale mm}
\psset{linewidth=0.3,dotsep=1,hatchwidth=0.3,hatchsep=1.5,shadowsize=1,dimen=middle}
\psset{dotsize=0.7 2.5,dotscale=1 1,fillcolor=black}
\psset{arrowsize=1 2,arrowlength=1,arrowinset=0.25,tbarsize=0.7 5,bracketlength=0.15,rbracketlength=0.15}
\begin{pspicture}(0,0)(33,22.79)
\rput{90}(5.3,6.32){\psellipse[linestyle=none,fillstyle=solid](0,0)(1.08,1)}
\psbezier(6.3,6.32)(10.3,7.4)(10.3,9.56)(10.3,11.72)
\psbezier(4.3,6.32)(0.3,7.4)(0.3,9.56)(0.3,11.72)
\psline(5.3,0.19)(5.3,6.02)
\rput(16.4,10.7){$=$}
\psline(33,2.8)(33,19.1)
\rput{90}(5.3,16.85){\psellipse[linestyle=none,fillstyle=solid](0,0)(1.05,-1)}
\psbezier(6.3,16.85)(10.3,15.8)(10.3,13.71)(10.3,11.62)
\psbezier(4.3,16.85)(0.3,15.8)(0.3,13.71)(0.3,11.62)
\psline(5.3,22.79)(5.3,17.13)
\end{pspicture}
 \end{center}
that is, symbolically, 
\[
\delta^\dagger\circ\delta=1_X\,.
\]
But by far, the most fascinating law which they obey are the \em Frobenius equations\em:
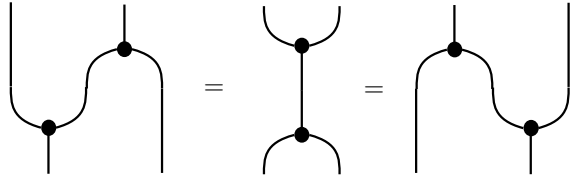
\begin{center} 
\ifx\JPicScale\undefined\def\JPicScale{1}\fi
\psset{unit=\JPicScale mm}
\psset{linewidth=0.3,dotsep=1,hatchwidth=0.3,hatchsep=1.5,shadowsize=1,dimen=middle}
\psset{dotsize=0.7 2.5,dotscale=1 1,fillcolor=black}
\psset{arrowsize=1 2,arrowlength=1,arrowinset=0.25,tbarsize=0.7 5,bracketlength=0.15,rbracketlength=0.15}
\begin{pspicture}(0,0)(74.45,23)
\rput{90}(5.3,6.32){\psellipse[linestyle=none,fillstyle=solid](0,0)(1.08,1)}
\psbezier(6.3,6.32)(10.3,7.4)(10.3,9.56)(10.3,11.72)
\psbezier(4.3,6.32)(0.3,7.4)(0.3,9.56)(0.3,11.72)
\psline(5.3,0.19)(5.3,6.02)
\rput(27.3,11.4){$=$}
\psline(20.4,0.3)(20.4,11.5)
\rput{90}(15.4,16.76){\psellipse[linestyle=none,fillstyle=solid](0,0)(1.05,-1)}
\psbezier(16.4,16.76)(20.4,15.71)(20.4,13.62)(20.4,11.53)
\psbezier(14.4,16.76)(10.4,15.71)(10.4,13.62)(10.4,11.53)
\psline(15.4,22.7)(15.4,17.05)
\psline(0.3,11.8)(0.3,23)
\rput{90}(69.43,6.26){\psellipse[linestyle=none,fillstyle=solid](0,0)(1.09,-1.01)}
\psbezier(68.42,6.26)(64.4,7.34)(64.4,9.5)(64.4,11.66)
\psbezier(70.43,6.26)(74.45,7.34)(74.45,9.5)(74.45,11.66)
\psline(69.42,0.13)(69.42,5.96)
\psline(54.18,0.3)(54.18,11.5)
\rput{90}(59.28,16.7){\psellipse[linestyle=none,fillstyle=solid](0,0)(1.05,1)}
\psbezier(58.27,16.7)(54.25,15.65)(54.25,13.56)(54.25,11.47)
\psbezier(60.28,16.7)(64.3,15.65)(64.3,13.56)(64.3,11.47)
\psline(59.27,22.64)(59.27,16.99)
\psline(74.45,11.74)(74.45,22.94)
\rput{90}(38.97,5.35){\psellipse[linestyle=none,fillstyle=solid](0,0)(1.05,1)}
\psbezier(37.97,5.35)(33.95,4.3)(33.95,2.21)(33.95,0.12)
\psbezier(39.98,5.35)(44,4.3)(44,2.21)(44,0.12)
\psline(38.98,11.29)(38.98,5.63)
\rput{90}(38.97,17.22){\psellipse[linestyle=none,fillstyle=solid](0,0)(1.06,-1)}
\psbezier(37.97,17.22)(33.95,18.28)(33.95,20.38)(33.95,22.49)
\psbezier(39.98,17.22)(44,18.28)(44,20.38)(44,22.49)
\psline(38.98,11.24)(38.98,16.93)
\rput(48.6,11.2){$=$}
\end{pspicture}
 \end{center}
that is, symbolically, 
\[
(1_X\otimes\delta^\dagger)\circ(\delta\otimes 1_X)=\delta\circ\delta^\dagger=
(\delta^\dagger\otimes 1_X)\circ(1_X\otimes\delta)\,.
\]
For a commutative dagger comonoid these two equations are easily seen to be equivalent.
We verify that these equations hold for the dagger comonoids in ${\bf FdHilb}$ and ${\bf Rel}$ discussed in  the previous section. 

In ${\bf FdHilb}$, we have
\[
\delta^\dagger: {\cal H}\otimes  {\cal H}\to {\cal H}::
|ij\rangle\mapsto \delta_{ij}\cdot|i\rangle
\quad\mbox{ and }\quad 
\epsilon^\dagger:\mathbb{C}\to{\cal H}::1\mapsto \sum_i |i\rangle
\]
so
\[
\begin{diagram}
|ij\rangle&\rMaps^{\delta\otimes 1_X}& |iij\rangle&\rMaps^{1_X\otimes\delta^\dagger}&|i\rangle\otimes (\delta_{ij}\cdot |i\rangle)=
\delta_{ij}\cdot|ii\rangle
\end{diagram}
\]
and
\[
\begin{diagram}
|ij\rangle&\rMaps^{\delta^\dagger}&\delta_{ij}\cdot |i\rangle&\rMaps^{\delta}
&\delta_{ij}\cdot |ii\rangle\,.
\end{diagram}
\]

In ${\bf Rel}$ we have 
\[
\delta^\dagger=   \{((x,x),x)\mid x\in X\}  \subseteq  (X\times X)\times X
\]
and 
\[
\epsilon ^\dagger= \{(*,x)\mid x\in X\}  \subseteq  \{*\}\times X
\]
so we obtain
\[
(1_X\otimes\delta^\dagger)\circ(\delta\otimes 1_X)=\delta\circ\delta^\dagger
=\{((x,x),(x,x))\mid x\in X\}\,.
\]

One can show that the Frobenius equation together with isometry 
guarantees a normal form for any connected picture made up of dagger Frobenius (co)monoids, identities and symmetry, and which only depends on the number of input and output wires  \cite{CP2006, Lack}.  As a result we can represent any such network as a `spider' e.g.,
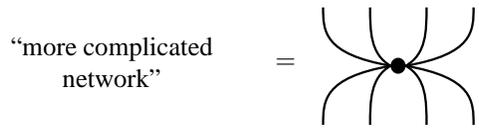
\begin{center} 
\ifx\JPicScale\undefined\def\JPicScale{1}\fi
\psset{unit=\JPicScale mm}
\psset{linewidth=0.3,dotsep=1,hatchwidth=0.3,hatchsep=1.5,shadowsize=1,dimen=middle}
\psset{dotsize=0.7 2.5,dotscale=1 1,fillcolor=black}
\psset{arrowsize=1 2,arrowlength=1,arrowinset=0.25,tbarsize=0.7 5,bracketlength=0.15,rbracketlength=0.15}
\begin{pspicture}(0,0)(63.75,17.5)
\rput(38.75,10){$=$}
\rput{90}(53.72,9.73){\psellipse[linestyle=none,fillstyle=solid](0,0)(1.06,-1)}
\psbezier(52.72,9.73)(50,11.29)(50,14.39)(50,17.5)
\psbezier(54.73,9.73)(63.75,11.29)(63.75,14.39)(63.75,17.5)
\psbezier(54.73,9.73)(57.5,11.29)(57.5,14.39)(57.5,17.5)
\psbezier(52.77,9.73)(43.75,11.29)(43.75,14.39)(43.75,17.5)
\psbezier(52.72,9.73)(50,8.15)(50,5.02)(50,1.88)
\psbezier(54.73,9.73)(63.75,8.15)(63.75,5.02)(63.75,1.88)
\psbezier(54.73,9.73)(57.5,8.15)(57.5,5.02)(57.5,1.88)
\psbezier(52.77,9.73)(43.75,8.15)(43.75,5.02)(43.75,1.88)
\rput(15.62,11.88){``more complicated}
\rput(15.62,8.12){network"}
\end{pspicture}
 \end{center}

Hence commutative dagger special Frobenius comonoids turn out to be structures which come with a very simple graphical calculus, but at the same time they are of key importance to quantum theory, as is exemplified by this theorem \cite{CPV}:

\begin{theorem}
In ${\bf FdHilb}$ there is a bijective correspondence between dagger special Frobenius comonoids and orthonormal bases.  Explicitly, each dagger special Frobenius comonoid in ${\bf FdHilb}$ is of the form 
\[
\delta: {\cal H}\to {\cal H}\otimes  {\cal H}::|i\rangle\mapsto |ii\rangle\quad
\mbox{ and }\quad \epsilon:{\cal H}\to\mathbb{C}::|i\rangle\mapsto1
\]
relative to some orthonormal basis $\{|i\rangle\}_i$.
\end{theorem}

In the category ${\bf 2Cob}$ we also encounter the Frobenius equation:
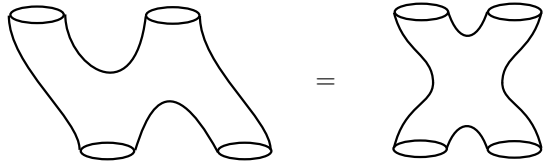
\begin{center} 
\ifx\JPicScale\undefined\def\JPicScale{1}\fi
\psset{unit=\JPicScale mm}
\psset{linewidth=0.3,dotsep=1,hatchwidth=0.3,hatchsep=1.5,shadowsize=1,dimen=middle}
\psset{dotsize=0.7 2.5,dotscale=1 1,fillcolor=black}
\psset{arrowsize=1 2,arrowlength=1,arrowinset=0.25,tbarsize=0.7 5,bracketlength=0.15,rbracketlength=0.15}
\begin{pspicture}(0,0)(72,20.5)
\rput(43.3,9.9){$=$}
\rput{0}(4.95,19){\psellipse[](0,0)(3.55,-1.1)}
\rput{0}(14.3,0.9){\psellipse[](0,0)(3.55,-1.1)}
\rput{0}(22.95,18.9){\psellipse[](0,0)(3.55,-1.1)}
\rput{0}(32.55,0.9){\psellipse[](0,0)(3.55,-1.1)}
\psbezier(36,1)(36.1,5.1)(26.6,12.1)(26.6,18.86)
\psbezier(10.6,1.04)(10.7,5.14)(1.2,12.14)(1.2,18.9)
\psbezier(19.4,18.8)(17.9,5.8)(8.7,12.3)(8.7,19.06)
\psbezier(29,0.7)(22.1,13.5)(19.3,5.04)(18,1)
\rput{0}(56.05,19.4){\psellipse[](0,0)(3.55,-1.1)}
\rput{0}(55.85,1.2){\psellipse[](0,0)(3.55,-1.1)}
\rput{0}(68.45,19.4){\psellipse[](0,0)(3.55,-1.1)}
\rput{0}(68.4,1.1){\psellipse[](0,0)(3.55,-1.1)}
\psbezier(66.8,9.9)(66.8,13.8)(70.63,13.9)(72,19.3)
\psbezier(59.6,19.4)(61,15.1)(63.4,15.19)(64.8,19.4)
\psbezier(59.4,1.2)(60.77,5.2)(63.43,5.2)(64.8,1.2)
\psbezier(66.8,10)(66.8,7.1)(70.63,7.1)(72,1.3)
\psbezier(57.6,9.9)(57.6,13.8)(53.77,13.9)(52.4,19.3)
\psbezier(57.6,10)(57.6,7.1)(53.77,7.1)(52.4,1.3)
\end{pspicture}
 \end{center}
but the (co)monoids involved are not special, since the two cobordisms
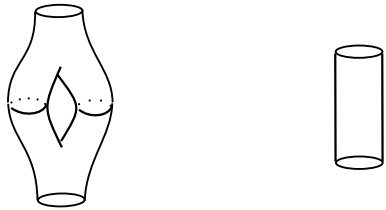
\begin{center} 
\ifx\JPicScale\undefined\def\JPicScale{1}\fi
\psset{unit=\JPicScale mm}
\psset{linewidth=0.3,dotsep=1,hatchwidth=0.3,hatchsep=1.5,shadowsize=1,dimen=middle}
\psset{dotsize=0.7 2.5,dotscale=1 1,fillcolor=black}
\psset{arrowsize=1 2,arrowlength=1,arrowinset=0.25,tbarsize=0.7 5,bracketlength=0.15,rbracketlength=0.15}
\begin{pspicture}(0,0)(50.5,27.61)
\rput{0.81}(7.79,1.61){\psellipse[linewidth=0.25](0,0)(3.11,0.86)}
\psbezier[linewidth=0.25](14.56,14.47)(14.49,10.19)(11.13,6.94)(10.9,1.5)
\psbezier[linewidth=0.25](0.73,14.4)(0.96,9.7)(4.52,8.52)(4.62,1.59)
\rput{0.81}(7.48,26.74){\psellipse[linewidth=0.25](0,0)(3.11,-0.82)}
\psbezier[linewidth=0.25](14.61,14.59)(14.56,19.09)(10.79,21.24)(10.59,26.93)
\psbezier[linewidth=0.25](0.7,14.58)(0.74,20.3)(4.41,20.02)(4.31,26.67)
\rput{0.13}(47.35,21.3){\psellipse[linewidth=0.25](0,0)(3.11,-0.82)}
\rput{0.13}(47.39,6.53){\psellipse[linewidth=0.25](0,0)(3.11,-0.83)}
\psline(50.46,21.1)(50.5,6.73)
\psline(44.21,21.09)(44.25,6.72)
\psbezier(7.72,19.33)(5.33,13.86)(5.33,13.86)(7.87,8.6)
\psbezier(7.28,18.27)(10.53,13.96)(10.53,13.96)(7.76,9.44)
\psbezier(10.14,13.63)(11.56,12.45)(14.05,12.69)(14.53,14.09)
\psbezier[linestyle=dotted](10.03,14.28)(11.42,15.2)(13.92,15.08)(14.43,14.04)
\psbezier(1.14,13.85)(2.65,12.67)(5.32,12.91)(5.83,14.32)
\psbezier[linestyle=dotted](1.03,14.5)(2.5,15.42)(5.17,15.31)(5.72,14.27)
\end{pspicture}
 \end{center}
are not homeomorphic. 
Therefore a normal form in ${\bf 2Cob}$ is of the form \cite{Kock}
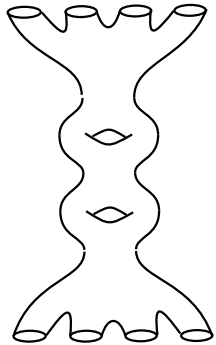
\begin{center}
\ifx\JPicScale\undefined\def\JPicScale{1}\fi
\psset{unit=\JPicScale mm}
\psset{linewidth=0.3,dotsep=1,hatchwidth=0.3,hatchsep=1.5,shadowsize=1,dimen=middle}
\psset{dotsize=0.7 2.5,dotscale=1 1,fillcolor=black}
\psset{arrowsize=1 2,arrowlength=1,arrowinset=0.25,tbarsize=0.7 5,bracketlength=0.15,rbracketlength=0.15}
\begin{pspicture}(0,0)(28.94,45.59)
\rput{0.77}(4.61,1.49){\psellipse[](0,0)(2.14,-0.72)}
\rput{0.6}(12.11,1.5){\psellipse[](0,0)(2.08,-0.72)}
\rput{0.76}(18.72,44.73){\psellipse[](0,0)(2.08,-0.72)}
\rput{0.75}(19.45,1.53){\psellipse[](0,0)(2.08,-0.72)}
\psbezier(18.52,33.26)(18.46,37.9)(25.56,38.52)(28.15,44.98)
\psbezier(13.54,44.66)(14.4,41.85)(15.81,41.93)(16.59,44.7)
\psbezier(14.19,1.52)(14.95,4.15)(16.51,4.17)(17.35,1.56)
\psbezier(18.77,12.81)(18.82,9)(26.18,9.1)(28.9,1.51)
\psbezier(11.54,33.59)(11.49,38.17)(4.35,38.2)(1.72,44.5)
\psbezier(11.9,12.73)(11.94,9.02)(4.9,8.93)(2.47,1.5)
\rput{0.7}(26.85,1.56){\psellipse[](0,0)(2.08,-0.72)}
\rput{0.75}(11.47,44.66){\psellipse[](0,0)(2.08,-0.72)}
\rput{0.65}(3.91,44.57){\psellipse[](0,0)(2.2,-0.72)}
\rput{0.76}(25.96,44.85){\psellipse[](0,0)(2.08,-0.72)}
\psbezier(6.86,1.62)(9.37,5.09)(10.46,6.16)(10.01,1.5)
\psbezier(6.24,44.56)(8.83,41.57)(9.94,40.66)(9.4,44.74)
\psbezier(24.65,1.84)(22.14,5.24)(21.07,6.3)(21.61,1.64)
\psbezier(23.78,44.78)(21.34,41.73)(20.3,40.79)(20.73,44.88)
\psbezier(21.7,28.3)(21.67,31.08)(18.54,31.04)(18.52,33.26)
\psbezier(18.64,23.26)(18.62,24.93)(21.74,24.97)(21.7,28.3)
\psbezier(21.83,18.2)(21.79,21.03)(18.67,20.99)(18.64,23.26)
\psbezier(18.77,13.05)(18.75,14.76)(21.87,14.8)(21.83,18.2)
\psbezier(8.58,28.14)(8.54,30.92)(11.67,30.96)(11.65,33.18)
\psbezier(11.76,23.18)(11.74,24.85)(8.62,24.8)(8.58,28.14)
\psbezier(8.7,18.03)(8.67,20.87)(11.79,20.91)(11.76,23.18)
\psbezier(11.89,12.97)(11.87,14.67)(8.75,14.63)(8.7,18.03)
\psbezier(18.38,17.96)(15.21,16.16)(15.21,16.16)(12.13,18.11)
\psbezier(17.5,17.61)(15.34,18.91)(15.34,18.91)(13.12,17.71)
\psbezier(18.25,28.32)(15.07,26.52)(15.07,26.52)(11.99,28.47)
\psbezier(17.36,27.96)(15.2,29.27)(15.2,29.27)(12.98,28.07)
\end{pspicture}
\end{center}

The commutative diagram in Definition \ref{internal_group} becomes
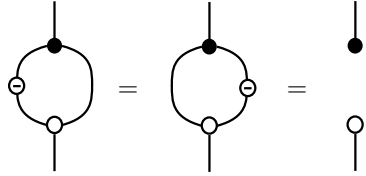
\begin{center} 
\ifx\JPicScale\undefined\def\JPicScale{1}\fi
\psset{unit=\JPicScale mm}
\psset{linewidth=0.3,dotsep=1,hatchwidth=0.3,hatchsep=1.5,shadowsize=1,dimen=middle}
\psset{dotsize=0.7 2.5,dotscale=1 1,fillcolor=black}
\psset{arrowsize=1 2,arrowlength=1,arrowinset=0.25,tbarsize=0.7 5,bracketlength=0.15,rbracketlength=0.15}
\begin{pspicture}(0,0)(47.6,22.7)
\rput{90}(6.6,6.19){\psellipse[](0,0)(1.08,1)}
\psbezier(7.6,6.2)(11.6,7.28)(11.6,9.44)(11.6,11.6)
\psbezier(5.6,6.2)(1.6,7.28)(1.6,9.44)(1.6,11.6)
\psline(6.6,0.07)(6.6,4.98)
\rput(16.4,10.7){$=$}
\rput{90}(6.6,16.73){\psellipse[linestyle=none,fillstyle=solid](0,0)(1.05,-1)}
\psbezier(7.6,16.73)(11.6,15.68)(11.6,13.59)(11.6,11.5)
\psbezier(5.6,16.73)(1.6,15.68)(1.6,13.59)(1.6,11.5)
\psline(6.6,22.67)(6.6,17.01)
\rput{90}(1.6,11.4){\psellipse[fillcolor=white,fillstyle=solid](0,0)(1.08,1)}
\psline(1.1,11.38)(2.1,11.38)
\rput{90}(27.2,6.1){\psellipse[](0,0)(1.08,1)}
\psbezier(28.2,6.1)(32.2,7.18)(32.2,9.34)(32.2,11.5)
\psbezier(26.2,6.1)(22.2,7.18)(22.2,9.34)(22.2,11.5)
\psline(27.2,-0.03)(27.2,4.88)
\rput{90}(27.2,16.63){\psellipse[linestyle=none,fillstyle=solid](0,0)(1.05,-1)}
\psbezier(28.2,16.63)(32.2,15.58)(32.2,13.49)(32.2,11.4)
\psbezier(26.2,16.63)(22.2,15.58)(22.2,13.49)(22.2,11.4)
\psline(27.2,22.57)(27.2,16.91)
\rput{90}(32.3,11.12){\psellipse[fillcolor=white,fillstyle=solid](0,0)(1.08,1)}
\psline(31.8,11.1)(32.8,11.1)
\rput(38.9,10.7){$=$}
\rput{90}(46.6,6.23){\psellipse[](0,0)(1.08,1)}
\psline(46.6,0.1)(46.6,5.01)
\rput{90}(46.6,16.76){\psellipse[linestyle=none,fillstyle=solid](0,0)(1.05,-1)}
\psline(46.6,22.7)(46.6,17.04)
\end{pspicture}
 \end{center}
when setting
\begin{center} 
\ifx\JPicScale\undefined\def\JPicScale{1}\fi
\psset{unit=\JPicScale mm}
\psset{linewidth=0.3,dotsep=1,hatchwidth=0.3,hatchsep=1.5,shadowsize=1,dimen=middle}
\psset{dotsize=0.7 2.5,dotscale=1 1,fillcolor=black}
\psset{arrowsize=1 2,arrowlength=1,arrowinset=0.25,tbarsize=0.7 5,bracketlength=0.15,rbracketlength=0.15}
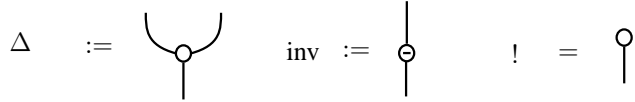
\begin{pspicture}(0,0)(87.2,16.9)
\rput{90}(27.6,9.95){\psellipse[](0,0)(1.08,1)}
\psbezier(28.6,9.96)(32.6,11.04)(32.6,13.2)(32.6,15.36)
\psbezier(26.6,9.96)(22.6,11.04)(22.6,13.2)(22.6,15.36)
\psline(27.6,3.83)(27.6,8.74)
\rput(16.4,10.7){$:=$}
\psline(57.2,4.23)(57.2,9.14)
\psline(57.3,16.9)(57.3,11.24)
\rput{90}(57.3,10.02){\psellipse[fillcolor=white,fillstyle=solid](0,0)(1.08,1)}
\psline(56.8,10)(57.8,10)
\rput(78.8,9.83){$=$}
\rput{90}(86.2,12.09){\psellipse[](0,0)(1.08,1)}
\psline(86.2,5.97)(86.2,10.88)
\rput(6.1,10.5){$\diag$}
\rput(71.7,9.87){$!$}
\rput(50.6,10.23){$:=$}
\rput(43.5,10.27){inv}
\end{pspicture}
 \end{center}
One refers to this equation typically as the \em Hopf law \em -- cf.~the Hopf algebras mentioned above. What also holds for these operations are the \em bialgebra laws\em:
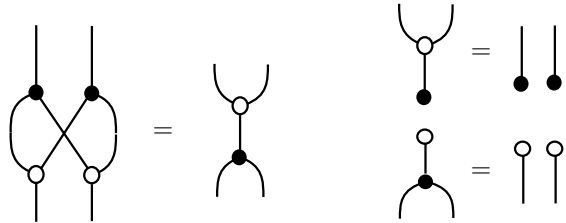
\begin{center} 
\ifx\JPicScale\undefined\def\JPicScale{1}\fi
\psset{unit=\JPicScale mm}
\psset{linewidth=0.3,dotsep=1,hatchwidth=0.3,hatchsep=1.5,shadowsize=1,dimen=middle}
\psset{dotsize=0.7 2.5,dotscale=1 1,fillcolor=black}
\psset{arrowsize=1 2,arrowlength=1,arrowinset=0.25,tbarsize=0.7 5,bracketlength=0.15,rbracketlength=0.15}
\begin{pspicture}(0,0)(76.2,29.49)
\psbezier(34.1,16.11)(37.05,17.17)(37.05,19.29)(37.05,21.41)
\psbezier(13.7,6.7)(16.85,7.78)(16.85,9.94)(16.85,12.1)
\psline(13.65,0.57)(13.65,5.48)
\rput{90}(6.25,17.68){\psellipse[linestyle=none,fillstyle=solid](0,0)(1.06,-1)}
\psbezier(6.3,17.68)(2.85,16.61)(2.85,14.49)(2.85,12.38)
\psline(6.25,26.64)(6.25,17.96)
\rput{90}(13.65,17.58){\psellipse[linestyle=none,fillstyle=solid](0,0)(1.06,-1)}
\psbezier(13.7,17.58)(16.85,16.51)(16.85,14.39)(16.85,12.28)
\psline(13.65,26.54)(13.65,17.86)
\psbezier(6.3,6.8)(2.85,7.91)(2.85,10.14)(2.85,12.36)
\psline(6.25,0.48)(6.25,5.54)
\psline(6.65,17.08)(13.05,7.68)
\psline(6.85,7.5)(13.05,16.98)
\rput{90}(6.25,6.79){\psellipse[fillcolor=white,fillstyle=solid](0,0)(1.12,1)}
\rput{90}(13.65,6.7){\psellipse[fillcolor=white,fillstyle=solid](0,0)(1.08,1)}
\psbezier(33.4,15.93)(29.95,17.04)(29.95,19.27)(29.95,21.49)
\psline(33.35,9.61)(33.35,14.67)
\rput{90}(33.35,15.92){\psellipse[fillcolor=white,fillstyle=solid](0,0)(1.12,1)}
\rput{90}(33.25,9.08){\psellipse[linestyle=none,fillstyle=solid](0,0)(1.06,-1)}
\psbezier(33.3,9.08)(36.45,8.01)(36.45,5.89)(36.45,3.78)
\psbezier(33.55,9.11)(30.3,8.04)(30.3,5.92)(30.3,3.81)
\rput(23.15,12.48){$=$}
\psbezier(58.65,24.11)(61.6,25.17)(61.6,27.29)(61.6,29.41)
\psbezier(57.95,23.92)(54.5,25.04)(54.5,27.26)(54.5,29.49)
\psline(57.9,17.61)(57.9,22.67)
\rput{90}(57.9,23.92){\psellipse[fillcolor=white,fillstyle=solid](0,0)(1.12,1)}
\rput{90}(57.8,17.07){\psellipse[linestyle=none,fillstyle=solid](0,0)(1.06,-1)}
\psline(70.8,19.44)(70.8,26.5)
\rput{90}(70.7,18.86){\psellipse[linestyle=none,fillstyle=solid](0,0)(1.06,-1)}
\psline(75.2,19.54)(75.2,26.6)
\rput{90}(75.1,19){\psellipse[linestyle=none,fillstyle=solid](0,0)(1.06,-1)}
\rput(65.4,22.9){$=$}
\psbezier(58.75,5.63)(61.7,4.7)(61.7,2.84)(61.7,0.98)
\psbezier(58.05,5.8)(54.6,4.82)(54.6,2.86)(54.6,0.9)
\psline(58,11.35)(58,6.9)
\rput{0}(58,5.8){\psellipse[linestyle=none,fillstyle=solid](0,0)(1,-0.98)}
\rput{0}(57.9,11.82){\psellipse[fillcolor=white,fillstyle=solid](0,0)(1,0.94)}
\psline(71,9.71)(71,2.9)
\rput{0}(70.9,10.2){\psellipse[fillcolor=white,fillstyle=solid](0,0)(1,0.94)}
\psline(75.3,9.73)(75.3,2.92)
\rput{0}(75.2,10.2){\psellipse[fillcolor=white,fillstyle=solid](0,0)(1,0.94)}
\rput(65.4,7.1){$=$}
\end{pspicture}
 \end{center}

There's lots more to say on the connections between algebraic structures and these pictures.  The reader may consult, for example, \cite{Kock, Selinger, StreetBook}.  A great place to find some very well-explained introductions to this is John Baez' This Week's Finds in Mathematical Physics \cite{TWF}, for example, weeks 174, 224, 268.

\section{Monoidal functoriality, naturality and TQFTs}\label{funcNatTQFT}

In this section we provide the remaining bits of theory required to state the definition of a topological quantum field theory.

\subsection{Bifunctors}\label{sec:BifunctAsFunct}

The category ${\bf Cat}$ which has categories as objects and functors as morphisms also comes with a monoidal structure:

\begin{definition} \em 
The {\em product} of categories $\cat$ and ${\bf D}$ is a category $\cat\times {\bf D}$: 
\begin{enumerate}
\item objects are pairs $(C,D)$ with $C\in|\cat|$ and $D\in|{\bf D}|$\,,
\item morphisms are pairs $( f,g):( C,D)\rTo( C',D')$ where $f:C\rTo C'$ is a morphism in ${\bf C}$ and $g:D\rTo D'$ is a morphism in ${\bf D}$\,,
\item composition is componentwise, that is, 
\[
( f',g')\circ( f,g)=( f'\circ f,g'\circ g)\,,
\]
and the identities are pairs of identities. 
\end{enumerate}
\end{definition}

This monoidal structure is Cartesian. The obvious projection functors 
\[
\cat\stackrel{P_1}\longleftarrow\cat\times{\bf D}\stackrel{P_2}\longrightarrow {\bf D}
\] 
provide the product structure:
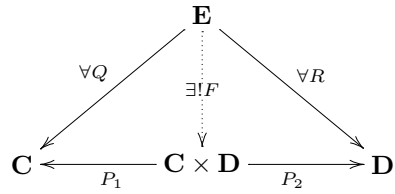
\[
\xymatrix@=.6in{&{\bf E} \ar[dl]_{\forall  Q}\ar[dr]^{\forall R}\ar@{..>}[d]|{\exists ! F}& \\ \cat & \cat\times {\bf D} \ar[l]^{P_1}\ar[r]_{P_2} & {\bf D}}
\]
This notion of product allows for a very concise definition of \em bifunctoriality\em. A \em bifunctor \em  is now nothing but an ordinary functor of type  
\[
F:\cat\times{\bf D}\longrightarrow{\bf E}\,.
\]
So, for instance, to say that a tensor is a bifunctor it now suffices to say that 
\[
-\otimes-:\cat\times\cat\longrightarrow \cat
\]
is a functor. Indeed, this implies that we have 
\[
\otimes(\varphi\circ\xi)=\otimes(\varphi)\circ\otimes(\xi)\quad\mbox{ and }\quad\otimes(1_\Xi)=1_{\otimes(\Xi)}
\]
for  all morphisms $\varphi,\xi$ and all objects $\Xi$ in $\cat\times\cat$, that is, 
\[
(g\circ f)\otimes(g'\circ f')=(g \otimes g')\circ(f\otimes f')\quad\mbox{ and }\quad 1_A \otimes 1_B=1_{A \otimes B}\,.
\]

We give another example of bifunctor which is contravariant in the first variable and covariant in the second variable.  This functor is key to the so-called \em Yoneda Lemma\em, which constitutes the core of many categorical constructs, for which we refer to the standard literature \cite{SML}.
For all $A\in|\cat|$ let
\[
\cat(A,-):\cat\longrightarrow{\bf Set}
\]
be the functor which maps
\begin{enumerate}
\item  each object $B\in|\cat|$ to the set $\cat(A,B)\in|{\bf Set}|$\,, and
\item  each morphism $g:B\rTo C$ to the function
\[
\cat(A,g):\cat(A,B)\rightarrow \cat(A,C)::f\mapsto g\circ f\,.
\]
\end{enumerate}
For all $C\in|\cat|$ let
\[
\cat(-,C):\cat^{\scriptsize \mbox{\em op}}\longrightarrow{\bf Set}
\]
be the functor which maps
\begin{enumerate}
\item  each object $A\in|\cat|$ to the set $\cat(A,C)\in|{\bf Set}|$\,, and
\item  each morphism $f:A\rTo B$ to the function
\[
\cat(f,C):\cat(B,C)\rightarrow \cat(A,C)::g\mapsto  g\circ f\,.
\]
\end{enumerate}
One verifies that given any pair $f:A\rTo B$ and $h: C\rTo D$ the diagram 
\[
\xymatrix@=.6in{ \cat(B,C)\ar[r]^{\cat(f,C)} \ar[d]_{\cat(B,h)} & \cat (A,C) \ar[d]^{\cat(A,h)}\\%
\cat (B,D)\ar[r]_{\cat(f,D)} & \cat (A,D)}
\]
commutes, sending a morphism $g:B\rTo C$ to the composite $h\circ g\circ f:A\rTo D$.  The bifunctor -- also called \em hom-functor \em --  which unifies the above two functors is
\[
\cat(-,-):\cat^{\scriptsize \mbox{\em op}}\times\cat\longrightarrow {\bf Set}
\]
which maps
\begin{enumerate}
\item  each pair of objects $(A,B)\in|\cat|$ to the set $\cat(A,B)\in|{\bf Set}|$\,, and
\item  each pair morphism $(f:A\rTo B,h: C\rTo D)$ to the function
\[
\cat(f,h):\cat(B,C)\rightarrow \cat(A,D)::g\mapsto  h\circ g\circ f\,.
\]
\end{enumerate}
We can now identify
\[
\cat(A,-):=\cat(1_A,-)\qquad\mbox{ and }\qquad\cat(-,A):=\cat(-,1_A)\,.
\]
These functors are called  \em representable functors\em. They enable us to represent objects and morphisms of any category as functors on the well-known category ${\bf Set}$.

 \subsection{Naturality}\label{sec:naturaltransformations} 
 
 We already encountered a fair number of examples of our restricted variant of natural isomorphisms, namely
 \[
 \II\otimes A\simeq A \simeq A\otimes \II \ \ \ , \, \  \ A\otimes B\simeq B\otimes A  \ \ \ , \, \ \
 A\otimes (B\otimes C)\simeq  (A\otimes B)\otimes C
 \]
 and
 \[
 A\times (B + C)\simeq(A\times B) + (A\times C)\,,
 \]
 as well as some proper natural transformations, namely 
 \[
A\stackrel{}{\leadsto} A \times A \ \ \ , \, \  \  A+A\stackrel{}{\leadsto} A
\ \ \ \mbox{and} \ \ \
 (A\times B) + (A\times C)\leadsto A\times (B + C)\,.
 \]
What makes all of these special is that all of the above expressions only involve objects of the category $\cat$ without there being any reference to morphisms.  This is not the case anymore for the general notion of natural transformations, which are in fact, structure preserving maps between functors.

\begin{definition} \em
Let $F,G:\cat\longrightarrow{\bf D}$ be functors.  A {\em natural transformation} 
\[
\tau:F\Rightarrow G
\] 
consists of a family of morphisms 
\[
\{\tau_A\in {\bf D}(FA, GA) \mid A\in|\cat|\}
\] 
which are such that  the diagram
\[
\xymatrix@=.5in{FA\ar[r]^{\tau_A} \ar[d]_{Ff} & GA\ar[d]^{Gf}\\ FB\ar[r]_{\tau_B} & GB}
\] 
commutes for any $A,B\in|\cat|$ and any $f\in \cat(A,B)$. 
\end{definition}

\begin{example} Given vector spaces $V$ and $W$, then two group representations 
\[
\rho_1:G\rightarrow\mbox{GL}(V)\ \ \ \ \mbox{ and }\ \ \ \ \rho_2:G\rightarrow\mbox{GL}(W)
\]
are {\em equivalent} if there exists an isomorphism $\tau:V\rightarrow W$ so that for all $g\in G$, 
\beq\label{groupnaturality}
\tau\circ\rho_1(g)=\rho_2(g)\circ\tau\,. 
\eeq
This isomorphism is a natural transformation.   Indeed, taking the functorial point of view for the two representations above, we get two functors 
\[
{\bf G}\stackrel{R_{\rho_1}}{\longrightarrow}\mathbf{FdVect}_\mathbb{K}\ \ \ \ \mbox{ and }\ \ \ \ {\bf G}\stackrel{R_{\rho_2}}{\longrightarrow} \mathbf{FdVect}_\mathbb{K}
\]
where $R_{\rho_1}$ maps $*$ on some vector space $R_{\rho_1}(*)$ and $R_{\rho_2}$ maps $*$ on some vector space $R_{\rho_2}(*)$. Naturality means  that the diagram
\[
\xymatrix@=.5in{R_{\rho_1}(*)\ar[r]^{\tau_*} \ar[d]_{R_{\rho_1}g} & R_{\rho_2}(*)\ar[d]^{R_{\rho_2}g}\\ R_{\rho_1}(*)\ar[r]_{\tau_*} & R_{\rho_2}(*)}
\]
commutes, which translates into eq.(\ref{groupnaturality}).
\end{example}

\begin{example}\label{exnatt} 
The family of canonical linear maps 
\[
\{\tau_V:V\rightarrow V^{**}\mid V\in\mathbf{FdVect}_\mathbb{K}\}
\]
from a vector space to its double dual is a natural transformation 
\[
\tau:1_{\mathbf{FdVect}_\mathbb{K}}\Rightarrow (-)^{**}
\]
from the identity functor to the double dual functor.  There is no natural transformation of type  $1_{\mathbf{FdVect}_\mathbb{K}}\Rightarrow (-)^{*}$. Indeed, while each finite dimensional vector space is isomorphic with its dual, there is no `natural choice' of an isomorphism, since constructing one depends on a choice of basis.  
\end{example}

The fact that for $\mathbf{FdVect}_\mathbb{K}$ naturality indeed means basis independence can immediately be seen from the definition of naturality. In
\[
\xymatrix@=.5in{FV\ar[r]^{\tau_V} \ar[d]_{Ff} & GV\ar[d]^{Gf}\\ FV\ar[r]_{\tau_V} & GV}
\] 
the linear map $f:V\to V$ can be interpreted as a change of basis, and then the linear maps $Ff:FV\to FV$ and $Gf:GV\to GV$ apply this change of basis to the expressions $FV$ and $GV$ respectively.  Commutation of the above diagram then means that it makes no difference whether we apply $\tau_V$ before the change of basis, or whether we apply it after the change of basis. Hence it asserts that $\tau_V$ is a basis independent construction.


\subsection{Monoidal functors and monoidal natural transformations}\label{sec:monoidal_functors}

A monoidal functor, unsurprisingly,  is a functor between two monoidal categories that preserves the monoidal structure `coherently'. 

\begin{definition} \em 
Let 
\[
( \cat,\otimes, \II,\alpha_\cat,\lambda_\cat,\rho_\cat)\quad \mbox{ and }\quad ( {\bf D}, \odot, 
{\rm J},\alpha_{\bf D},\lambda_{\bf D},\rho_{\bf D})
\]
be monoidal categories. Then a {\em monoidal functor} is a functor 
\[
F:\cat\longrightarrow {\bf D}\,,
\]
together with a natural transformation 
\[
\phi_{-,-}:(F-)\odot (F-)\Rightarrow F(-\otimes -)
\]
with components
\[
\{\phi_{A,B}:FA\odot FB\rTo F(A\otimes B)\mid A,B\in |\cat|\}\,,
\]
and a morphism
\[
\phi:{\rm J}\rTo F\II\,,
\]
which are such that for every $A,B,C\in |\cat|$ the diagrams
\[
\xymatrix{
(FA\odot FB)\odot FC\ar[r]^{\alpha_{\bf D}^{-1}} \ar[d]_{\phi_{A,B}\odot 1_{FC}} & FA\odot (FB\odot FC) \ar[d]^{1_{FA}\odot\phi_{B,C}}\\
F(A\otimes B)\odot FC\ar[d]_{\phi_{A\otimes B,C}} & FA\odot F(B\otimes C)\ar[d]^{\phi_{A,B\otimes C}} \\
F((A\otimes B)\otimes C)\ar[r]_{F\alpha_{\cat}^{-1}} & F(A\otimes (B\otimes C))
}
\]
and
\[
\xymatrix{
FA\odot {\rm J}\ar[r]^{1_{FA}\odot\phi}\ar[d]_{\rho_{\bf D}^{-1}} & FA\odot F\II\ar[d]^{\phi_{A,\II}} & \ar @{}[d]|{\mbox{,}} & {\rm J}\odot FB\ar[d]_{\lambda_{\bf D}^{-1}} \ar[r]^{\phi\odot 1_{FB}} & F\II\odot FB\ar[d]^{\phi_{\II,B}}\\
FA & F(A\otimes \II)\ar[l]^{F\rho_{\cat}^{-1}} & & FB & F(\II\otimes B)\ar[l]^{F\lambda_{\cat}^{-1}}
}
\]
commute in ${\bf D}$.  Moreover, a monoidal functor between symmetric monoidal categories is {\em symmetric} if, in addition, for all $A,B\in |\cat|$ the diagram
\[
\xymatrix{FA\odot FB\ar[r]^{\sigma_{FA,FB}}\ar[d]_{\phi_{A,B}}& FB\odot FA\ar[d]^{\phi_{B,A}}\\
F(A\otimes B)\ar[r]_{F\sigma_{A,B}} & F(B\otimes A)
}
\]
commutes in ${\bf D}$. A monoidal functor is {\em strong} if the components of the natural transformation $\phi_{-,-}$ as well as the morphism $\phi$ are isomorphisms, and it is {\em strict} if they are identities. In this case the equational requirements simplify to
\[
F(A\otimes B)=FA\odot FB \quad\qquad\mbox{and}\qquad\quad F\II={\rm J}\,,
\]
and 
\[
F\alpha_{\cat}=\alpha_{\bf D}\quad , \quad  F\lambda_{\cat}=\lambda_{\bf D}\quad , \quad F\rho_{\cat}=\rho_{\bf D}\quad \mbox{and} \quad F\sigma_{\cat}=\sigma_{\bf D}\,.
\]
Hence a strict monoidal functor between strict monoidal categories just means that the tensor is preserved by $F$.
\end{definition}

\begin{example} 
The functor $\dagger:\cat^{op}\longrightarrow\cat$ is a strict monoidal functor. In a compact category $\cat$, the functor $(-)^*:\cat^{op}\longrightarrow\cat$ which maps any object $A$ on $A^*$ and any morphism $f$ on its transpose $f^*$ is a strong monoidal functor. 
\end{example}

\begin{definition} \em
A {\em monoidal natural transformation}
\[
\theta:\left(F,\{\phi_{A,B}\!\mid\! A,B\in|\cat|\},\phi\right)\Rightarrow \left(G,\{\psi_{A,B}\!\mid\! A,B\in\!|\cat|\},\psi\right)
\]
between two monoidal functors is a natural transformation such that
\[
\xymatrix{ FA\odot FB\ar[d]_{\phi_{A,B}}\ar[r]^{\theta_A\odot\theta_B} & GA\odot  GB\ar[d]^{\psi_{A,B}} & \ar@{}[d]|{\mbox{and}} & & {\rm J}\ar[dr]^\psi\ar[dl]_\phi & \\
F(A\otimes B)\ar[r]_{\theta_{A\otimes B}} & G(A\otimes B) & & F\II\ar[rr]_{\theta_\II} & & G\II}
\]
commute.  A monoidal natural transformation is {\em symmetric} if the two monoidal functors which constitute its domain and codomain are both symmetric. 
\end{definition}

\subsection{Equivalence of categories}

In Example~\ref{excat4} we defined the category ${\bf Cat}$ which has  categories as objects and functors as morphism. Definition \ref{def:isomorphic} on isomorphic objects, when applied to this special category ${\bf Cat}$, tells us that two categories $\cat$ and ${\bf D}$ are isomorphic if there exists two functors $F:\cat\longrightarrow{\bf D}$ and $G:{\bf D}\longrightarrow\cat$ such that 
\[
G\circ F=1_{\cat}\qquad\mbox{ and }\qquad F\circ G=1_{\bf D}\,. 
\]
Thus, the functor $F$ defines a bijection between the objects \underline{as well as} between the hom-sets  of $\cat$ and ${\bf D}$. However, many categories that are --- for most practical purposes --- equivalent are not isomorphic.  For example, 
\bit
\item the category ${\bf FSet}$ which has all finite sets as objects, and functions between these sets as morphisms, and, 
\item a category which has for each $n\in \mathbb{N}$ exactly one set of that size as objects, and functions between these sets as morphisms.  
\eit
Therefore, it is useful to define some properties for functors that are weaker than being isomorphisms. For instance, the two following definitions describe functors whose morphism assignments are injective and surjective respectively.

\begin{definition}\em 
A functor $F:\cat\longrightarrow{\bf D}$ is {\em faithful} if for any $A,B\in|\cat|$ and any $f,g:A{\rTo} B$ we have that
\[
Ff=Fg:FA{\rTo} FB\qquad\mbox{ implies }\qquad f=g:A{\rTo} B\,.
\]
\end{definition}

\begin{definition}\em 
A functor $F:\cat\longrightarrow{\bf D}$ is {\em full} if for any $A,B\in|\cat|$ and for any $g:FA\rightarrow FB$ there exists an $f:A{\rTo} B$ such that $Ff=g$.
\end{definition}

A {\em subcategory} ${\bf D}$ of a category $\cat$ is a collection of objects  of $\cat$ as well as a collection of  morphisms of $\cat$ such that
\begin{itemize}
\item for every morphism $f:A{\rTo} B$ in ${\bf D}$, both $A$ and $B\in|{\bf D}|$\,,
\item for every $A\in|{\bf D}|$, $1_A$ is in ${\bf D}$\,, and
\item for every pair of composable morphisms $f$ and $g$ in ${\bf D}$, $g\circ f$ is in ${\bf D}$.
\end{itemize}
These conditions  entail that ${\bf D}$ is itself a category. Moreover, if ${\bf D}$ is a subcategory of $\cat$, the inclusion functor $F:{\bf D}\longrightarrow \cat$ which maps every $A\in|{\bf D}|$ and $f\in{\bf D}$ to itself in $\cat$ is automatically faithful. If in addition $F$ is full, then we say that ${\bf D}$ is a {\em full subcategory} of ${\bf C}$.  A full and faithful functor is in general  {\em not} an isomorphism, as we shall see in Theorem \ref{Thm:fullfaithiso} below. 

\begin{definition} \em
Two categories ${\bf C}$ and ${\bf D}$ are \em equivalent \em if there is a pair of functors $F:\cat\longrightarrow {\bf D}$ and $G:{\bf D}\longrightarrow\cat$ and natural isomorphisms 
\[
G\circ F\stackrel{_\sim}{\Rightarrow} 1_{\bf C}\qquad \mbox{and} \qquad F\circ G\stackrel{_\sim}{\Rightarrow}1_{\bf D}\,.
\]
\end{definition}
An equivalence of categories is weaker than the notion of isomorphism of categories. It captures the essence of what we can do with categories without using concrete descriptions of objects: if two categories $\cat$ and ${\bf D}$ are equivalent then any result following from the categorical structure in $\cat$ remains true in ${\bf D}$, and vice-versa. 

\begin{theorem}\label{Thm:fullfaithiso}
{\rm\cite[p.~93]{SML}} A functor $F:\cat\longrightarrow{\bf D}$ is an equivalence of categories if and only if it is both full and  faithful, and if each object $B\in|{\bf D}|$ is isomorphic to an object $FA$ for some $A\in|\cat|$.
\end{theorem}


\begin{example} 
A {\em skeleton} ${\bf D}$ of a category $\cat$ is any full subcategory of $\cat$ such that each $A\in|\cat|$ is isomorphic   in $\cat$ to exactly one $B\in|{\bf D}|$. 
An equivalence between a category ${\bf C}$ and one of its skeleton ${\bf D}$ is defined as follows: 
\begin{enumerate}
\item As ${\bf D}$ is a full subcategory of ${\bf C}$, there is an inclusion functor $F:{\bf D}\longrightarrow {\bf C}$.
\item By the definition of a skeleton, every $A\in|{\bf C}|$ is isomorphic to an $A'\in|{\bf D|}$, so we can set $GA:=A'$ and pick an isomorphism $\tau_A:A\rTo GA$.
\item From the preceding point, there is a unique way to define a functor $G:{\bf C}\longrightarrow{\bf D}$ such that we have $FG\stackrel{\sim}{\Rightarrow} 1_{\bf C}$ and $GF\stackrel{\sim}{\Rightarrow} 1_{\bf D}$.
\end{enumerate} 
Particular instances of this are:
\begin{itemize}
\item The two categories with sets as objects and functions as morphisms discussed at the beginning of this section.
\item ${\bf FdHilb}$ is equivalent to the category with $\mathbb{C}^0,\mathbb{C}^1,\mathbb{C}^2,\cdots,\mathbb{C}^{n},\cdots$ as objects and linear maps between these as morphisms. This category is isomorphic to  the category ${\bf Mat}_\mathbb{C}$ of matrices with entries in $\mathbb{C}$ of Example~\ref{excat2_5}. 
\end{itemize}
\end{example}

\subsection{Topological quantum field theories}\label{sec:TQFT}


TQFTs are primarily used in condensed matter physics to describe, for instance, the fractional quantum Hall effect. Perhaps more accurately, TQFTs are quantum field theories that compute topological invariants. In the context of this paper, TQFTs are our main example of monoidal functors. Defining a TQFT as a monoidal functor is very elegant, however, the seemingly short definition that we will provide is packed with subtleties. In order to appreciate it to its full extent, we will first give the non-categorical axiomatics of  a generic $n$-dimensional TQFTs as given in \cite{Turaev}. We then derive the categorical definition from it. The bulk of this section is taken from~\cite{Kock} to which the reader is referred for a more detailed discussion on the subject.  

An {\em $n$-dimensional TQFT}  is a rule $\mathcal{T}$ which associates to each closed oriented $(n-1)$-dimensional manifold $\Sigma$ a vector space $\mathcal{T}(\Sigma)$ over the field $\mathbb{K}$, and to each oriented cobordism $M:\Sigma_0\rightarrow\Sigma_1$ a linear map $\mathcal{T}(M):\mathcal{T}(\Sigma_0)\rightarrow \mathcal{T}(\Sigma_1)$, subject to the following conditions:
\begin{enumerate}
\item if $M\simeq M'$ then $\mathcal{T}(M)=\mathcal{T}(M')$\,;
\item each cylinder $\Sigma\times [0,1]$ is sent to the identity map of $\mathcal{T}(\Sigma)$\,;
\item If $M=M'\circ M''$ then
\[
\mathcal{T}(M)=\mathcal{T}(M')\circ\mathcal{T}(M'')\,;
\]
\item the disjoint union $\Sigma=\Sigma' + \Sigma''$ is mapped to 
\[
\mathcal{T}(\Sigma)=\mathcal{T}(\Sigma')\otimes\mathcal{T}(\Sigma''),
\]
and the disjoint union $M=M' + M''$ is mapped to 
\[
\mathcal{T}(M)=\mathcal{T}(M')\otimes\mathcal{T}(M'')\,;
\]
\item the empty manifold $\Sigma=\emptyset$ is mapped to  the ground field $\mathbb{K}$ and  the empty cobordism is sent to the identity map on $\mathbb{K}$\,.
\end{enumerate}
All of this can be written down in one line.

\begin{definition} \em
An {\em $n$-dimensional TQFT} is a symmetric monoidal functor 
\[
\mathcal{T}:( {\bf nCob}, +, \emptyset, T)\rightarrow ({\bf FdVect}_\mathbb{K},\otimes,\mathbb{K},\sigma)
\]
where  $T$ are the `twist' cobordisms e.g.~$T_1=\ensuremath{\vcenter{\hbox{
\ifx\JPicScale\undefined\def\JPicScale{1}\fi
\psset{unit=\JPicScale mm}
\psset{linewidth=0.3,dotsep=1,hatchwidth=0.3,hatchsep=1.5,shadowsize=1,dimen=middle}
\psset{dotsize=0.7 2.5,dotscale=1 1,fillcolor=black}
\psset{arrowsize=1 2,arrowlength=1,arrowinset=0.25,tbarsize=0.7 5,bracketlength=0.15,rbracketlength=0.15}
\begin{pspicture}(0,0)(7.96,8.73)
\rput{-0}(6.62,2){\psellipse[linewidth=0.25](0,0)(1.27,0.55)}
\rput{-0}(2.97,2.04){\psellipse[linewidth=0.25](0,0)(1.26,0.56)}
\rput{-0}(6.7,8.14){\psellipse[linewidth=0.25](0,0)(1.26,0.55)}
\rput{-0}(3.06,8.18){\psellipse[linewidth=0.25](0,0)(1.26,0.56)}
\psbezier[linewidth=0.25](1.74,2.09)(1.74,4.51)(5.46,5.9)(5.46,8.21)
\psbezier[linewidth=0.25](4.24,1.99)(4.24,4.41)(7.96,5.79)(7.96,8.1)
\psbezier[linewidth=0.25](5.35,2.02)(5.3,4.9)(1.74,5.29)(1.74,8.21)
\psbezier[linewidth=0.25](7.91,2.05)(7.86,4.94)(4.29,5.33)(4.29,8.25)
\end{pspicture}
}}}$.
\end{definition}

The rule that maps manifolds to vector spaces and cobordisms to linear maps gives the domain and the codomain of the functor. Condition 1 says that we consider homeomorphism classes of cobordisms. Conditions 2 and 3 spell  out that the TQFT is a functor. Conditions 4 and 5 say that it is a monoidal functor.  

We now construct such a functor. In the case of 2-dimensional quantum field theories, it turns out that this question can be answered with the material we introduced in the preceding sections. 

We have the following result \cite{Kock}:

\begin{proposition}\label{generate}  
The monoidal category ${\bf 2Cob}$ is generated by
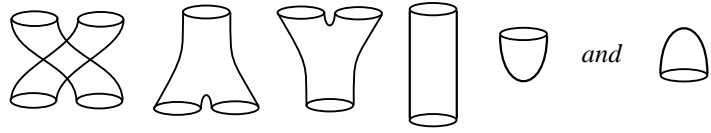
\begin{center}
\ifx\JPicScale\undefined\def\JPicScale{1}\fi
\psset{unit=\JPicScale mm}
\psset{linewidth=0.3,dotsep=1,hatchwidth=0.3,hatchsep=1.5,shadowsize=1,dimen=middle}
\psset{dotsize=0.7 2.5,dotscale=1 1,fillcolor=black}
\psset{arrowsize=1 2,arrowlength=1,arrowinset=0.25,tbarsize=0.7 5,bracketlength=0.15,rbracketlength=0.15}
\begin{pspicture}(0,0)(94.26,17.54)
\rput{0}(13.42,4.5){\psellipse[linewidth=0.25](0,0)(3.04,0.98)}
\rput{0}(4.68,4.56){\psellipse[linewidth=0.25](0,0)(3.04,0.98)}
\rput{0}(13.6,15.38){\psellipse[linewidth=0.25](0,0)(3.03,0.98)}
\rput{0}(4.86,15.44){\psellipse[linewidth=0.25](0,0)(3.04,0.98)}
\psbezier[linewidth=0.25](1.7,4.66)(1.7,8.94)(10.63,11.41)(10.63,15.51)
\psbezier[linewidth=0.25](7.71,4.47)(7.71,8.76)(16.64,11.22)(16.64,15.32)
\psbezier[linewidth=0.25](10.38,4.53)(10.26,9.64)(1.7,10.33)(1.7,15.51)
\psbezier[linewidth=0.25](16.52,4.59)(16.4,9.7)(7.83,10.4)(7.83,15.57)
\rput{0}(31.51,3.58){\psellipse[linewidth=0.25](0,0)(3.11,0.86)}
\rput{0}(23.83,3.52){\psellipse[linewidth=0.25](0,0)(3.11,0.86)}
\rput{0}(27.67,16.35){\psellipse[linewidth=0.25](0,0)(3.11,0.86)}
\psbezier[linewidth=0.25](26.94,3.46)(26.94,5.89)(28.34,5.89)(28.34,3.46)
\psbezier[linewidth=0.25](20.72,3.58)(25.05,11.16)(24.44,10.57)(24.56,16.5)
\psbezier[linewidth=0.25](34.62,3.76)(30.66,10.69)(30.84,9.39)(30.84,16.32)
\rput{0}(47.91,16.18){\psellipse[linewidth=0.25](0,0)(3.11,-0.82)}
\rput{0}(40.23,16.24){\psellipse[linewidth=0.25](0,0)(3.11,-0.82)}
\rput{0}(44.07,3.94){\psellipse[linewidth=0.25](0,0)(3.11,-0.83)}
\psbezier[linewidth=0.25](43.34,16.3)(43.34,13.97)(44.74,13.97)(44.74,16.3)
\psbezier[linewidth=0.25](37.12,16.19)(41.45,8.92)(40.84,9.49)(40.96,3.8)
\psbezier[linewidth=0.25](51.02,16.01)(47.06,9.37)(47.24,10.62)(47.24,3.97)
\rput{0}(57.8,1.94){\psellipse[linewidth=0.25](0,0)(3.11,-0.82)}
\rput{0}(57.8,16.71){\psellipse[linewidth=0.25](0,0)(3.11,-0.82)}
\psline(54.69,2.15)(54.69,16.52)
\psline(60.94,2.15)(60.94,16.52)
\rput{0}(69.83,13.42){\psellipse[linewidth=0.25](0,0)(3.11,-0.82)}
\psbezier(72.94,13.22)(72.94,5.1)(66.69,5.1)(66.69,13.22)
\rput(80.2,10.9){and}
\rput{0}(91.15,7.9){\psellipse[linewidth=0.25](0,0)(3.11,0.82)}
\psbezier(94.26,8.09)(94.26,16.21)(88.01,16.21)(88.01,8.09)
\end{pspicture}

\end{center}
That means, any cobordism in ${\bf 2Cob}$ can be written in terms of these generators when using composition and tensor.
\end{proposition}

Following the discussion of Section~\ref{sec:comonoids}, it is easily seen that these generators satisfy the axioms of a Frobenius comonoid. Moreover, since $\mathcal{T}$ is a monoidal functor, it is sufficient to give the image of the generators of ${\bf 2Cob}$ in order to specify it completely. Hence we can map this Frobenius comonoid in ${\bf 2Cob}$ on a Frobenius comonoid in ${\bf FdVect}_\mathbb{K}$:
\begin{center} 
\begin{tabular}{ccccl}
Objects: & $n$ & $\mapsto$ &\ \ \  & $\underbrace{V\otimes V\otimes ...\otimes V}_{n\mbox{ times}}$\vspace{.5em}\\
Identity: & \ensuremath{\vcenter{\hbox{
\ifx\JPicScale\undefined\def\JPicScale{1}\fi
\psset{unit=\JPicScale mm}
\psset{linewidth=0.3,dotsep=1,hatchwidth=0.3,hatchsep=1.5,shadowsize=1,dimen=middle}
\psset{dotsize=0.7 2.5,dotscale=1 1,fillcolor=black}
\psset{arrowsize=1 2,arrowlength=1,arrowinset=0.25,tbarsize=0.7 5,bracketlength=0.15,rbracketlength=0.15}
\begin{pspicture}(0,0)(3.73,11.9)
\rput{-0}(2.16,1.8){\psellipse[linewidth=0.25](0,0)(1.55,-0.54)}
\rput{-0}(2.16,11.36){\psellipse[linewidth=0.25](0,0)(1.56,-0.53)}
\psline(0.6,1.93)(0.6,11.23)
\psline(3.73,1.93)(3.73,11.23)
\end{pspicture}
}}} & $\mapsto$ & &$1_V:V\rightarrow V$ \vspace{.5em}\\

Twist: & \ensuremath{\vcenter{\hbox{}}} & $\mapsto$ & & $\sigma_{V,V}:V\otimes V\rightarrow V\otimes V$\vspace{.5em} \\

$e$: &  \ensuremath{\vcenter{\hbox{
\ifx\JPicScale\undefined\def\JPicScale{1}\fi
\psset{unit=\JPicScale mm}
\psset{linewidth=0.3,dotsep=1,hatchwidth=0.3,hatchsep=1.5,shadowsize=1,dimen=middle}
\psset{dotsize=0.7 2.5,dotscale=1 1,fillcolor=black}
\psset{arrowsize=1 2,arrowlength=1,arrowinset=0.25,tbarsize=0.7 5,bracketlength=0.15,rbracketlength=0.15}
\begin{pspicture}(0,0)(4.4,4.7)
\rput{0.9}(2.77,4.23){\psellipse[linewidth=0.25](0,0)(1.56,-0.45)}
\psbezier(4.33,4.15)(4.4,-0.3)(1.27,-0.35)(1.2,4.1)
\end{pspicture}
}}} & $\mapsto$ & & $e:\mathbb{K}\rightarrow V$\vspace{.5em}\\

$\mu$: & \ensuremath{\vcenter{\hbox{
\ifx\JPicScale\undefined\def\JPicScale{1}\fi
\psset{unit=\JPicScale mm}
\psset{linewidth=0.3,dotsep=1,hatchwidth=0.3,hatchsep=1.5,shadowsize=1,dimen=middle}
\psset{dotsize=0.7 2.5,dotscale=1 1,fillcolor=black}
\psset{arrowsize=1 2,arrowlength=1,arrowinset=0.25,tbarsize=0.7 5,bracketlength=0.15,rbracketlength=0.15}
\begin{pspicture}(0,0)(7.85,8.44)
\rput{0}(6.41,1.65){\psellipse[linewidth=0.25](0,0)(1.44,0.43)}
\rput{0}(2.85,1.62){\psellipse[linewidth=0.25](0,0)(1.45,0.43)}
\rput{0}(4.63,8.01){\psellipse[linewidth=0.25](0,0)(1.44,0.42)}
\psbezier[linewidth=0.25](4.29,1.59)(4.29,2.8)(4.94,2.8)(4.94,1.59)
\psbezier[linewidth=0.25](1.4,1.65)(3.41,5.43)(3.13,5.14)(3.19,8.09)
\psbezier[linewidth=0.25](7.85,1.74)(6.01,5.2)(6.1,4.55)(6.1,8)
\end{pspicture}
}}} & $\mapsto$ & & $\mu:V\otimes V\rightarrow V$\vspace{.5em}\\

$\epsilon:$ & \ensuremath{\vcenter{\hbox{
\ifx\JPicScale\undefined\def\JPicScale{1}\fi
\psset{unit=\JPicScale mm}
\psset{linewidth=0.3,dotsep=1,hatchwidth=0.3,hatchsep=1.5,shadowsize=1,dimen=middle}
\psset{dotsize=0.7 2.5,dotscale=1 1,fillcolor=black}
\psset{arrowsize=1 2,arrowlength=1,arrowinset=0.25,tbarsize=0.7 5,bracketlength=0.15,rbracketlength=0.15}
\begin{pspicture}(0,0)(4.7,6.19)
\rput{-0}(3.14,1.28){\psellipse[linewidth=0.25](0,0)(1.55,0.49)}
\psbezier(4.7,1.4)(4.7,6.19)(1.57,6.19)(1.57,1.4)
\end{pspicture}
}}} & $\mapsto$ & & $\epsilon:V\rightarrow\mathbb{K}$ \vspace{.5em}\\

$\delta:$ & \ensuremath{\vcenter{\hbox{
\ifx\JPicScale\undefined\def\JPicScale{1}\fi
\psset{unit=\JPicScale mm}
\psset{linewidth=0.3,dotsep=1,hatchwidth=0.3,hatchsep=1.5,shadowsize=1,dimen=middle}
\psset{dotsize=0.7 2.5,dotscale=1 1,fillcolor=black}
\psset{arrowsize=1 2,arrowlength=1,arrowinset=0.25,tbarsize=0.7 5,bracketlength=0.15,rbracketlength=0.15}
\begin{pspicture}(0,0)(7.22,8.07)
\rput{-1.44}(5.94,7.49){\psellipse[linewidth=0.25](0,0)(1.28,-0.44)}
\rput{-1.44}(2.79,7.59){\psellipse[linewidth=0.25](0,0)(1.27,-0.45)}
\rput{-1.44}(4.2,0.89){\psellipse[linewidth=0.25](0,0)(1.27,-0.45)}
\psbezier[linewidth=0.25](4.06,7.59)(4.03,6.33)(4.61,6.31)(4.64,7.57)
\psbezier[linewidth=0.25](1.52,7.59)(3.19,3.62)(2.95,3.92)(2.92,0.84)
\psbezier[linewidth=0.25](7.21,7.36)(5.5,3.8)(5.59,4.48)(5.5,0.88)
\end{pspicture}
}}} & $\mapsto$ & & $\delta:V\rightarrow V\otimes V$ \vspace{.5em}
\end{tabular}
\end{center} 
The converse is also true, that is, given a Frobenius comonoid on $V$, then we can define a TQFT with the preceding prescription, so there is a one-to-one correspondence between commutative Frobenius comonoids and 2-dimensional TQFTs. This is interesting in itself but we can go a step further. 

We can now define the category ${\bf 2TQFT}_\mathbb{K}$ of 2-dimensional TQFTs and symmetric monoidal natural transformations between them. 
Given two TQFTs $\mathcal{T},\mathcal{T}'\in|{\bf 2TQFT}_\mathbb{K}|$, then the components of the natural transformation $\theta$ must be --- by the definition above --- of the form 
\[
\theta_n:\underbrace{V\otimes V\otimes ...\otimes V}_{n\ \mbox{times}}\rightarrow\underbrace{W\otimes W\otimes ...\otimes W}_{n\ \mbox{times}}.
\]
Since this natural transformation is monoidal, it is completely specified by the map $\theta_1:V\rightarrow W$. The morphism $\theta_\mathbb{K}$ is the identity mapping from the trivial Frobenius comonoid on $\mathbb{K}$ to itself. Finally, naturality of $\theta$ means that the components must commute with the morphisms of ${\bf 2Cob}$. Since the latter  can be decomposed into the generators listed in Proposition~\ref{generate}, we just have to consider these cobordisms. E.g.
\[
\xymatrix{%
V\otimes V\ar[r]^{\theta_{2}} \ar[d]_{\mu_V} & W\otimes W\ar[d]^{\mu_W} \\ V\ar[r]_{\theta_1} & W}
\]

We can now define the category ${\bf CFC}_\mathbb{K}$ of commutative Frobenius comonoids and morphisms of Frobenius comonoids, that is, linear maps that are both comonoid homomorphisms and monoid homomorphisms. 


\begin{theorem}{\rm\cite{Kock}}
The category ${\bf 2TQFT}_\mathbb{K}$ is equivalent to the category ${\bf CFC}_\mathbb{K}$.
\end{theorem}  
 
\section{Further reading}\label{Further_reading}

This concludes our tutorial of  (a small fraction of) category theory.  We particularly focussed on monoidal categories, given that we expect their role to grow within physics.  We indicated how the monoidal structure encodes the nature of physical systems, e.g.~classical versus quantum.  Admittedly, the distinction as presented here requires substantial qualification, and by no means characterizes what quantum theory is truly about.  A recent more elaborated categorical comparison of classical vs.~quantum theories is in \cite{BES}.  All of this is part of a novel vastly growing research area, and we hope that this chapter may help the interested reader to take a bite of it.

We end this chapter by pointing in the direction of other important categorical concepts, for which we refer the reader to other sources.  A good place to start are the YouTube postings by the Catsters \cite{Catsters}.
   
\em Adjoint functors \em  are, at least from a mathematical perspective, the
greatest achievement of category theory thus far: it essentially unifies
all known mathematical constructs of a variety of areas of mathematics such
as algebra, geometry, topology, analysis and combinatorics within a single
mathematical concept.
  
The restriction of adjoint functors to posetal categories, that is, those
discussed in Examples \ref{excat6}, \ref{excat6_5}, \ref {ex:lattice} and
\ref{Quant_log}, is the concept of \em Galois adjoints\em .  These play an
important role in computer science when reasoning about \em computational
processes\em.  Let  $P$ be a partial order which represents the properties
one wishes to attribute to the input data of a process, with `$a\leq b$'
if and only if `whenever $a$ holds, then $b$ must hold too', and let $Q$ be
the partial order which represents the properties one wishes to attribute
to the output data of that process.  So the process is an order preserving
map $f:P\to Q$.  The order preserving map $g: Q\to P$, which maps a
property $b$ of the output to the `weakest' property (i.e.~highest in the
partial ordering) which the input data needs to satisfy in order to
guarantee that the output satisfies $b$, is then the \em left Galois
adjoint \em to $f$.  One refers to $g(b)$ as the \em weakest
precondition\em. Formally $f$ is left Galois adjoint to $g$ if and only if
for all $a\in P$ and all $b\in Q$ we have
\[
f(a)\leq b \Longleftrightarrow a\leq g(b)\,.
\]

The \em orthomodular law \em of quantum logic \cite{Piron}, that is, in
the light of Example \ref{Quant_log}, a weakening of the distributive law
which $L({\cal H})$ does satisfy, is an example of such an adjunction of
processes, namely
\[
{\rm P}_c (a)\leq b \Longleftrightarrow a\leq [c\rightarrow](b)
\]
where:
\bit
\item ${\rm P}_c$ is an order-theoretic generalization of the linear
algebraic notion of an `orthogonal projector on subspace $c$', formally
defined to be
\[
{\rm P}_c: L\to L:: a\mapsto c\wedge(a\vee c^\perp)\,,
\]
where $(-)^\perp$ stands for the orthocomplement\,;
\item $[-\rightarrow](-)$ is referred to as \em Sasaki hook\em, or
unfortunately, also sometimes referred to as `quantum implication', and
is formally defined within
\[
[c\rightarrow]: L\to L:: a\mapsto c^\perp\vee(a\wedge c)\,.
\]
\eit
Heyting algebras, that is, the order-theoretic incarnation of
intuitionistic logic, and which play an important role in the recent work
by Doering and Isham \cite{DoeringIsham} are, by definition, Galois
adjoints now defined within
\[
[c\wedge ](a)\leq b \Longleftrightarrow a\leq [c\Rightarrow](b)\,.
\]
So these Galois adjoints relate logical conjunction to logical
implication.

The general notion of adjoint functors involves, instead of an `if and
only if' between statements $f(a)\leq b$ and $a\leq g(b)$, a `natural
equivalence' between hom-sets ${\bf D}(FA, B)$ and $\cat(A, GB)$, where
$F:\cat \longrightarrow {\bf D}$ and $G:  {\bf D}\longrightarrow\cat$ are now functors.  We refer
to \cite{AbramskyT, BaezStay} in these volumes for an account on adjoint
functors and the role they play in logic.  We also recommend
\cite{LambekScott} on this topic.

The composite $G\circ F:\cat\longrightarrow\cat$ of a pair of adjoint functors is a
\em monad\em, and each monad arises in this manner.  The posetal
counterpart to this is a \em closure operator\em, of which the linear span
in a vector space is an example.

The composite $F\circ G:{\bf D}\longrightarrow{\bf D}$ of a pair of adjoint functors
is a \em comonad\em.  Comonads are an instance of the research area of \em
coalgebra\em, of which comonoids are also an instance.  The study of
coalgebraic structures has become increasingly important both in computer
science and physics.  These structures are very different from algebraic
structures: while algebraic structures typically would take
two pieces of data $a$ and $b$ as input, and produce the composite
$a\bullet b$, coalgebraic structures would do the opposite, that is, take
one piece of data as input and produce two pieces of data as output,
cf.~a copying operation.  Another example of a coalgebraic concept is quantum measurement. Quantum measurements take a quantum state as input and produces another quantum state together with classical data \cite{CPav2006}.

There also is the area of \em higher-dimensional category theory\em, after which the $n$-category cafe is named \cite{Cafe}.  Monoidal categories are a special case of \em bicategories\em, since we
can compose the objects with the tensor, as well as the processes between
these objects.  There is currently much activity  on the study of \em
$n$-categories\em, that is, categories in which the hom-sets are
themselves categories, and the hom-sets of these categories are again
categories etc.  Why would we be interested in that?   If one is interested
in processes then one should also be in modifying processes, and that
is exactly what these higher dimensional categorical structures enable to
model.  An excellent book on  higher-dimensional category theory is
\cite{Leinster}.

We end by recommending  the other chapters in these volumes entitled New Structures for Physics, which, among many other things,  contain complementary tutorials on category theory and its graphical calculus \cite{AbramskyT,BaezStay,Selinger}.

\section*{Acknowledgements}

We very much appreciated the feedback from the $n$-category cafe on a previous draft of this paper, by  John Baez, Hendrik Boom, Dave Clarke, David Corfield and  Aaron Lauda.  We in particular thank Frank Valckenborgh for proofreading the final version.

\end{document}